\documentclass[twocolumn]{aastex63}

\usepackage{amsmath}

\usepackage{tikz}

\usetikzlibrary{shapes,arrows}

\begin{document}

\shortauthors{Guy et al.}
\shorttitle{The DESI Spectroscopic Data Processing Pipeline}
\title{
The Spectroscopic Data Processing Pipeline for the Dark Energy Spectroscopic Instrument
}


\author{J.~Guy}
\affiliation{Lawrence Berkeley National Laboratory, 1 Cyclotron Road, Berkeley, CA 94720, USA}

\author{S.~Bailey}
\affiliation{Lawrence Berkeley National Laboratory, 1 Cyclotron Road, Berkeley, CA 94720, USA}

\author{A.~Kremin}
\affiliation{Lawrence Berkeley National Laboratory, 1 Cyclotron Road, Berkeley, CA 94720, USA}

\author{Shadab Alam}
\affiliation{Institute for Astronomy, University of Edinburgh, Royal Observatory, Blackford Hill, Edinburgh EH9 3HJ, UK}

\author{D.~M.~Alexander}
\affiliation{Centre for Extragalactic Astronomy, Department of Physics, Durham University, South Road, Durham, DH1 3LE, UK}

\author{C.~Allende~Prieto}
\affiliation{Instituto de Astrof\'{i}sica de Canarias, C/ Vía L\'{a}ctea, s/n, E-38205 La Laguna, Tenerife, Spain \\  Universidad de La Laguna, Dept. de Astrof\'{\i}sica, E-38206 La Laguna, Tenerife, Spain}

\author{S.~BenZvi}
\affiliation{Department of Physics \& Astronomy, University of Rochester, 206 Bausch and Lomb Hall, P.O. Box 270171, Rochester, NY 14627-0171, USA}

\author{A.~S.~Bolton}
\affiliation{NSF's NOIRLab, 950 N. Cherry Ave., Tucson, AZ 85719, USA}

\author{D.~Brooks}
\affiliation{Department of Physics \& Astronomy, University College London, Gower Street, London, WC1E 6BT, UK}

\author{E.~Chaussidon}
\affiliation{IRFU, CEA, Universit\'{e} Paris-Saclay, F-91191 Gif-sur-Yvette, France}

\author{A.P.~Cooper}
\affiliation{Institute of Astronomy and Department of Physics, National Tsing Hua University, 101 Kuang-Fu Rd. Sec. 2, Hsinchu 30013, Taiwan}

\author{K.~Dawson}
\affiliation{Department of Physics and Astronomy, The University of Utah, 115 South 1400 East, Salt Lake City, UT 84112, USA}

\author{A.~de la Macorra}
\affiliation{Instituto de F\'{\i}sica, Universidad Nacional Aut\'{o}noma de M\'{e}xico,  Cd. de M\'{e}xico  C.P. 04510,  M\'{e}xico}

\author{A.~Dey}
\affiliation{NSF's NOIRLab, 950 N. Cherry Ave., Tucson, AZ 85719, USA}

\author{Biprateep~Dey}
\affiliation{Department of Physics \& Astronomy and Pittsburgh Particle Physics, Astrophysics, and Cosmology Center (PITT PACC), University of Pittsburgh, 3941 O'Hara Street, Pittsburgh, PA 15260, USA}

\author{G.~Dhungana}
\affiliation{Department of Physics, Southern Methodist University, 3215 Daniel Avenue, Dallas, TX 75275, USA}

\author{D.~J.~Eisenstein}
\affiliation{Center for Astrophysics $|$ Harvard \& Smithsonian, 60 Garden Street, Cambridge, MA 02138, USA}

\author{A.~Font-Ribera}
\affiliation{Institut de F\'{i}sica d’Altes Energies (IFAE), The Barcelona Institute of Science and Technology, Campus UAB, 08193 Bellaterra Barcelona, Spain}

\author{J.~E.~Forero-Romero}
\affiliation{Departamento de F\'isica, Universidad de los Andes, Cra. 1 No. 18A-10, Edificio Ip, CP 111711, Bogot\'a, Colombia}

\author{E.~Gaztañaga}
\affiliation{Institut d'Estudis Espacials de Catalunya (IEEC), 08034 Barcelona, Spain}
\affiliation{Institute of Space Sciences, ICE-CSIC, Campus UAB, Carrer de Can Magrans s/n, 08913 Bellaterra, Barcelona, Spain}

\author{S.~Gontcho A Gontcho}
\affiliation{Lawrence Berkeley National Laboratory, 1 Cyclotron Road, Berkeley, CA 94720, USA}

\author{D.~Green}
\affiliation{Department of Physics and Astronomy, University of California, Irvine, 92697, USA}

\author{K.~Honscheid}
\affiliation{Center for Cosmology and AstroParticle Physics, The Ohio State University, 191 West Woodruff Avenue, Columbus, OH 43210, USA}
\affiliation{Department of Physics, The Ohio State University, 191 West Woodruff Avenue, Columbus, OH 43210, USA}
\affiliation{The Ohio State University, Columbus, 43210 OH, USA}

\author{M.~Ishak}
\affiliation{None}

\author{R.~Kehoe}
\affiliation{Department of Physics, Southern Methodist University, 3215 Daniel Avenue, Dallas, TX 75275, USA}

\author{D.~Kirkby}
\affiliation{Department of Physics and Astronomy, University of California, Irvine, 92697, USA}

\author{T.~Kisner}
\affiliation{Lawrence Berkeley National Laboratory, 1 Cyclotron Road, Berkeley, CA 94720, USA}

\author{Sergey~E.~Koposov}
\affiliation{Institute for Astronomy, University of Edinburgh, Royal Observatory, Blackford Hill, Edinburgh EH9 3HJ, UK}

\author{Ting-Wen Lan}
\affiliation{Graduate Institute of Astrophysics and Department of Physics, National Taiwan University, No. 1, Sec. 4, Roosevelt Rd., Taipei 10617, Taiwan}

\author{M.~Landriau}
\affiliation{Lawrence Berkeley National Laboratory, 1 Cyclotron Road, Berkeley, CA 94720, USA}

\author{L.~Le~Guillou}
\affiliation{Sorbonne Universit\'{e}, CNRS/IN2P3, Laboratoire de Physique Nucl\'{e}aire et de Hautes Energies (LPNHE), FR-75005 Paris, France}

\author{Michael E.~Levi}
\affiliation{Lawrence Berkeley National Laboratory, 1 Cyclotron Road, Berkeley, CA 94720, USA}

\author{C.~Magneville}
\affiliation{IRFU, CEA, Universit\'{e} Paris-Saclay, F-91191 Gif-sur-Yvette, France}

\author{Christopher~J.~Manser}
\affiliation{Department of Physics, University of Warwick, Coventry, CV4 7AL, UK}

\author{P.~Martini}
\affiliation{Center for Cosmology and AstroParticle Physics, The Ohio State University, 191 West Woodruff Avenue, Columbus, OH 43210, USA}
\affiliation{Department of Astronomy, The Ohio State University, 4055 McPherson Laboratory, 140 W 18th Avenue, Columbus, OH 43210, USA}
\affiliation{The Ohio State University, Columbus, 43210 OH, USA}

\author{Aaron M. Meisner}
\affiliation{NSF's NOIRLab, 950 N. Cherry Ave., Tucson, AZ 85719, USA}

\author{R.~Miquel}
\affiliation{Instituci\'{o} Catalana de Recerca i Estudis Avan\c{c}ats, Passeig de Llu\'{\i}s Companys, 23, 08010 Barcelona, Spain}
\affiliation{Institut de F\'{i}sica d’Altes Energies (IFAE), The Barcelona Institute of Science and Technology, Campus UAB, 08193 Bellaterra Barcelona, Spain}

\author{J.~Moustakas}
\affiliation{Department of Physics and Astronomy, Siena College, 515 Loudon Road, Loudonville, NY 12211, USA}

\author{Adam~D.~Myers}
\affiliation{Department of Physics \& Astronomy, University  of Wyoming, 1000 E. University, Dept.~3905, Laramie, WY 82071, USA}

\author{Jeffrey A.~Newman}
\affiliation{Department of Physics \& Astronomy and Pittsburgh Particle Physics, Astrophysics, and Cosmology Center (PITT PACC), University of Pittsburgh, 3941 O'Hara Street, Pittsburgh, PA 15260, USA}

\author{Jundan Nie}
\affiliation{National Astronomical Observatories, Chinese Academy of Sciences, A20 Datun Rd., Chaoyang District, Beijing, 100012, P.R. China}

\author{N.~Palanque-Delabrouille}
\affiliation{IRFU, CEA, Universit\'{e} Paris-Saclay, F-91191 Gif-sur-Yvette, France}
\affiliation{Lawrence Berkeley National Laboratory, 1 Cyclotron Road, Berkeley, CA 94720, USA}

\author{W.J.~Percival}
\affiliation{Department of Physics and Astronomy, University of Waterloo, 200 University Ave W, Waterloo, ON N2L 3G1, Canada}
\affiliation{Perimeter Institute for Theoretical Physics, 31 Caroline St. North, Waterloo, ON N2L 2Y5, Canada}
\affiliation{Waterloo Centre for Astrophysics, University of Waterloo, 200 University Ave W, Waterloo, ON N2L 3G1, Canada}

\author{C.~Poppett}
\affiliation{Lawrence Berkeley National Laboratory, 1 Cyclotron Road, Berkeley, CA 94720, USA}
\affiliation{Space Sciences Laboratory, University of California, Berkeley, 7 Gauss Way, Berkeley, CA  94720, USA}
\affiliation{University of California, Berkeley, 110 Sproul Hall \#5800 Berkeley, CA 94720, USA}

\author{F.~Prada}
\affiliation{Instituto de Astrof\'{i}sica de Andaluc\'{i}a (CSIC), Glorieta de la Astronom\'{i}a, s/n, E-18008 Granada, Spain}

\author{A.~Raichoor}
\affiliation{Lawrence Berkeley National Laboratory, 1 Cyclotron Road, Berkeley, CA 94720, USA}

\author{C.~Ravoux}
\affiliation{IRFU, CEA, Universit\'{e} Paris-Saclay, F-91191 Gif-sur-Yvette, France}

\author{A.~J.~Ross}
\affiliation{Center for Cosmology and AstroParticle Physics, The Ohio State University, 191 West Woodruff Avenue, Columbus, OH 43210, USA}
\affiliation{Department of Astronomy, The Ohio State University, 4055 McPherson Laboratory, 140 W 18th Avenue, Columbus, OH 43210, USA}
\affiliation{The Ohio State University, Columbus, 43210 OH, USA}

\author{E.~F.~Schlafly}
\affiliation{Space Telescope Science Institute, 3700 San Martin Drive, Baltimore, MD 21218, USA}

\author{D.~Schlegel}
\affiliation{Lawrence Berkeley National Laboratory, 1 Cyclotron Road, Berkeley, CA 94720, USA}

\author{M.~Schubnell}
\affiliation{Department of Physics, University of Michigan, Ann Arbor, MI 48109, USA}
\affiliation{University of Michigan, Ann Arbor, MI 48109, USA}

\author{Ray~M.~Sharples}
\affiliation{Centre for Advanced Instrumentation, Department of Physics, Durham University, South Road, Durham DH1 3LE, UK}
\affiliation{Institute for Computational Cosmology, Department of Physics, Durham University, South Road, Durham DH1 3LE, UK}

\author{Gregory~Tarl\'{e}}
\affiliation{University of Michigan, Ann Arbor, MI 48109, USA}

\author{B.~A.~Weaver}
\affiliation{NSF's NOIRLab, 950 N. Cherry Ave., Tucson, AZ 85719, USA}

\author{Christophe~Yèche}
\affiliation{IRFU, CEA, Universit\'{e} Paris-Saclay, F-91191 Gif-sur-Yvette, France}

\author{Rongpu Zhou}
\affiliation{Lawrence Berkeley National Laboratory, 1 Cyclotron Road, Berkeley, CA 94720, USA}

\author{Zhimin~Zhou}
\affiliation{National Astronomical Observatories, Chinese Academy of Sciences, A20 Datun Rd., Chaoyang District, Beijing, 100012, P.R. China}

\author{H.~Zou}
\affiliation{National Astronomical Observatories, Chinese Academy of Sciences, A20 Datun Rd., Chaoyang District, Beijing, 100012, P.R. China}


\begin{abstract}
  We describe the spectroscopic data processing pipeline of the Dark Energy Spectroscopic Instrument (DESI), which is conducting a redshift survey of about 40 million galaxies and quasars using a purpose-built instrument on the 4-m Mayall Telescope at Kitt Peak National Observatory. The main goal of DESI is to measure with unprecedented precision the expansion history of the Universe with the Baryon Acoustic Oscillation technique and the growth rate of structure with Redshift Space Distortions. Ten spectrographs with three cameras each disperse the light from 5000 fibers onto 30 CCDs, covering the near UV to near infrared (3600 to 9800\,\AA) with a spectral resolution ranging from 2000 to 5000. The DESI data pipeline generates wavelength- and flux-calibrated spectra of all the targets, along with spectroscopic classifications and redshift measurements. Fully processed data from each night are typically available to the DESI collaboration the following morning. We give details about the pipeline's algorithms, and provide performance results on the stability of the optics, the quality of the sky background subtraction, and the precision and accuracy of the instrumental calibration. This pipeline has been used to process the DESI Survey Validation data set, and has exceeded the project's requirements for redshift performance, with high efficiency and a purity greater than 99\% for all target classes.
\end{abstract}

\keywords{cosmology, large scale structure, surveys spectroscopy}



\section{Introduction}
\label{sec:Introduction}

The Dark Energy Spectroscopic Instrument (DESI) is a project
whose primary objective is to measure with unprecedented precision the expansion history of the universe and the growth rate of large scale structure~\citep{DESI2016a}. The goal is to constrain better or detect a deviation from the standard cosmological model which  relies on a puzzling cosmological constant or an unknown source of ``dark energy'' to explain the recent acceleration of the expansion. DESI should achieve a measurement of cosmological distances with the baryon acoustic oscillation (BAO) technique with an aggregate precision\footnote{Precision on the BAO peak position from the 2-point statistics of a galaxy sample covering a large redshift range, which requires assuming a fiducial cosmology to convert angles on the sky and redshifts to distances.} better than 0.3\% for redshifts below 1.1, 0.4\% in the range $1.1 < z < 1.9$, and 1\% at higher redshifts. It will use a 3D catalog of Luminous Red Galaxies (LRG), Emission Line Galaxies (ELG) and quasars (or QSO) as tracers of the matter density field, along with Lyman-alpha forests in the spectra of quasars at redshifts $z>1.9$. This result will be obtained with the spectra and redshift of about 40 million galaxies and quasars covering a 14,000 square degree footprint. About 10 million stellar spectra will also be acquired during the survey.
These improved statistics, about 10 times larger than the SDSS~\citep{SDSS-DR16}, are made possible by an instrument specifically optimized for redshift surveys, with a large multiplex factor with 5020 robotically actuated fibers in a 3.2-deg field of view, a large aperture 4-m telescope, and high-throughput spectrographs with a spectral resolution adapted to detect the [\ion{O}{2}]~$\lambda \lambda 3726,3729$ doublet of faint emission line galaxies.

In this paper we present the DESI spectroscopic pipeline, which consists of converting the raw CCD images of the 30 cameras from the 10 spectrographs into 5000 wavelength and flux calibrated spectra, along with a spectroscopic identification and redshift for all the observed targets.
This pipeline inherits from the experience gathered over the years in the processing of SDSS BOSS and eBOSS survey data~\citep{Bolton2012,Dawson2013,SDSS-DR16}. However, the code was entirely rewritten and includes several conceptual improvements. The main one is related to the spectral extraction technique which consists of a forward model of the CCD image~\citep{BoltonSchlegel2010}, using a precise model of the 2D point spread function instead of a projected 1D cross-dispersion profile. This approach is more complex, it requires a specific linear system solver to account for the large number of correlated parameters to extract, and a technique to decorrelate the output flux values. Its advantages are a minimal variance, a common wavelength grid for all fibers, and a well defined resolution matrix that improves the sky subtraction, the spectro-photometric calibration and the redshift estimation. Another notable difference with the SDSS pipeline is the treatment of the noise where we rely on a model of the CCD image to estimate the CCD pixel Poisson noise, which ensures a linearity of the fluxes. We also pay attention to the error propagation down to the final calibrated spectra.

After an overview of the instrument and the observations in Sections~\ref{sec:instrument} and~\ref{sec:observations}, we provide a detailed description of the algorithms that were developed and a first evaluation on their performance in Section~\ref{sec:algorithms-and-performance}. We describe the collaborative software development methods we have used in Section~\ref{sec:methodology}. We then present in Section~\ref{sec:data-processing} the data processing, including the real-time analysis, the daily updates, which comprise a quality assurance to identify failures and monitor the survey progress, and the large reprocessing runs with an homogeneous software version for the astrophysical and cosmological analyses. Finally in Section~\ref{sec:data-products} we give an overview of the data products before concluding in Section~\ref{sec:summary} with prospects for further improvements both for the data quality and the efficiency of the processing.

Machine-readable tables of the data shown in some of the figures of this paper are available at \url{https://doi.org/10.5281/zenodo.7087815}.
This paper documents the version of the spectroscopic pipeline used for the
``Fuji'' data processing run to be included with the DESI Early Data Release (EDR), covering Survey Validation and ancillary observations from December 2020 through May 13 2021.  The software codes used for this run are available on GitHub\footnote{\url{https://github.com/desihub}}, with the tags (or versions)
desispec/0.51.13, specex/0.8.4, specter/0.10.0,
redrock/0.15.4, redrock-templates/0.7.2,
desimodel/0.17.0, desitarget/2.4.0, and desiutil/3.2.5 (see also \S\ref{sec:methodology}).
Algorithmic updates for future data releases will be documented as part of those releases.

\section{Instrument}
\label{sec:instrument}

The DESI instrument is presented in detail in \cite{desi-collaboration22a} and references therein; we provide here only a brief overview. DESI is a multi object spectroscopic system installed at the Mayall 4-m telescope at Kitt Peak in Arizona. It features a prime focus instrument with a new corrector, a focal plane composed of 5000 fiber robots, and ten 3-arm spectrographs in the Coud\'e room connected to the focal plane with 50-m long fiber cables.

The high throughput of the instrument, the fiber positioning accuracy, and the stability of the spectrograph optics allows us to achieve an excellent redshift efficiency.

\subsection{Corrector}

The DESI corrector is composed of six lenses of about 1-m diameter, converting the telescope focal ratio from f/2.8 to f/3.9 over a 3.2$^\circ$ diameter field of view, while achieving an image quality of 0.6$\arcsec$ FWHM for the best nights~\citep{miller22a}. The corrector incorporates an atmospheric dispersion corrector (ADC), which compensates for the wavelength-dependent atmospheric dispersion in the field of view up to 60\,$\deg$ from Zenith. The lenses are hold by a barrel that is connected to the rest of the telescope structure by an hexapod which provides 6 degrees of freedom to adjust the position and orientation of the corrector and the focal plane system.

\subsection{Focal plane}

The DESI focal plane is segmented in 10 petals, each of them composed of 500 fiber positioners, up to 12 fiducial sources of light, and one Guide/Focus/Alignment (GFA) detector system~\citep{silber22a}. Six of the ten GFAs are used for guiding and four are used for the auto-focus and optical alignment. The fibers can be back-illuminated from the spectrographs. A fiber view camera (FVC) placed in an opening at the center of the primary mirror is used as a feed-back system to obtain an accurate positioning of the fibers relative to the fiducials. Each fiber positioner is composed of two rotating 3-mm arms, the second being placed at the end of the first, providing a 12-mm diameter patrol area for each fiber. As the separation between adjacent positioners is of 10.4\,mm, most of the petal area is within the reach of a positioner.
The positioning loop is composed of a first ``blind'' move, where a typical positioning precision of about 50\,$\mu$m is obtained, followed by a ``correction'' move, where a positioning accuracy better than 10\,$\mu$m is achieved after the analysis of a FVC image. This is much smaller that the 107\,$\mu$m diameter of a fiber, such that less than 3\% of the light is lost because of positioning errors (see more details in \S\ref{sec:fluxcalibration}).
The complete positioning loop takes less than a minute. A last FVC image is taken to record the final position of the fibers.

\subsection{Spectrographs}

\begin{figure}
\centering
\includegraphics[width=0.98\columnwidth,angle=0]{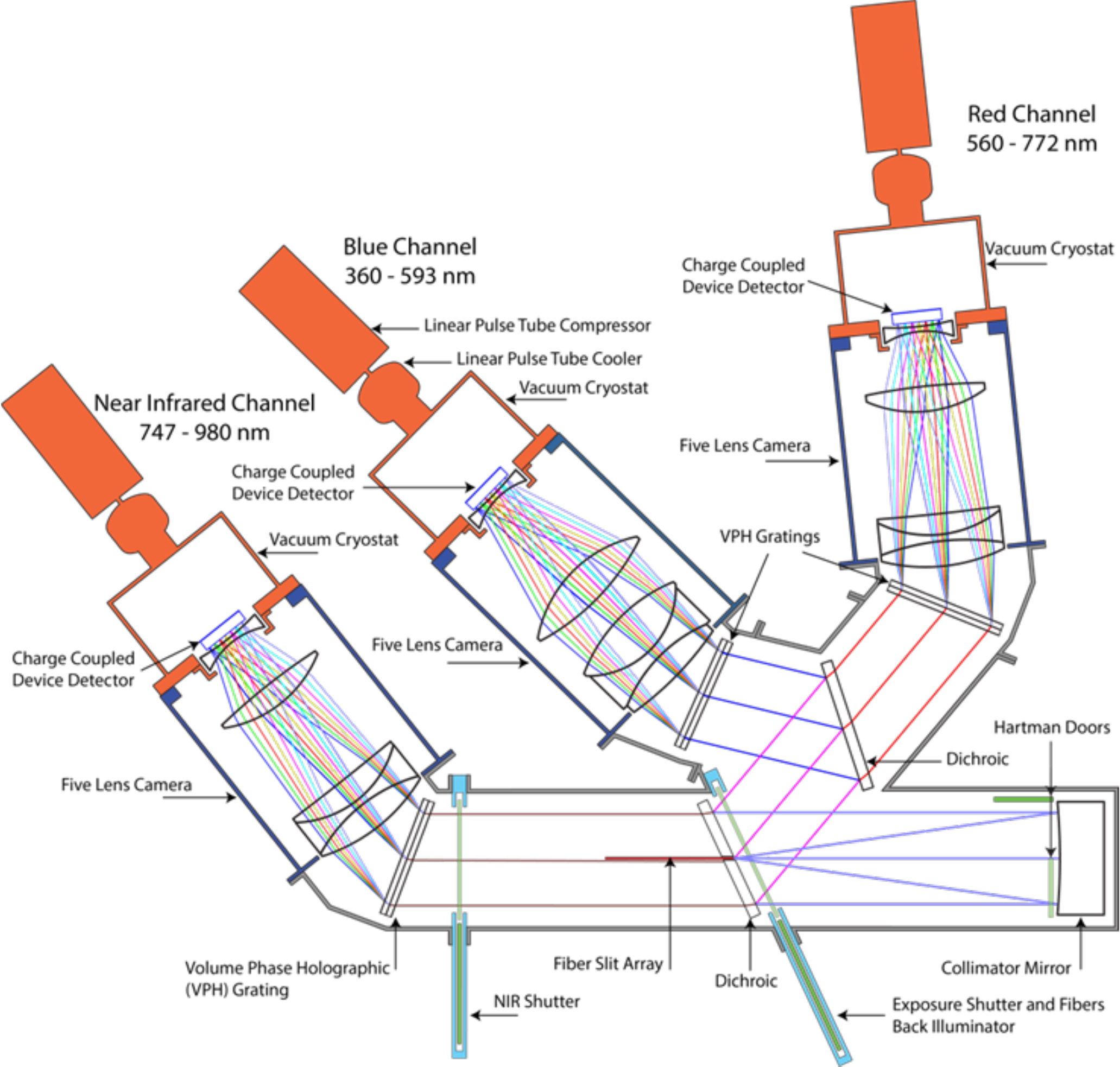}
\caption{Schematic view of one of the 10 DESI spectrographs. See text for details.}
\label{fig:spectrograph}
\end{figure}

50-m long fiber cables connect each petal in the focal plane to one spectrograph in the Coud\'e room. The fiber system and the spectrographs are described in detail in~\cite{poppett22a}  and \cite{jelinsky22a}.
On the spectrograph end of the cable, the 500 fibers are aligned to form a pseudoslit in a mechanical assembly called a slithead. The latter is inserted in the spectrograph with the pseudoslit precisely positioned in a slot in a NIR dichroic (see Figure~\ref{fig:spectrograph}). The light emitted from the fibers is reflected on a collimator mirror back to the NIR dichroic which is transparent to NIR wavelengths but reflects optical wavelengths at an angle. In front on the NIR dichroic is the main shutter which also contains a fiber back illumination system. A second shutter is located behind the dichroic to protect the NIR camera from this back illumination. The spectrograph is also equipped with a pair of Hartman doors placed in front of the collimator. Optical wavelengths are further split with a second dichroic, resulting in three arms for blue (3600-5930\,\AA), red (5600-7720\,\AA), and NIR (7470-9800\,\AA) wavelengths. Each arm utilizes a volume phase holographic (VPH) grating and a five lens camera, the last lens being the entrance window of a cryostat. Each cryostat hosts a CCD cooled down to 163 K for the blue channel, and 140 K for the red and NIR channels.


\subsection{CCDs}
\label{sec:ccdimage}

\begin{figure}
  \centering
  \includegraphics[width=0.9\columnwidth,angle=0]{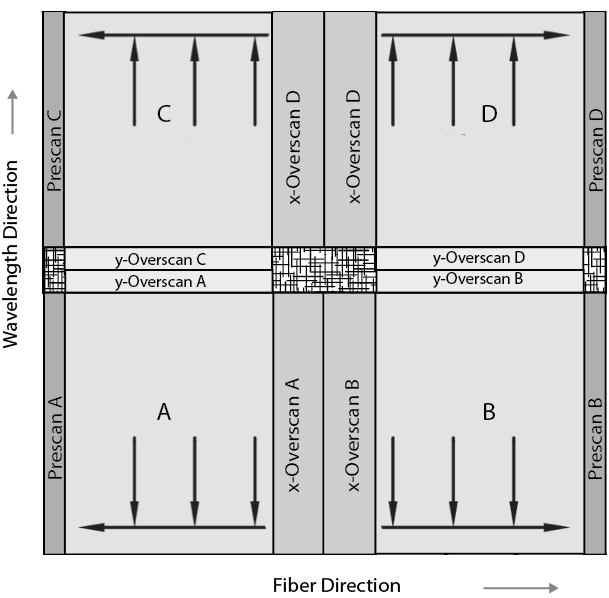}
  \caption{DESI CCD image layout with four amplifiers. The CCD images for all cameras have the same orientation after transformation of the data array by the instrument control system; the column index (AXIS1 in the {\it fits} files headers, often labeled $X$ in this paper) increases with the fiber number, and the row index (AXIS2 in {\it fits}, often label $Y$ in this paper) increases with increasing wavelength. The black arrows indicate the direction of readout, with the parallel clock in the wavelength direction ($Y$) and the serial clock along the fiber number ($X$). Also shown are the prescan and overscan regions used to measure and remove the bias level during pre-processing (see \S\ref{sec:preprocessing}).}
  \label{fig:ccd-layout}
\end{figure}

The blue CCDs are $4096 \times 4096$ STA4150 CCDs from Semiconductor Technology Associates (STA), that were processed and packaged by the University of Arizona Imaging Technology Laboratory (ITL). The red and NIR CCDs are 250\,$\mu$m thick, fully-depleted p-channel CCDs of $4114 \times 4128$ pixels.
Both types of CCDs are read with four amplifiers. Figure~\ref{fig:ccd-layout} shows the CCD layout as it appears in the files on disk, after some transformation performed by the instrument control system. The figure also shows the direction of the parallel and serial clocks, along with the location of over-scan and pre-scan regions. The measured CCD read noise (see~\S\ref{sec:preprocessing}) is in the range of 2.8 to 4.2 electrons r.m.s for the blue CCD amplifiers, with one outlier close to 5 electrons, and 2.2 to 3.5 electrons for red and NIR CCD amplifiers again with one outlier at 4.5 electrons.

\subsection{Calibration system}
\label{sec:calibration-system}

\begin{figure}
\centering
\includegraphics[width=0.8\columnwidth,angle=0]{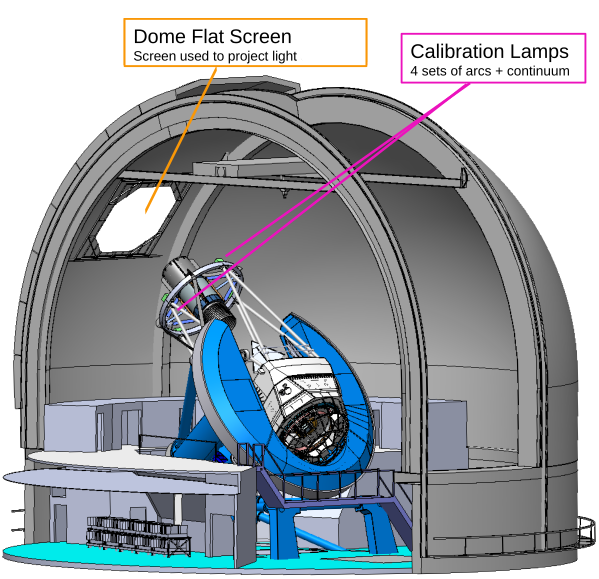}
\caption{Illustration of the DESI calibration system with the location of the screen and calibration lamps. The diameter of the dome screen is 5\,m.}
\label{fig:mayall}
\end{figure}

The calibration system for DESI consists in a dome screen and a set of lamps installed on the upper ring of the telescope structure that can illuminate the screen when the telescope is pointing at it (see Figure~\ref{fig:mayall}).
The screen is larger than the telescope mirror to account for the range of incidence angles and is covered with a coating of nearly Lambertian and achromatic reflectance. The lamps are placed in four boxes evenly placed along the upper ring. LED arrays are used for flat-fielding (see~\S\ref{sec:fiberflat}) and a combination of Mercury, Argon, Cadmium, Neon, Xenon, and Krypton lamps are used for the wavelength calibration and the point spread function modeling (see~\S\ref{sec:psf}).

\begin{figure}
  \centering
  \includegraphics[width=0.99\columnwidth,angle=0]{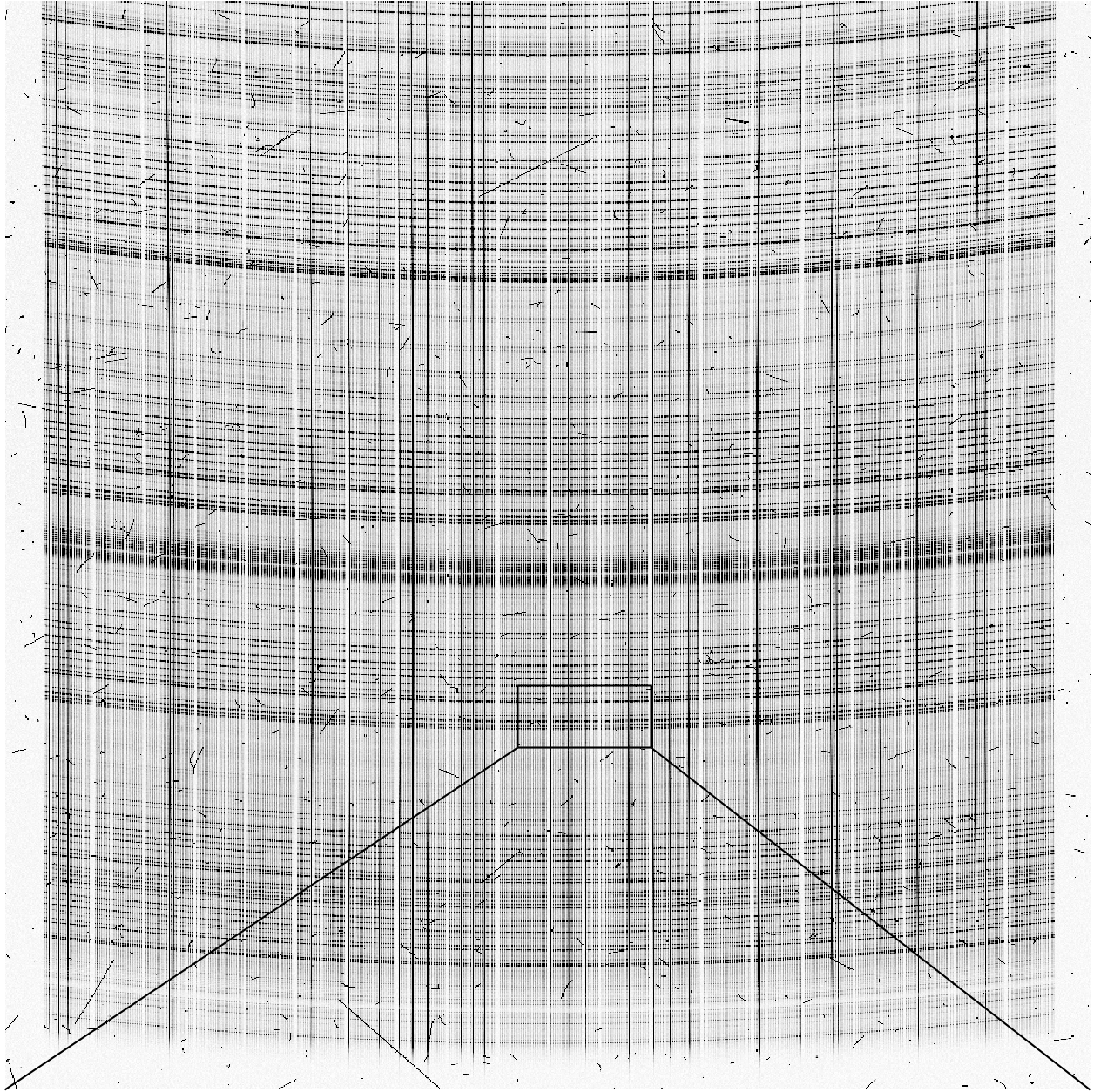}
  \includegraphics[width=0.99\columnwidth,angle=0]{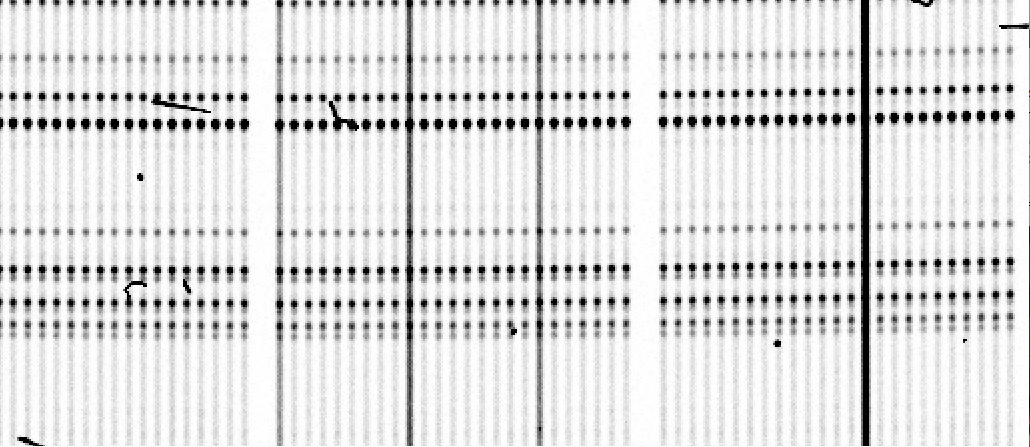}
  \caption{Example CCD image after preprocessing. This is a NIR CCD image of spectrograph SM10 after a 900s exposure. On the top panel, one can see the 500 fibers organized in 20 blocks of 25 fibers each. The mostly horizontal curved lines are sky lines. Bright fibers (appearing as vertical dark bands in this negative color scale) are the spectral traces from standard star fibers. One can also note many cosmic ray hits. The lower panel is a zoom highlighting the clear separation of the fiber traces, the space between blocks, and the spectrograph resolution (see~\S\ref{sec:psf} for more details on the resolution).}
\label{fig:ccd-image}
\end{figure}


\section{Observations}
\label{sec:observations}

We present in this section the sequence of observations
and the data set used by the spectroscopic pipeline, starting with the afternoon calibrations and then providing some information about a typical exposure of the main survey.

\subsection{Spectroscopic calibration observations}
\label{sec:calib-obs}

The instrument control system allows us to operate all 30 cameras simultaneously and control the calibration lamps. Each action is registered in a queue that can be filled interactively or with predefined scripts that are used for routine operations such as the ones described here.

The afternoon calibration sequence starts with a broadcast call to a specific stabilization routine applied to all the CCDs. This stabilization consists in running the CCDs with a series of non-standard configurations of the clocks voltages to efficiently remove charges and reach rapidly a stable state (for most CCDs, and as long as they are continually clocked).
We then acquire 25 ``zero'' exposures (zero exposure time and shutters closed) that are used to build a nightly master bias frame. This sequence is ended with a 300 sec ``dark'' exposure (still with the spectrographs' shutter closed) which is used to monitor the quality of the pre-processing (see \S\ref{sec:preprocessing}).

Calibration exposures using the dome screen are acquired later in the afternoon, when the dome lights can be switched off after the maintenance operations by the day crew. They consists of
\begin{itemize}
\item 5 exposures of 5 seconds with all the arc lamps on (Hg, Ar, Cd, Ne, Xe, Kr)
\item 5 exposures of 30 seconds with only the Cd and Xe lamps in order to fill wavelength gaps with low S/N
\item 4 series of 3 exposures of 120 seconds with the LEDs arrays from each of the four calibration boxes, one at a time.
\end{itemize}

  The first series of five exposures is used to fit the point spread function and determine the wavelength calibration (see arc lamp spectra in Figure~\ref{fig:psf1}). We use five exposures in order to be robust to cosmic rays that sometimes compromise the PSF fit by hitting regions of the CCD with useful emission lines (see \S\ref{sec:psf}). The set of LED lamps exposures is used for the fiber flat fielding. Three exposures per illumination configuration are needed to mitigate the effect of cosmic rays and four configurations with one lamp at a time in order to inter-calibrate the lamps (see \S\ref{sec:fiberflat} for details).

  A few month after the beginning of the main survey, we started acquiring two additional dark exposures of 1200 sec at the end of the night to build statistics for master dark frames (more darks are also acquired when on-sky observations are not possible).

\subsection{Standard exposure sequence}
\label{sec:std-expo-seq}

The standard exposure sequence is described in detail in \cite{desi-collaboration22a}.
We summarize here the information we need for the spectroscopic pipeline.

A survey tile with assigned fibers consists of the sky coordinates of the tile center, a field rotation angle, and the sky positions of the 5000 fibers. During the telescope slew towards the tile center and the hexapod moves to rotate the field, the first blind moves of the positioners are operated. After the telescope slew and the field rotation, a 15 sec acquisition exposure is obtained to precisely point the telescope and perform a residual field rotation. After this move (telescope + hexapod), guide stars are starting to be measured in 6 of the 10 GFAs to maintain the telescope pointing, the fiber back-illumination is turned on, and a FVC image is taken to adjust the location of positioners. After the positioners are adjusted, a last FVC image is acquired, the back-illumination turned off, and the spectrographs' shutters are opened.

The hexapod is regularly rotated during the exposure to compensate for the residual field rotation. Also during the exposure a dynamic exposure time calculator estimates the remaining time needed to achieve a pre-set effective exposure time which is a function of the sky brightness, the sky transparency, the airmass and the image quality. The sky brightness is estimated from sky monitor fibers placed on the edge of the focal plane and read regularly with a dedicated CCD camera. The sky transparency and image quality are determined from the GFA images.

\subsection{Exposure data set}
\label{sec:expo-data}

At the end of the exposure sequence, the shutters are closed, the CCD read, and the instrument control system collects the individual images and saves them in a single fits files with multiple HDUs, with one for each of the 30 cameras, while adding many header keywords about the telescope and its pointing, environmental parameters in the dome, along with monitoring parameters for the CCDs, cryostat, readout electronics, and temperatures and humidity in the vicinity of the cameras. All of the guide star images are saved for offline analysis. This is useful to model the fraction of light in the fiber aperture given the current image quality and the performance of the guiding. Those images also provide redundant information about the sky transparency. The FVC images and in particular the coordinates of the fiber tips are recorded along with their expected positions for a perfect alignment on the targets. This information is used for the flux calibration, the estimation of the expected $S/N$ in the spectra, and for general quality assurance. The data from the dynamic exposure time calculator is saved, along with the input fiber assignment table associated with the observed tile, and a table of fiber positioning offsets, determined from the analysis of the FVC control image. All of this data is saved in one directory per exposure (see the DESI data model referenced in \S\ref{sec:data-products} for more details). This directory is copied to NERSC within a few minutes, and additional copies are stored at NOIRLab and the NERSC archival tape backup system within one day.

The fiber assignment table contains important information about the targets for the spectroscopic pipeline.
The fibers pointing to blank sky coordinates are separately identified; they are used to model the sky spectrum (see \S\ref{sec:skysubtraction}). The table also contains the various target bitmasks among which standard stars can be found for the flux calibration (\S\ref{sec:starfit}). It provides useful information from the imaging catalogs. This includes in particular the total flux in the $g$, $r$, and $z$ DECam pass-bands\footnote{See \citealt{Dey2019} for a description of the surveys and the pass-bands, and \url{https://www.legacysurvey.org/dr9/} for the imaging catalogs.}, along with the fiber fluxes, which are the fluxes one would have collected in a 1.5$\arcsec$ diameter fiber for a seeing or image quality of 1$\arcsec$ FWHM for the target, given its surface density profile as determined from the imaging data.

The majority of tiles (or pointings) are observed with a single exposure, with an average exposure time of about 800 sec in bright time, and 1100~sec in dark time. Tiles that require exposure times longer than 1800~sec because of poor observing conditions are split into several exposures to minimize the impact of cosmic rays and readjust the fiber positioners. This exposure time threshold was decided after studying the effect of unmasked cosmic ray hits on the redshift success rate of emission line galaxies. Observing systematically tiles with several shorter exposures would further reduce the effect of cosmics rays but at the cost of an increased read noise and a loss of exposure time when reading the CCDs (60 sec).

\section{Algorithms and performance}
\label{sec:algorithms-and-performance}

\subsection{Overview}

The spectroscopic pipeline deliverables are wavelength and flux calibrated spectra of the observed targets (with flux variance, bit mask, spectral resolution for each wavelength and fiber), and a redshift catalog with a spectroscopic classification of the targets, their redshift uncertainty and a confidence level (see \S\ref{sec:data-products} for a more complete description of the data products).
We review here the data flow. The algorithms, fitting procedures and performances are presented in more details in the following sections.
The data flow is graphically presented in Figure~\ref{fig:data-flow}.


\tikzstyle{block} = [rectangle, draw, text centered]
\tikzstyle{choice} = [ellipse, draw, text centered]
\tikzstyle{line} = [draw, -latex']
\tikzstyle{cloud} = [rectangle, draw,fill=gray!20, text centered, rounded corners]

\begin{figure}
  \centering
  \begin{tikzpicture}[node distance=0.5cm]
    \tiny
    \node [block] (raw) {raw CCD image};
    \node [choice, below of=raw, node distance=0.55cm] (biaschoice) {bias?};
    \node [cloud, left of=biaschoice, node distance=3.5cm] (biasfit) {bias averaging};
    \node [block, below of=biasfit] (bias) {Night bias frame};
    \node [cloud, below of=biaschoice, node distance=0.8cm] (preprocessing) {pre-processing (\S\ref{sec:preprocessing})};
    \node [block, below of=preprocessing] (preproc) {pre-processed image};
    \node [choice, below of=preproc, node distance=0.55cm] (arclampchoice) {arc lamps?};
    \node [cloud, left of=arclampchoice, node distance=3.5cm] (psffit) {PSF fit and averaging (\S\ref{sec:psf})};
    \node [block, below of=psffit] (psf) {Night PSF};
    \node [cloud, below of=arclampchoice, node distance=0.8cm] (wavecalib) {trace and wavelength calibration (\S\ref{sec:wave-calib-trace-coords})};
    \node [block, below of=wavecalib] (exposurepsf) {Exposure PSF};
    \node [cloud, below of=exposurepsf] (extraction) {spectral extraction (\S\ref{sec:extraction})};
    \node [block, below of=extraction] (frame) {uncalibrated spectral frame};
    \node [choice, below of=frame, node distance=0.55cm] (fiberflatchoice) {LED lamps?};
    \node [cloud, left of=fiberflatchoice, node distance=3.5cm] (computefiberflat) {compute flat-field (\S\ref{sec:fiberflat})};
    \node [block, below of=computefiberflat] (fiberflat) {Night fiber flat-field};
    \node [cloud, below of=fiberflatchoice, node distance=0.8cm] (applyfiberflat) {apply flat-field};
    \node [block, below of=applyfiberflat] (fframe) {flat-fielded spectral frame};
    \node [cloud, below of=fframe] (skysub) {sky subtraction (\S\ref{sec:skysubtraction})};
    \node [block, below of=skysub] (sframe) {sky-subtracted spectral frame};
    \node [cloud, below of=sframe] (starfit) {stellar model fit (\S\ref{sec:starfit})};
    \node [block, left of=starfit, node distance=3.5cm] (startemplates) {stellar models};
    \node [block, below of=starfit] (stdstar) {standard star calibrated spectra};
    \node [cloud, below of=stdstar] (fluxcalib) {flux calibration (\S\ref{sec:fluxcalibration})};
    \node [block, left of=fluxcalib, node distance=3.5cm] (throughputmodel) {throughput model};
    \node [cloud, below of=fluxcalib] (xtalk) {fiber cross-talk correction (\S\ref{sec:fibercrosstalk})};
    \node [block, below of=xtalk] (cframe) {calibrated spectral frame};
    \node [cloud, below of=cframe] (coadd) {spectral grouping and co-addition (\S\ref{sec:coaddition})};
    \node [block, below of=coadd] (spectra) {spectra group};
    \node [cloud, below of=spectra] (redrock) {identification and redshift fitting  (\S\ref{sec:redshift})};
    \node [block, left of=redrock, node distance=3.5cm] (galaxytemplates) {spectral templates};
    \node [block, below of=redrock] (catalog) {redshift catalog};

    \path [line] (raw) -- (biaschoice);
    \path [line] (biaschoice) -- node[left] {no} (preprocessing);
    \path [line] (biaschoice) -- node[above] {yes} (biasfit);
    \path [line] (biasfit) --  (bias);
    \path [line] (bias) |-  (preprocessing);
    \path [line] (preprocessing) -- (preproc);
    \path [line] (preprocessing) -- (preproc);
    \path [line] (preproc) -- (arclampchoice);
    \path [line] (arclampchoice) -- node[above] {yes} (psffit);
    \path [line] (psffit) --  (psf);
    \path [line] (psf) |- (wavecalib);
    \path [line] (arclampchoice) -- node[left] {no} (wavecalib);
    \path [line] (wavecalib) -- (exposurepsf);
    \path [line] (exposurepsf) -- (extraction);
    \path [line] (extraction) -- (frame);
    \path [line] (frame) -- (fiberflatchoice);
    \path [line] (fiberflatchoice) -- node[above] {yes} (computefiberflat);
    \path [line] (computefiberflat) -- (fiberflat);
    \path [line] (fiberflat) |- (applyfiberflat);
    \path [line] (fiberflatchoice) -- node[left] {no} (applyfiberflat);
    \path [line] (applyfiberflat) -- (fframe);
    \path [line] (fframe) -- (skysub);
    \path [line] (skysub) -- (sframe);
    \path [line] (sframe) -- (starfit);
    \path [line,dashed] (startemplates) -- (starfit);
    \path [line] (starfit) -- (stdstar);
    \path [line] (stdstar) -- (fluxcalib);
    \path [line,dashed] (throughputmodel) -- (fluxcalib);
    \path [line] (fluxcalib) -- (xtalk);
    \path [line] (xtalk) -- (cframe);
    \path [line] (cframe) -- (coadd);
    \path [line] (coadd) -- (spectra);
    \path [line] (spectra) -- (redrock);
    \path [line,dashed] (galaxytemplates) -- (redrock);
    \path [line] (redrock) -- (catalog);
\end{tikzpicture}
\caption{Spectroscopic pipeline data flow}
\label{fig:data-flow}
\end{figure}
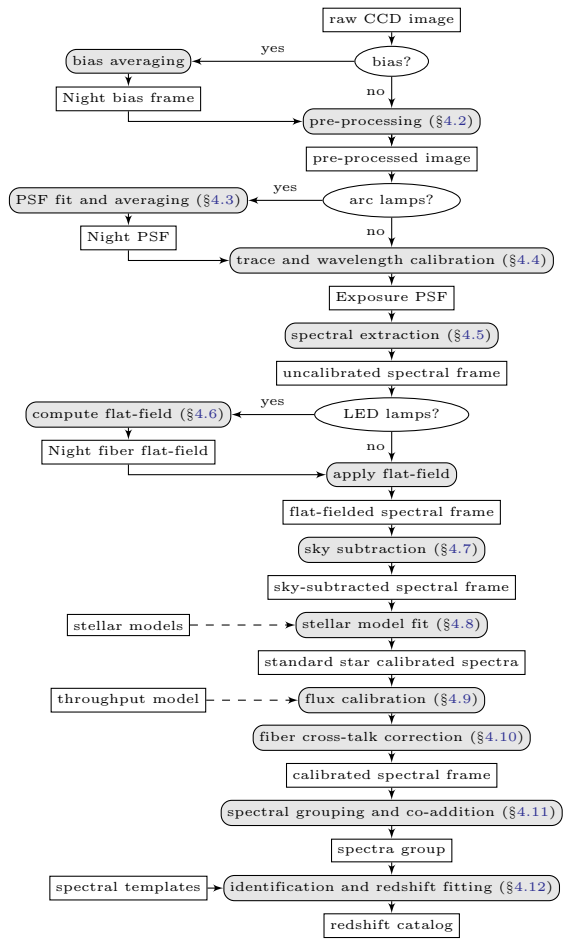


\begin{enumerate}
\item A first step for all incoming images is the pre-processing (\S\ref{sec:preprocessing}) of the CCD images, where ADC counts are converted to electrons per pixel, with an estimate of their variance, a subtraction of dark current, a flat field correction, and a mask indicating bad pixels (due to CCD defects or cosmic rays); this process also incorporates other corrections of the imperfections of the hardware.

\item  Then arc lamp calibration images are analyzed to perform a precise wavelength calibration, fit the spectral traces CCD coordinates and measure the spectroscopic Point Spread Function (hereafter PSF) shape (\S\ref{sec:psf}). In general, an average PSF is computed from the calibration data acquired during an afternoon to process the following flat-fielding calibration exposures and then the scientific exposures from the night that follows. The pipeline can also use a PSF model derived from previous nights if calibration data is missing.

\item The wavelength calibration and trace coordinates vary by a fraction of a CCD pixel during the night because of small variations of environmental parameters in the spectrographs. For this reason an adjustment is required for each science exposure (\S\ref{sec:wave-calib-trace-coords}).

\item The next step is the spectral extraction (see \S\ref{sec:extraction}). It consists in fitting a full model of the CCD image, using the known 2D PSF. The fit is a least square minimization with inverse variance weights ($\chi^2$). It is very close to be the statistically optimal estimator because the pixel noise is nearly Gaussian and uncorrelated. The drawback of this method is the computation challenge and the need for a specific treatment of the high frequency noise resulting from the deconvolution.

\item We then proceed with the calibration of these spectra. We first use the extracted spectra from calibration exposures on the white screen illuminated with LED lamps to determine a fiber flat-field correction (\S\ref{sec:fiberflat}). It also corrects for residual CCD non-uniformities and accounts for the difference of throughput from one spectrograph to another.

\item Once the spectra are flat-fielded, we use fibers intentionally pointing to empty regions of the sky (hereafter called ``sky fibers'') to compute a sky spectrum model and subtract this sky model from all the target spectra (\S\ref{sec:skysubtraction}). It makes use of the precise estimate of the spectral resolution determined during the extraction.

\item The flat-fielded and sky subtracted spectra of standard stars are used to fit stellar spectral models, normalized by the photometric fluxes from the input imaging catalog. The fit uses data from several exposures from the same target. It does not require a prior knowledge of the instrument throughput. We detail the fit algorithm and the stellar templates in \S\ref{sec:starfit}.

\item The spectral models of the standard stars can then be used to derive the instrument throughput as a function of wavelength by comparing the measured counts in electrons with the expected incoming flux (\S\ref{sec:fluxcalibration}). We use for this purpose our past knowledge of the instrument response while accounting for the variable atmospheric conditions and the optical properties of the telescope (focal plane PSF, plate scale, vignetting).

\item We apply a fiber cross-talk correction (\S\ref{sec:fibercrosstalk}) to remove the contamination of spectra from neighboring bright fiber traces in the CCD images.

\item Calibrated spectra from several exposures and nights are grouped together according to their location on the sky, and optionally co-added (or averaged), while avoiding any resampling of the spectral data arrays (see \S\ref{sec:coaddition}).

\item Finally, regrouped spectra are analyzed to classify targets and measure their redshifts. We give an overview of the methods used in~\S\ref{sec:redshift}.

\end{enumerate}

\subsection{CCD calibration and pre-processing}
\label{sec:preprocessing}

Pre-processing is the task that consists in converting the 'raw' CCD images, consisting of an array of Analog Digital Units (hereafter ADUs) per pixel into a number of electrons per pixel, with a flat-field correction, an estimate of the variance in each pixel, and a mask to discard some pixels from the subsequent spectroscopic analysis.

\subsubsection{Bias, dark and over-scan subtraction}
\label{sec:bias-subtraction}

The first step of the pre-processing is the subtraction of  a bias frame to the raw image in ADUs. The bias frame is either a yearly master bias obtained independently from past observations or a nightly master bias. The latter is obtained from the series of 25 bias exposures taken during the afternoon as part of the daily calibration sequence (see \S\ref{sec:calib-obs}). This master bias is simply the clipped average per pixel of the raw CCD images after over-scan subtraction, including the pre-scan and over-scan columns and rows. The clipped average is an iterative evaluation of the mean and dispersion of the data with a 5-$\sigma$ outlier rejection. The initial values for the mean and dispersion are the median and the normalized median absolute deviation (NMAD) of the pixel values. The choice of using the yearly or nightly bias frame is based on the reduction of a 300 second dark exposure taken during the afternoon calibration. This dark frame is fully pre-processed (including dark current subtraction) with both bias frames, and the bias that gives the smallest residuals in the processed image is selected for the following night of data. This procedure allows us to account for night to night variations of the bias levels while protecting us from unsettled CCD biases that happen in some rare sequences when one or several CCDs had been turned off earlier during the day.

After subtracting the reference bias frame we proceed with the over-scan subtraction. We use by default the clipped mean of all of the pixels in the over-scan columns (``x-overscan'' regions in Figure~\ref{fig:ccd-layout}) , per CCD amplifier (pre-scan columns are not used). We however apply a more complex treatment in two situations. First, some CCD quadrants have bias fluctuations with a correlation length exceeding the length of a CCD line. Some rare nights for some specific CCD amplifiers, the amplitude of the correlated fluctuations exceeds a threshold for which it becomes advantageous to perform a bias subtraction per row (i.e. using the clipped average of each row independently in the over-scan area). This decision is made automatically in the pre-processing with a measure of the amplitude of the variation from row to row (recorded under the {\it OSTEP\{A,B,C,D\}} keyword in the pre-processed CCD image headers). When doing so, we have to first detect and mask over-scan rows affected by residual trailing  charges (positive or negative, see \S\ref{sec:darktrails}) from large cosmic ray deposits in the last columns of the CCD active region.
For problematic CCD amplifiers (only one among 120 for the first DESI data release), we apply an additional correction derived from the average of the over-scan rows (``y-overscan'' regions in Figure~\ref{fig:ccd-layout}) as a function of the column number.

In addition to a standard dark current subtraction, proportional to the exposure time, we also correct for residual bias levels depending non-linearly on the exposure time. This extra bias level in the readout electronics builds up during the exposure (and so is not present in the master bias), and then decays as the CCD is read, leaving an exponential profile as function of the CCD row, with a maximum in the row that is read first for each amplifier (the top or bottom row on the CCD). This effect is largely mitigated by pre-reading the equivalent of 300 'blank' CCD rows (while running the serial clocks but not the parallel clocks, which results in disabling the row to row charge transfer and reading dummy values) but a residual effect of a few electrons at maximum is left and corrected with calibration data obtained from the analysis of many dark exposures with various exposure times, from 1 second to 1200 seconds. The dark exposures are taken during nights with poor atmospheric conditions or during the day when possible. As already mentioned above, two additional 1200s darks are taken at the end of each night.

The procedure we just described is not sufficient to obtain for all exposures a residual bias level below one electron per pixel for the blue CCDs. Such a residual bias can lead to spectral discontinuities of about 5 electrons per Angstrom (the effective width of the cross-dispersion is about 3 pixels, and a pixel corresponds to a wavelength variation of 0.6\,\AA\ along the dispersion axis). For this purpose we evaluate a residual background in the space between the fiber traces of adjacent fiber blocks in each science exposure, and then interpolate the values across the whole CCD. This background estimate is obtain with a median filter across CCD rows of the mean bias level in each row in the space between blocks. We have reprocessed a large number of science exposures to verify that this new approach improves the bias level estimation and the redshift performances.

\subsubsection{Gains}

\begin{figure}
  \centering
  \includegraphics[width=0.99\columnwidth,angle=0]{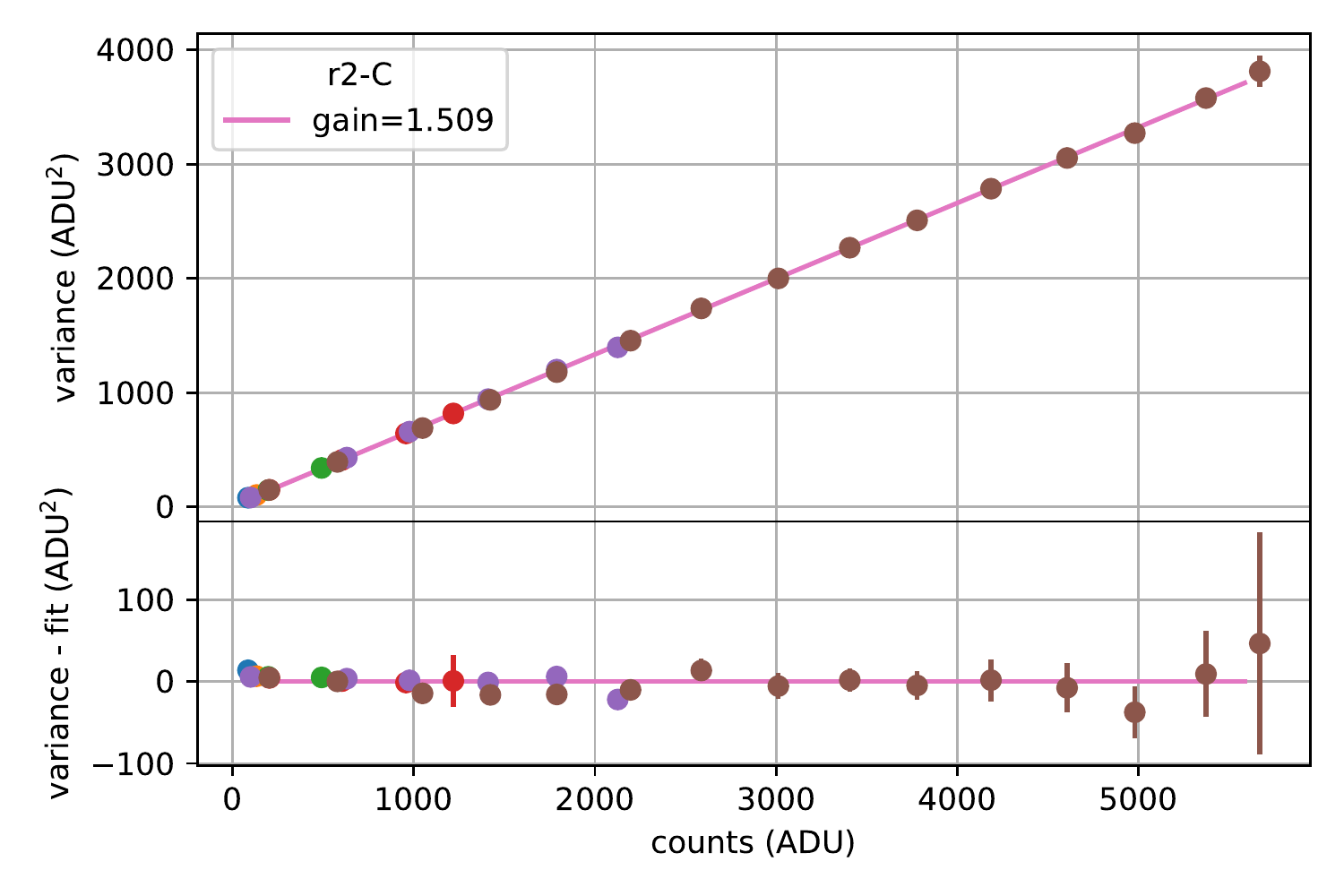}
  \caption{Photon transfer curve for spectrograph SM5, camera r, amplifier C, obtained with pairs of exposures of 1, 2, 5, 10, 20 and 50 seconds with the LED lamps illuminating the dome screen. The units on the figure are ADU (or ADC counts). A gain of 1.509\,electron/ADU is determined from the inverse of the slope.}
  \label{fig:gain}
\end{figure}

Gains in electrons/ADU are applied to the bias subtracted images, per amplifier. The gain values have been obtained from photon transfer curves (i.e. variance as a function of mean counts) derived from exposures with the telescope pointed towards the dome screen, the latter being illuminated with LED lamps.
We take advantage of the excellent temporal stability of the LEDs to estimate the variance from a comparison of spectra obtained in consecutive exposures. A boxcar extraction is used for this purpose to avoid potential biases introduced by variations in the position of spectral traces. Pairs of observations of different exposure times are used to cover the dynamic range from 0 to 5,000 electrons. The gains measured with this technique are reproducible and consistent with measurement obtained in very different conditions on a test bench at CEA, Saclay, France. We also do not detect discontinuities at the boundary of amplifiers after applying the gains. Figure~\ref{fig:gain} shows an example of such a transfer curve. 

\subsubsection{Electronic cross-talk correction}

An electronic cross-talk between amplifiers of the same CCD has been detected for some chips. The measurement was done with a sparse fiber slithead where only a few isolated fiber traces were used to detect a flipped ghost trace on another amplifier. It is not possible to perform this study with the science slit as the true fiber traces and the ghosts from other amplifiers overlap.
A few pairs of amplifiers have cross-talk amplitudes exceeding $10^{-4}$, the maximum value being $6 \times 10^{-4}$ with a contamination of amplifier A into B for the blue CCD of the spectrograph SM2, but the signal is much smaller or undetected for the majority of CCDs. Although the cross-talk effect is arguably negligible, we nevertheless correct for the effect in image pre-processing. It consists simply in subtracting from each image quadrant the flipped image of the other quadrants multiplied by the corresponding cross-talk amplitude.

\subsubsection{Cosmic ray identification}

\begin{figure}
  \centering
  \includegraphics[width=0.99\columnwidth,angle=0]{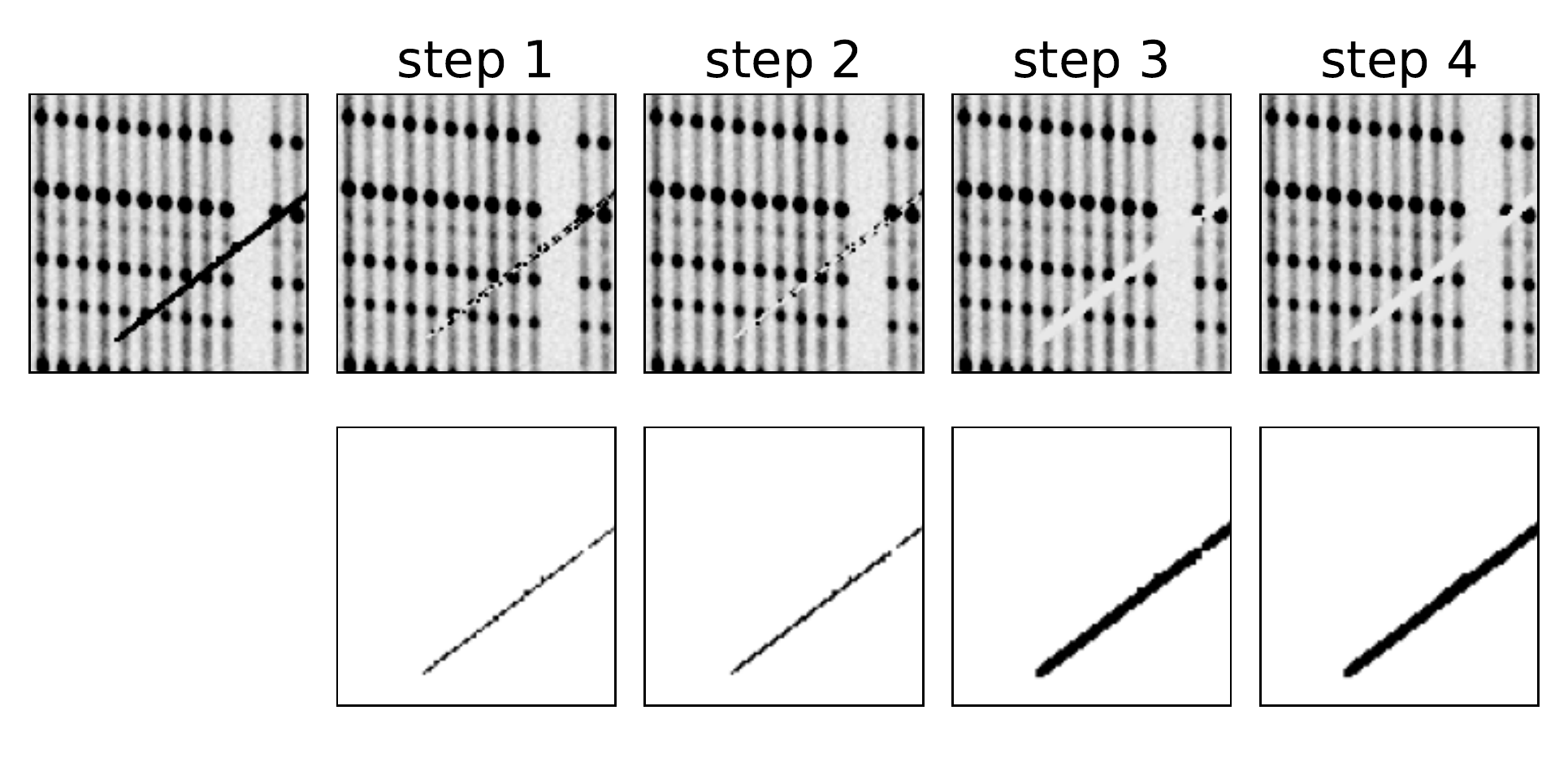}
  \caption{A $100\times100$-pixel region of an image from a red camera. The upper panels show the pixel values with the pixels identified as cosmic ray masked out. The bottom panels show the masked pixels at each step of the cosmic ray identification procedure.}
  \label{fig:cosmicrays}
\end{figure}

Cosmic ray hits have to be identified and masked with a very high efficiency to avoid spurious emission lines in the spectra that could lead to incorrect spectroscopic identification and redshifts for emission line galaxies.
We are rejecting cosmic rays with the following steps during the pre-processing (see illustration on Figure~\ref{fig:cosmicrays}).

1) We first mask pixels using the spatial gradient of the intensity in the CCD. We mask a pixel if its value is significantly above its neighbors along at least two axes out of four, counting diagonals (Eq~\ref{eq:cosmicray_criterion1}), and if the intensity gradient relative to the peak value is significantly larger than the expectation given the spectrograph point spread function along one of the four axes (Eq~\ref{eq:cosmicray_criterion2}).

\begin{subequations}
\label{eq:cosmicray}
\begin{equation}    P > P_{neigh} + N \, \sigma(P_{neigh}) \label{eq:cosmicray_criterion1} \end{equation}
\begin{equation}    c_2 ( P - c \, \sigma(P) ) \delta_{PSF} > P_{neigh} + c \, \sigma(P_{neigh}) \label{eq:cosmicray_criterion2}\end{equation}
\end{subequations}
where $P$ is the value of the pixel being tested, $\sigma(P)$ its uncertainty, $P_{neigh}$ is the average of the two pixels neighboring the pixel of interest along one of the four axis, $\sigma(P_{neigh})$ its uncertainty, $\delta_{PSF}$ is the maximum PSF variation along this axis, and  $N=6$, $c=3$, and $c_2=0.5$ coefficients that have been adjusted on the data. We follow here the algorithm implemented for SDSS imaging\footnote{See the Photo Pipelines documentation, \url{https://www.astro.princeton.edu/~rhl/photo-lite.pdf}}, with minor modifications. The gradients $\delta_{PSF}$ have been derived from the measured point spread function. There is one set of values for each camera arm (blue, red, and NIR).

2) After a first pass, we repeat the test with a lower significance threshold $N=3$, but only applying it to the pixels neighboring the ones that have already been masked. This last step is repeated as long as new pixels are masked.

3) We then broaden the masked regions in all four directions by 1 pixel.

\begin{figure}
  \centering
  \setlength{\tabcolsep}{0pt}
  \renewcommand{\arraystretch}{0}
  \begin{tabular}{ccccccc}
  \includegraphics[width=0.142\columnwidth]{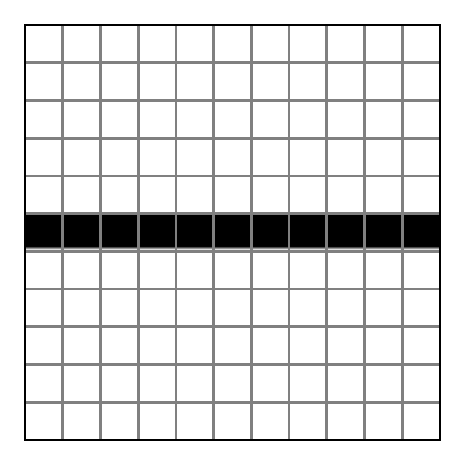} &
  \includegraphics[width=0.142\columnwidth]{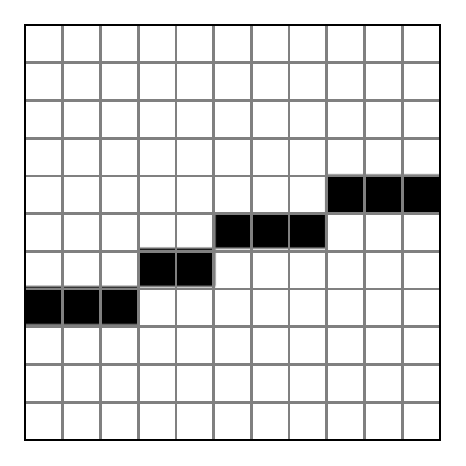} &
  \includegraphics[width=0.142\columnwidth]{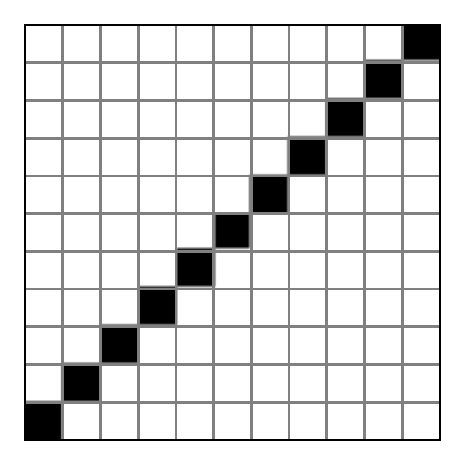} &
  \includegraphics[width=0.142\columnwidth]{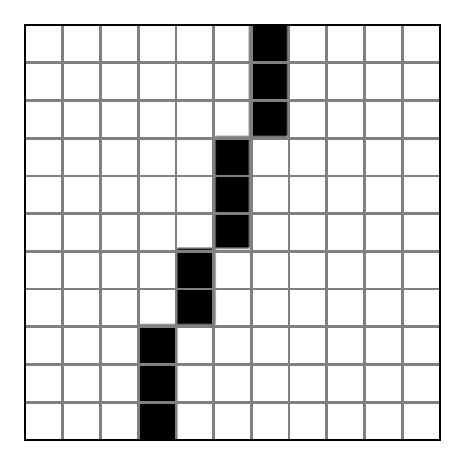} &
  \includegraphics[width=0.142\columnwidth]{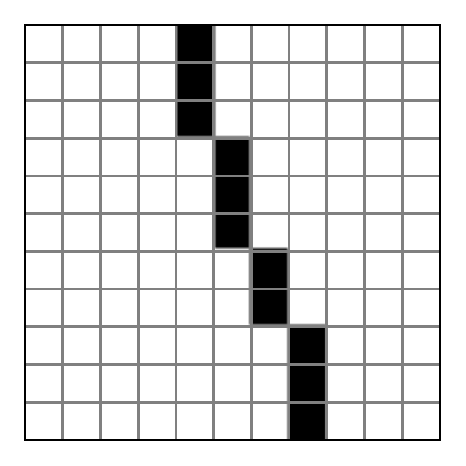} &
  \includegraphics[width=0.142\columnwidth]{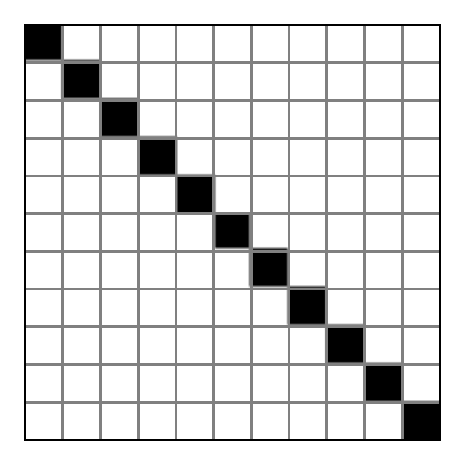} &
  \includegraphics[width=0.142\columnwidth]{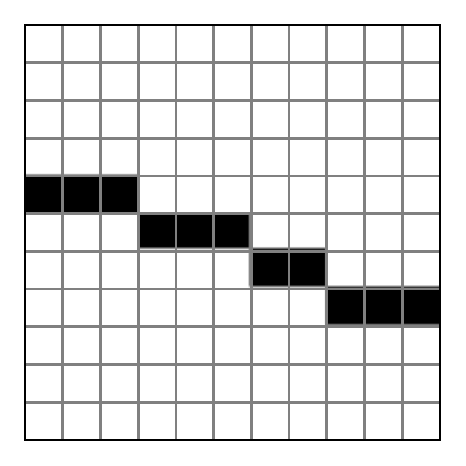}
  \end{tabular}
  \setlength{\tabcolsep}{6pt}
  \renewcommand{\arraystretch}{1}
  \caption{Seven $11\times11$-pixel binary ``structuring elements'' used to
  merge disconnected particle tracks with binary closure during the final step
  of cosmic-ray identification.}
  \label{fig:binaryclosure}
\end{figure}

4) Finally, to connect separated clusters of pixels belonging to individual cosmic-ray tracks, we perform a morphological binary closure \citep{Dougherty:2003}. The closure applies a sequence of seven ``structuring elements'' shown in Figure~\ref{fig:binaryclosure} to the cosmic-ray mask, with the elements chosen to align with possible track orientations while discarding the vertical element that would align with the spectral traces. The closure operation produces seven masks which are combined with a bitwise OR.

The outcome of cosmic-ray masking, broadening, and binary closure is shown in Figure~\ref{fig:cosmicrays}. Note that the residual signal from cosmic rays is further detected in later steps of processing (during the spectral extraction, at each step of the spectral calibration, and finally when spectra from several exposures are co-added).

\subsubsection{Negative trail correction}
\label{sec:darktrails}

\begin{figure}
  \centering
  \includegraphics[width=0.9\columnwidth,angle=0]{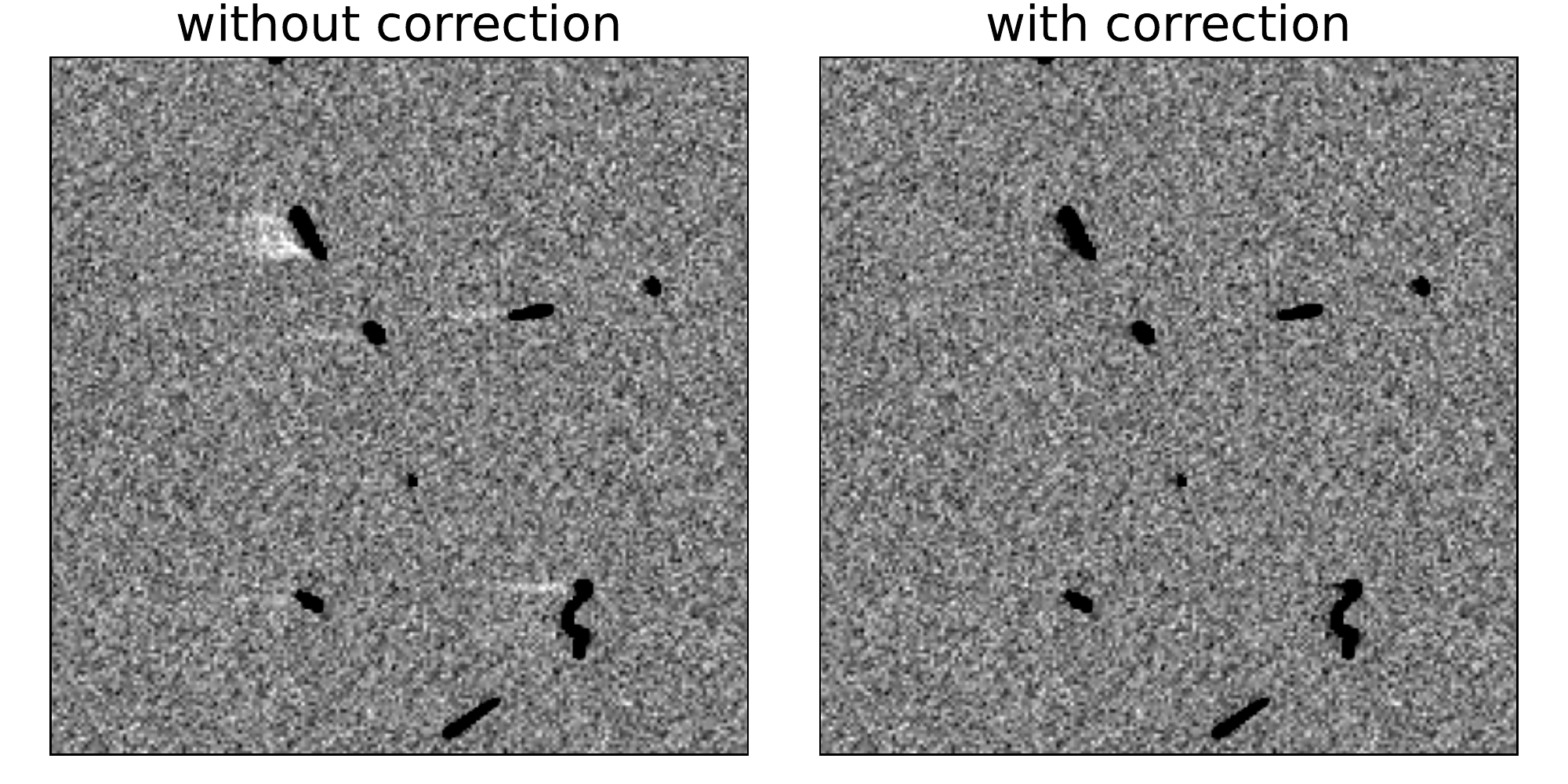}
  \caption{A 200x200 pixel region of a dark image from the red camera of spectrograph SM6, without and with the negative trail correction. Note the bright trails corresponding to negative values in this inverted gray scale on the left of cosmic ray hits on the left panel. The serial clock moves charges to the right in this region of the CCD read with amplifier B.}
  \label{fig:negative-trails}
\end{figure}

Negative trails have been found for bright illuminations (typically cosmic rays) along the serial transfer direction for some amplifiers in the red and NIR CCDs. We have not been able to remove those with changes to the parameters of the CCD readout (clocks voltages and profiles, digital correlated double sampling parameters, namely the position and size of the readout time window and the number of digital samples). Fortunately this effect has been found to be linear, of small amplitude (0.1\% of the charge for the worst amplifier, see Table~\ref{table:negative-trails}), and easy to model with a simple spatial convolution of the CCD image. We have used images from bright illuminations with a continuum lamp to fit the parameters of an exponential convolution kernel along the direction of serial transfer (CCD rows). The parameters for the few affected amplifiers are given in Table~\ref{table:negative-trails}. We subtract from the original image the convolved image as part of pre-processing. This process should in principle be iterative but this is not necessary given the amplitude of the effect. It also introduces correlated noise between pixels but this is also fortunately a negligible effect. An example of an image before and after this correction is shown on Figure~\ref{fig:negative-trails}.

\begin{table}
  \begin{tabular}{ccccc}
    Spectrograph, & $A$ & $L$ \\
    Camera, Amplifier &  &  (pixels) \\
    \hline
    SM1 r B & 0.00024 & 13 \\
    SM2 r D & 0.00021 & 18 \\
    SM2 z D & 0.00006 & 28 \\
    SM3 r B & 0.00020 & 23 \\
    SM6 r B & 0.00120 & 6 \\
    SM6 z C & 0.00005 & 20 \\
    SM6 z D & 0.00008 & 35 \\
    SM7 r B & 0.00041 & 4 \\
    \hline
  \end{tabular}
  \caption{Parameters of the negative trails convolution kernel $K(x) = -(A/L) \exp(-x/L)$ for the CCD amplifiers where this effect has been detected.}
  \label{table:negative-trails}
\end{table}

\subsubsection{CCD flat field}

\begin{figure}
  \centering
  \includegraphics[width=0.9\columnwidth,angle=0]{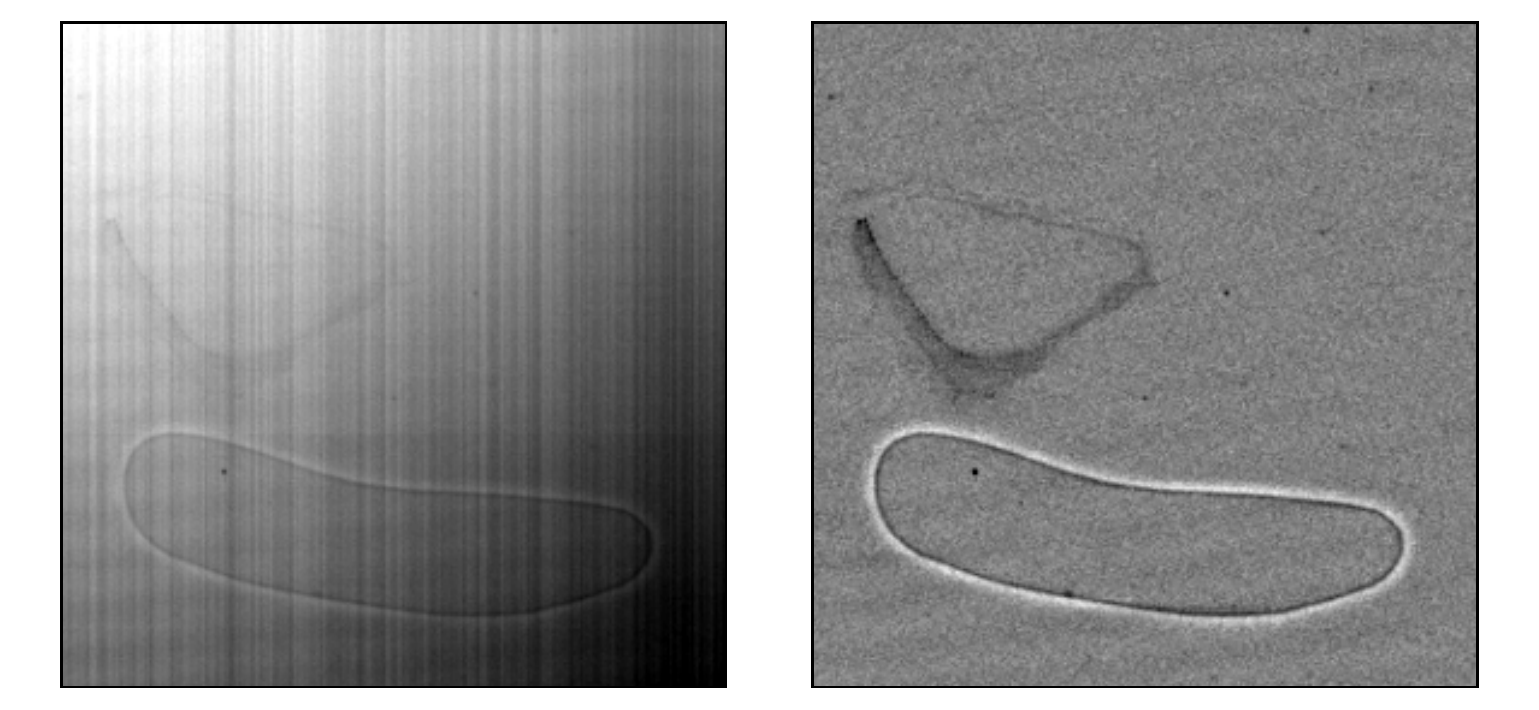}
  \caption{Left: a 300x300 pixel region showing the average of a series of images obtained with the flat field slit for the red camera of spectrograph SM8. Right: the pixel flat field correction derived from the image. The range of scales is $\pm$20\% in the left panel and only $\pm$5\% in the right panel. A lighter gray corresponds to a higher efficiency. The vertical strips on the left panel are caused by small variations in the slit width. The gradient from top to bottom is due to spectral variations in the illumination.}
  \label{fig:pixel-flat-field}
\end{figure}

The CCD flat fielding has been obtained on site at Kitt Peak using a spectrograph slit head specifically designed for this purpose. During calibration runs, the science slit is removed from the spectrograph and the flat field slit inserted. It consists of a fiber, on the tip of which is located a lens that spread the light to a diffusing sheet in front of which is placed a thin slit. The thin slit is purposefully slightly offset from the spectrograph best focus position in order to smooth out the image on the spectrograph. A special purpose software is used to correct for the non-uniformities of the images due to the illumination pattern while preserving to the best possible extent the variation of efficiency of the CCD pixels. The algorithm consists in iteratively dividing the image by the same image convolved with a 1D Gaussian kernel, one axis at a time, while masking pixels identified as outliers at each iteration. An example is given on Figure~\ref{fig:pixel-flat-field}.

\subsubsection{Scattered light}

During the first few months of the spectrographs' commissioning, we have found a continuous increase of scattered light in the CCD images. This was caused by an accumulation of contaminants on the cryostat windows due to a polluted source of dry air. This has since then been corrected, the windows have been cleaned, and we no longer detect a significant amount of scattered light.

In order to process this early data set, we have modeled this scattered light contribution as a convolution of the input image with a 2D kernel derived from the profile of scattered light around bright spots of arc lamp images. The model image is then modulated with a 2D smooth correction whose coefficients are determined by comparing the model to the background level measured in the space between blocks of fiber traces. This is similar to the residual background subtraction presented in \S\ref{sec:bias-subtraction}. This algorithm was efficient when the amount of scattered light was small. It is disabled for the processing of recent data for which there is no detectable scattered light.

\subsubsection{Bad pixel mask}

A 2D map of bad pixels in each of the CCDs is provided as part of the pre-processing which we incorporate into a bitmask, each bit flagging a specific issue. The pixels flagged during pre-processing have large dark currents, are saturated, affected by cosmic rays, in regions of dead columns or large cosmetic defects, or have a low flat field value.

A static bitmask marking some permanently flagged pixels is used as an input to the pre-processing step. This is obtained flagging outliers in the 2D distribution of the median and inter-quartile range (IQR) of pixels measured from a series of bias subtracted and gain applied set of dark exposures. The algorithm then uses a binary closing operation to flag CCD pixels completely surrounded by masked pixels. A standard set of thresholds for median and IQR work well for most CCDs but for some CCDs, it was necessary to manually adjust them based on a visual inspection of the mask and 2D distribution of the median and IQR values. These pixel masks need to be regularly updated to reflect any changes to the hardware.

We do not provide here the list of bits and their meaning but invite the interested reader to find this information in the DESI data model documentation\footnote{See~\url{https://desidatamodel.readthedocs.io} for the current version, but note this documentation will evolve with data releases.} which will vary from one data release to the next.

\subsubsection{Estimating the variance}
\label{sec:ccd-variance}

The read noise is determined from the dispersion in the over-scan regions (the average read noise of the blue, red and NIR camera CCDs are respectively of 3.4, 2.7 and 2.5 electrons per pixel r.m.s.).

A method used in many spectroscopic pipelines to estimate the variance in pixels in the active region of the CCD has been to use the sum of the read noise variance and an estimate of the Poisson variance directly inferred from the individual pixel values, once converted in electrons.

\begin{equation}
  \widetilde{\sigma}_P^2 = \sigma_R^2 + P \ \mathrm{if} \ P>0 \label{eq:bad-variance}
\end{equation}

As mentioned in~\citet{Horne1986}, this method introduces a correlation between the pixel value and its estimated variance that propagates to spectra and leads to biases in the estimation the sky level, the calibration, and the final calibrated spectra.
For instance the weighted mean of $P$, with weights equal to the estimated inverse variance $\widetilde{\sigma}_P^{-2}$ is off by $-1$ at the leading order. The variance itself is also biased at low counts, because of the truncation to positive values.

\begin{figure}
  \centering
  \includegraphics[width=0.9\columnwidth,angle=0]{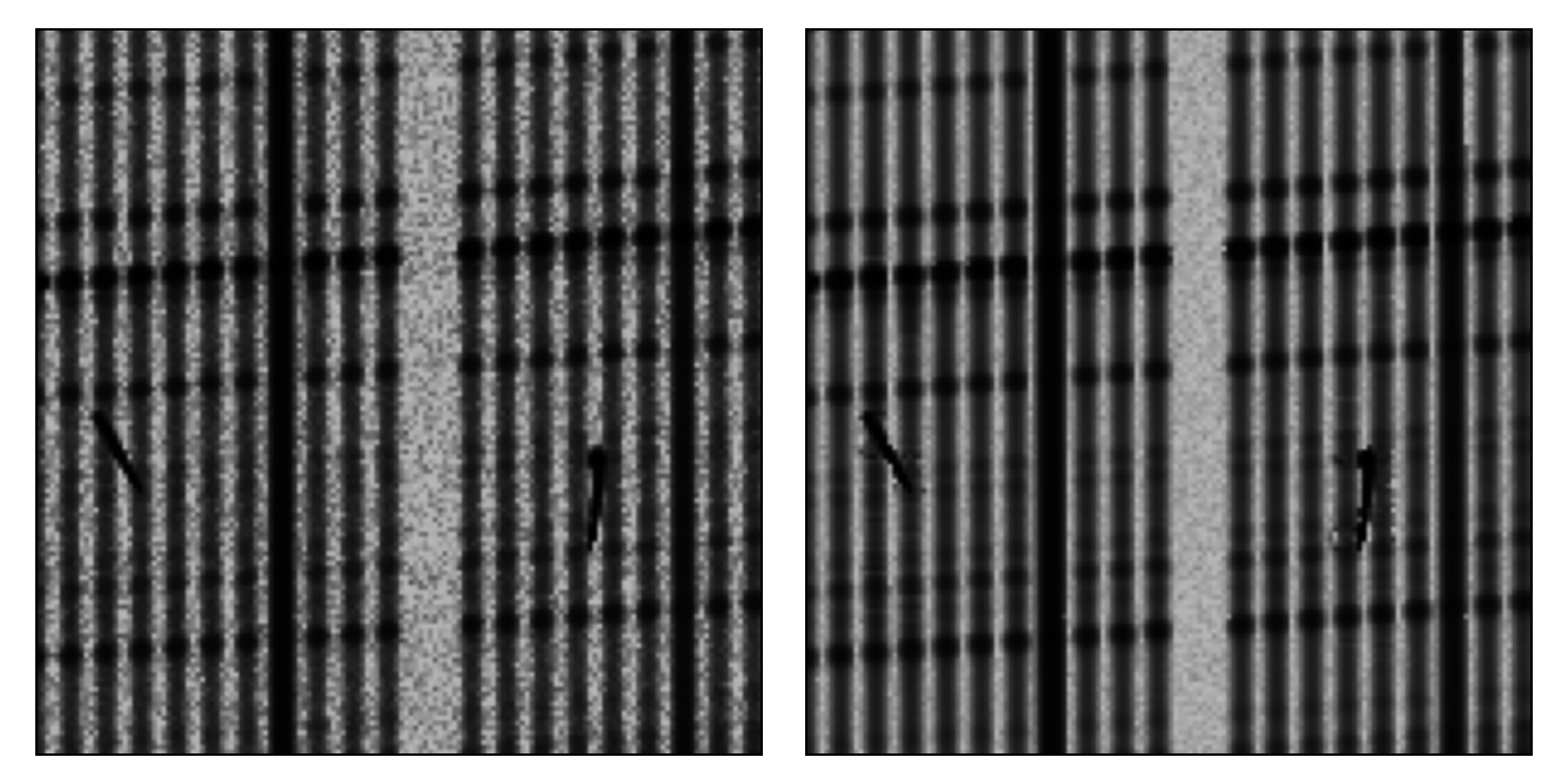}
  \caption{Left: a 150x150 pixel region showing the inverse variance derived from the individual pixel values for a 1200 second dark time exposure in a red camera CCD. Right: same pixel area with the inverse variance model used for the data processing. Spectral traces appear dark as the Poisson noise is larger and so the inverse of the variance smaller. Note the model has much lower noise. The residual spatial noise is coming from the pixel flat field. Note also that the inverse variance on cosmic rays and bright fibers is preserved.}
  \label{fig:inverse_variance_model}
\end{figure}

In order to circumvent this problem, we have resorted to use a full model of the CCD image to estimate the Poisson noise. We proceed as follows. i) We first extract the spectra with a fast boxcar extraction, after having adjusted the spectral traces (see \S\ref{sec:adjustcalib}). ii) After fiber flat fielding based on predetermined calibration data (see \S\ref{sec:fiberflat}), we compute an average sky model with a simple wavelength resampling and averaging of the sky fiber spectra. iii) We determine a smoothed version of the sky subtracted spectrum of each target, obtained with a convolution using a Gaussian kernel of parameter $\sigma=10$\,\AA, with outlier rejection (to limit the impact of residual cosmic rays), and after interpolating over values affected by masked pixels. iv) The final spectral model is obtained by adding back the sky model to the smoothed spectrum and applying the inverse of the fiber flat field correction v) We then project the spectra back on the CCD, assuming a Gaussian cross-dispersion profile. We keep the original CCD pixel values that are inconsistent with the model by more than 5 standard deviation and use the model for the others to derive the Poisson noise. An example of the inverse variance model image compared to the traditional estimate $\widetilde{\sigma}_P^{-2}$ is shown on Figure~\ref{fig:inverse_variance_model}.

The variance model ensures the linearity of the spectroscopic fluxes. It is important for analyses like the Lyman-alpha forest \citep{Bautista2017}. The propagation of the pixel variance to the extracted spectral flux variance is described in \S\ref{sec:extraction}. This variance estimate is then combined with other sources of uncertainties and compared to the scatter in the sky background flux in \S\ref{sec:sky-residual-variance}.

\subsubsection{Automated detection of bad columns}

The daily calibration sequence includes a 300 second dark exposure that can be used to verify that the pre-processing accurately subtracts the bias level and dark current. As part of the automated daily data reduction, we pre-process this exposure, and measure the median value per column of each CCD amplifier. CCD columns with values exceeding a threshold of 0.005 electron per second per pixel, positive or negative, are recorded in a calibration file for each camera. This file is subsequently used for each exposure to flag fiber spectra affected by the bad columns. The spectral contamination is estimated using the distance between fiber spectral trace (see~\S\ref{sec:adjustcalib}) and the bad column, using a pre-determined cross-dispersion profile. Spectral fluxes with contamination larger than 0.005 electron per Angstrom per second are flagged.

\subsection{Spectrograph point spread function}
\label{sec:psf}

We present in this section the point spread function (PSF) used for the spectral extraction (see \S\ref{sec:extraction}). We first describe the model and qualitative properties of the PSF, we then
describe the method we use to fit this PSF model, and finally we discuss the stability of the PSF as a function of time, temperature, telescope pointing and fiber positioner moves.

\subsubsection{Model}

\begin{figure}
  \centering
  \includegraphics[width=\columnwidth,angle=0]{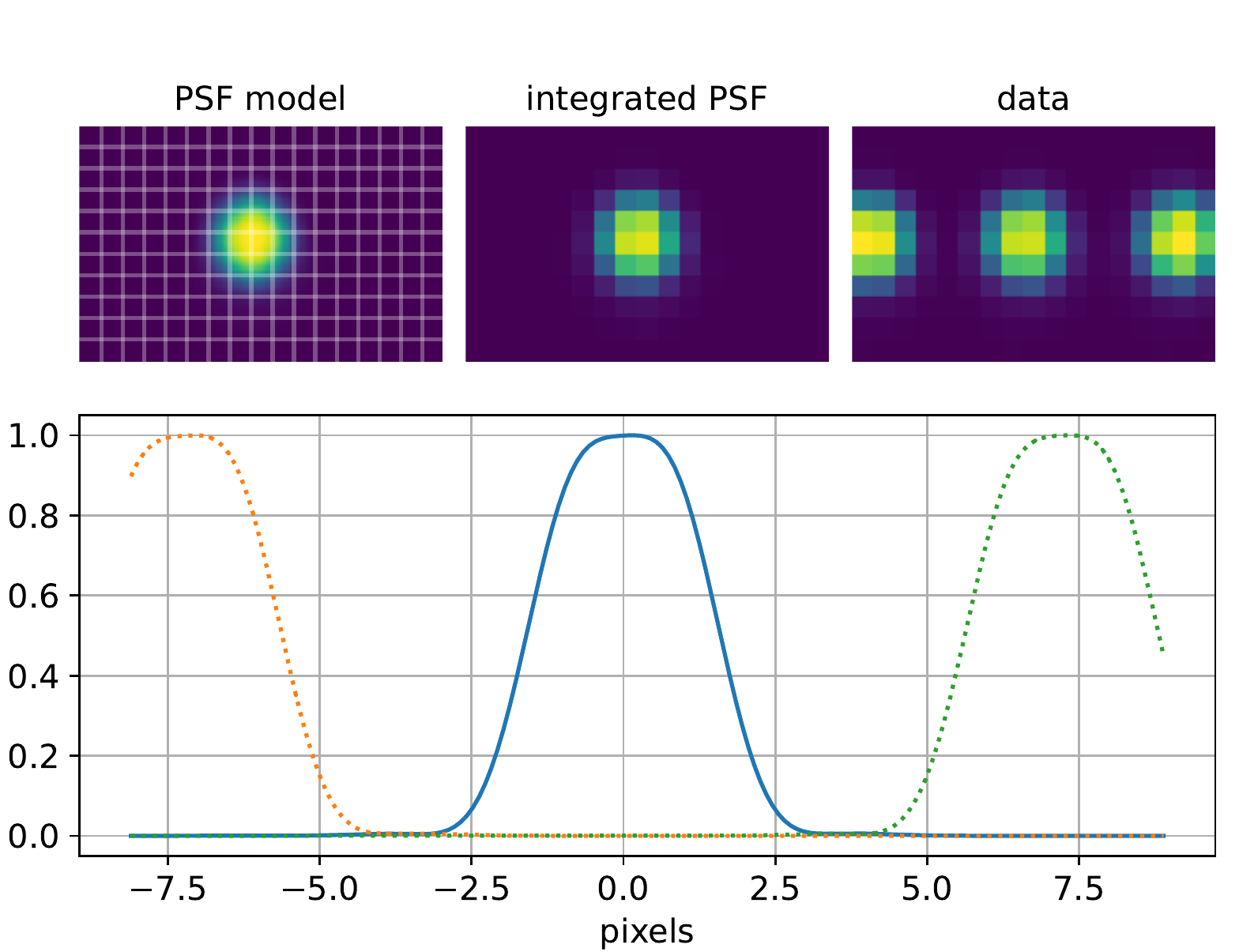}
  \caption{Example PSF of a central fiber (\#260) at 7034.3\,\AA\ in a red camera CCD. From top left to bottom right: 2D PSF model, PSF model integrated in pixels, arc lamp image where PSF from neighboring fibers is visible, and cross-dispersion profile of the PSF.}
  \label{fig:psf-r1}
\end{figure}

We consider here the PSF of the system composed of the fiber and spectrograph, such that the PSF is the convolution of the image
of the fiber tip (near field) convolved with the optical blur from the spectrograph cameras.
The diameter of the fiber image on the CCD is of 51\,$\mu$m, or 3.4~CCD pixels, after accounting for the camera demagnification of 0.48 (see \citealt{DESI2016b}).
The PSF appears as a blured disk presenting a central plateau because the fiber image is partially resolved (see Fig.~\ref{fig:psf-r1}).

We have considered an empirical model of the central part of the PSF consisting in a linear combination of Gauss-Hermite functions.
\begin{eqnarray}
  \small
  PSF_H&&(x,y)  = \sum_{i,j}  a_{i,j} \mathrm{He}_i \left( \frac{x-x_c}{\sigma_x} \right) \mathrm{He}_j \left( \frac{y-y_c}{\sigma_y} \right) \nonumber \\
  && \times  \frac{1}{2 \pi \sigma_x  \sigma_y} \exp \left( -\frac{(x-x_c)^2}{2 \sigma_x^2}-\frac{(y-y_c)^2}{2 \sigma_y^2} \right) \label{eq:psf}
\end{eqnarray}
where $\mathrm{He}_i$ is a Hermite polynomial\footnote{We use the probabilist's and not the physist's definition of the Hermite polynomials.}  of degree $i$ (from 0 to 6),  $x_c$ and $y_c$ are the coordinates of the spectral traces in CCD pixels, and $\sigma_x$ and $\sigma_y$ the Gaussian parameters along both axis. The coefficients $(a_{i,j})$ and the parameters $x_c,y_c,\sigma_x,\sigma_y$ are allowed to vary continuously with wavelength and from fiber to fiber. We model them as Legendre polynomials of the wavelength, per fiber.

The primary motivation for this modeling approach is its flexibility to describe most PSF shapes. Having the complex PSF shape described by a linear combination of components makes the fit convergence easier. Also the basis is orthogonal which means uncorrelated best fit $(a_{i,j})$ coefficients in the limit of small pixel size and constant pixel weights. Finally the integral of each component is analytic, so the integral of the PSF in the pixels is analytic as are all the derivatives of the model with respect to its parameters. This reduces considerably the computation time for the fit. This choice of basis is however not a perfect match to model a partially resolved disk: a Hermite degree as high as 6 is needed to model reasonably well the PSF shape.

The spectrograph PSF is also composed of a faint extended tail due to the roughness of the optical surfaces and impurities that scatter a fraction of the light at large angles. A power law component has been implemented in the model to account for this effect. It is however not used in the current version of the pipeline as the Gauss-Hermite terms appear to be sufficient to describe the PSF tails up to a scale corresponding to the spacing between neighboring fiber traces (about 7 pixels). We will see however in \S\ref{sec:fibercrosstalk} that we still need to correct for a fiber cross-talk induced by those tails.
We chose to ignore extended PSF tails in the spectral extraction (and PSF modeling) to mitigate the effect of unmasked cosmic ray hits that would contaminate many spectra across many fibers and wavelengths if extended PSFs were used for extraction. Reducing the PSF size also helps the computing performance, and finally, we have found that the cross-talk correction in post-processing was precise enough.

Figure~\ref{fig:psf-r1} illustrates the PSF shape for a central fiber of a red camera CCD, at 7034.3\,\AA\ which is a bright Neon line. The cross-dispersion profile highlights the central plateau that shows the fiber image is nearly resolved. The profile also illustrates that there is very little overlap between the spectral traces of adjacent fibers. It is also made clear on the figure that the sharpness the PSF profile requires that we integrate precisely the PSF in the pixels instead of sampling its value at the pixel center.

\begin{figure}
  \centering
  \includegraphics[width=\columnwidth,angle=0]{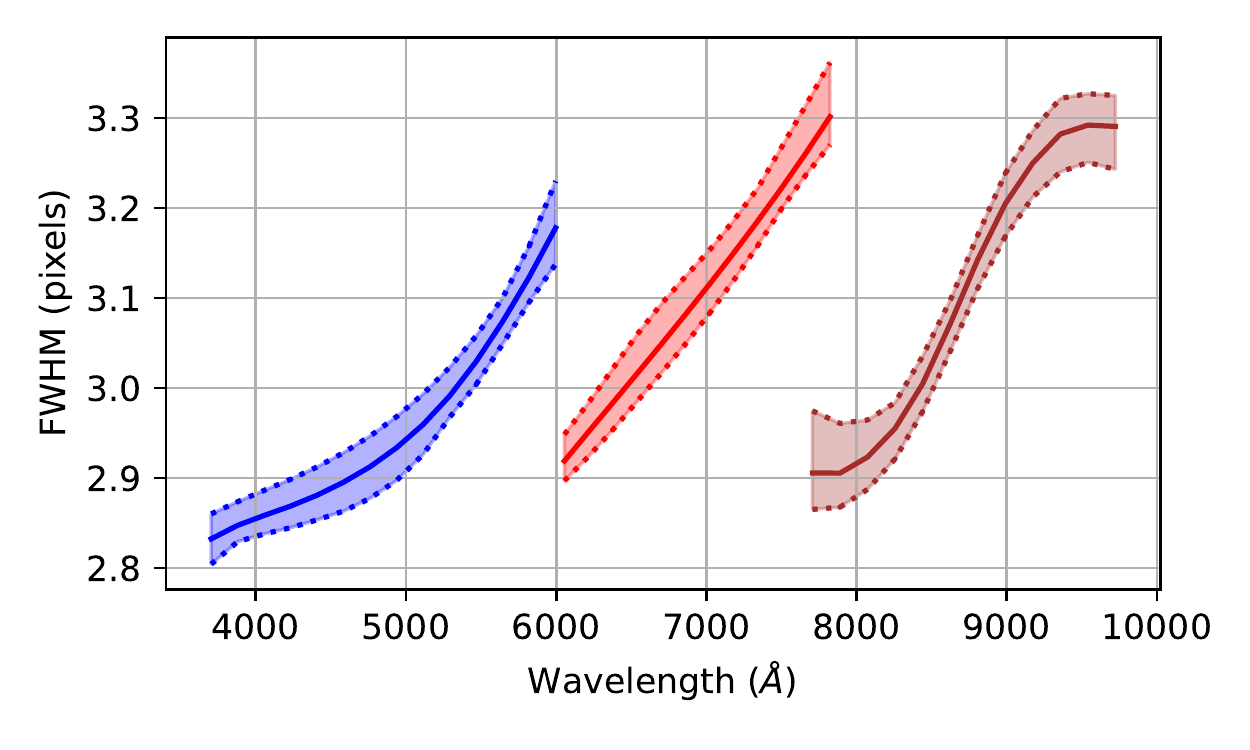}
  \caption{Spectrograph PSF FWHM in the blue, red and NIR cameras, in units of CCD pixel (the pixel size is 15\,$\mu$m). This is the width of a slice of the PSF along the dispersion axis and it is not convolved with the pixel size, so this is different from the line spread function FWHM. This PSF width as to be compared with the diameter the fiber tip image of 3.4 pixel. The solid curves are the average PSF widths and the dotted curves the minimum and maximum PSF widths.}
  \label{fig:psf-fwhm}
\end{figure}

The average PSF FWHM (full width half max) is represented in Figure~\ref{fig:psf-fwhm} as a function of wavelength for the three cameras of a spectrograph.

\subsubsection{Fit procedure}

The PSF parameters are fit for each arc lamp exposure. The fit is performed with the {\it specex}\footnote{{\it specex:} \url{https://github.com/desihub/specex/}} C++ package that is part of the DESI software suite. The parameter estimation consists in a classic least square minimization of the model parameters using the Gauss-Newton algorithm. The inputs to the code are a pre-processed image (see~\S\ref{sec:preprocessing}) with the pixel counts, their variance and a mask, a first approximate solution of the trace coordinates, wavelength calibration, and PSF, and a list of arc lamp lines with their wavelength in vacuum (given in Table~\ref{table:arc-lamp-lines-list} in Appendix~\ref{sec:arc-lamp-lines-list}).

The parameters are the Legendre coefficients of the parameters $(a_{i,j}),x_c,y_c,\sigma_x,\sigma_y$, and the intensity of all the arc lines from all the fibers. We note here that we use the arc lamp exposures to adjust the cross-dispersion coordinates of the fiber traces ($x_C$) when we could have used the continuum lamp exposures which contain more information. Our approach has the advantage to use in a consistent way the 2D PSF model for both $x$ and $y$ directions. We have verified that we have enough emission lines to get precise trace coordinates for all wavelength. As we will see in~\S\ref{sec:adjustcalib}, we also readjust on the actual data the fiber-trace coordinates with low order polynomial corrections.

The fit actually determines the Legendre coefficients of two dimensional polynomials of the CCD coordinates ($x,y$), which allows us to account for spatially continuous changes of the resolution in the camera. The final output is presented in the form of Legendre polynomials of wavelength and fiber number which is more convenient for the spectral extraction. The fit is iterative because the model is a non-linear function of its parameters. Derivatives are analytic but the code is still CPU intensive because we integrate the PSF for all of the lines in all of the fibers at each minimization step. The balance between precision and computing time can be adjusted with varying the number of Gauss-Hermite terms, the PSF stamp size, and the number of emission lines entering in the fit. The fit is performed in several steps as follows: i) first the Legendre coefficients of trace coordinates and wavelength solution ($x_c,y_c$) polynomials are adjusted using the input PSF, ii) the Gaussian $(\sigma_x,\sigma_y)$ Legendre coefficient are fit, iii) the Legendre coefficients  of the Gauss-Hermite terms $(a_{i,j})$ are fit. The line intensities are fit at the same time as the coefficients for all three steps. The code contains additional steps to optionally fit the Lorentzian tails, account for masked spots and dead fibers. The PSF fit is performed independently per group of 25 fibers, following the fiber pseudo-slit layout composed of 20 blocks of 25 tightly packed fibers each, with an extra spacing of about 2 inter fiber distance between the blocks. There is no constraint of non-negativity for the (pixel-deconvolved) PSF model during the PSF fit. However, for extractions, the PSF values integrated in CCD pixels are forced to be positive and the PSF renormalized accordingly.

\begin{figure}
  \centering
  \includegraphics[width=\columnwidth,angle=0]{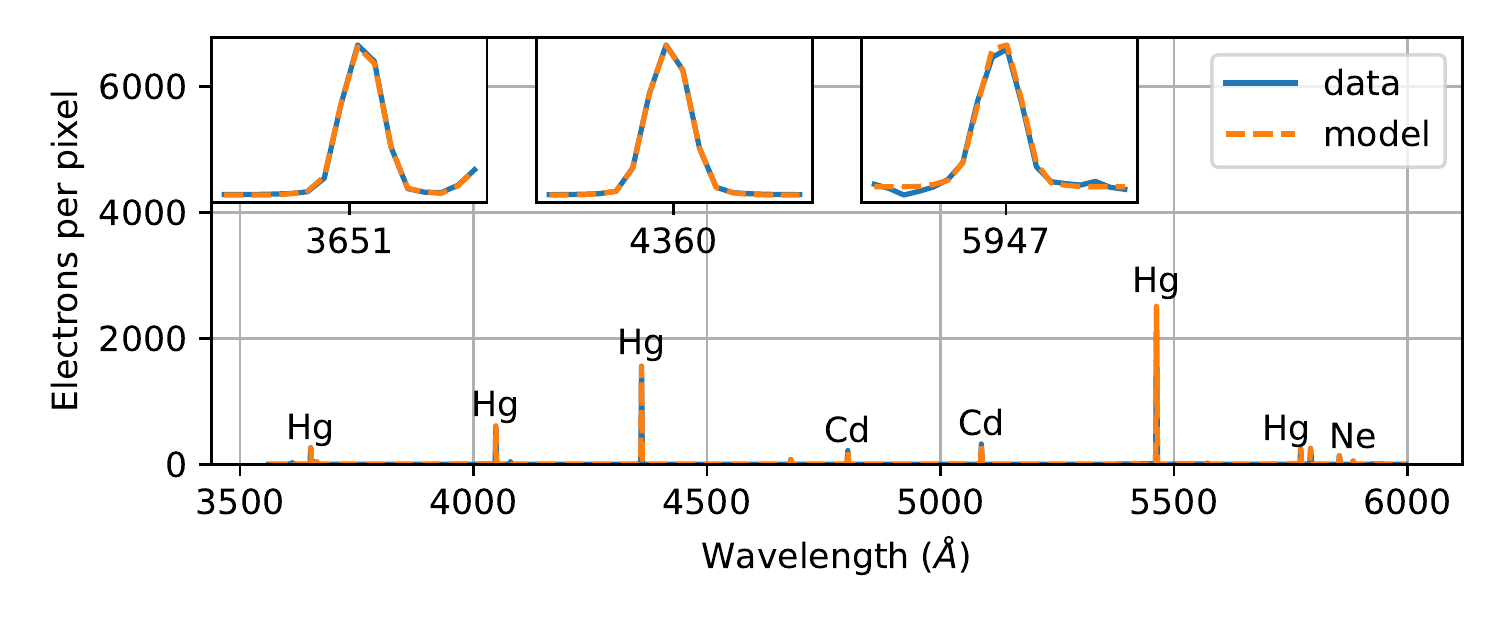}
  \includegraphics[width=\columnwidth,angle=0]{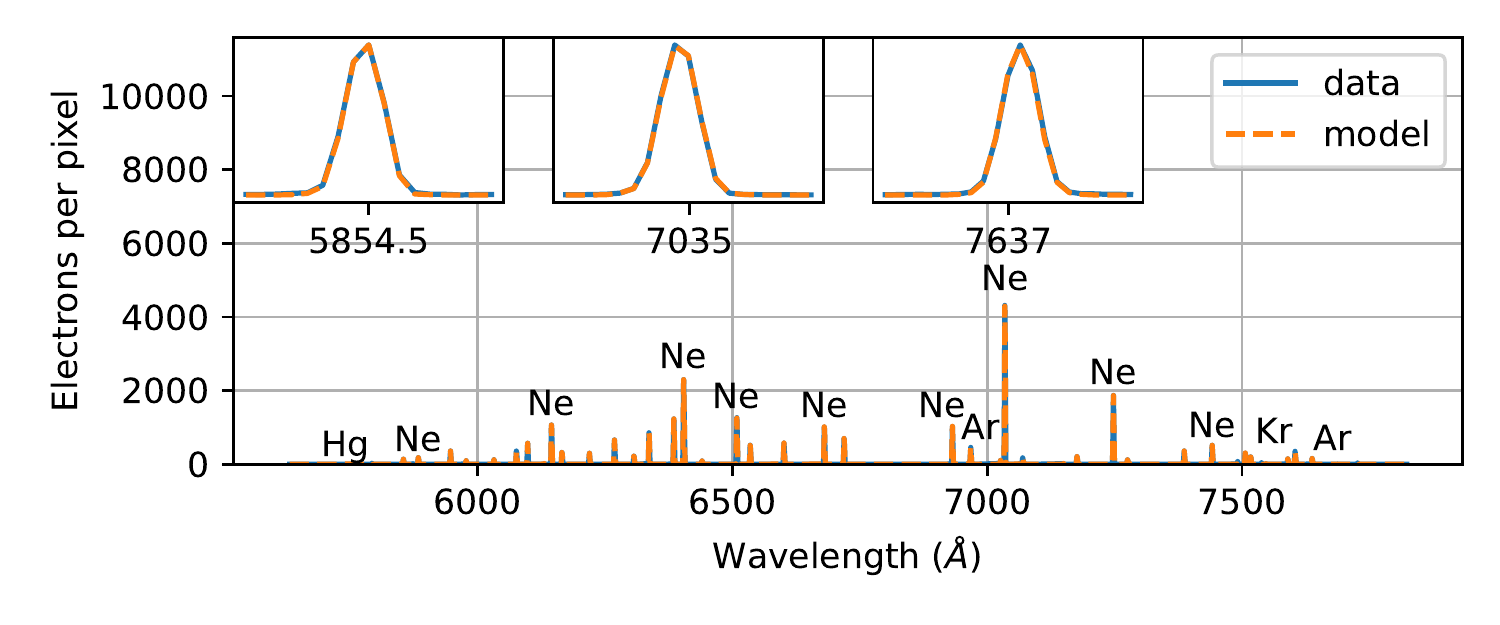}
  \includegraphics[width=\columnwidth,angle=0]{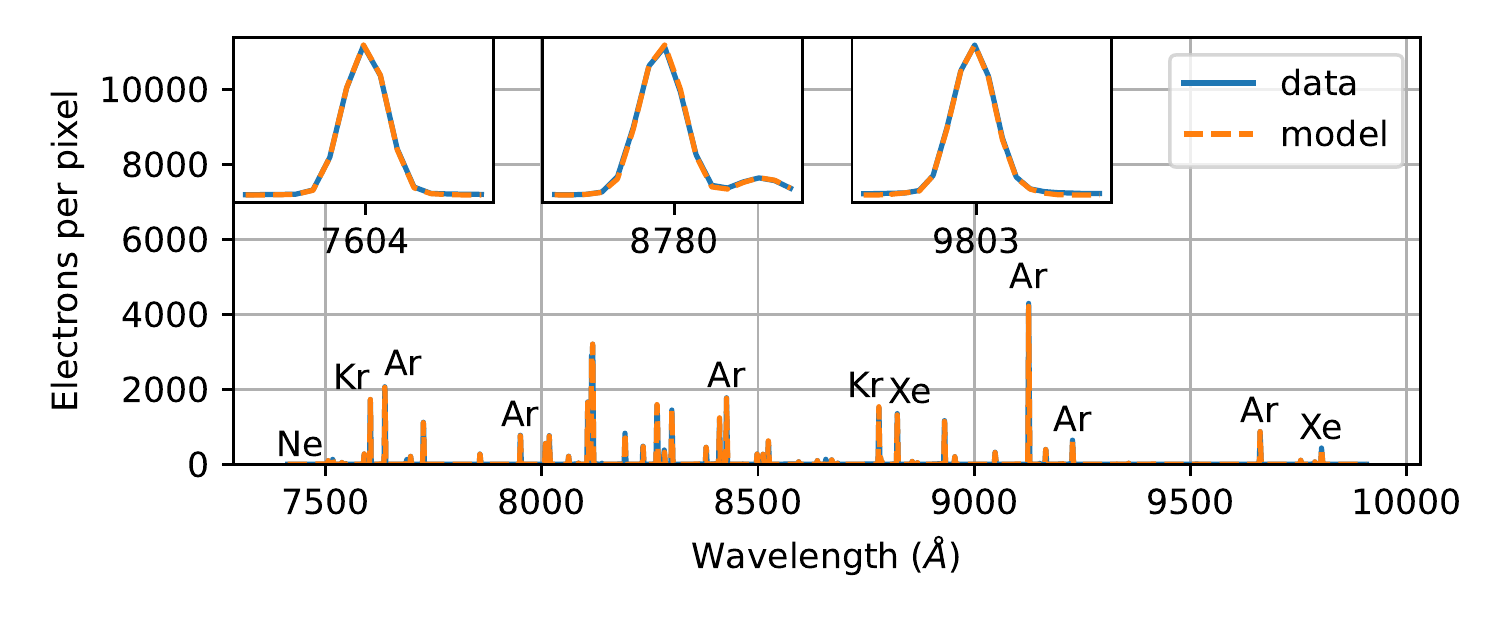}
  \caption{Spectra from arc-lamp exposures in the blue, red and NIR camera CCDs, from top to bottom. The y-axis gives the average number of electron per pixel, for each row of a vertical band covering fibers 250 to 274. The blue curve is the data and the orange dashed curve the best fit model from the PSF fit. The full wavelength range accessible for each camera is shown.}
  \label{fig:psf1}
\end{figure}

\begin{figure}
  \centering
  \includegraphics[width=\columnwidth,angle=0]{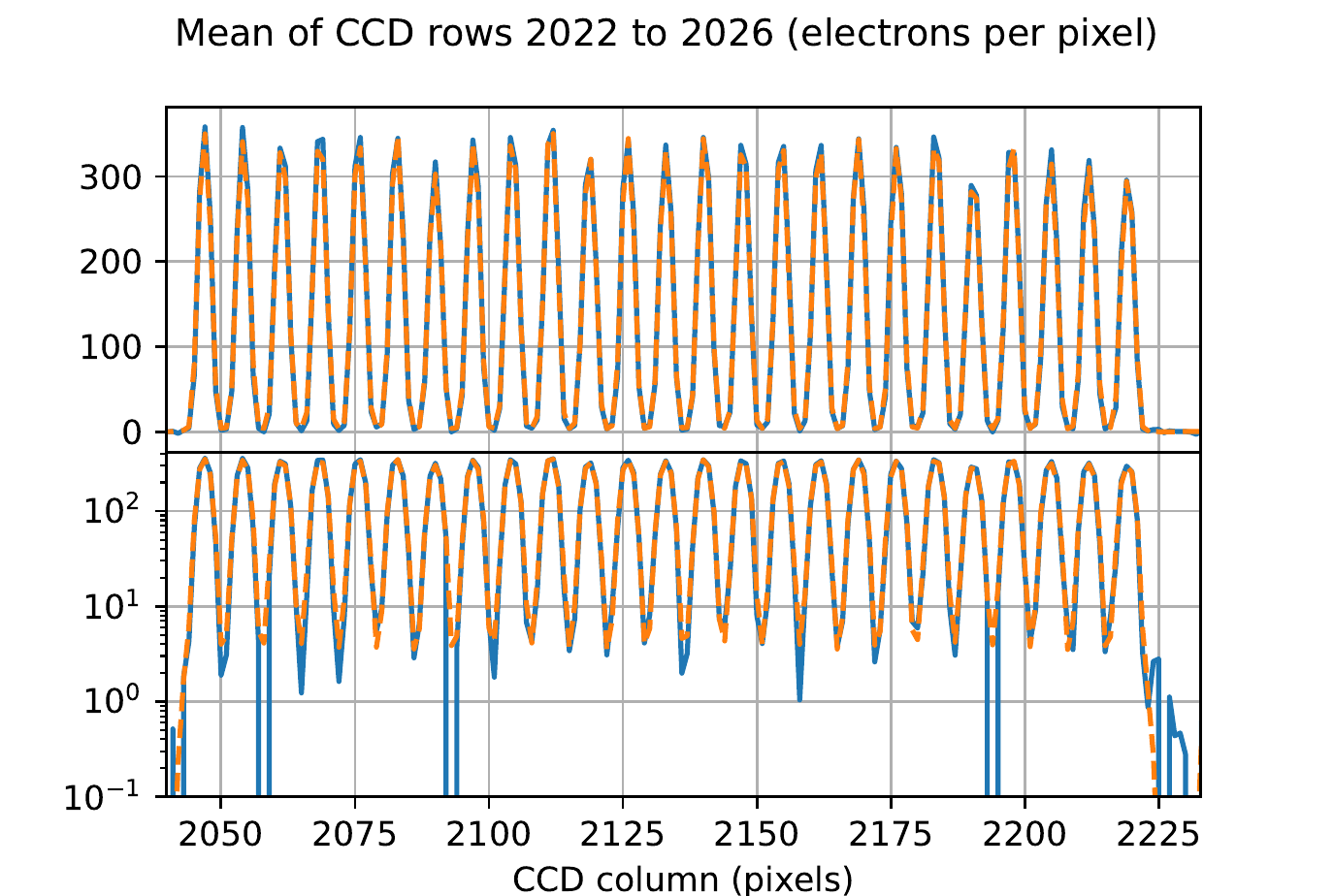}
  \includegraphics[width=\columnwidth,angle=0]{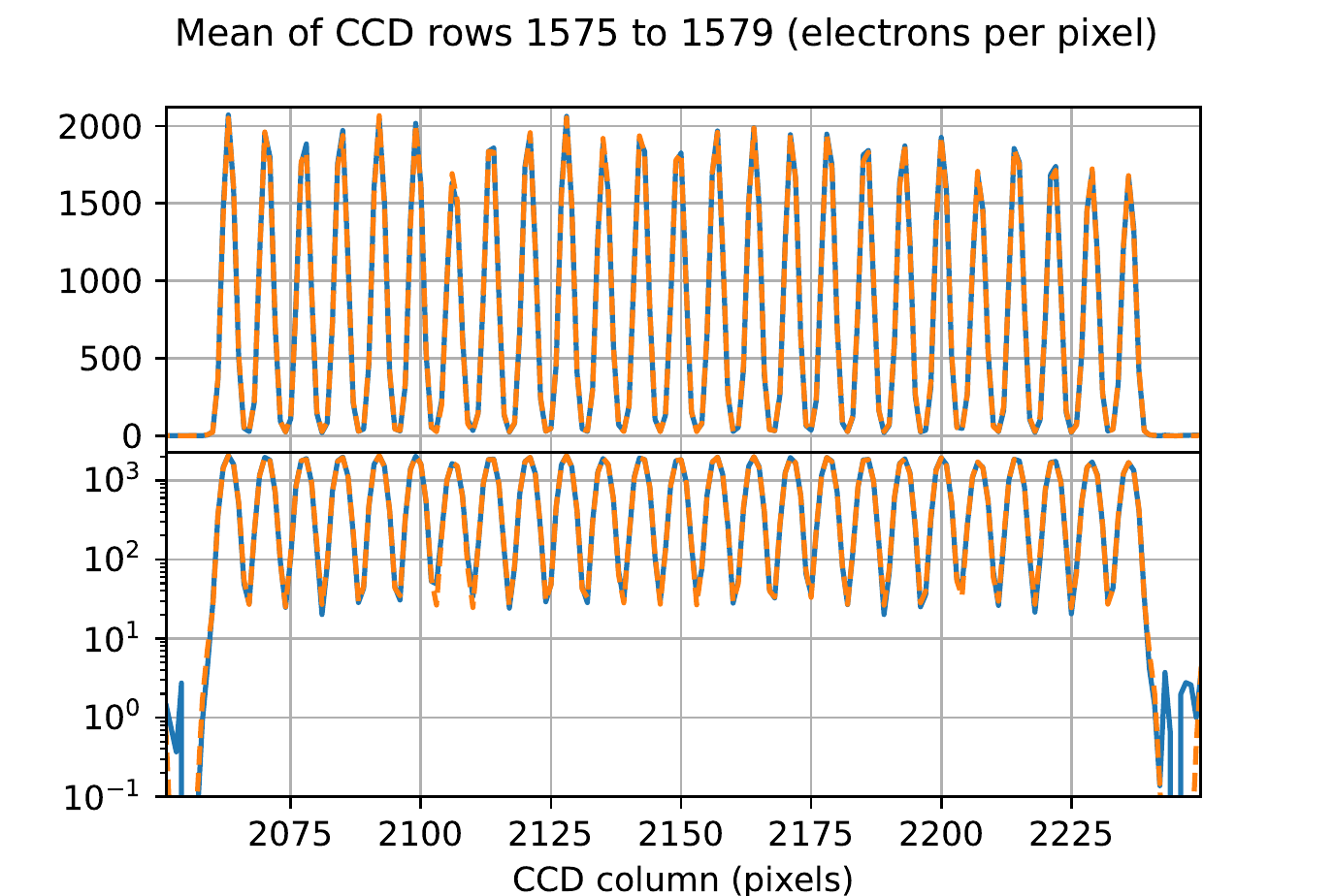}
  \includegraphics[width=\columnwidth,angle=0]{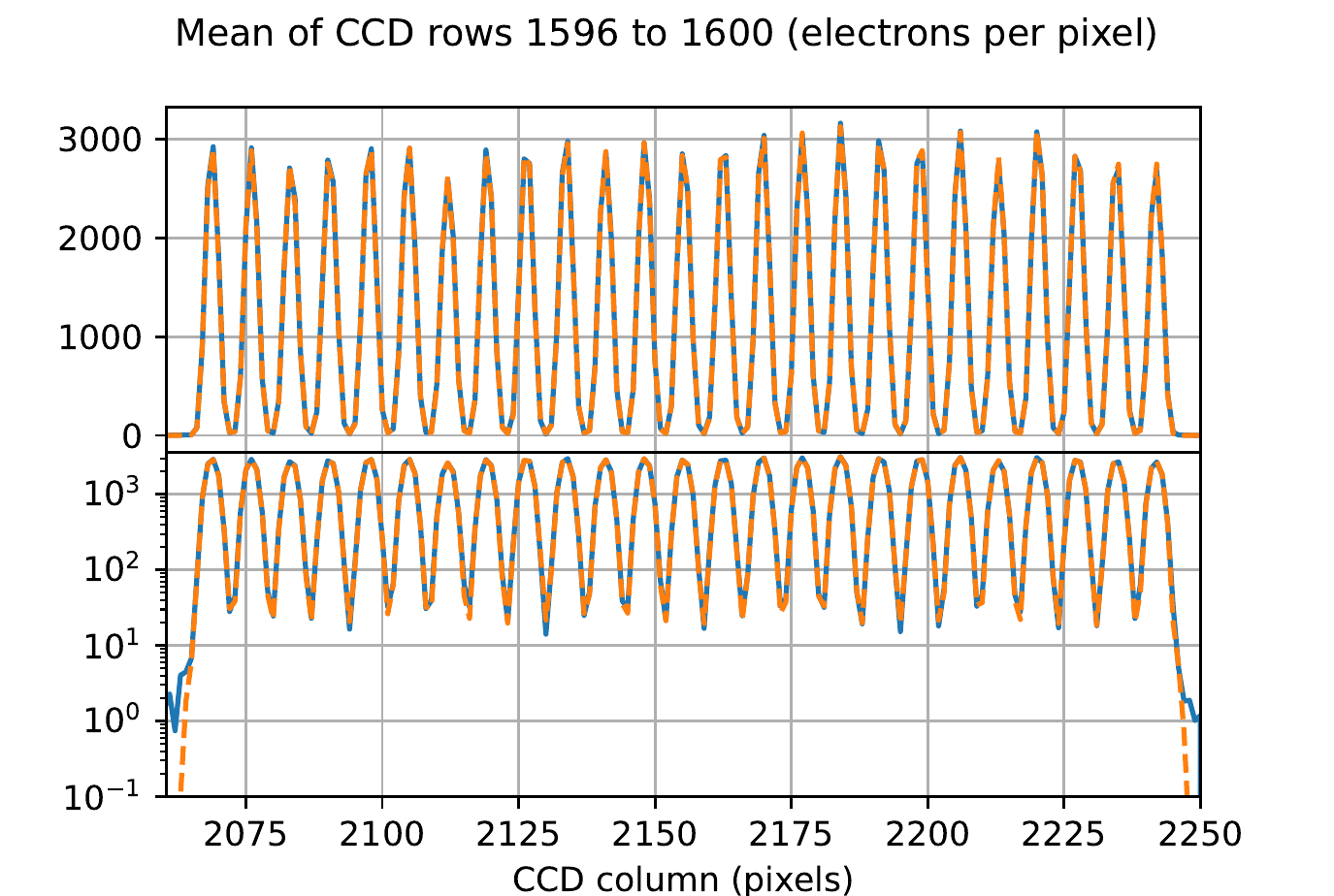}
  \caption{Cross-dispersion profile from arc-lamp exposures in the blue, red and NIR camera CCDs, from top to bottom. The y-axis gives the average number of electron per pixel, for each column of a narrow band centered on the brightest emission line in the central region of the CCD. The blue curve is the data and the orange dashed curve the best fit model from the PSF fit. Both linear and log-scale graphs are shown to highlight the agreement of the PSF model with data in the core and tails of the PSF.}
  \label{fig:psf2}
\end{figure}

The agreement between the best fit PSF model and the spectroscopic data is presented in Figures~\ref{fig:psf1} and~\ref{fig:psf2}.
In the first figure are shown the data and model spectra from a typical arc lamp calibration exposures with Mercury, Argon, Cadium, Neon, Krypton and Xenon lines. The quantity on the $y$-axis is the average pixel counts per row of a band covering the spectral traces of a central fiber block (fiber block with fibers 250 to 274). Both the average signal from the pre-processed image and the model image are computed and displayed. The spectra covering the whole CCD range are shown, along with zooms on some emission lines.
There is an excellent agreement between the model and the data.
The second figure (Fig.~\ref{fig:psf2}) represents cross-dispersion profiles from the CCD images of the same exposure and fiber block. The signal being displayed is now the average per column of an horizontal band centered on the brightest central emission line for each camera. Again a good agreement between the data and model is found. In particular, the logarithmic scaled figures show that the PSF tails are well described with the Gauss-Hermite model. 
As mentioned above, PSF Lorentzian tails are not part of the model used for extraction.

\subsubsection{Stability}

The stability of the spectroscopic PSF is critical for the quality of the spectroscopic data reduction. Because the PSF model is fit on the arc lamp calibration images in the afternoon preceeding the night of observation (see \S\ref{sec:calib-obs}), the spectroscopic PSF has to be stable for a duration of about 12 hours, and should not vary too much with the fiber flexure induced by the positioner moves and the change of telescope pointing.

A quantitative criterion for the PSF stability has to be defined as there are several scalar numbers one could use: full width half maximum, second moments, flux measurement biases (for a continuum spectrum or emission lines), line fit bias. The most stringent scientific driver is the sky subtraction accuracy because it affects the false identification rate of emission line galaxies, which is a level 2 requirement of the project. This sky subtraction bias is affected by the relative error on the PSF shape from fiber to fiber, and not by a global systematic PSF error. Indeed any average PSF error which is shared among all fibers is compensated by a change of the average sky spectral shape and intensity during the sky subtraction procedure (see \S\ref{sec:skysubtraction}).

In the following we will present the stability based on the most stringent criterion which is the emission line flux bias induced by a change of PSF shape. The effect on this multiplicative bias is maximum when the CCD pixel noise is dominated by readout noise and not Poisson noise. In this limit, assuming a perfect centering of the PSF, the same readout noise for all pixels, and identical pixel quantum efficiency and pixel size, this bias is
\begin{equation}
  \delta F / F  = \frac{ \sum_{i} P_i P^{truth}_i}{ \sum_j P_j^2 } - 1 \label{eq:flux-bias}
\end{equation}
where the sum on the numerator and denominator run on pixels, $P_i$ is the PSF model integrated in a pixel and $P^{truth}_i$ is the same quantity for the true PSF profile. In this formula it is assumed that both PSF profiles (model and truth) are normalized by their integral. The derivation of this formula is given in Appendix~\ref{sec:emission-line-bias}.

We will also report the more standard variation of second moments.

In the following we discuss the PSF stability with time, telescope move and positioner moves.

\begin{figure}
  \centering
\includegraphics[width=0.99\columnwidth,angle=0]{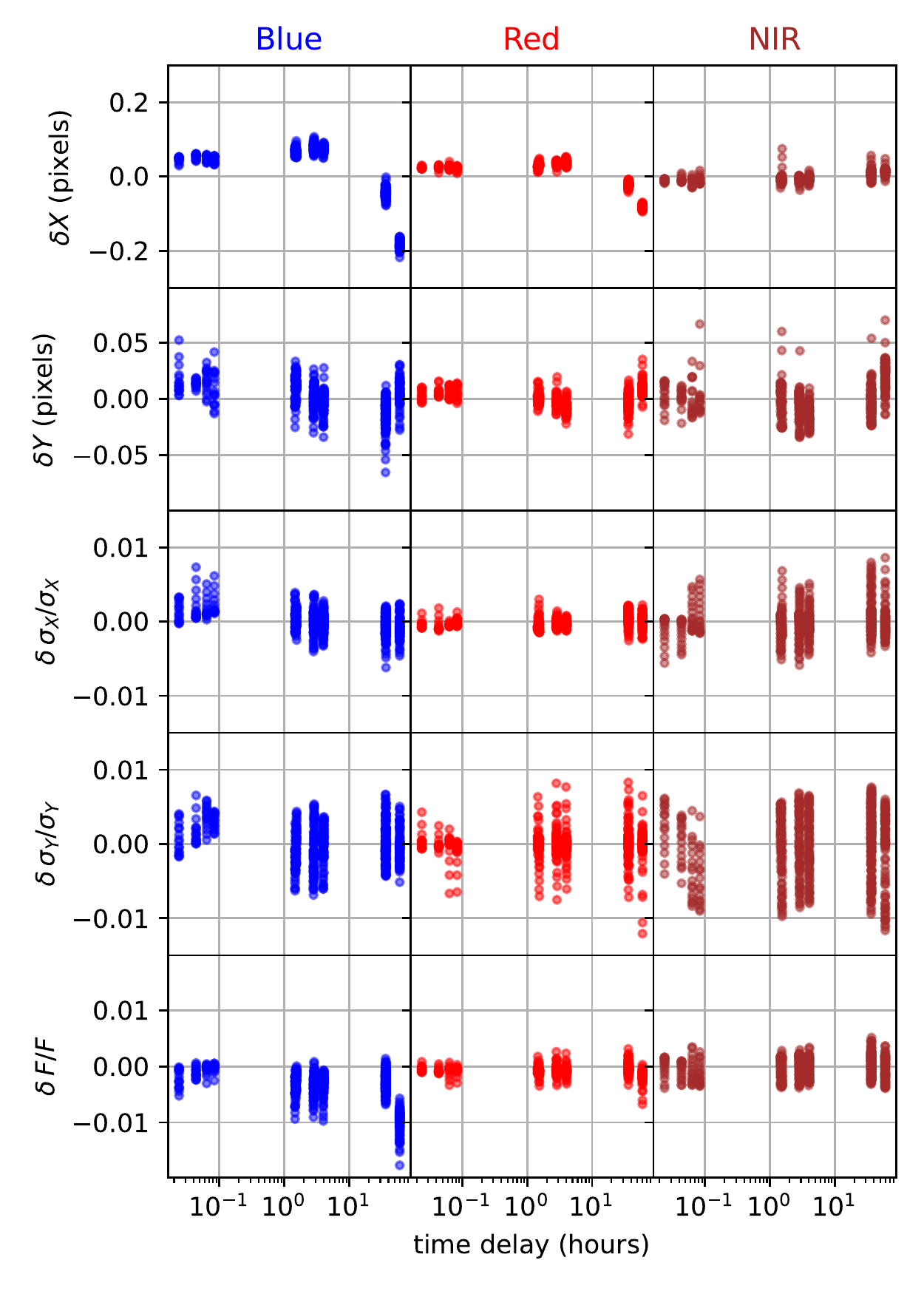}

\caption{PSF stability measurement. Each dot on the figures is the variation of a property of the PSF from exposure to exposure (i.e. current value minus average over exposures) for a given exposure, fiber and wavelength. The properties are, from top to bottom, the variation of the PSF center along $X$ (fiber direction) or $Y$ (wavelength direction), the relative change of PSF width along $X$ and $Y$, and finally the variation of emission line flux bias (Eq~\ref{eq:flux-bias}). The fibers and wavelength were chosen to cover evenly the active region of the CCD, excluding regions at short and long wavelength for each camera where the statistical noise is too large. Fibers 10, 100, 200, 300, 400, 490 are shown, and the wavelength are respectively in the range 3860-5650\,\AA, 5900-7500\,\AA, and 7700-9500\,\AA, for the blue, red and NIR cameras. The PSF model of the spectrograph SM10 are shown. The calibration exposures are from the nights of March 13, 14 and 15, 2020. Note that the fiber positioners were moved during the nights, so the variations from night to night are a test of the stability with positioner moves. Also, the dome was moved between the calibration observations of March 13, so the stability of the PSF during this sequence is a test of the stability with the change of telescope pointing (because the calibration screen the telescope is pointing to is attached to the dome).}
\label{fig:psf-stability}
\end{figure}

{\bf Stability with time}:
Figure~\ref{fig:psf-stability} shows the PSF stability over several days for the three cameras of a spectrograph. Each dot on the figure is the PSF model for one fiber and one specific wavelength. The fiber numbers and wavelength were chosen to sample evenly the CCDs.
The primary conclusion is that the PSF shape is remarkably stable. Looking at the most sensitive quantity which is the flux bias induced by the change of PSF ($\delta F/F$), we find a maximum variation of about 0.6\% within several hours. We note that this is an upper limit because part of this number is due to the statistical noise of the PSF parameters (which is reduced when averaging the result of several calibration exposures for each night). The variation from night to night is also small. It however exceeds the requirement of 1\%  in the blue camera for this particular set of observations.
One can note a significant drift of the PSF center along the cross-dispersion axis ($X$ on the figure), in other words from fiber to fiber, which physically corresponds to a vertical axis as the pseudo slit is standing vertically on the spectrograph optical bench.
We explain our approach to correct for this effect in $\S$\ref{sec:wave-calib-trace-coords}.

During these three days the temperatures recorded on the spectrograph camera body have varied by less than 0.1 degree C thanks to the temperature control while the humidity level, which is monitored but not controlled, has varied from 40\% to 28\%. The atmospheric pressure was stable with maximum variation of 3\,mbar (the maximum being on March 14).

{\bf Stability with telescope moves}:
The PSF calibration run of March 13, which is the first night of the 3 shown on Fig.~\ref{fig:psf-stability} is composed of 4 sequences with different azimuthal angles of the telescope dome: 253, 359, 107 and 180 degrees. This is the best possible test of the PSF stability with changes of the telescope orientation, because the telescope which needs to point to the white screen was also moved to follow the dome rotation. This is testing the possibility that the flexure of the fiber cables along the equatorial mount and in the cable chains affect the PSF. The figures demonstrate this is not a concern; we do not see a variation of the PSF shape larger than 0.6\%.

{\bf Stability with positioner moves}:
Figure~\ref{fig:psf-stability} also shows to some extent the PSF stability with positioner moves, because the fibers were positioned to targets during the nights following each afternoon calibration sequence. We see that the PSF is mostly unaffected by positioner moves.

This has been confirmed with another data set where the positioners from petal 0 were moved between series of arc lamps exposures with the telescope pointing to the dome screen.
The spectra have been obtained with a boxcar extraction and then combined per fiber over the exposures of each series (with a median to reduce the effect of cosmic rays). We then measured the intensity of each emission line in each fiber and computed their variation from one series to the next. In order to reduce further the effect of residual cosmic rays, we finally considered the median of the ratio of emission line flux for each fiber.
Among the fibers from positioners that were moved, the maximum variation in flux of the emission lines was found to be of $0.015$, and the mean standard deviation of $0.0035$. This is marginally larger than the dispersion obtained from non-moving positioners. Considering the quadradic difference for the moving and non-moving positioners, the excess scatter caused by the moves was found to be only of $0.002$ (0.2\%). We note that at the time of this tests, positioners were moved in a restricted range to avoid collisions. More recent data based on the relative variation of sky line intensities show a larger variation of 0.009. This latter dispersion however includes other effects (anisotropy of the sky background, wavelength calibration or flat field systematic errors).

\subsection{Wavelength calibration and trace coordinates}
\label{sec:wave-calib-trace-coords}

Trace coordinates and wavelength calibration are determined at three different phases of the spectroscopic data analysis.
The first one occurs when a new spectrograph is calibrated with no prior measurement of its optical properties. This procedure has to be robust but does not need to be very precise, it is described in~\S\ref{sec:bootcalib}. The second one is on the contrary a precise fit of the trace and wavelength calibration which requires a first solution. It is performed at the same time as the PSF model fit and is based on a forward model of the emission line spots in the CCD images. The model and fit procedure are described in detail in section~\ref{sec:psf} and we do not come back to it in this section. The third and final phase consists in tuning the trace coordinates and wavelength calibration for each scientific exposure to compensate for fine changes in the optics. We present our approach to this important adjustment in~\S\ref{sec:adjustcalib}.

In all cases, the output are the coefficients of Legendre polynomials as a function of wavelength, per fiber, providing the $X$ and $Y$ coordinates of the PSF center.
As described in \S\ref{sec:ccdimage}, $X$ is indexing CCD columns, and increases with increasing fiber number, and $Y$ is indexing CCD rows, and increases with increasing wavelength.

\subsubsection{Initial wavelength calibration and trace coordinates fit}
\label{sec:bootcalib}

This initialisation is performed for a new hardware setup. It has been used for three different configurations of the spectrographs: first during the spectrograph tests at the vendor in France, second for the functional verification tests at Kitt Peak with a test slit and a dedicated illumination setup in the spectrographs enclosure, and finally when the science slits were inserted in the spectrographs, with calibration exposures obtained with the telescope pointing to the dome screen.

Trace coordinates are first fit on preprocessed images obtained with continuum lamp exposures.
The continuum light source is either a Tungsten lamp or an array of LEDs. A Tungsten lamp was used for tests at the vendor and for functional verification, whereas LED lamps shining on the dome screen were preferred for the initial calibration of the spectrographs with the science slits.

First a median cross-dispersion profile is extracted from a central band of 100 pixel rows in the CCD. Isolated peaks are found and the $X$ coordinates of their maximum is recorded. This defines the fiber array. Then, starting from the coordinates at the center, the barycenter of the cross-dispersion profile of each CCD row and fiber is determined, walking up and then down the CCD rows, starting from the center and using as initial guess the fit from the previous row.
Those coordinates are fit with a Legendre polynomial for each fiber. A Gaussian cross-dispersion profile is also fit per fiber and CCD row. This fast fit consists in an iterative weighted second moment computation where the weight at each iteration is a Gaussian profile whose $\sigma$ is given by the previous iteration (with a $\sqrt{2}$ scaling factor due to the Gaussian weight).

The wavelength calibration makes use of preprocessed images obtained with arc lamps exposures taken shortly after the continuum exposures. Not all lamps are used simultaneously to avoid confusion between neighboring lines. For the blue cameras,  Mercury-Argon and Cadmium lamps are used together. For the red and near infrared (hereafter NIR) cameras, only the Neon lamp is used (see Table~\ref{table:arc-lamp-lines-list} in Appendix~\ref{sec:arc-lamp-lines-list} for the list of emission lines used for the wavelength calibration).

The spectral extraction is a profile fit based on the trace $X$ coordinate and Gaussian $\sigma$ derived with the continuum lamp data. The resulting spectra are then analysed to find the most significant peaks.

We then use the following algorithm (inspired by~\citealt{Valdes1995}) to match the detected peaks with the known emission lines from the lamps, which is a challenge because of a poor initial knowledge of the wavelength calibration and the relative strength of the emission lines.
For each of the two arrays to match (the known emission line wavelengths and the list of $Y$ coordinates from the detected peaks), we record all possible combinations of ordered triplets. We then compute all possible pairs of triplets from one array to the other and derive for each pair the parameters of a second order polynomial transformation from the wavelength to the $Y$ coordinates. A four-dimensional histogram is then filled with, for each triplet pair, the first and second degree terms of the transformation, the indices the longest wavelength in one triplet and the largest $Y$ coordinate in the other. The bin with the highest number of counts gives us a pair of matching lines and a good first guess of the transformation between wavelength and CCD coordinates. Then, we proceed to find additional matching pairs of lines and detected peaks and subsequently refine the transformation. This procedure is performed independently for each fiber. A last step consists in comparing the solutions among fibers to identify the set of good matches, enforcing this match among all fibers and refitting the transformation.

This procedure works well for the combination of lamps given above and for a specific list of lines per lamp. One also has to predefine the bins of the 4D histogram which requires some prior on the first and second degree of the polynomial converting wavelength to CCD coordinates.

\subsubsection{Adjustment of wavelength calibration and trace coordinates per exposure}
\label{sec:adjustcalib}

\begin{figure}
  \centering
  \includegraphics[width=\columnwidth,angle=0]{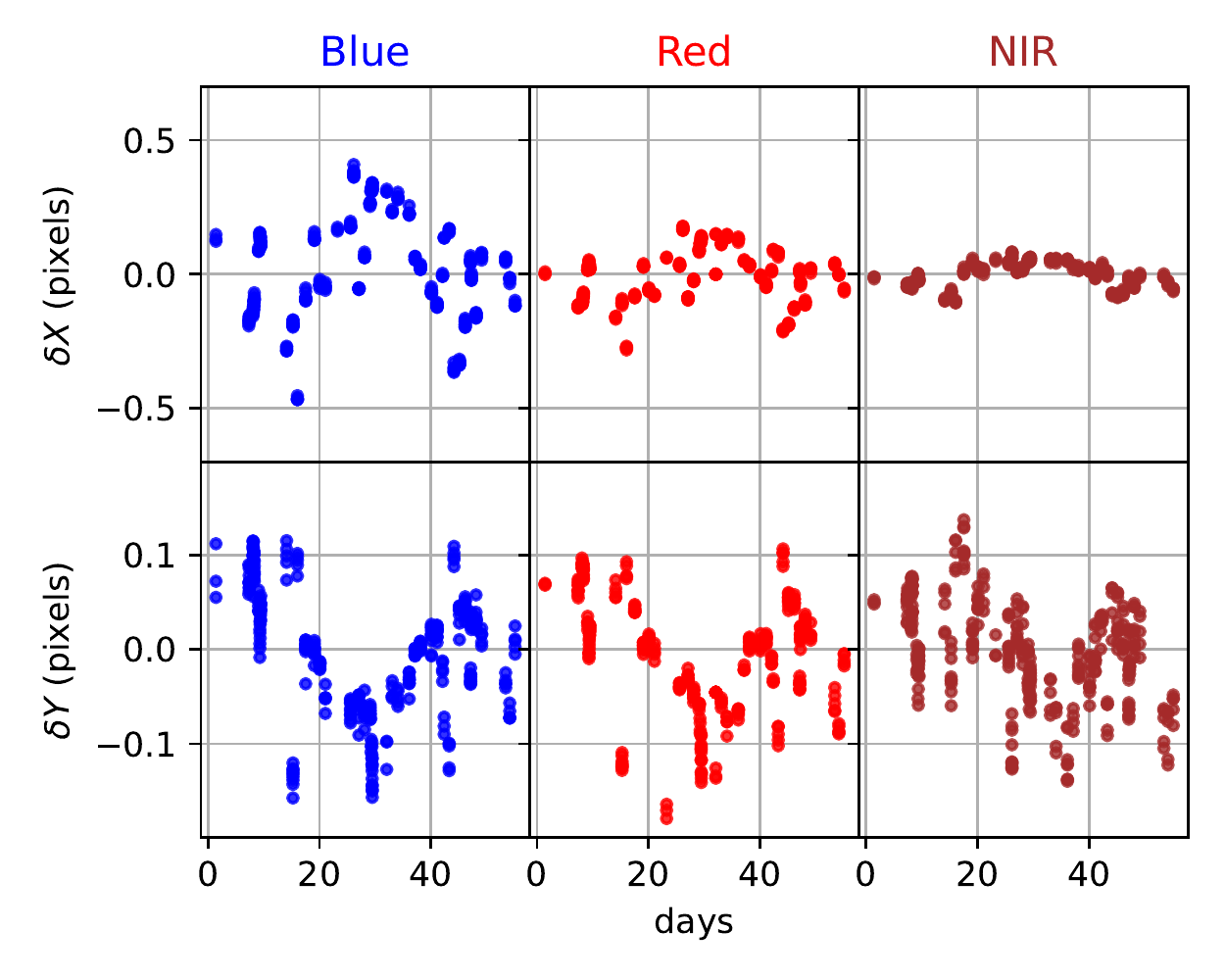}
  \caption{Variations of the spectral traces coordinates with time for the three cameras of spectrograph SM2. Variation along $X$ (direction of fiber number, left panels) for the blue cameras are the largest with peak to peak variations of $ \pm 0.7$ pixel and an r.m.s of 0.3 pixel. The variations are smaller in the red and NIR cameras (0.15 and 0.03 pixels rms for $X$, respectively). The rms variation along the $Y$ axis (direction of wavelength dispersion, right panels)  are about 0.05 pixel rms for all cameras. This would represent a wavelength calibration error of about 0.03\,\AA\ if not corrected.}
  \label{fig:dxdy}
\end{figure}

Despite a fine control of the temperature in the spectrographs enclosure, we have noticed drifts of trace coordinates with time as shown in Figure~\ref{fig:dxdy} for the spectrograph SM2. The other spectrographs have similar variations.

These trace-coordinate variations have been correlated with environmental parameters, and a large correlation was found between the $X$ coordinates for the blue and red cameras with the humidity in the spectrographs' room. It is suspected that this correlation is due to a humidity dependent tilt of the NIR dichroic, possibly due to water absorption by the sealant that bonds the dichroics in their cells. This would explain why no correlation is found with the NIR camera data (as the dichroic is transmitting NIR light) while a correlation is found for the two other cameras (reflection on the dichroic). This would also explain the relative amplitude between the blue and red cameras.

While further investigation may allow the development of a predictive model for the variations of trace coordinates, we have resorted to adjust the trace coordinates and wavelength calibration on the data itself.

For each exposure, prior to the 2D extraction, we estimate the $\delta X$ offset of each fiber trace, for values of $Y$ coordinates, using cross-dispersion profiles averaged over several $Y$ rows. The $\delta X$ values across the whole CCD are then fit with a low order 2D polynomial of $X$ and $Y$, while making sure to have an efficient outlier rejection (because of cosmic rays, broken fibers, spectral regions with low signal).

For the wavelength calibration adjustment ($\delta Y$), we rely on a cross-correlation of spectra. We first extract the spectra of all fibers with a fast algorithm (simple row-by-row sum of pixels around each trace) and then cross-correlate the spectra of all fibers with the average fiber spectrum to detect relative shifts of wavelength calibration. This simple approach works because of the strong sky lines that are present in all spectra. We also perform this cross-correlation for several wavelength intervals which makes it easier to discard erroneous measurements due to bright target spectra or cosmic rays. As for the other axis, a low order polynomial is fit across the CCD. Finally, we rely on an external sky spectrum to get the final wavelength adjustment for science exposures.
This external sky spectrum is derived from a set of high-resolution spectra from~\citet{Hanuschik2003}, convolved to the resolution of the DESI spectrographs. The spectra were obtained at the Paranal Observatory with UVES, ESO's echelle spectrograph at the 8-m UT2 telescope of the Very Large Telescope (VLT).

The performance of this method can be evaluated by studying the spectral residuals after the sky subtraction (see~\S\ref{sec:skysubtraction}), on sky lines to test the wavelength calibration, and in the continuum to test the stability of the flux calibration from fiber to fiber which is a function of the accuracy of the trace coordinates $\delta X$.

\subsubsection{Solar system barycentric velocity correction}

The wavelength are corrected for the shift due the relative velocity of the Earth with respect to the solar system barycenter.
We apply the same correction to all fibers of an exposure, using the velocity shift calculated for the center of the field of view. Our approach is to apply the opposite correction to the wavelength array used for the spectral extraction (see \S\ref{sec:extraction}), and then simply correct the reported wavelength. This avoids any resampling of the extracted spectra and in consequence does not introduce any extra correlation between the flux values.

\subsubsection{Radial velocity measurements }

\begin{figure}
  \centering
  \includegraphics[width=\columnwidth,angle=0]{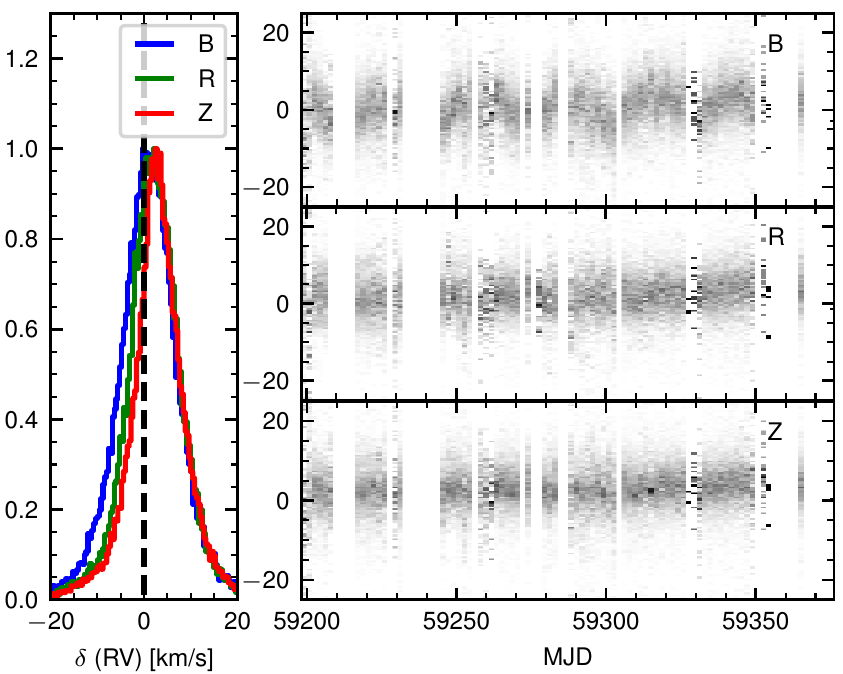}
  \caption{Radial velocity offsets of stars with respect to the measurements in the SoS catalog \citep{Tsantaki2022} computed separately for each arm of the instrument. The left panel shows the histogram of radial velocity offsets in DESI Survey Validation data. The B (blue), R (red) and Z (for NIR) median offsets are of 0.97, 2.13, and 2.88\,km\,s$^{-1}$ respectively, and the median absolute deviations are of 4.2, 3.7 and 3.5\,km\,s$^{-1}$. The right panel shows the radial velocity offsets on the y-axis, in km\,s$^{-1}$, as a function of the MJD of observation.}
  \label{fig:rv_offset}
\end{figure}

As the Milky Way Survey (hereafter MWS) of DESI will be observing millions of stars in the Galaxy, we are interested in extracting stellar radial velocities that are accurate down to the photon-noise limits and are not biased with respect to other surveys. The detailed comparison of the radial velocities with external surveys is presented in the MWS overview paper~\citep{cooper22a}, while here we focus on issues related to the wavelength calibration and spectral reduction performance.

The radial velocities of stars in the MWS are determined by chi-square fitting of the spectra by interpolated templates \citep{Koposov2011} from the PHOENIX stellar library \citep{Husser2013} using the rvspecfit\footnote{\url{https://github.com/segasai/rvspecfit}} code \citep{Koposov2019rvspecfit}. The radial velocity uncertainties are determined from the posterior on radial velocity conditional on best-fit stellar atmospheric parameters values. The accuracy and precision of radial velocities derived from DESI spectra was assessed through several tests described below.

We looked at repeated observations of stars observed multiple times as these could be used to determine whether the observed scatter between radial velocities match the formal radial velocity uncertainties computed based on the pixel noise. The analysis of large number of repeats done during the science verification shows that the radial velocity scatter between different exposures can be approximated as $\sigma_{RV} \approx \sqrt{ \sigma_{RV,0}^2 + (0.8\,\rm{km}\,\rm{s}^{-1}) ^2}$ where the $\sigma_{RV,0}$ is the formal radial velocity error determined by rvspecfit. This implies that there are radial velocity systematics that are different from exposure to exposure and are likely associated with the inaccuracies of the wavelength calibration. We note however that for most but the bluest stars most of the radial velocity information comes from the NIR arm.

To study the radial velocity systematics in further detail we look at the radial velocity measurements of all the stars observed during the DESI Survey Validation (hereafter SV, see~\citealt{sv22a}) that are part of the survey of surveys (SoS) catalog \citep{Tsantaki2022}. This catalog is based on stellar radial velocity measurements from multiple surveys such as APOGEE, SDSS, LAMOST, Gaia-ESO and {\it Gaia} that have been brought together to the same radial velocity zero-point and thus can be used as radial velocity reference catalog. To understand the wavelength calibration for different arms of the instrument separately, we separately measured the radial velocity from the blue, red and NIR arm spectra of each star from the SoS observed by DESI in SV. We then analyzed the radial velocity offsets with respect to measurements in the SoS catalog. The distribution of the offsets for stars with DESI RV errors smaller than 3\,km\,s$^{-1}$ is shown on Figure~\ref{fig:rv_offset}. It shows that the offset distribution is not the same for different arms, with the mean offset of $\sim 3$\,km\,s$^{-1}$ for the NIR arm and smaller offset for the blue arms. The blue arms show also larger spread of offsets than the red arm. This can be explained by some time dependence of those offsets, shown in the right-hand panels of Figure~\ref{fig:rv_offset}. These residuals seem to be also correlated with the wavelength corrections described in \S\ref{sec:adjustcalib} and the sky brightness. These correlations suggest that the likely cause of the radial velocity offsets in the blue arm is the wavelength adjustment based on sky lines. The offsets in the red and the NIR arm do not show much correlation with time of observation and observational conditions thus their cause remains to be determined.

\subsection{Spectral extraction}
\label{sec:extraction}

\subsubsection{2D PSF spectral extraction}
\label{sec:extraction-2d}

Extracting 1D spectra from the 2D CCD images follows the ``spectroperfectionism''
methodology described in \cite{BoltonSchlegel2010}.
CCD pixels $p$ are modeled as a linear combination of the input
spectral flux $F$ on a discrete wavelength grid such that
\begin{equation}
    p = A F + \text{noise} \label{eq:projection}
\end{equation}
The PSF model is encoded in the matrix $A$, {\it i.e.}~modeling how
the input flux $F$ (interpreted as a series of $\delta$-functions) is distributed across the CCD pixels $p$, including the
fact that an individual pixel can have contributions from multiple wavelengths
and multiple fibers.  In this notation the 2D array of pixels are flattened
into a single 1D array of pixels $p$, and the 2D array of fluxes vs.~fiber
and wavelength are flattened into a single 1D array $F$.
The $A_{ij}$ element is the PSF for fiber wavelength $\lambda_j$ integrated over
CCD pixel $i$.

The optimal estimator of $F$ is obtained by maximizing the likelihood $L$ of $p$ or minimizing
\begin{equation}
  \chi^2 = -2 \ln L + \mathrm{const} = (p - A F)^T W ( p - A F) \label{eq:extractionchi2}
\end{equation}
where $W$ is the inverse of the covariance of the pixels $p$. The pixel
noise is uncorrelated to a good approximation so $W$ is diagonal and its elements are the inverse of the pixel variance (see \S\ref{sec:ccd-variance} for how the CCD pixel-level variance is estimated).

Minimizing Eq.~\ref{eq:extractionchi2} with respect to $F$ consists of solving the following linear system
\begin{equation}
    (A^T W A) F = A^T W p
    \label{eq:extract-f}
\end{equation}

The quantity $(A^T W A) = C^{-1}$ is the inverse covariance matrix of $F$,
which in general has off-diagonal covariance, {\it i.e.} the elements of
$F$ are correlated.

Following \cite{BoltonSchlegel2010}, with a eigen decomposition
$C^{-1} = P^T D^{-1} P$, one can define the square root matrix $Q = P^T D^{-1/2} P$, the diagonal matrix $S$ with $S_{i,i}= 1/\sum_j Q_{ij}$, and construct a {\it resolution matrix} $R$,

\begin{equation}
\label{eq:resolution-matrix}
R = S Q = S P^T D^{-1/2} P
\end{equation}

This matrix can be used to recombine the correlated flux values into an array of uncorrelated noise,
\begin{equation}
\tilde F \equiv R F \label{eq:ftilde}
\end{equation}

Indeed the covariance $\tilde C$ of $\tilde F$ is diagonal:
\begin{eqnarray}
\tilde C &\equiv& \left< (\delta \tilde F) (\delta \tilde F)^T \right> =  R C R^T \nonumber \\
&=& S P^T D^{-1/2} P \left(  P^T D P \right) P^T D^{-1/2}  P S \nonumber \\
&=& S^{2} \label{eq:cov-c-tilde}
\end{eqnarray}

The final outputs are:
the resolution convolved flux $\tilde F$ which has by construction uncorrelated noise between wavelengths;
the variances $\sigma_\lambda^2$ which are the diagonals of $\tilde C$ (off-diagonal elements are 0);
and the resolution matrix $R$ that was used to decorrelate the data.

These can be combined to compare a high resolution model spectrum $M$ to the resolution
convolved extracted spectra $\tilde F$ via $R M$. In the form of a $\chi^2$, this gives:
\begin{equation}
\label{eq:extract-chi2-1D}
\chi^2_{\mathrm{1D}} = \sum_{\mathrm{wavelengths}~\lambda} \left( {\tilde{F} - (R M)_\lambda \over \sigma_\lambda}\right)^2
\end{equation}

The fluxes of Eq.~\ref{eq:ftilde} obtained with the resolution matrix of Eq.~\ref{eq:resolution-matrix}
are not only decorrelated across wavelength but also across fibers. What seems a good thing at first glance has the undesired effect of mixing the true underlying flux of neighboring fibers, and introducing fiber cross-talk. In order to avoid this, we have decided to compute the resolution matrix per fiber individually, such that the noise is decorrelated across wavelength but not fibers. This is obtained by considering independently the blocks of the covariance matrix $C$ that address each fiber spectrum when computing its corresponding resolution matrix.

\subsubsection{Resolution matrix is not flux conserving}

It is important to note that $R$ is not a flux-conserving convolution and its normalization vs.~wavelength depends upon the input noise model,
i.e.~it is not simply a model of the instrument resolution.
Appendix~\ref{sec:spectroperfconvolution} discusses the interpretation of the $R$ tranformation from $F$ to $\tilde F$ as a convolution, which is only
valid in the limit of constant noise vs.~wavelength, which in general is
not the case.
When the input noise varies with wavelength, for instance near a sky emission line, then the normalization of $R$ can vary by up to 20\% for DESI spectra.

\begin{figure}
\includegraphics[width=0.99\columnwidth,angle=0]{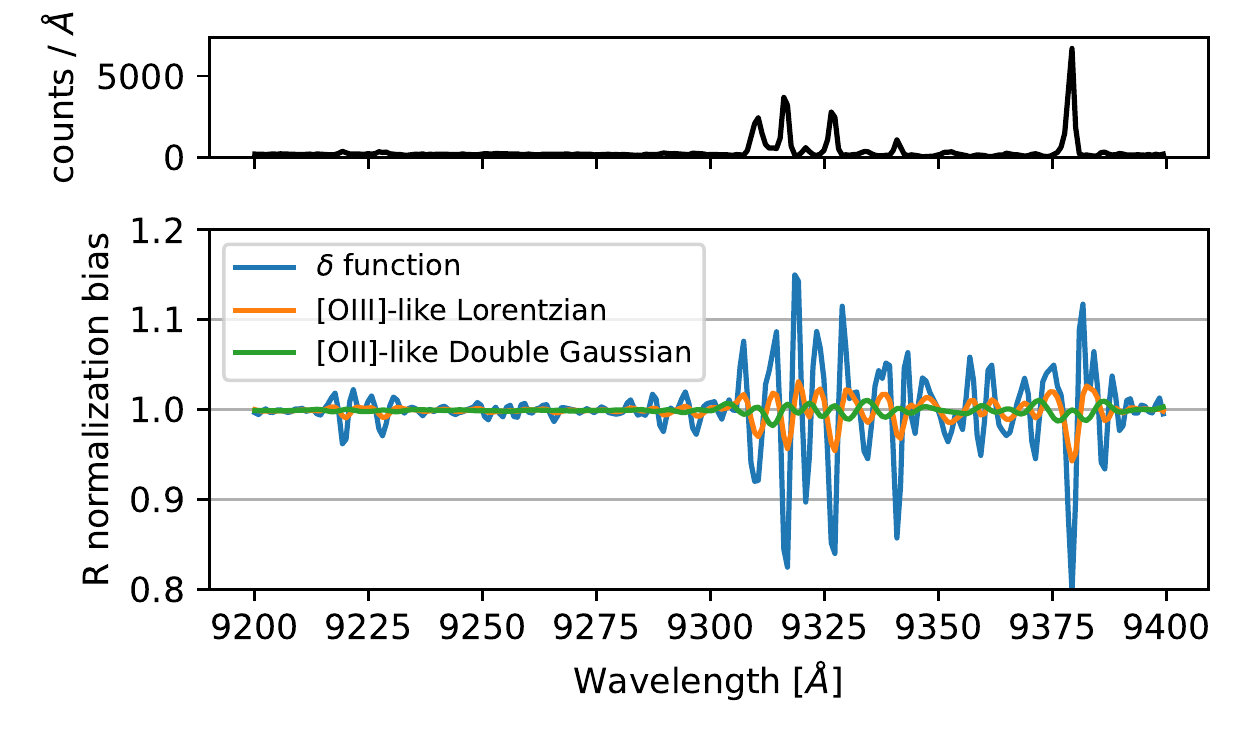}
  \caption{Measurement bias if fitting a model directly to data without using the resolution matrix.  The upper plot shows a 200\,\AA\ region of a raw extracted sky spectrum.  The bottom plot blue line shows the resolution matrix $R$ normalization vs.~wavelength.  Orange and green lines show the normalization convolved with an input [OIII]-like Lorentzian or a [OII]-like double Gaussian.}
  \label{fig:Rnorm-bias}
\end{figure}

As a consequence, any model fits to data
should use a metric such as Equation~\ref{eq:extract-chi2-1D}, where the
model is multiplied by the resolution matrix before comparing to data.
Directly fitting a line amplitude to data can lead to wavelength dependent systematic biases as shown in Figure~\ref{fig:Rnorm-bias}.
The top plot shows a 200\,\AA\ region of a raw extracted sky spectrum.  The bottom plot blue line shows the resolution matrix $R$ normalization vs.~wavelength, exhibiting strong ringing where the input noise is varying
due to the Poisson noise of the input sky.
For instance if an input signal $F$ was a true $\delta$-function, the output
extracted flux $\tilde F = R F$ could vary by up to 20\% depending upon
wavelength.  Broader lines are less affected since they sample more of the
$R$ normalization oscillations. The Figure~\ref{fig:Rnorm-bias} bottom orange line shows the normalization bias of an [OIII]-like Lorentzian emission
line with full-width half-max (FWHM) of 2.7\,\AA, and the green line shows
the normalization bias of an [OII]-like double Gaussian with $\sigma=$2\,\AA.
These broader lines have less normalization variation than the $\delta$-function, but still can vary by several percent.

We emphasize, however, that this normalization ``bias'' disappears if one multiplies an input model by $R$ before comparing to data, as will be shown in \S\ref{sec:extraction-noise-model}.

\subsubsection{Extraction regularization and masking}

Masking of CCD pixels for cosmic rays, hot columns, or defects can result in wavelength bins with few or no unmasked input pixels, thus requiring regularization to be added to Equation~\ref{eq:extractionchi2} to avoid a singular matrix or extracting ringing due to ill-constrained bins.
At the same time, overly aggressive regularization can lead to biases in the extracted spectra, especially at the high flux limit where bright pixels have a relatively low weight due to the Poisson noise.
For each wavelength bin, we sum the weights of the input pixels contributing to that bin (masked pixels have zero weight), and set a threshold of
$10^{-4}$ of the maximum summed weight of any flux bin.  Bins whose summed
weight is below that threshold receive a regularization term towards 0-flux
with that threshold weight.  This approach was empirically tuned to avoid ringing from ill-constrained bins while minimizing bias on bright bins.

Cosmic rays that are not completely masked can result in poor PSF fits which
bias the extracted flux.  To check for this, a CCD image model $A F$ is calculated and used to measure the $\chi^2$ of pixels contributing to each extracted flux bin.  Bins with $\chi^2>100$ are flagged with a \textsc{bad2dfit} mask.  Additionally, bins with $>50\%$ masked input pixels
get a \textsc{somebadpix} mask, while those with completely masked inputs get
a \textsc{allbadpix} mask bit.

\subsubsection{Subdividing extractions for computational efficiency}

Since the solution for $R$ requires an eigen-decomposition of $C^{-1}$,
it is not practical to solve all wavelengths ($\sim$2500) for all fibers (500)
simultaneously since this would require an ${\cal O}(n^3)$ calculation with $n\sim 1.25 \times 10^6$ per CCD per exposure.  Instead, the problem
is decomposed into multiple overlapping subregions which are extracted
independently and then recombined, thus turning the calculation into a large
number of small matrices to solve instead of a small number of very large matrices.

Each subregion has a core range of wavelengths to extract, which could
contribute photons to a contiguous region of CCD pixels.  It is not
sufficient to simply extract those wavelengths using those pixels, since
those pixels also have photons from wavelengths outside of the core
wavelengths of interest.  The impact on the core wavelengths can be minimized
by including additional wavelengths and additional pixels, solving the extraction for the full set, and then keeping the solution for only the core wavelengths of interest for each subregion.
The core wavelengths extracted for each subregion are unique, but the CCD pixels used and the buffer wavelengths extracted do overlap.

The PSF spot for a given (fiber,wavelength) is modeled over a
$11\times17$ rectangular grid of CCD pixels.
This size is large enough to include the
peak of the PSF from the neighboring fibers, and extend in wavelength enough
to represent $>99.95\%$ of the light.

The amount of padding in pixels and wavelengths was empirically determined
to minimize the extraction bias while maintaining pragmatic computational runtime.
Considering the first fiber in the patch, the pixels are extended in $y$ (wavelength direction) by 8 pixels, and then the wavelength grid is extended to include any wavelengths whose CCD coordinates are within 2 pixels of that extended CCD patch.
Analogous to padding in the wavelength direction, neighboring fibers for each subregion are also extracted and then discarded.

\begin{figure}
  \centering
  \includegraphics[width=0.9\columnwidth,angle=0]{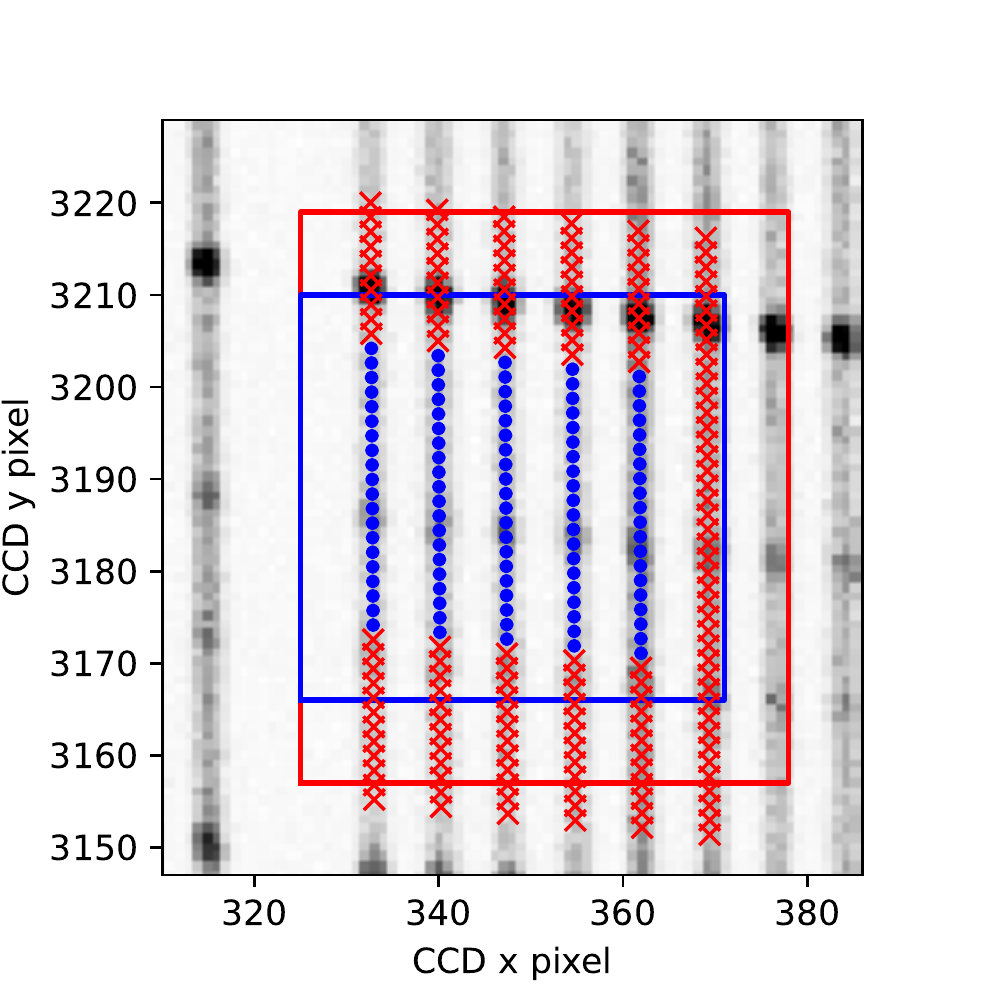}
  \caption{Illustration of a subregion of 2D extractions. See text for details.}
  \label{fig:extraction-ccd-patch}
\end{figure}

This is shown in Figure~\ref{fig:extraction-ccd-patch} for a portion of
CCD r1 fibers 25-29 for wavelengths 7300\,\AA\ to 7315.2\,\AA.
Blue dots show the centroids of the PSFs on a 0.8\,\AA\ grid and the blue box shows the pixels
receiving light from those fibers and wavelengths using a $11\times 17$ spot size.
Red $\times$ marks and the red box indicate the
extra wavelengths and pixels extracted to minimize the bias on the blue region.  In this case, the leftmost fiber (25) is at the edge of a bundle of 25 fibers with a gap before
the left-neighboring fiber, thus it isn't necessary to include a padding fiber on the left, but an extra padding fiber is included on the right.

In practice, the extractions use a core region of 50 wavelengths (compared to 20 in Figure~\ref{fig:extraction-ccd-patch}).

\begin{figure}
  \centering
   \includegraphics[width=0.99\columnwidth,angle=0]{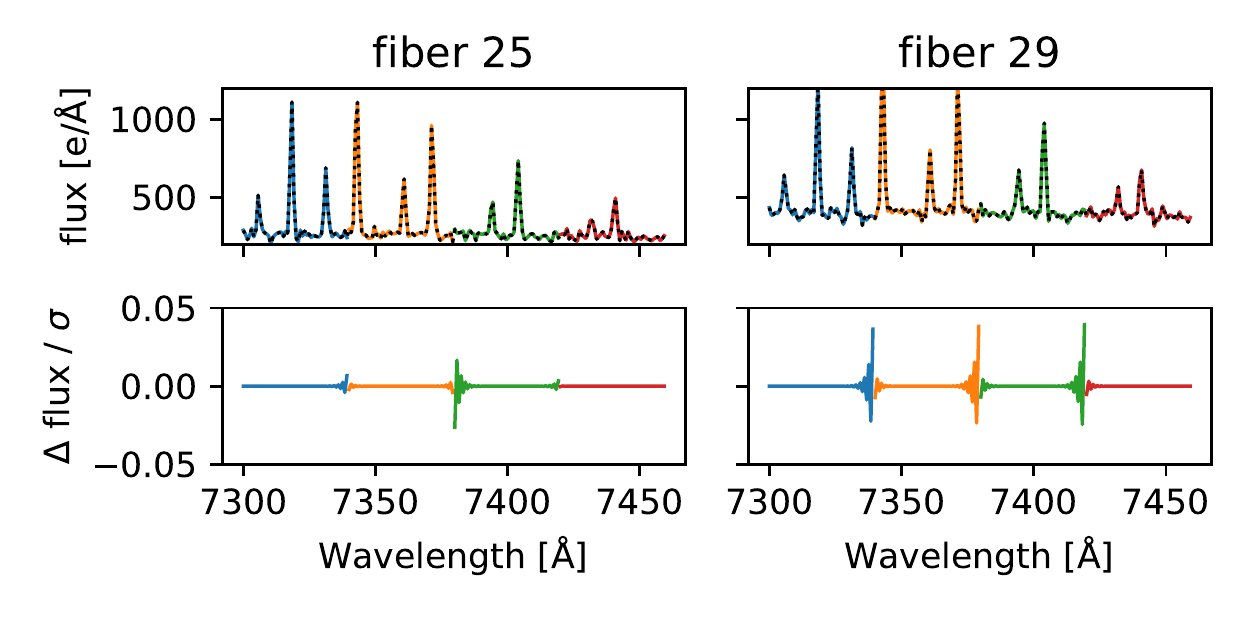}
  \caption{Extraction bias from using overlapping subregions for fibers
  25 and 29 of the r1 CCD.  The top plot shows the subregion extracted flux (solid colored) overlayed with the full extraction (dotted black).
  Bottom show the difference normalized by the statistical error of the extractions.  Although there is systematic ringing at the subregion boundaries,
  it is below 5\% of the statistical error.
  }
  \label{fig:extraction-bias}
\end{figure}

The impact of the biases from extracting overlapping subregions is shown in
Figure~\ref{fig:extraction-bias} for fibers 25 and 29 (the left and right
blue fibers in Figure~\ref{fig:extraction-ccd-patch}).  On the top, the
colored solid lines show the extractions from the subregions, while the overlayed dotted line shows the full solution without using subregions.
The bottom figures show the difference between the two, normalized by the
statistical error on the extractions.  Although there is a residual systematic
ringing due to the use of subregion extractions, it is less than 5\% of the
statistical error.  If needed in the future, this could be further reduced
by extending the padding.

The mapping from wavelength to CCD row varies with fiber number, thus using a rectangular region of pixels to extract results in an asymmetric amount of padding from one fiber to another.
This can be seen in Figure~\ref{fig:extraction-ccd-patch}, comparing the leftmost fiber with symmetric top/bottom pixel padding, vs. the rightmost fiber having more padding at higher (upper) wavelengths than lower.
This results in slightly more bias at the subregion boundaries for the rightmost fiber (see~Figure~\ref{fig:extraction-bias}, right).
In practice this effect is still quite small compared to the statistical errors and we have not optimized the padding to be the same for all fibers, and have
kept the use of rectangular CCD regions to simplify the code.

\subsubsection{Wavelength grid}

One of the key conveniences of the 2D extraction algorithm is that
it allows one to choose a common grid of extraction wavelengths for every
fiber, regardless of how those align with the CCD pixel grid.  By construction
the output wavelength grid is the same for every fiber and the flux bins
have uncorrelated noise, unlike row-by-row extraction methods that either have a
different wavelength grid for every fiber (tied to the CCD rows), or
introduce correlations by resampling to a common wavelength grid.

DESI extracts flux using a linear wavelength grid from 3600\,\AA--9824\,\AA\ in 0.8\,\AA\ wavelength steps. This choice of extraction resolution was chosen to be slightly larger than the native CCD pixel scale ($\sim$0.6\,\AA/row) to avoid numerical artifacts.

$b$~cameras extract 3600--5800\,\AA, $r$-cameras use 5760--7620\,\AA, and
$z$-cameras use 7520--9824\,\AA.  Note that the extraction wavelengths for the
individual cameras overlap, but are phased to be on a common 0.8\,\AA\ grid across
all 3 cameras.

\begin{figure}
  \centering
  \includegraphics[width=0.99\columnwidth,angle=0]{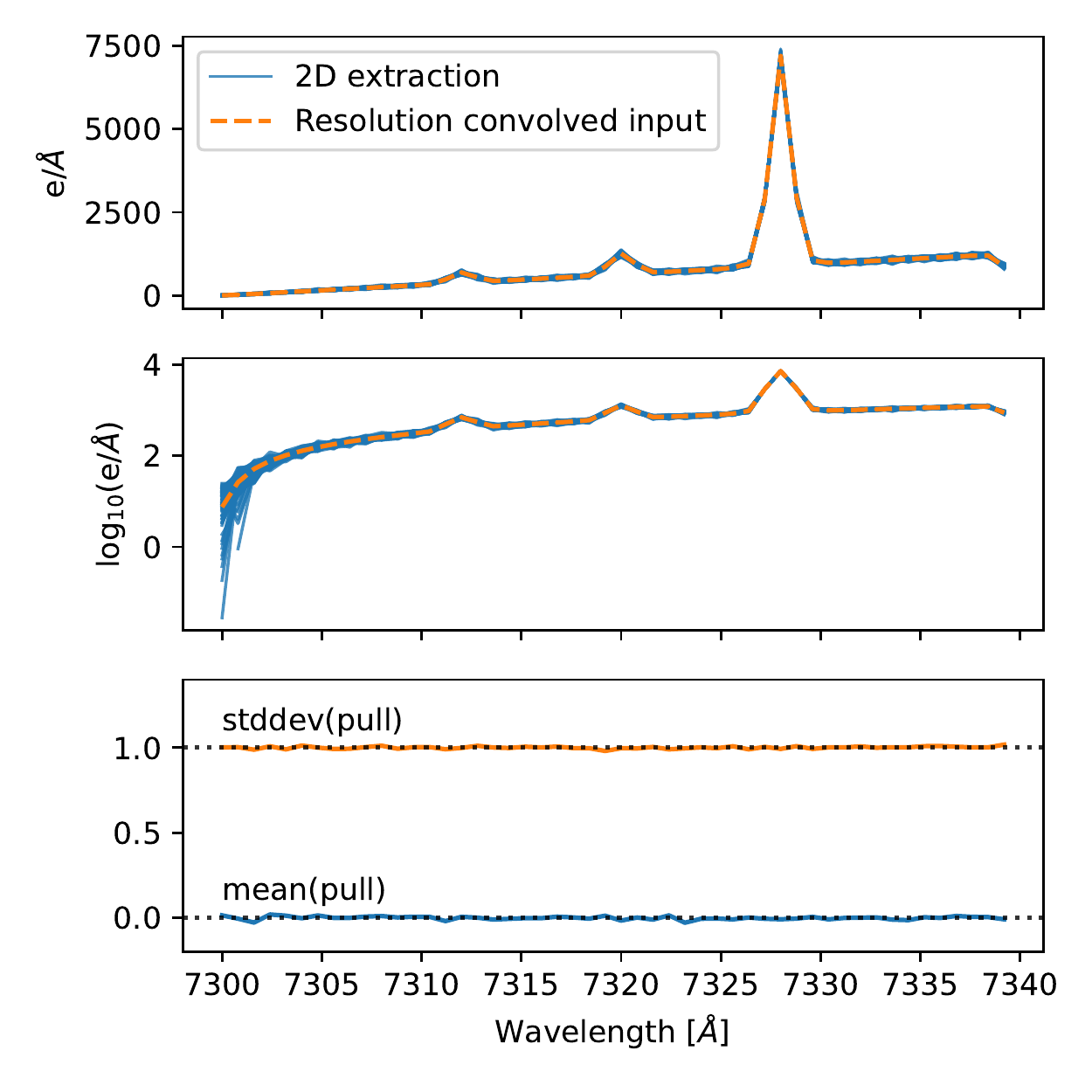}
  \includegraphics[width=0.8\columnwidth,angle=0]{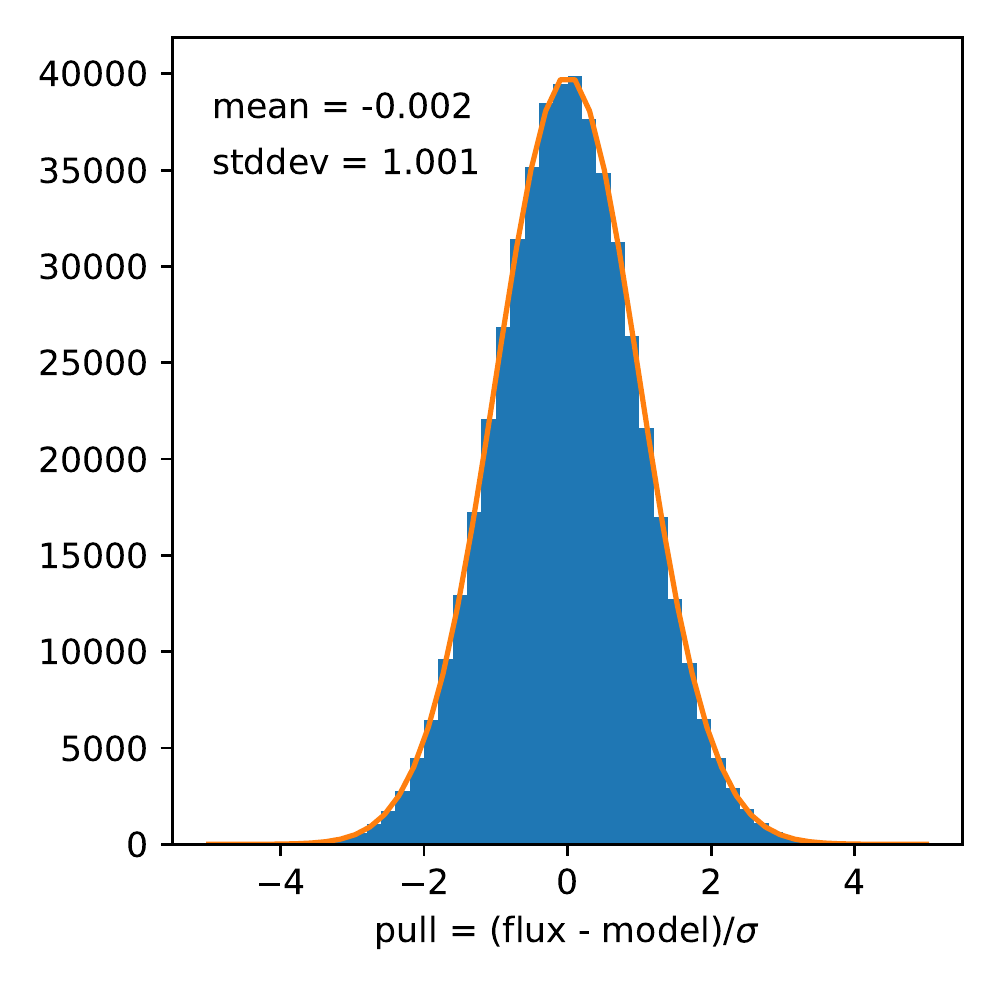}
  \caption{2D flux extractions compared to resolution convolved input for
  10000 noise realizations of simulated CCD images.}
  \label{fig:extraction-pull}
\end{figure}

\subsubsection{Validity of extraction noise model}
\label{sec:extraction-noise-model}

To study extraction bias and the validity of the variance model, we simulated
10000 CCD images of a spectrum with a smooth sloping continuum ranging from
readnoise limited to photon shot noise limited, plus 3 emission lines of varying strengths.
Each realization has the same truth spectrum but a different noise realization
of the CCD pixels (Gaussian readnoise of 3 electrons/pixel plus Poisson signal variations), and then is processed with the 2D extraction code.
The results are shown in Figure~\ref{fig:extraction-pull}.  The extracted flux compared with the resolution-convolved input spectrum is very consistent with the reported noise model, with a pull distribution (the distribution of the variable (data-model)/$\sigma$, sometimes called $\chi$) very consistent with a mean of 0 and standard deviation of 1 at all wavelengths.

\begin{figure}
  \centering
 \includegraphics[width=0.99\columnwidth,angle=0]{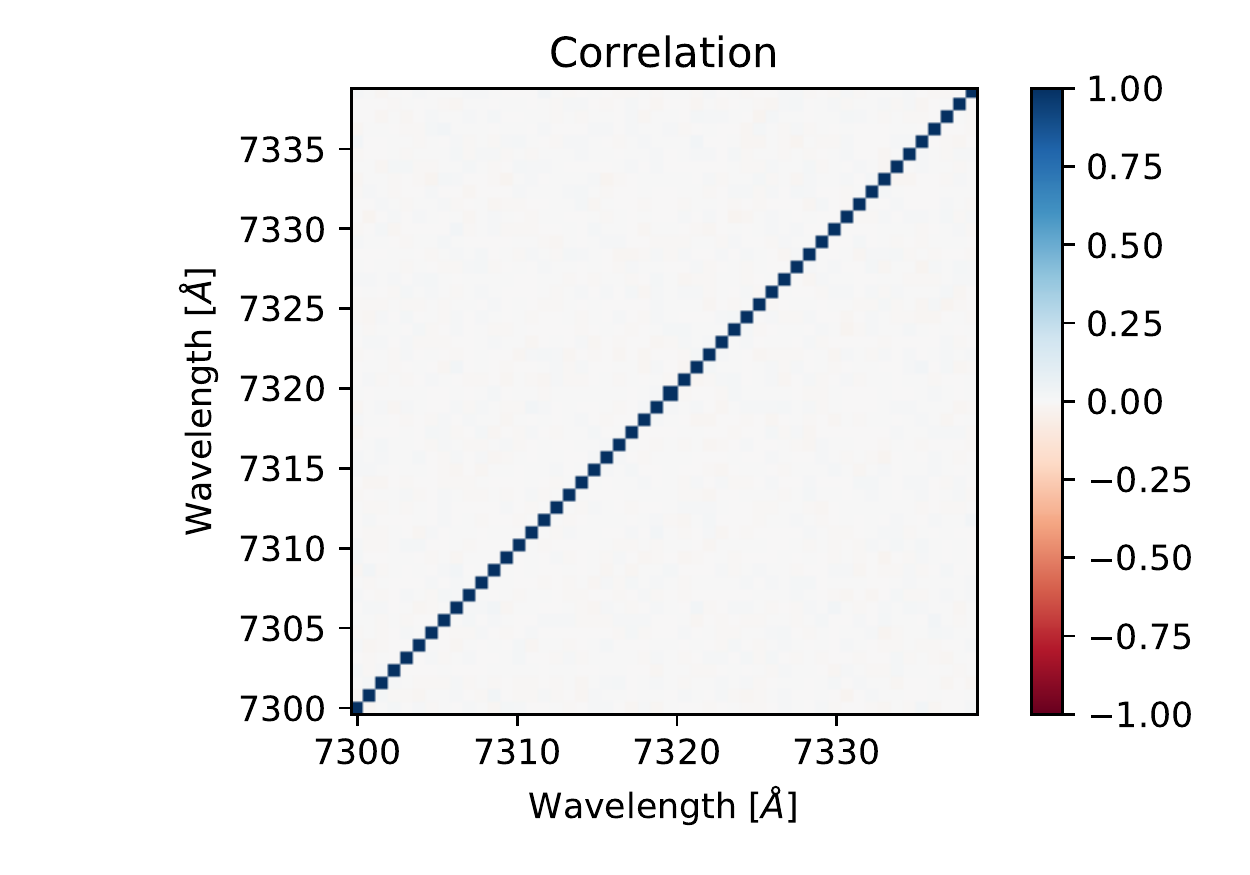}
 \includegraphics[width=0.99\columnwidth,angle=0]{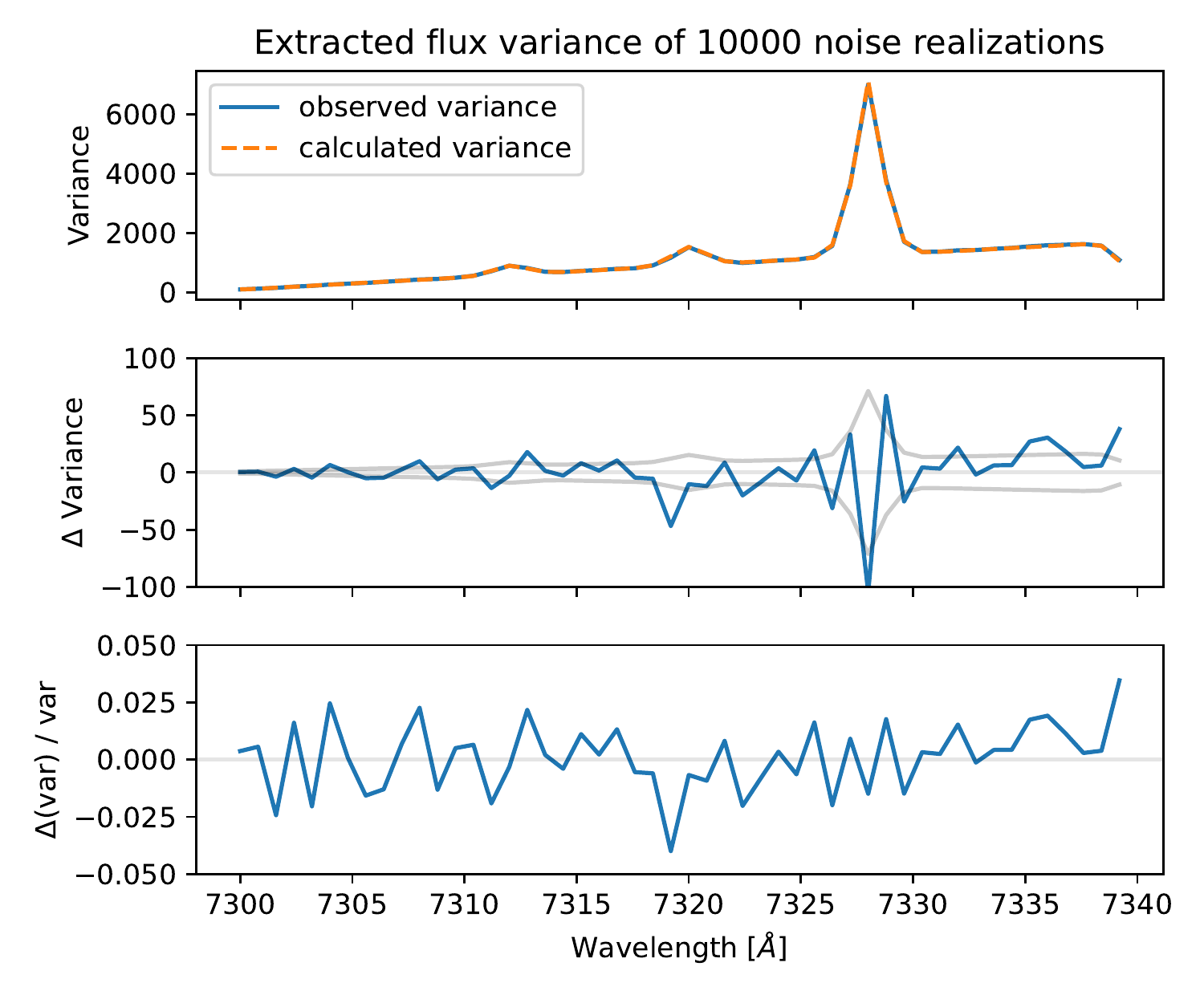}
  \caption{Variance and correlation of 2D extractions of 10000 noise realizations of CCD images of a spectrum.  The upper plot shows that the wavelengths bins have uncorrelated noise, while the lower 3 plots show the excellent agreement with the reported variance.
The second from the bottom plot shows the statistical uncertainty of the
variance of 10000 simulations in thin gray lines.
}
  \label{fig:extraction-covcorr}
\end{figure}

Figure~\ref{fig:extraction-covcorr} shows the variance and correlation of the wavelength bins, confirming that they are uncorrelated between wavelength bins and in agreement with the reported variance to within the statistical precision of the 10000 realizations.

\subsubsection{Key features of the extraction}

We conclude this section by emphasizing some of the features of the spectra
extracted using a full 2D PSF model.  On the positive side, spectra of every
fiber use a common wavelength grid and have by construction uncorrelated noise across
wavelengths.  Use of the full 2D PSF model maximizes the information extracted from the raw data and has excellent validity of the resulting uncorrelated noise model. The resolution matrix $R$ models the per-fiber per-wavelength effective resolution of each spectrum with more fidelity than a simple Gaussian vs.~wavelength line-spread-function (LSF) model.  On the negative side, the non-unity normalization of $R$ results in more complex model fitting since one can no longer directly fit a model $M$ to the spectrum $F$, but rather one must fit $R M$ to $F$.  However, the uncorrelated bins of $F$ on a common wavelength grid are a counterbalancing convenience for analyses.

\subsection{Fiber flat fielding}
\label{sec:fiberflat}

The fiber flat fielding consists of determining a correction for the variations of throughput from fiber to fiber, as a function of wavelength. This correction is essential to homogenize the response of the fibers before the sky background subtraction (see~\S\ref{sec:skysubtraction}). It is also needed to propagate the spectro-photometric calibration obtained from the standard star fibers to the other fibers of the focal plane (see~\S\ref{sec:fluxcalibration}).
This throughput variation is principally due the relative variation of fiber transmission, but it is also affected by the variation of transmission in the spectrograph optics (in particular the change of transmission of the dichroics with the incidence angle which is correlated with the position of the fibers on the slit), the change of solid angle seen by a fiber (due to the change of plate scale in the field of view), the vignetting (mostly the shadow of the cylindrical focal plane instrument), and the residual variations of CCD pixel efficiency that have not been perfectly corrected by the pixel flat field (see \S\ref{sec:preprocessing}).

Because the primary goal is to homogenize the response of fibers for the sky background subtraction, one could think about using directly night sky observations for this purpose. We do not use this approach in practice because the sky spectrum is composed of bright emission lines that would leave an imprint on the estimated transmission, and spectral regions with a faint continuum that would require a prohibitive exposure time to acquire enough signal and reduce statistical fluctuations. The twilight sky is better in that respect but still not ideal; a dedicated calibration system with stable and smoother spectral features is preferred. We use for this purpose an array of LEDs lamps shining on a white screen in the dome as described in \S\ref{sec:calib-obs}. The drawback of this approach is that the spatial and angular distribution of the illumination is not exactly the same as for the night sky.

The flat-field correction is computed in several steps. We first compute a correction independently for each camera and exposure (\S\ref{sec:fiberflat1}). We then combine several exposures obtained with different lamps to get a more homogeneous illumination, closer to the night sky, and we combined the data from the cameras of all spectrographs to inter-calibrate them (\S\ref{sec:fiberflat2}). We evaluate the geometrical differences between the dome screen illumination and the night sky using twilight sky data in \S\ref{sec:twilight}, and finally we present the method used to correct for variation of the flat field with humidity in~\S\ref{sec:fiberflat_vs_humidity}.

\subsubsection{Fiber flat fielding algorithm}
\label{sec:fiberflat1}

The algorithm consists in fitting simultaneously an average spectrum and the relative transmission of each fiber from a camera, for each wavelength of the spectral extraction grid. We take into account the different resolution of each fiber and take advantage of the fact that spectral bins are not correlated.

We minimize the following quantity

\begin{equation}
\chi^2 = \sum_{k} \sum_{i} w_{k,i} \left( \tilde{F}_{k,i} - T_{k,i} \sum_j R_{k,i,j} S_j \right)^2 \label{eq:fiberflat}
\end{equation}
where $k$ is the fiber number, $i$ the wavelength index, $w_{k,i}$ the inverse variance, $\tilde{F}_{k,i}$ the data, $T_{k,i}$ the relative transmission we want to measure, $R_{k,i,j}$ an element of the resolution matrix defined in~\S\ref{sec:extraction} for the fiber $k$ and $S_j$ a flux value of the deconvolved spectrum (from an hypothetical average fiber). The only constraint is that the average transmission is one for all wavelength, $\left< T_{k,i} \right>_k=1$.
We note we should in principle have considered the transmission before applying the resolution to the mean spectrum (with $T$ to the right of $R$ in Eq.~\ref{eq:fiberflat}), but in practice this is not important because we do not expect sharp variations of the relative fiber transmission over the typical scale of the spectral resolution\footnote{Figure 26 of \cite{desi-collaboration22a} shows the DESI throughput as a function of wavelength, which includes the fiber transmission. The only sharp feature in the blue channel at 4400\,\AA\ is caused by the collimator mirror reflectivity. The other absorption features at larger wavelength are due to the atmospheric transmission.}.

The challenge in this fit are the cosmic ray hits. We first fit iteratively for a smooth mean spectrum and then a smooth fiber transmission, ignoring variations of resolution, while rejecting outlier spectral pixels. When all the outliers are removed, we fit for the exact solution. For both fits we use the Gauss-Newton algorithm. The fit is iterative because it is non-linear (we have to determine the product of $T$ and $S$).

\begin{figure}
  \centering
  \includegraphics[width=0.99\columnwidth,angle=0]{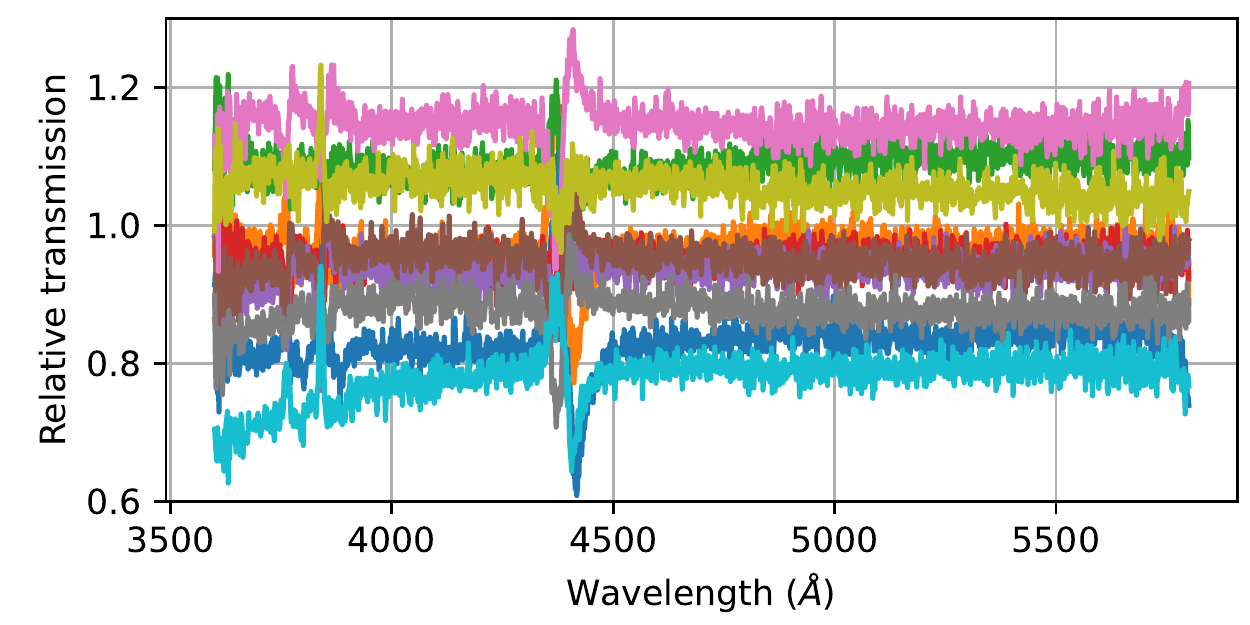}
  \includegraphics[width=0.99\columnwidth,angle=0]{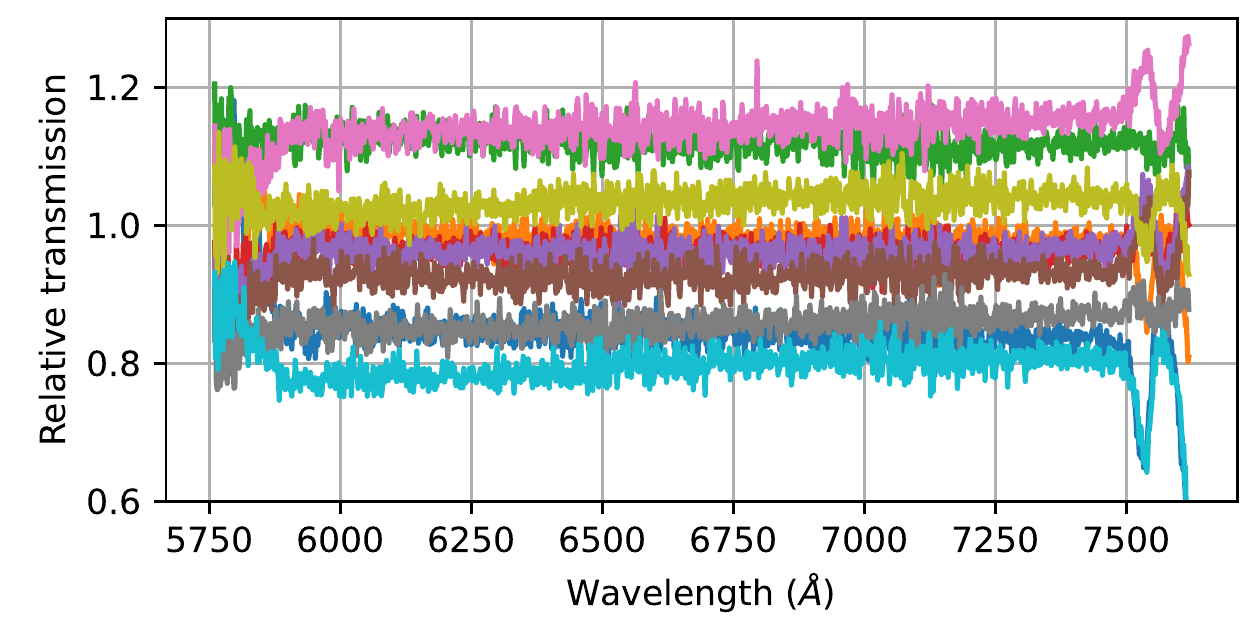}
  \includegraphics[width=0.99\columnwidth,angle=0]{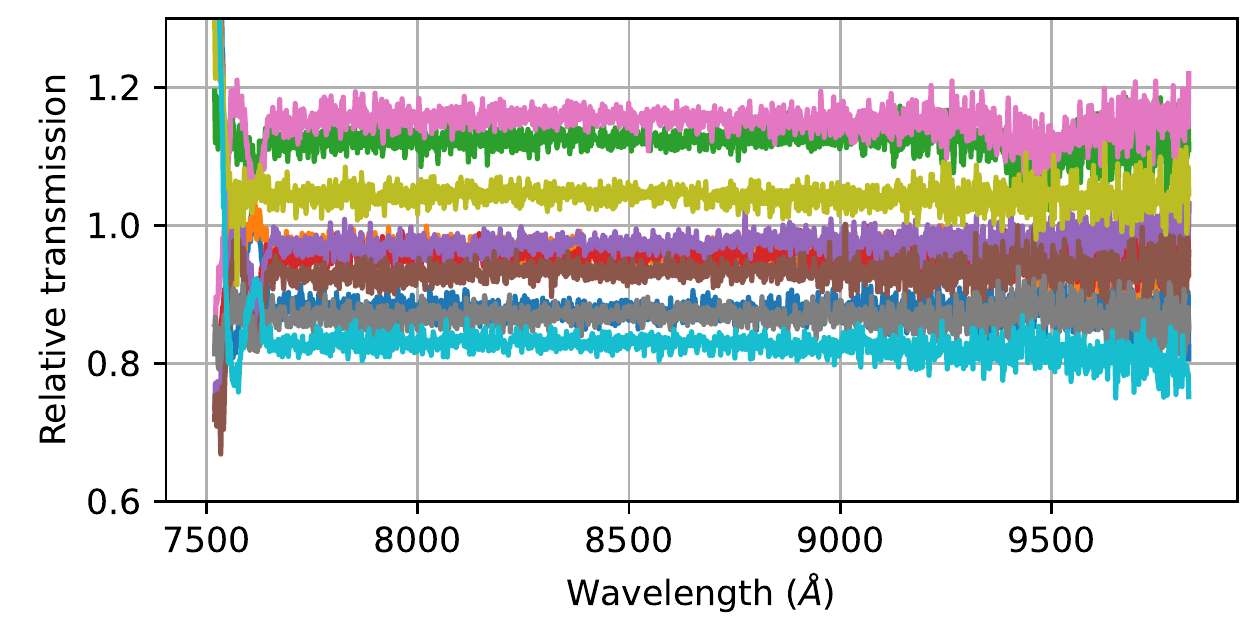}
  \caption{Relative transmission of a sub-sample of 10 fibers from the spectrograph SM2 obtained with one exposure of 2\,min on the dome screen with LED lamps in the blue, red and NIR cameras. Each colored curve corresponds to a fiber. The measurement appear noisy but 12 such exposures are combined to determine the final correction. The spectral features in the blue camera at 3800 and 4400\,\AA\ are due to a dip in the reflectivity of the spectrograph collimator mirror at these wavelength, with a profile that varies with the location in the mirror and with the light incidence angle. The variations around 5750\,\AA\ and 7500\,\AA\ in the red and NIR camera spectra are due to the change of transmission and reflection of the dichroics again with the light incidence angle that varies with the fiber location in the pseudo-slit. Spectral regions affected by cosmic ray hits have been masked and are interpolated over in the figures.}
  \label{fig:fiberflat}
\end{figure}

Examples of flat field corrections are shown in Figure~\ref{fig:fiberflat} for one 120\,sec exposure obtained with the LED array of one calibration lamp.

\subsubsection{Combining the fiber flat fields from various lamps and cameras}
\label{sec:fiberflat2}

\begin{figure}
  \centering
  \includegraphics[width=0.49\columnwidth,angle=0]{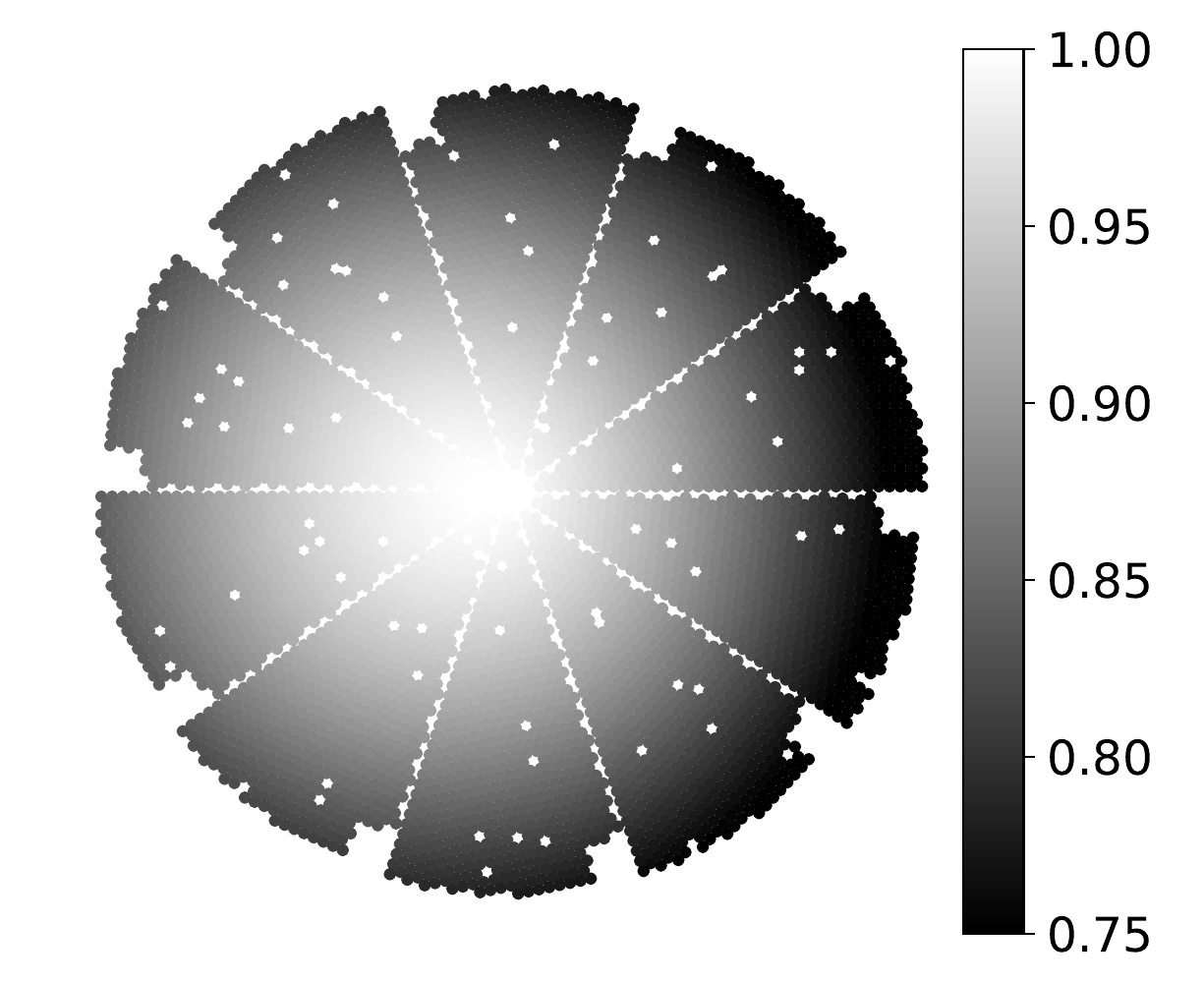}
  \includegraphics[width=0.49\columnwidth,angle=0]{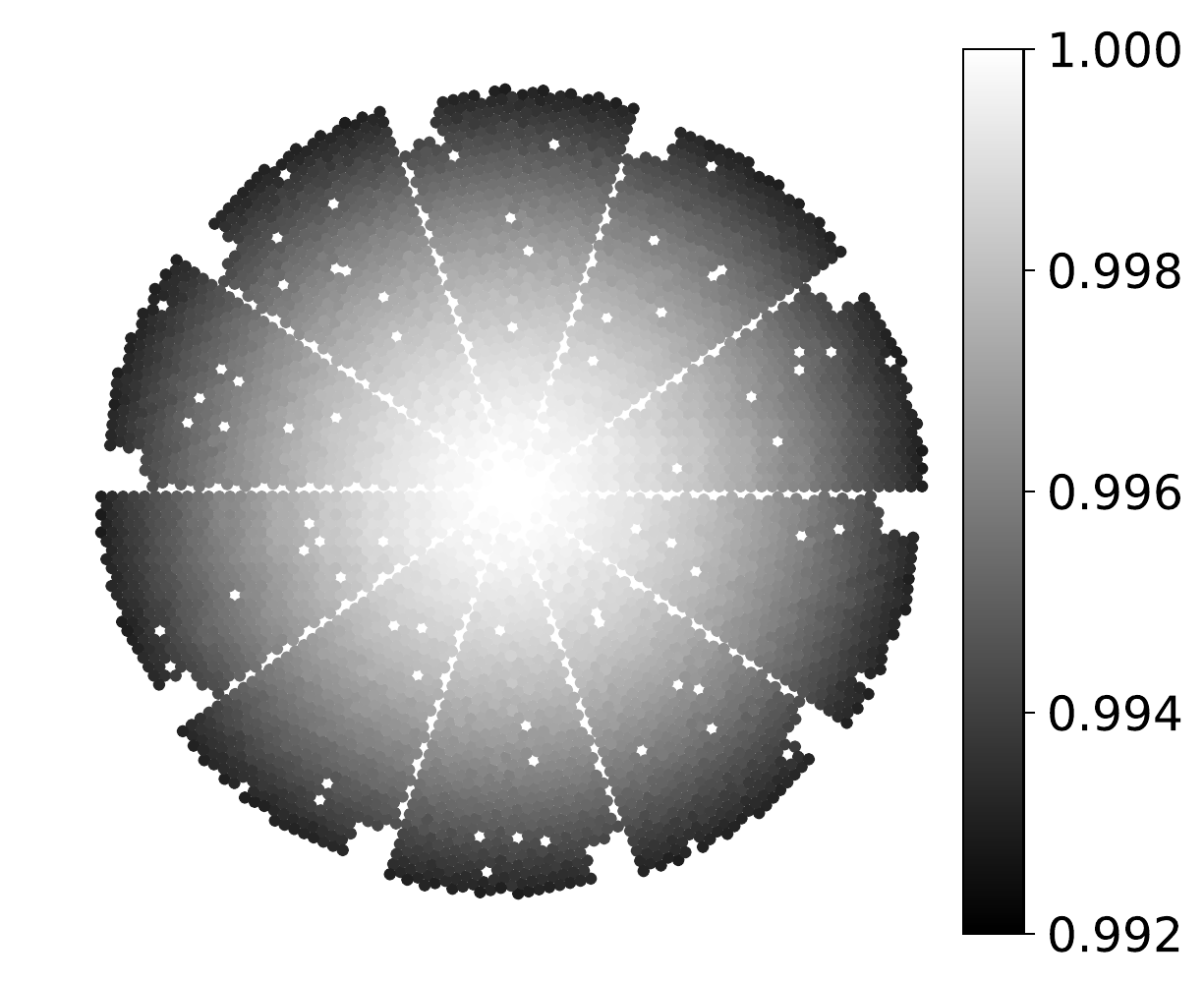}
  \caption{Left: illumination pattern on the focal plane for one calibration lamp. Right: illumination pattern  obtained with four calibration lamps divided by the illumination from the sky. The non-homogeneity of the illumination ratio is purely radial, with a variation of about 1\% from the center to the edge of the focal plane.}
  \label{fig:lamp-illumination}
\end{figure}

The fiber flat fields obtained with each of the four lamps placed on the upper ring of the telescope (see Figure~\ref{fig:mayall}) are combined to get a more homogeneous pattern. A numerical computation has shown that the illumination is azimuthally homogeneous to better than 0.1\% with four lamps, with an expected residual radial gradient of about 1\% when compared to the night sky illumination (see Figure~\ref{fig:lamp-illumination}). This residual difference is due to differences in the vignetting by the focal plane instrument.

We combine the fiber flats from exposures obtained with each lamp one at a time instead of all the lamps together in order to mitigate the possible variations of lamp intensity. Indeed a variation of intensity of one lamp has no effect on the flat field correction obtained from an exposure with this individual lamp as it is absorbed in the mean spectrum term in Eq.~\ref{eq:fiberflat} while it would lead to a gradient in the flat field correction if all the lamps were used together.

\begin{figure}
  \centering
  \includegraphics[width=0.99\columnwidth,angle=0]{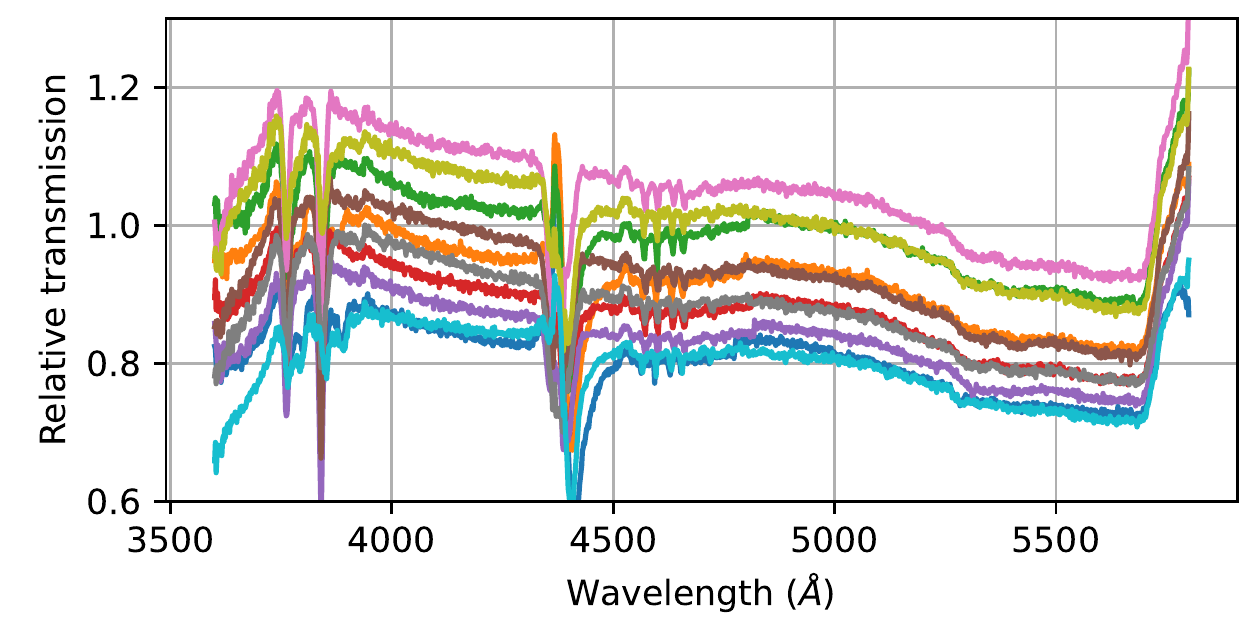}
  \includegraphics[width=0.99\columnwidth,angle=0]{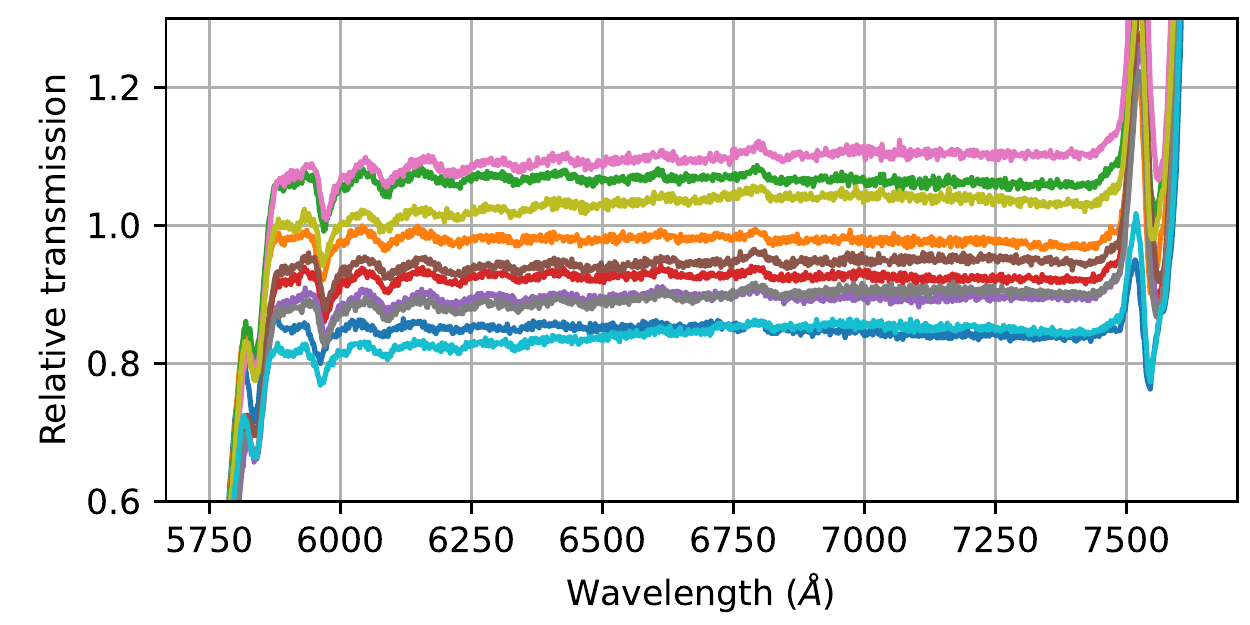}
  \includegraphics[width=0.99\columnwidth,angle=0]{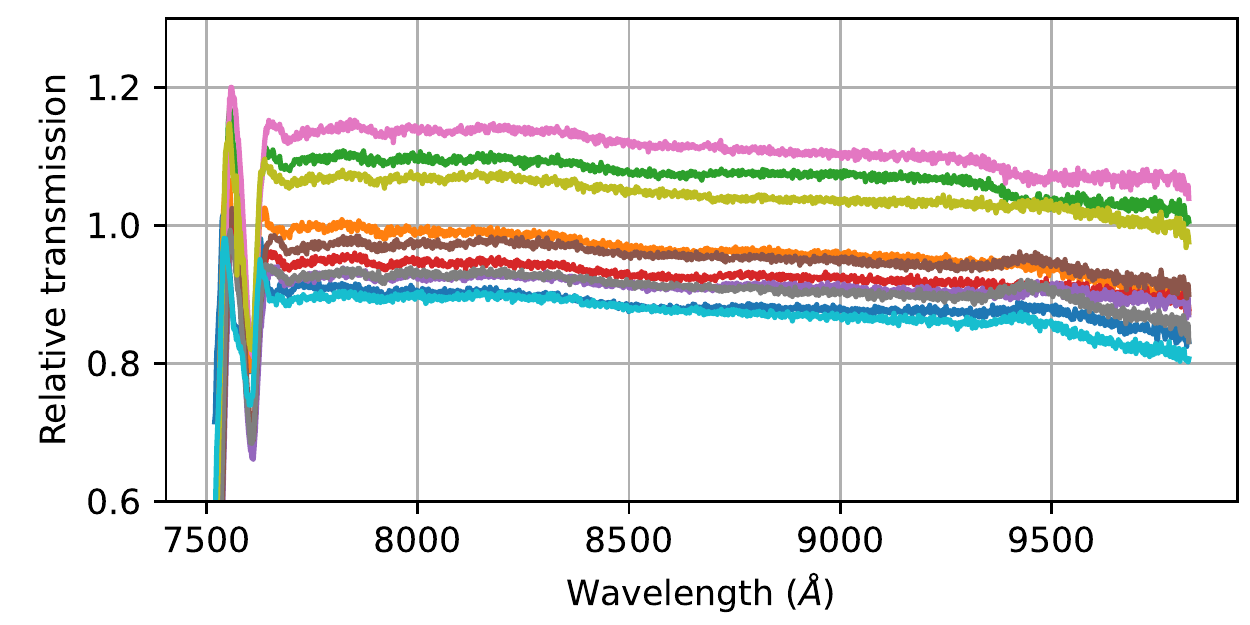}
  \caption{Transmission of a sub-sample of fibers from spectrograph SM2 normalized to the average transmission of all the fibers from the 10 spectrographs. These curves are the result of the combination of 12 individual fiber flats as shown in Figure~\ref{fig:fiberflat}, but with a different normalization (average of 10 spectrographs instead of one).}
  \label{fig:fiberflatnight}
\end{figure}

We also combine fiber flats from all cameras of the same type (blue, red, or NIR) while normalizing them to the same mean spectrum. This results in an inter-calibration of the spectrographs. We show these fiber flats in Figure~\ref{fig:fiberflatnight}.

\subsubsection{Validation with sky background measurements}
\label{sec:twilight}

The expected radial pattern in the ratio of dome flat to sky flat (Fig.~\ref{fig:lamp-illumination}) has been verified with sky background observations.

We expect this residual anisotropy to be stable and achromatic as it is the combination of purely geometric terms (location of dome screen, lamps, shadows), and the reflectivity of the screen which is quasi-Lambertian over the whole wavelength range.

\begin{figure}
  \centering
  \includegraphics[width=0.99\columnwidth,angle=0]{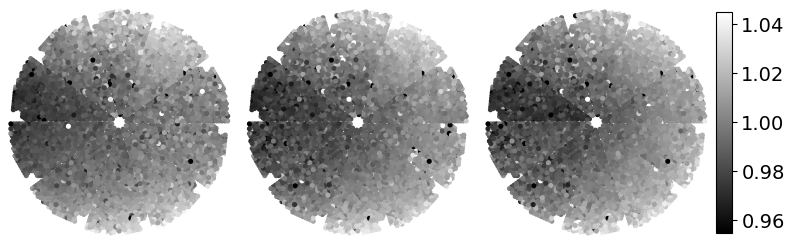}
  \caption{Focal plane view of the median flux in fibers from a twilight observation (exposure \#00055559 from 2020/03/15) after applying the nightly averaged fiber flat correction (see \S\ref{sec:fiberflat2}) and a normalization factor per camera (blue, red and NIR cameras from left to right). The color scales have been saturated to enhance the residual gradient. This figure highlights the agreement between cameras, demonstrating this twilight flat is very nearly achromatic.}
  \label{fig:twilightflat}
\end{figure}

Figure~\ref{fig:twilightflat} presents the focal plane views of the median flux in fibers for a twilight sky observation conducted on March 15, 2020, after flat fielding.
We measure a gradient of a few percent along one axis and a quadratic term only along the other. More quantitatively, noting $x$ and $y$ the cartesian focal plane coordinates of fibers in units of 400\,mm, we have
\begin{eqnarray}
\mathrm{blue\ flat} & \propto & 0.991 -0.003 	\, x +0.012 	\, y -0.002 \, x^2 +0.041 \, y^2 \nonumber \\
\mathrm{red\ flat} & \propto & 0.988 +0.000 	\, x +0.019 	\, y +0.003 \, x^2 +0.041 \, y^2 \nonumber \\
\mathrm{NIR\ flat} & \propto & 0.987 -0.001 	\, x +0.022 	\, y +0.010 \, x^2 +0.039 \, y^2 \nonumber \\
\label{eq:flatfield-polynomial}
\end{eqnarray}

 As visually evident on the figure and the polynomial coefficients, the twilight flat is consistent in all three cameras, blue, red, and NIR, confirming it consists in a purely geometric correction term. The gradient is due to the variation of the twilight sky brightness with elevation (the parallactic angle was close to 90\,$\deg$).

\begin{figure}
  \centering
  \includegraphics[width=0.9\columnwidth,angle=0]{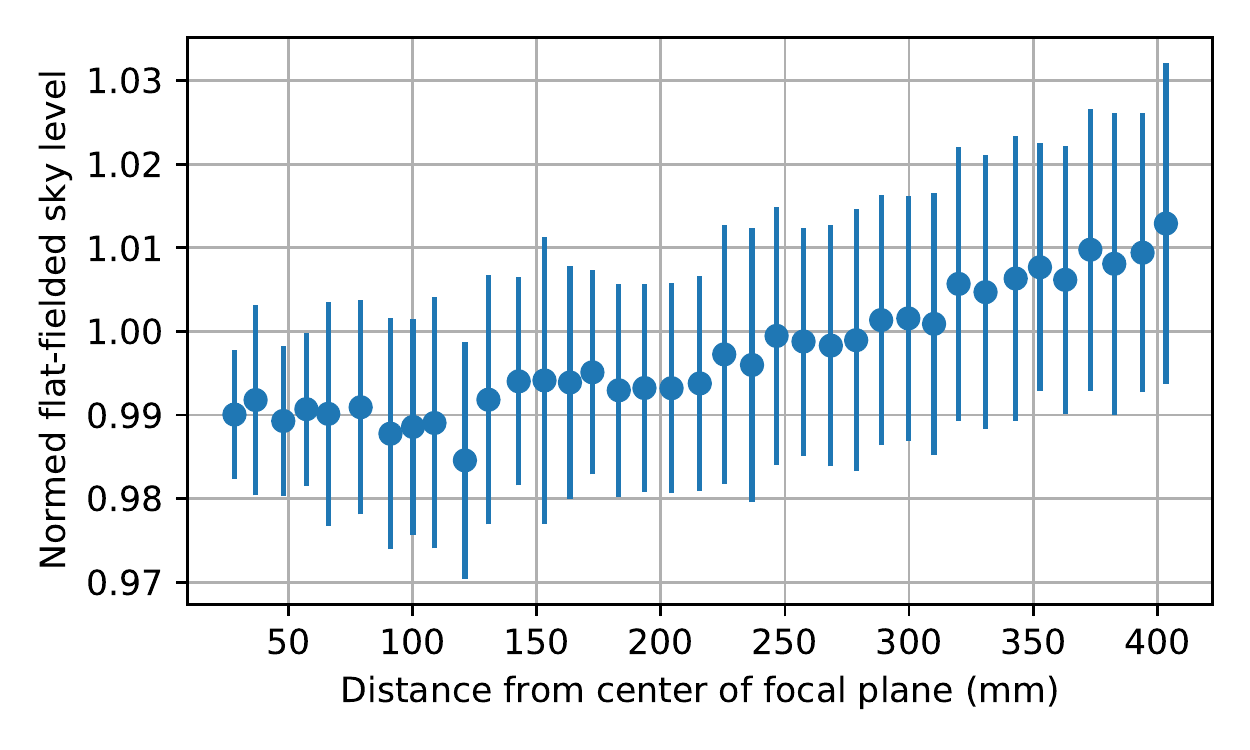}
  \caption{Flat-fielded night sky background variation in the r-camera as a function of the distance from the focal plane center. This figure has been obtained by averaging the sky spectra of all sky fibers from 4 nights of observations. The rms of 1 to 1.5\% comprises the measurement statistical errors, the residual contamination for stars, and systematic errors in the spectroscopic measurements. The radial variation is due to a difference in the vignetting between the sky and the calibrations with the dome screen.} \label{fig:skyflat}
\end{figure}

This study is complemented with a measurement of the dark sky during science observations.
The flat-fielded flux in the r-camera of all the sky fibers of 4 nights has been averaged as a function of the fiber coordinates in the focal plane. The normalized and average flux is shown as a function of the focal plane distance on Figure~\ref{fig:skyflat}. We measure a radial pattern with a variation of 2\% which is a bit larger than the expectations from simulations (see Figure~\ref{fig:lamp-illumination}, where the expected inverse ratio is shown). Note that this residual anisotropy is absorbed by a normalization coefficient estimated for each exposure and fiber (see~\S\ref{sec:skynormalization}).

\subsubsection{Humidity correction}
\label{sec:fiberflat_vs_humidity}

\begin{figure}
  \centering
  \includegraphics[width=0.99\columnwidth,angle=0]{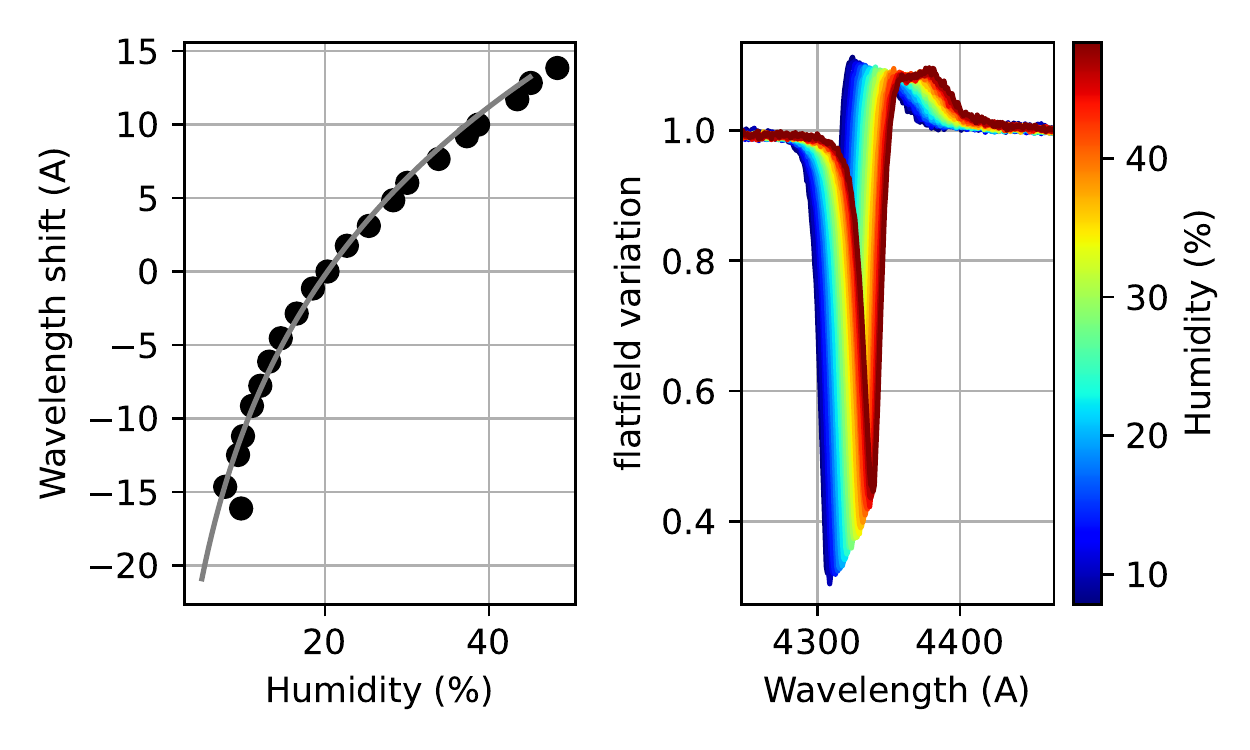}
  \caption{Flat field variation of the central fiber of the SM10 blue camera as a function of the humidity in the spectrographs enclosure. Left: wavelength shift of the feature; right: fiber flat as a function of wavelength (normalized to the average of the 500 fibers for this camera).}
  \label{fig:fiberflat-vs-humidity}
\end{figure}

There are absorption features at 3800 and 4400\,\AA\ in the collimator mirror reflectivity of most spectrographs. The wavelength of those features are a strong function of humidity. Figure~\ref{fig:fiberflat-vs-humidity} shows this variation for one fiber. It was obtained by averaging the flat field measurements obtained in 2021, in bins of humidity, from the minimum of about 8\% obtained in winter to the maximum of 50\% that is reached in summer during the moonsoon season (this maximum value is enforced with a dehumidifier controlled by the environmental system of the spectrographs enclosure). We use this variation as a model to correct for the change of humidity between the calibration runs in the afternoon and the observations during the night.

\subsection{Sky subtraction}
\label{sec:skysubtraction}

The sky subtraction algorithm takes advantage of the spectral extraction method, which provides fiber spectra on the same wavelength grid, with uncorrelated noise (within a spectrum), along with a resolution matrix that provides the transform to apply from a high resolution spectral model to a lower resolution fiber spectrum.

The sky subtraction is performed in two steps. First a sky spectrum model is fit to the sky fibers spectra for each camera independently and then this model is subtracted from each fiber of the camera.

\subsubsection{Sky model fit}
\label{sec:skymodelfit}

We fit for a deconvolved sky spectrum by minimizing the following.

\begin{equation}
\chi^2 = \sum_{k} \sum_{i} w_{k,i} \left( \tilde{F}_{k,i} - T_{k,i} \sum_j R_{k,i,j} P(x_k,y_k,\lambda_j) S_j \right)^2 \label{eq:skymodel}
\end{equation}
where $k$ is a sky fiber number, $i$ the wavelength index, $w_{k,i}$ the inverse variance, $\tilde{F}_{k,i}$ the data, $T_{k,i}$ the fiber flat field correction (see~\S\ref{sec:fiberflat}), $R_{k,i,j}$ an element of the resolution matrix defined in~\S\ref{sec:extraction} for the fiber $k$, $P$ an optional polynomial correction and $S_j$ a flux value of the average sky model spectrum. Because of the resolution term, $S$ is a deconvolved spectrum.

The polynomial correction is optional, it is intended to model angular variations of the sky spectrum across the focal plane. We can consider either an isotropic sky spectrum (in which case $P = 1$), an anisotropic sky intensity with a same spectral shape, or an anisotropic and color-varying sky spectrum. In that later case, the most complex, the polynomial is 3D, it is a function of the fiber coordinates in the focal plane (i.e. angular sky coordinates) and wavelength. We have not considered in this version\footnote{Code version numbers are given at the end of the introduction.} of the code the possible relative variations of the OH emission line intensities with respect to the sky continuum. We are presently developing corrections for variations across the field of view due to diffuse moonlight and the twilight sky.

The fit is performed iteratively. Each iteration starts with a fit of the mean sky spectrum $S$ given the polynomial $P$ (simple linear system), then optionally  $P$ given $S$, and then each iteration ends with an outlier rejection which aims at discarding flux bins that are affected by cosmic rays (and have not been flagged at the extraction level).
The loop ends when the $\chi^2$ decrement between iterations falls below a given threshold and no more outliers are found.
After the last iteration the covariance matrix of $S$ is computed (inverse of half of the Hessian of the $\chi^2$), assuming constant $P$, which is an approximation.
We then compute the sky model for all fibers of a camera, by reconvolving the model using the resolution matrix of each fiber, and possibly applying the polynomial correction. We also save the sky model covariance assuming the average resolution and transmission of all fibers.

\subsubsection{Normalization of the sky background}
\label{sec:skynormalization}

The sky model determined from the sky fibers is then subtracted from each flat-fielded target spectrum of the same camera. In this process, an achromatic scale factor is adjusted for each fiber and camera. This scale factor can absorb various calibration errors like a flux extraction bias due to a variation of PSF shape or more likely an imperfect centering of fiber traces, or an error in the flat field corrections like the percent variations revealed by the twilight flats (see \S\ref{sec:fiberflat}). It can also correct for a genuine variation of fiber throughput between the calibration exposures and the current one due to either fiber flexure, which changes the fiber focal ratio degradation\footnote{The Focal Ratio Degradation (FRD) characterizes the broadening of the light beam angular distribution between the fiber input and output. A larger FRD leads to a reduced fiber throughput because the light loss in the spectrograph}, or a change of vignetting with the fiber position for the fibers on the edge of the focal plane.

\begin{figure}
\centering
  \includegraphics[width=0.99\columnwidth,angle=0]{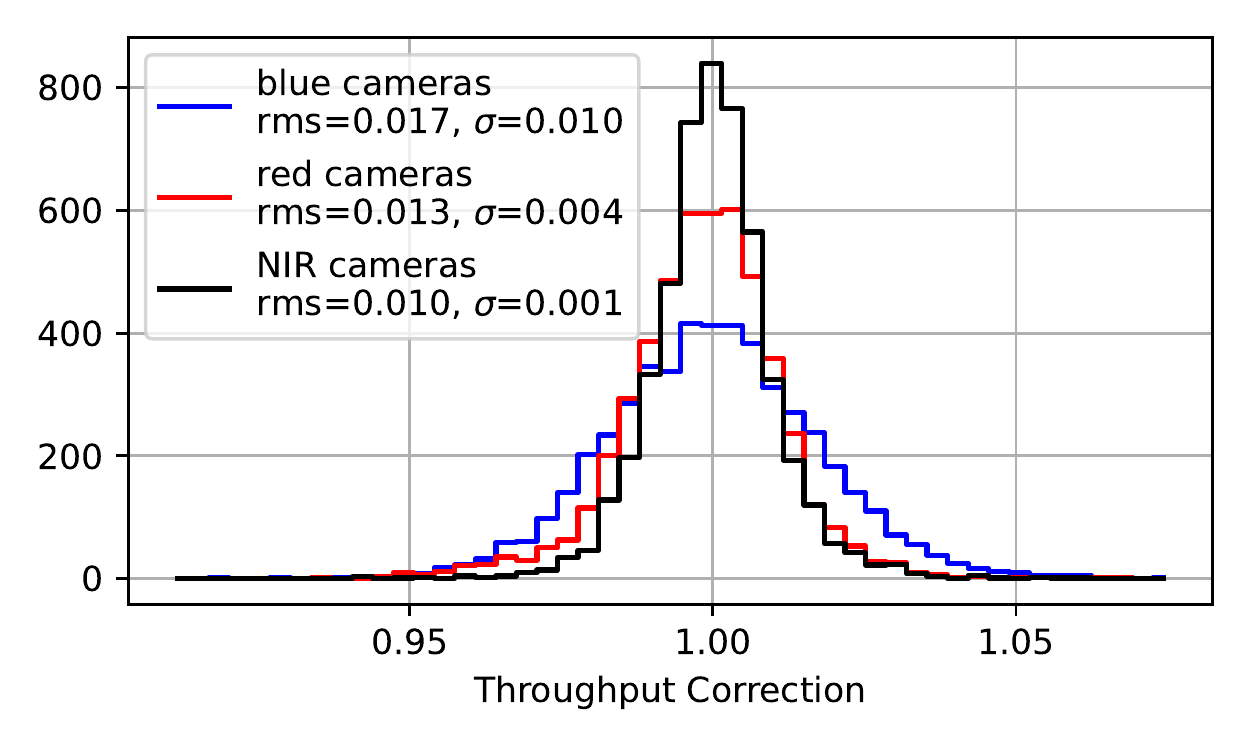}
  \caption{Histogram of the multiplicative correction term applied in the sky subtraction per camera for a 900 second sky exposure. There is one entry per fiber. Measurements with statistical uncertainties ($\sigma$ in the legend) larger than 0.03 were discarded.}
  \label{fig:sky-throughput-correction}
\end{figure}

This scale factor is estimated as follows.
We first fit for the amplitude of each sky line in a predefined list along with a local background. The fit of the amplitude and background is performed simultaneously in a narrow spectral range of 9\,\AA\ centered on the line in order to account for the unknown target flux density. We use a least square minimization accounting for the flux variance and the known sky spectrum line profile from the model.
We then combined the amplitudes of all the lines within a spectrum into a single scale factor, per fiber, with inverse variance weights, and while rejecting 3 sigma outliers.
Figure~\ref{fig:sky-throughput-correction} shows that the typical correction is of order 1\%.

\subsubsection{Corrections with a principal component analysis}

The model described above was successfully tested on simulations, however systematic sky residuals on bright sky lines were found when applied to real DESI data, in particular for the bright NIR lines at 9000\,\AA\ and above. The residuals can be described in first approximation as wavelength calibration errors of about 0.02\,\AA\ (about 3\% of the size of a CCD pixel), and errors on the Line Spread Function (hereafter LSF, which is the 1D spectral point spread function) width of a few 0.1\,\AA.
Those errors are quite stable between exposures, but vary from fiber to fiber, with a correlation length of about 10-20 fibers (not fully correlated with the fiber bundles of 25 fibers).
We have not understood the source of those systematic residuals. In particular, we could not relate them to PSF fit residuals. The working hypothesis is that they are the result of differences in the illumination pattern between the white screen in the dome and the night sky, causing a different distribution of light in the spectrographs and hence different PSFs.

Our approach to correct those effects is purely empirical. For each exposure, each sky fiber and each bright sky line, we fit for 3 parameters: an amplitude correction, a wavelength calibration offset ($\Delta \lambda$), and a LSF width correction ($\Delta LSF$). The amplitude correction is ignored but the $\Delta \lambda$ and $\Delta LSF$ arrays are recorded.

In a first calibration step, we computed those arrays for a large sample of sky observations conducted in 2020.
For each camera/exposure, the $\Delta \lambda$ and $\Delta LSF$ values obtained for a finite set of sky lines in the sky fibers are linearly interpolated (or extrapolated) to the full 2D array that comprise the 500 fibers and the wavelength grid of the spectral extraction. The resulting 2D frames are then used to perform a principal component analysis (hereafter PCA, see also~\citealt{WildHewett2005} for another PCA approach to sky subtraction). The resulting four first components (also 2D frames) are saved as part of the calibration product for each of the 30 cameras of DESI spectrographs.

Then, for each science exposure, the limited set of  $\Delta \lambda$ and $\Delta LSF$ on the sky fibers are used to compute linear coefficients for the saved principal components, and the resulting linear combination is then applied to all of the fibers and wavelength from the frame.

Presented in the form of an equation, the complete sky model for the flat-fielded spectrum in fiber $f$ for the wavelength index $i$ is :

\begin{eqnarray}
  S_{k,i} &=& \alpha_k \sum_j R_{k,i,j} P(x_k,y_k,\lambda_j) S_j \nonumber \\
  && + \sum_l X^{(\lambda)}_l (\partial_{\lambda} \tilde{S})_i  C^{(\lambda)}_{l,k,i} \nonumber \\
  && + \sum_p X^{(LSF)}_p (\partial_{LSF} \tilde{S})_i  C^{(LSF)}_{p,k,i} \label{eq:fullskymodel}
\end{eqnarray}

Here, $\alpha_k$ is the scale factor described in \S\ref{sec:skynormalization}. $R_{k,i,j}$, $P(x_k,y_k,\lambda_j)$, and $S_j$ are the terms contributing to the initial sky model described in \S\ref{sec:skymodelfit}. $X^{(\lambda)}_l$ and $C^{(\lambda)}_{l,k,i}$ are respectively the PCA coefficients and components for the wavelength correction, and $\partial_{\lambda} \tilde{S}$ is the derivative of the reconvolved sky model ($\tilde{S} \equiv R S$) with respect to the wavelength $\lambda$. The terms in the last row are the equivalent for the LSF width correction. $\partial_{\lambda} \tilde{S}$ and  $\partial_{LSF} \tilde{S}$ are computed numerically. The latter is derived from the difference between a (second) convolution of sky spectrum and the original spectrum; we use Gaussian kernel with $\sigma=0.3$\AA\ for this purpose.

\subsubsection{Estimating the variance of the sky subtracted spectra}
\label{sec:sky-residual-variance}
The sky model covariance (hereafter $C_S$) is the inverse of the Fisher matrix (half of the Hessian of the $\chi^2$ given by Eq.~\ref{eq:skymodel}). It is not diagonal, the neighboring  flux values in the vector $S$ have large anti-correlations as a result of the deconvolution process (the sky model $S$ is first multiplied by the resolution matrix $R$ before comparison with data).
Conversely, the flux values of the sky model convolved with the resolution of a fiber ($\tilde{S_k} \equiv R_k S$) have uncorrelated noise to a good approximation. In consequence, rather than keeping the full covariance $C_S$, and evaluating the reconvolved model variance for each fiber, we estimate and save the diagonal of the covariance of the convolved sky model for the average resolution matrix $\bar{R}$ (average of all the valid fibers of the camera). We have the covariance $C_{\tilde{S}} \equiv \bar{R} C_S \bar{R}^T$, and we note $\sigma^2_{\tilde{S}}$ the diagonal terms of this covariance matrix in the following.

\begin{figure}
\centering
  \includegraphics[width=0.95\columnwidth,angle=0]{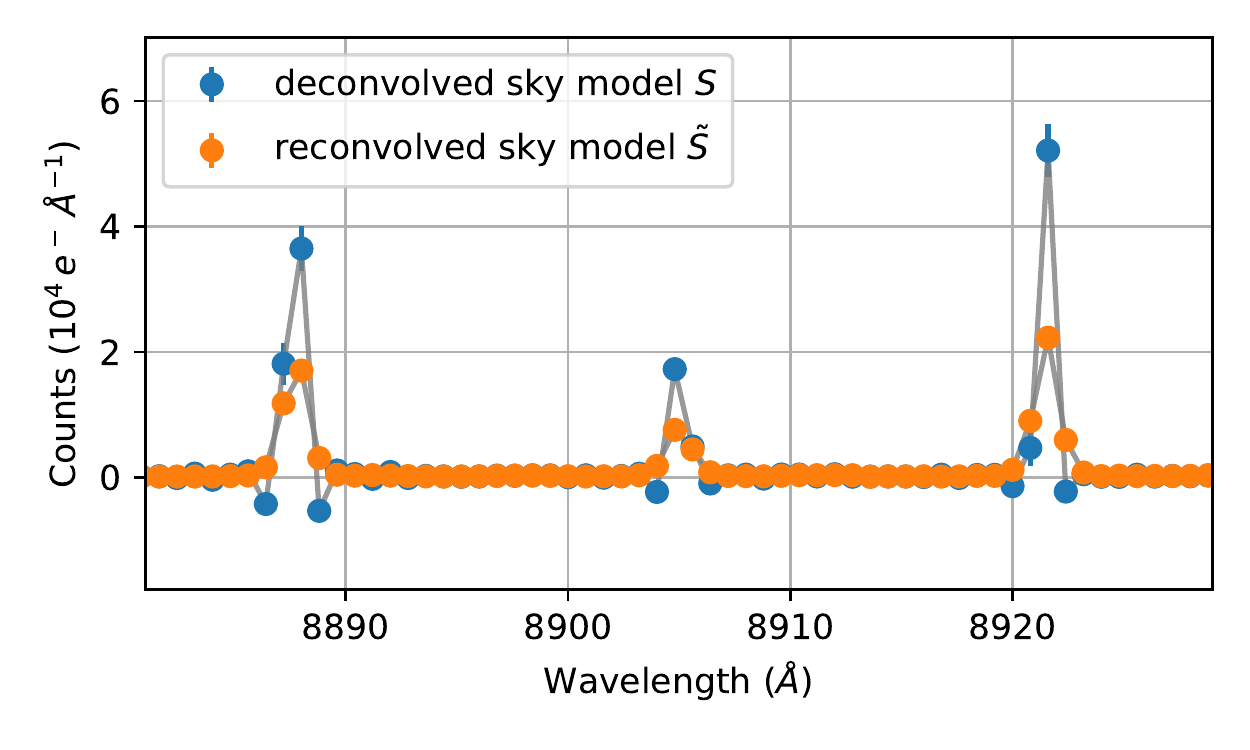}
  \includegraphics[width=\columnwidth,angle=0]{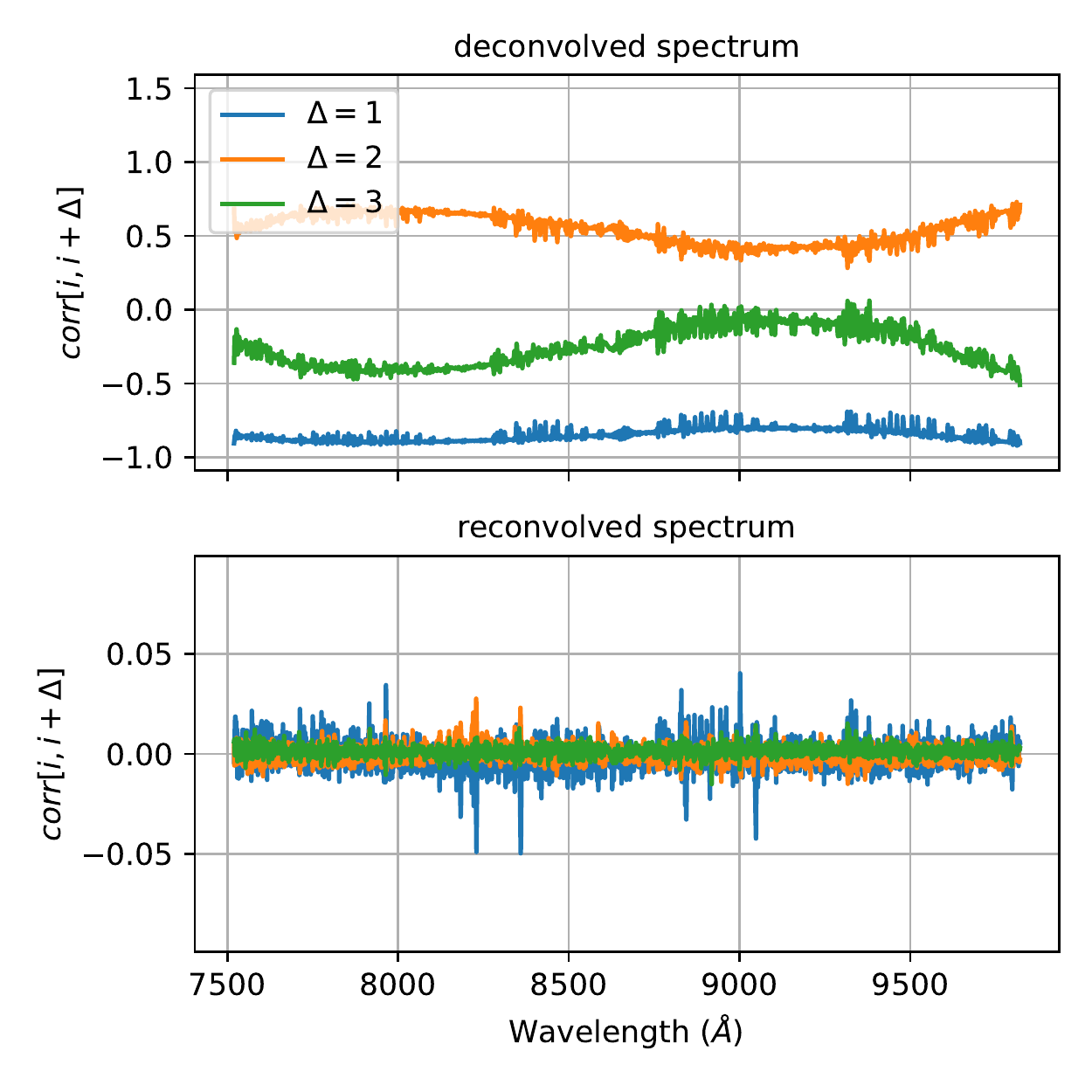}
  \caption{Top: zoom on a few NIR lines of the sky model spectrum derived from a 900 second science exposure with 80 sky fibers. The blue curve is the deconvolved spectrum ($S$) and the orange the reconvolved spectrum $\tilde{S} = R S$. Bottom: correlation coefficients ($ \equiv C_{i,i+\Delta} / \sqrt{ C_{i,i} C_{i+\Delta,i+\Delta}}$ with $C$ the covariance matrix)  of neighboring flux values in the same spectrum. The neighboring fluxes ($\Delta=1$) on the deconvolved spectrum are highly anti-correlated with a coefficient of about -0.8, whereas the correlations in the reconvolved spectrum never exceed a few percent.}
  \label{fig:skymodel}
\end{figure}

Figure~\ref{fig:skymodel} shows a few model sky lines before and after reconvolution ($S$ and $\tilde{S}$). Also shown are the correlation coefficients between neighboring fluxes in the same spectra. The correlation between neighboring flux bins in the reconvolved spectrum does not exceed a few percent.

The variance of the target spectra after sky subtraction is modeled as
\begin{eqnarray}
  \sigma^2_{\mathrm{corr},k,i} &=& \sigma^2_{\mathrm{extract},k,i} \, T_{k,i}^{-2} + \sigma^2_{\tilde{S},i} \\ \nonumber
  && + \left( \sigma_{T,i} \, \tilde{S}_{k,i} \right)^2 + \left( \epsilon_{\lambda}  \, (\partial_{\lambda} \tilde{S}_{k})_i \right)^2 \label{eq:spectralvariance}
\end{eqnarray}
where $\sigma^2_{\mathrm{extract},k,i}$ is the variance from the spectral extraction for the fiber $k$ and wavelength $i$ (the diagonal of $\tilde{C}$ defined in \S\ref{sec:extraction}), $T_{k,i}^{-1}$ is the fiber flat field correction (inverse of the relative fiber throughput, see Eq.~\ref{eq:fiberflat}), $\sigma^2_{\tilde{S}}$ the sky model error, $\tilde{S_{k}}$ the resolution convolved sky for the fiber $k$ and $\partial_{\lambda} \tilde{S}_k $ its derivative with respect to wavelength. $\sigma_{T}$ is the flat field error and $\epsilon_{\lambda}$ is a term used to quantify the wavelength calibration error.

\begin{figure}
\centering
  \includegraphics[width=0.95\columnwidth,angle=0]{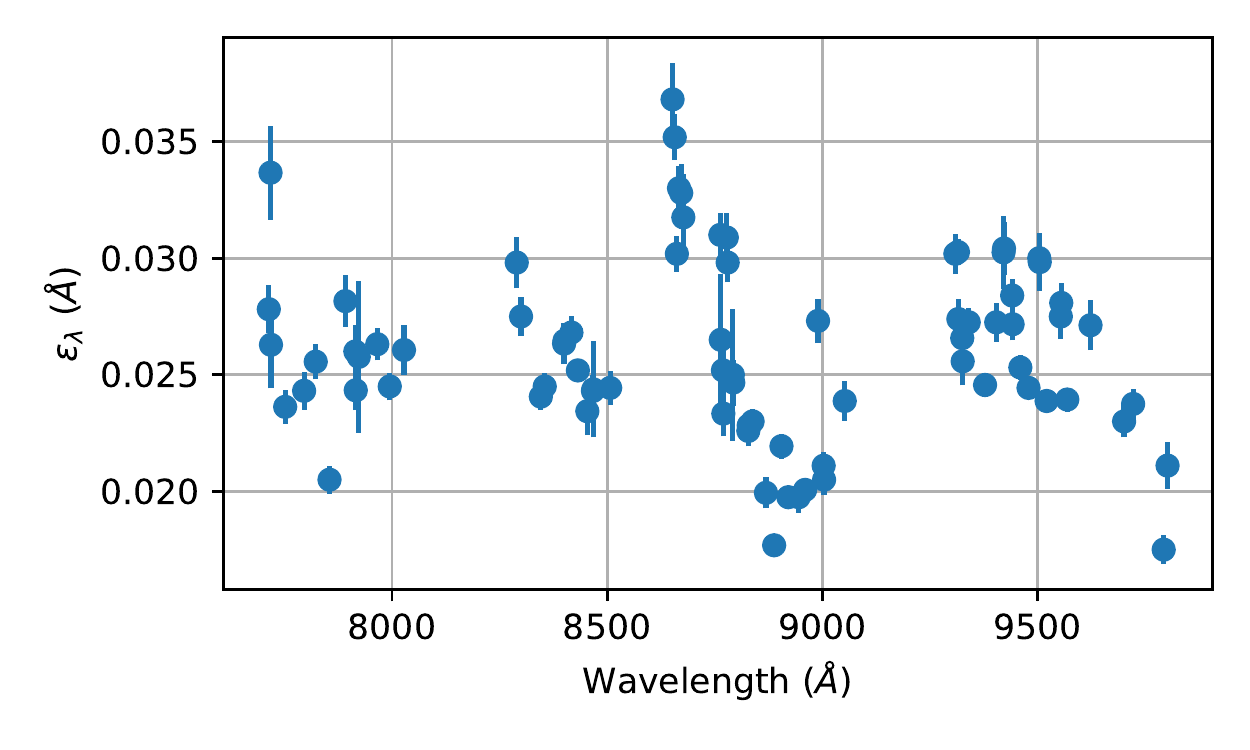}
  \caption{Wavelength calibration error term $\epsilon_{\lambda}$ determined from the sky line fit (one value per sky line, averaged over cameras and exposures, for typical dark time exposures of about 1000 sec).}
  \label{fig:sky-epsilon-lambda}
\end{figure}

The term $\epsilon_{\lambda}$ is adjusted for each sky line independently, but considering all sky fibers at once. Its value is chosen to get a reduced  $\chi^2/ndf = 1$ in a spectral range of $\pm 3$\,\AA\ around each sky line ($ndf$ is the number of degrees of freedom of the fit). It is evaluated only on the sky fibers but is then applied to the variance of all the target fibers. We stress however that it applied only to the wavelength affected by sky lines (in a range of $\pm 3$\,\AA). Figure~\ref{fig:sky-epsilon-lambda} shows the median value $\epsilon_{\lambda}$ as a function of wavelength for several science exposures in NIR. Typical values are of 0.025\,\AA.

\subsubsection{Results}

\begin{figure}
\centering
  \includegraphics[width=\columnwidth,angle=0]{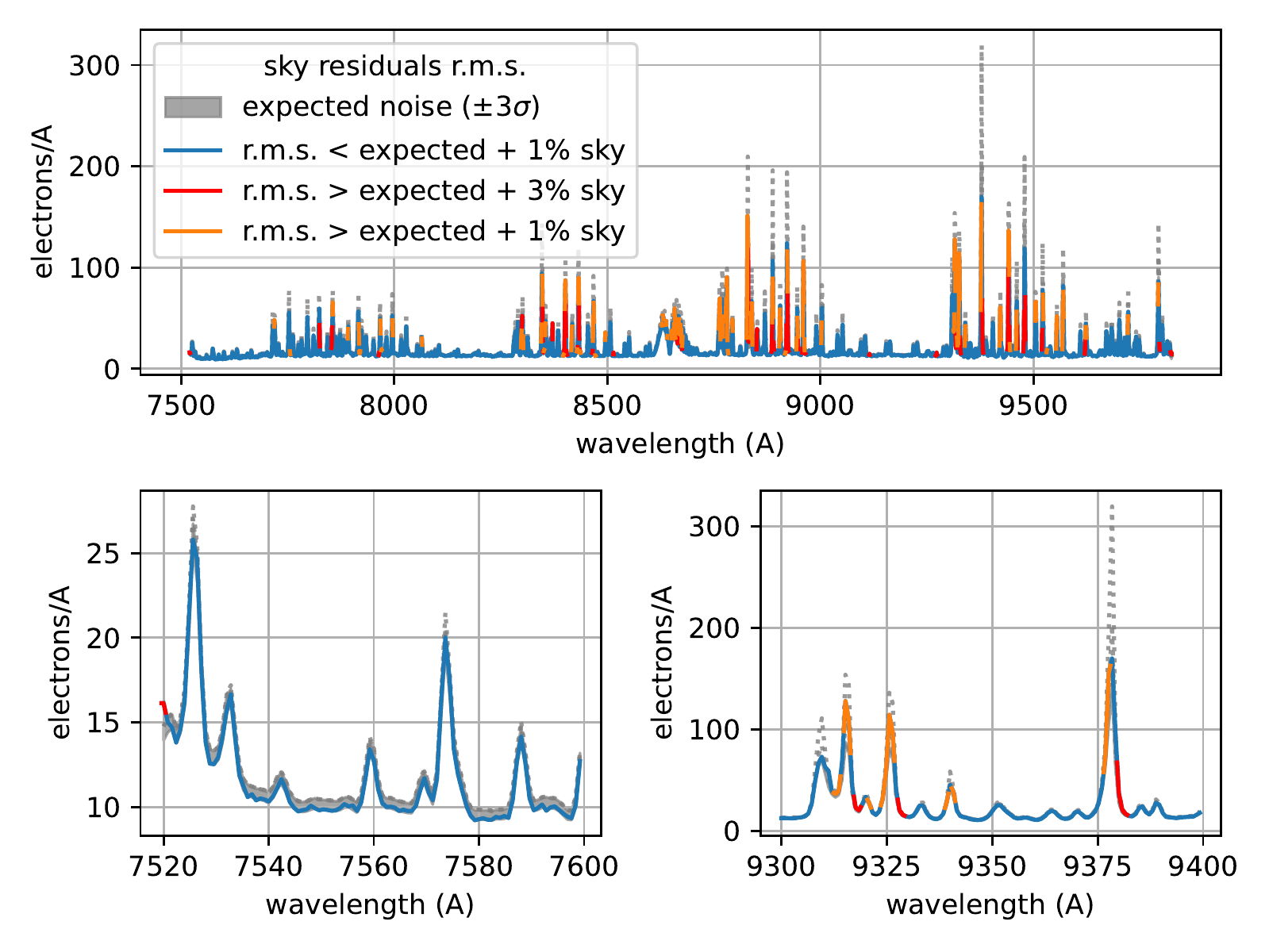}
  \caption{Dispersion of the spectral residuals in sky fibers as a function of wavelength. The value at each wavelength is the RMS of the residuals in all the sky fibers from  all spectrographs, and for several exposures of 1000 $\pm$ 10 sec. The gray band is the expected noise RMS from Eq.~\ref{eq:spectralvariance}, in the ideal case with $\epsilon_\lambda=0$ (but accounting for the flat field and sky model statistical uncertainties). The colored curves are the measured RMS, the colors indicating when the variance exceeds the expected noise. The top panel is the full spectral range of the NIR camera, whereas the lower panels focus on some narrower wavelength ranges. There is naturally a larger rms on the lines because of the Poisson noise, but the measurements also exhibit deviations due to systematic errors in the flat field, wavelength calibration or PSF shape.}
  \label{fig:skyres}
\end{figure}

We evaluate here the performance on the sky subtraction. Figure~\ref{fig:skyres} shows a comparison of the r.m.s. (root mean square) of the spectral residuals in the sky fibers after sky subtraction with the expected noise given by Eq.~\ref{eq:spectralvariance} (with $\epsilon_\lambda=0$), for the NIR cameras. Spectral regions where the rms of residuals exceeds the expected noise are highlighted in orange and red colors, when the extra scatter exceeds 1\% and 3\% of the sky level.

One can see on the figure that the realized noise is consistent with the expectation in the continuum. This is an important validation of the estimation of the noise at the CCD level and its propagation to the spectra.

On bright sky lines however, one can see that the noise is underestimated. The excess noise is typically between 1 and 3\% of the sky level on the brightest lines. This is of course dependent on the sky brightness as the relative contribution from electronic noise, Poisson noise, and systematics in the PSF model depend on the signal level. The data we are presenting here are from science exposures of 1000 $\pm$ 10 sec.  The exposure time, the number of sky fibers used and the observation conditions are typical of the ones found for the main dark time survey of DESI.

\begin{figure}
\centering
  \includegraphics[width=\columnwidth,angle=0]{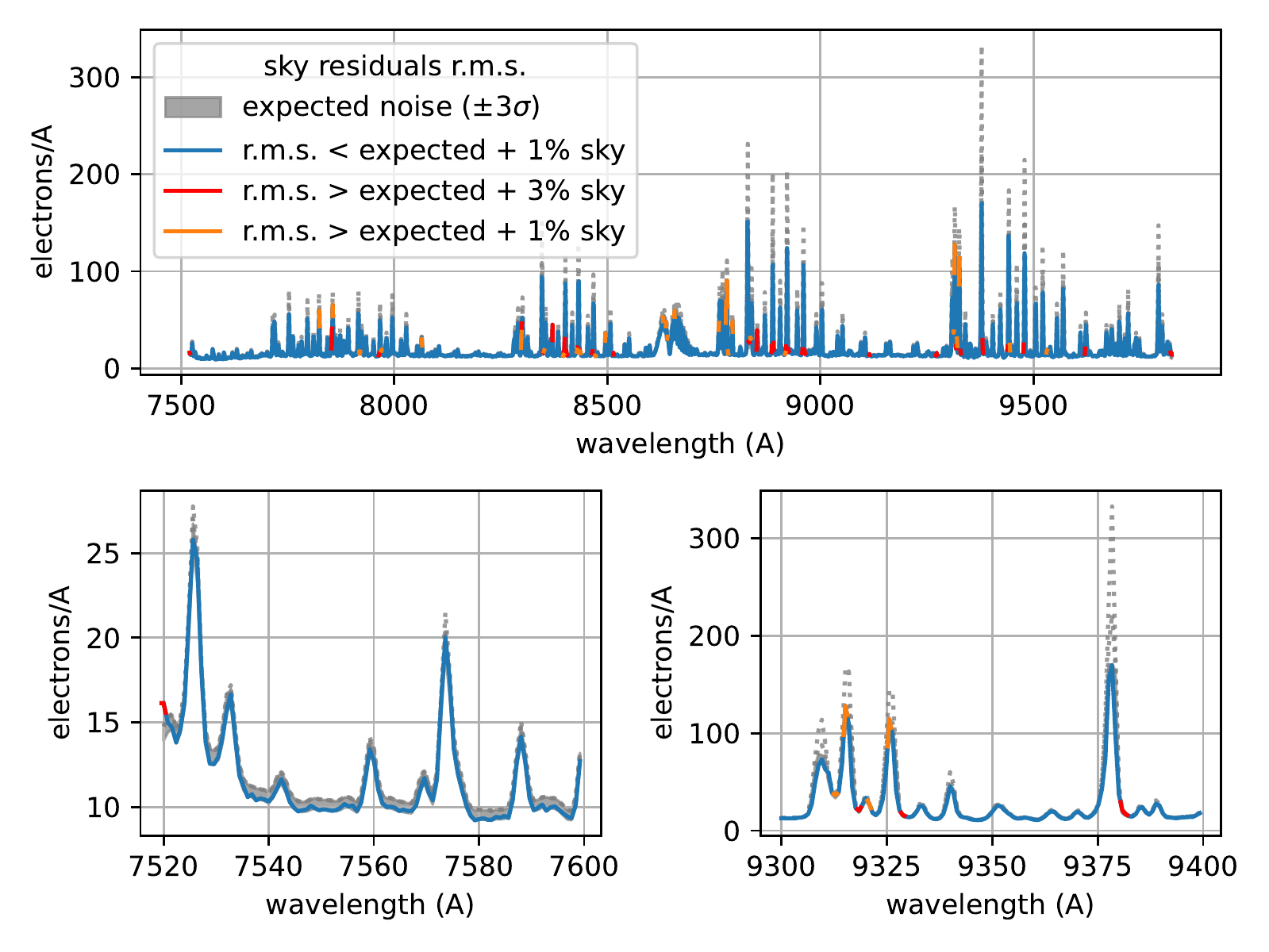}
  \caption{Same as Figure~\ref{fig:skyres} with a variance model including the $\epsilon_\lambda$ term (see Eq.~\ref{eq:spectralvariance}).}
  \label{fig:skyres-extra-variance}
\end{figure}

Figure~\ref{fig:skyres-extra-variance} shows the same dispersion now compared to a variance model where the extra term $\epsilon_\lambda$ has been adjusted. The measured scatter is now consistent with the noise model for most wavelength. This is an important result for the optimization of the redshift fitting.

\subsection{Stellar model fit}
\label{sec:starfit}

The flux calibration relies on a comparison of the measured and expected spectra of standard stars observed during each science exposure. About ten fibers per petal are allocated to standard stars for the main survey, but we benefited from a larger number of standard stars per pointing during commissioning.


The  standard stars are selected from the imaging catalogs based on their $g$, $r$ and $z$ AB magnitudes from MzLS+BASS surveys in the north, and the DECaLS survey in the south (see~\citealt{Dey2019} for a detailed description of the surveys and the effective band-passes). The color cuts designed to select main-sequence F stars are $0 < g-r < 0.35$ and $r-z < 0.2$. An additional magnitude cut was used to select a bright ($15<g<18$) and a faint ($16<g<19$) sample designed for short and long observations. Finally, lower metallicity halo stars are preferentially selected with a cut on parallax  ($<1$\,milliarcsecond, hereafter mas), and peculiar stars with cool kinematics are avoided with a lower limit on the proper motion ($>2$\,mas\,yr$^{-1}$). Both astrometric quantities are extracted from the GAIA DR2 catalog~\citep{GAIADR2}. A comprehensive description of the selection of standard star targets can be found in Section 4.2 of~\citet{Myers22a} (in particular the selection of stars based on GAIA data only in some parts of the sky).

A stellar model is fit to each observed standard star. This fit relies on a grid of stellar models. We describe the templates, the fitting technique, and present early results in the subsequent sections.

\subsubsection{Stellar templates}

We use a theoretical stellar models grid from \citet{AllendePrieto2018}, spanning a wide range of stellar effective temperature ($T_{\rm eff}$), surface gravity ($\log_{10} g$), and iron abundance ([Fe/H]). The spectral coverage of the models ranges from 0.13-6.5\,$\mu$m at a resolution of $R= 10,000$, and the dispersion is constant in $\log \lambda$ with a pixel size of 9.9\,km\,s$^{-1}$.

\begin{figure}
  \centering
\includegraphics[width=0.95\columnwidth,angle=0]{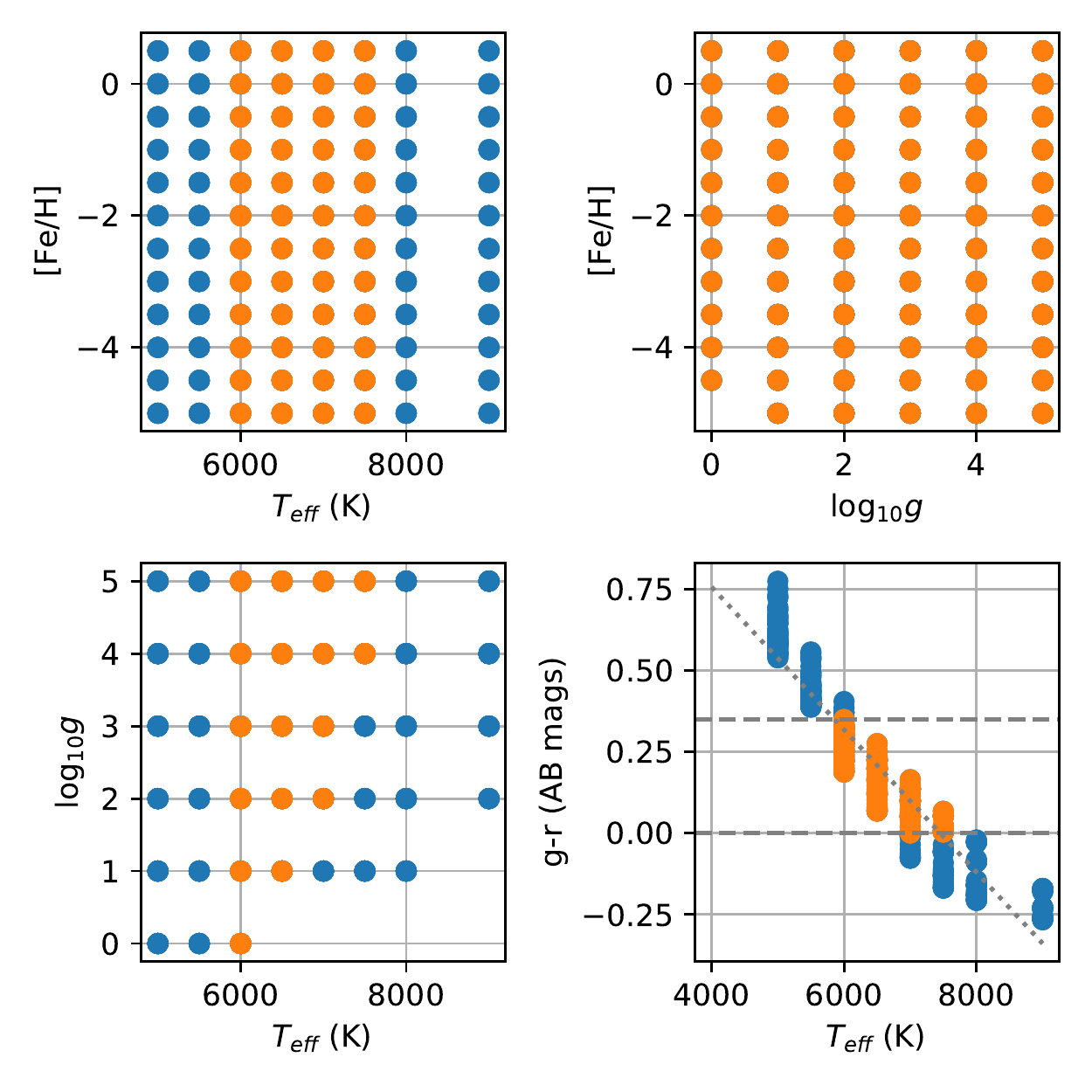}
\caption{Projections of the tri-dimensional stellar parameter grid as a function of effective temperature ($T_{\rm eff}$), surface gravity ($\log_{10} g$), and iron abundance ([Fe/H]). Models satisfying the color selection cuts for the DESI standard stars are marked as orange dots while the others are in blue.}
\label{fig:startemplategrid}
  \end{figure}

Figure~\ref{fig:startemplategrid} illustrates the stellar template grid, focusing on the surface temperature range of the standard stars. The temperature resolution of the grid is 500\,K, $[Fe/H]$ are in steps of 0.5 dex, from -5 to +1, and the surface gravity in steps of 1 dex from 1 to 10$^5$\,cm\,s$^{-2}$. One can see that the $g-r$ color cut selects stars in the temperature range from 5500 to 8000\,K.

\subsubsection{Fit procedure}

The stellar model fit is based on a comparison of the uncalibrated observed spectrum of a star with a model, after both the data and model have been divided by a smooth continuum, effectively circumventing the need for a prior knowledge on the flux calibration.
The smooth continuum is obtained with a sliding median filtering of the original spectrum with a filter width of 160\,\AA. This width was chosen empirically to efficiently erase the variations of throughput while retaining most of the information on the stellar absorption lines.
We perform the fit a second time after a first calibration of the data in order to avoid any systematic bias on the selection of stars due to the absence of original calibration.

The input data comprises the observed spectra of a star from the three cameras (blue, red and NIR) and from one or several exposures when available. The spectroscopic pipeline variance from Eq.~\ref{eq:spectralvariance} is used, with the proper normalization from the smooth continuum, and with an additional variance term of 0.1$^2$ introduced to improve the relative weights in the fit while accounting for the potential modeling errors.
Bright sky lines and atmospheric (telluric) absorption lines are masked (inverse variance set to zero), along with the Ca H\&K absorption lines which receive contributions from the interstellar medium\footnote{We did not mask the Na D-lines which also vary with the interstellar gas column density but may consider this potential improvement for a future data release.}.

The stellar model is a function of five parameters, a normalization based on the $r$-band magnitude from the targeting catalog (set independently and not affecting the fit), a radial velocity, and the three stellar parameters from the templates, $T_{\rm eff}$, $\log_{10} g$, and [Fe/H].

The radial velocity is first fit by cross-correlating the data with a canonical spectral model at $T_{\rm eff}=6000$\,K, $\log_{10} g=4$, and [Fe/H]=-1.5. We have found that cross-correlating the data with the final best fit model does not significantly improve the velocity measurement. For this radial velocity fit, the canonical model is first convolved with the resolution from the data (assuming no radial velocity), then both the model and data are resampled on a fine logarithmic wavelength grid where a change of velocity becomes a simple translation of indices, and both are divided by a smooth continuum before being compared. The scan is performed in the velocity range of $\pm 1500$\,km\,s$^{-1}$ with a step of $15$\,km\,s$^{-1}$. The best fit and uncertainty is given by a parabolic fit around the $\chi^2$ minimum in the scan.

Once the radial velocity is known, the model templates are blue or red-shifted, resampled to the wavelength grid of the data, then multiplied by the corresponding resolution matrix, and finally divided by their smooth continuum.
Because this procedure is CPU intensive, we first select the possible models from the template grid that have a $g-r$ color consistent with the measured color of the star (from the targeting catalog), within a color range of $\pm 0.2$. Prior to this selection, we first correct the measured color for the Milky-Way dust reddening, using \cite{SFD98} map with corrections from \cite{SchlaflyFinkbeiner2011}, and \cite{Fitzpatrick99} dust extinction law. Also, for practical interpolation purposes, we extend the selection to the models in the smallest rectangular region of the 3D parameter space that includes all of the color selected models. The model for a given set of parameters $T_{\rm eff}$, $\log_{10} g$, and [Fe/H] is a trilinear interpolation of spectra from the 8 closest nodes in the 3D parameter grid. The fit is a least square minimization with inverse variance weight. We start from the best fit node in the grid and refine the parameters with a linear combination of its neighbors in the grid.

\begin{figure}
  \centering
\includegraphics[width=\columnwidth,angle=0]{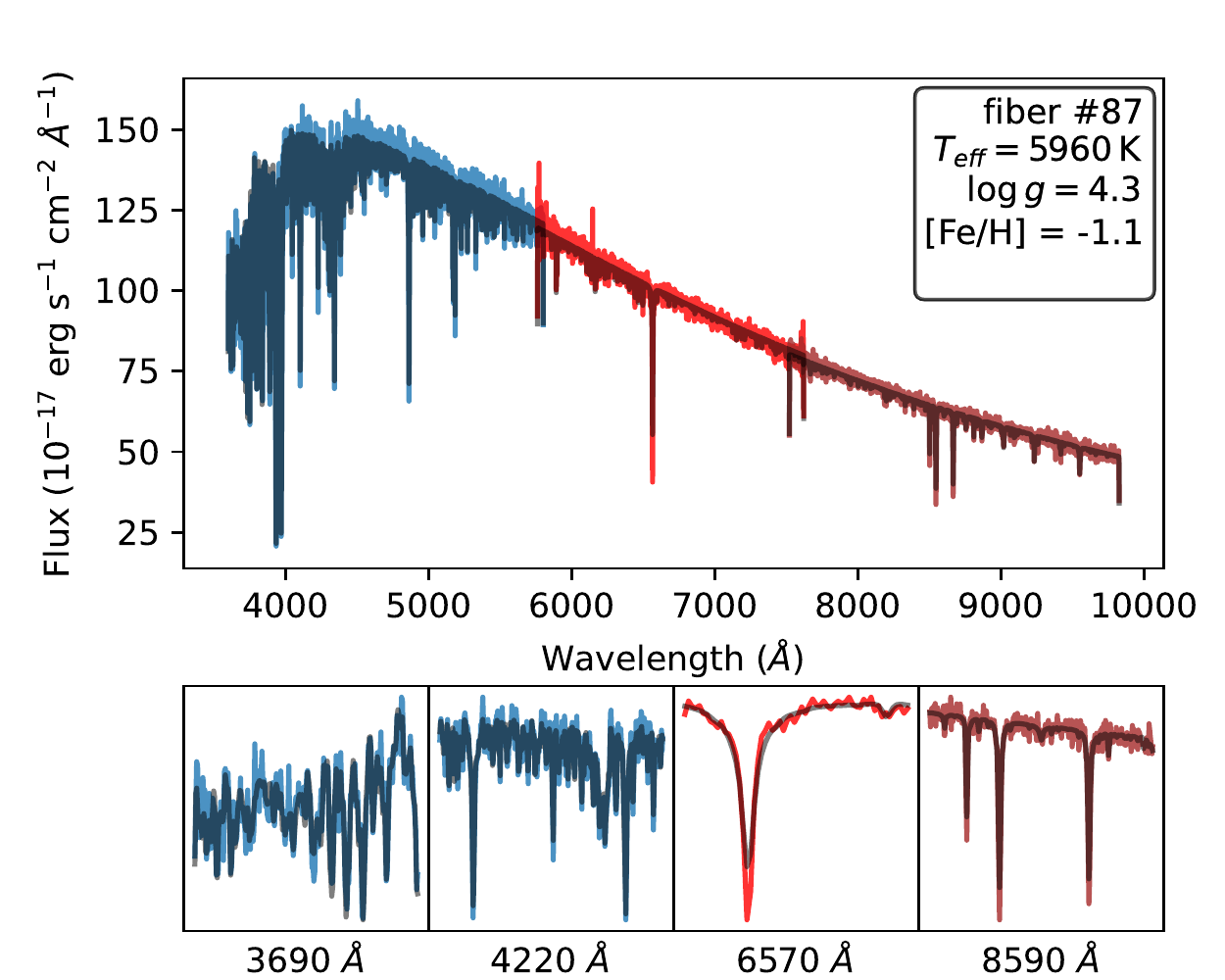}
\caption{A typical standard star spectrum from an exposure of 900s (colored curves) along with the best fit model (black curve). The upper panel shows the full spectrum while the lower panels present zooms in various wavelength range. For these, the model has been renormalized to better fit the data and highlight the agreement on the absorption lines.}
\label{fig:starspectrum}
\end{figure}

Once the best fit linear coefficients are found, the best fit spectrum is recomputed from the templates (with red-shifting, but without the resampling, resolution convolution and continuum normalization), dust extinction is applied to the model and it is then normalized to get the same $r$-band magnitude as the measured one from the photometric catalog.
An example fit is shown on Figure~\ref{fig:starspectrum}.

\begin{figure}
  \centering
\includegraphics[width=0.95\columnwidth,angle=0]{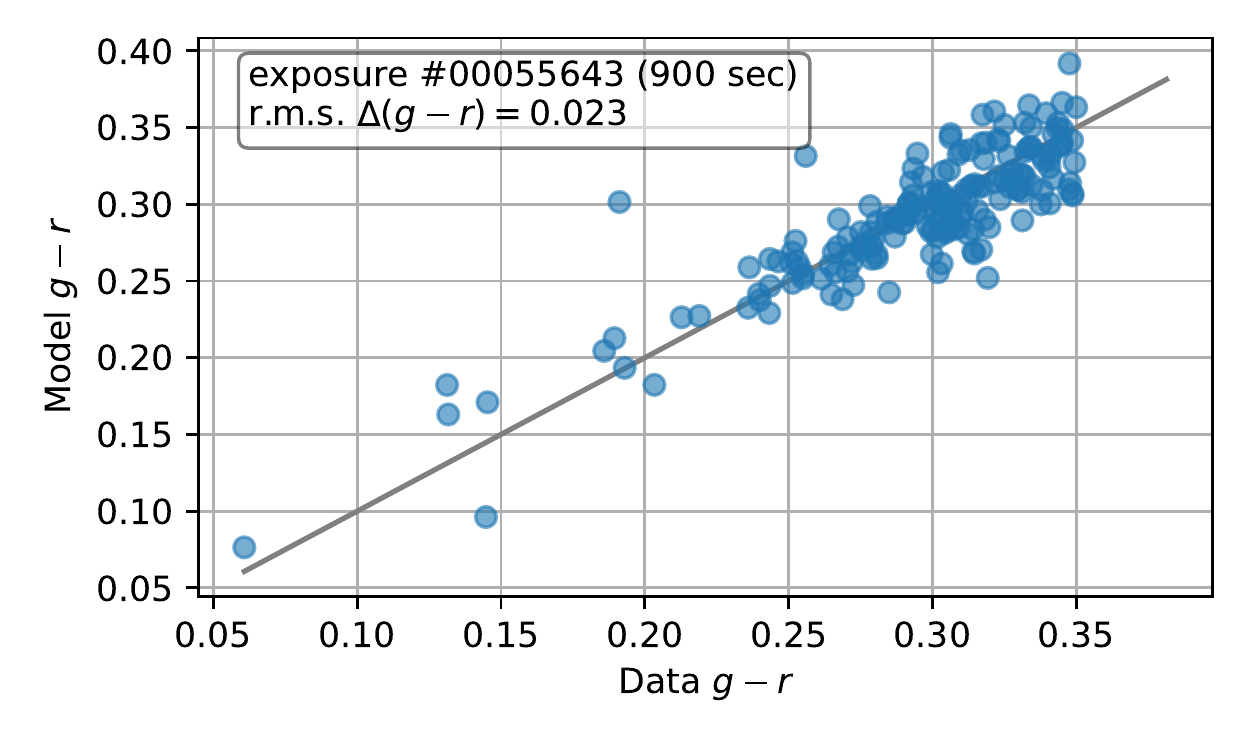}
\caption{Comparison of the $g-r$ colors from the targeting catalog with the colors inferred from the stellar spectrum fit. Each dot is a standard star from a typical 900sec exposure.}
\label{fig:stdstar-color-color}
\end{figure}

An objective criterion to validate the quality of the temperature fit is to compare the measured $g-r$ color of the stars from the targeting catalog with the $g-r$ color computed from the best fit spectral model.
These colors are compared in Figure~\ref{fig:stdstar-color-color} for one of the science exposures (a 900 sec exposure). The agreement between the model and measured color is excellent, with an r.m.s. of 0.023, corresponding approximately to a precision of approximately 100\,K on the surface temperature of the stars.

\subsection{Flux calibration}
\label{sec:fluxcalibration}

In this section is described the flux calibration of each exposure. It consists in converting counts in electrons per unit wavelength into spectral energy distributions. The default calibration that is applied to all spectra provides an estimate of the total flux for point sources. It is however only a lower limit on the total flux for extended sources. In order to obtain a more useful calibration for those, we also provide a coefficient (\texttt{PSF\_TO\_FIBER\_SPECFLUX}, see \S\ref{sec:fiber-aperture-loss-correction}) to convert the measurements to ``fiber fluxes'', corresponding to the flux one would record in a fiber of angular diameter 1.5$\arcsec$ observed for a seeing of $1\arcsec$ FWHM\footnote{Same fiber flux definition as in the Legacy Surveys Data Release 9 catalogs, see \url{https://www.legacysurvey.org/dr9/catalogs/}}. Fiber fluxes are designed to be valid both for point sources and extended sources. This flux calibration is performed as follows.

\subsubsection{Selecting the stars}
We first measure for each standard star from the 10 petals the ratio between the measured fiber flat-fielded flux in r-band and the expectation from the stellar model obtained in the previous section. The numerator is the average inverse variance weighted flux in the wavelength range [6000-7300]\,\AA\ and the denominator is the average model flux in the same range. The distribution of those ratios is used to exclude outliers at three standard deviation. Note that the fiber flat field includes a correction from spectrograph to spectrograph so that we can compare the fluxes from different spectrographs.

Stars for which the difference between the model and observed $g-r$ color from the imaging catalogs exceed $0.1 + 0.2 \times E(B-V)$ are also excluded. The second term relaxes the color criterion in regions of large Galactic extinction because of uncertainties in the dust reddening law.

\subsubsection{Average flux calibration over fibers, per camera}
Once the stars are selected, the calibration is performed for each spectrograph camera individually.

The camera flux calibration vector as a function of wavelength (hereafter $C_j$) is computed as the average ratio over fibers between the flat-fielded but otherwise uncalibrated measured spectra and the model of standard stars. Each stellar model spectrum (hereafter $M_{k,j}$, with $k$ and $j$ being respectively the fiber and wavelength indices) is first resampled to the wavelength grid of the data.

The calibration vector is obtained by minimizing the following.
 \begin{eqnarray}
   \chi^2 &=& \sum_{k} \sum_{i} w_{k,i} \left( \tilde{F}_{k,i} -  T_{k,i} \sum_j R_{k,i,j} M'_{k,j} \right)^2 \\  \nonumber
   &&\mathrm{with} \ M'_{k,j} \equiv P(x_k,y_k,\lambda_j) S_j + \alpha_k C_j M_{k,j} \label{eq:fluxcalchi2}
 \end{eqnarray}

The notations are the same as for Eq.~\ref{eq:skymodel}, with the additional terms $C_j$, $M_{k,j}$ defined above, and a scale factor $\alpha_k$ per standard star (the sum also applies this time to the standard star fibers and not the sky fibers).

We iteratively i) fit the mean calibration vector, ii) compute the scale factors $\alpha_k$ to correct for possible offsets in the fiber positioning, iii) reject outlier fluxes as a protection against residual spikes from unmasked cosmic ray hits. For step ii), we fit for a ``deconvolved'' mean calibration vector that we multiply by the model flux before convolving the product using the resolution matrix of each standard star fiber. At the end of this iterative procedure we use the resolution matrix of each of the 500 fibers of the camera to obtain a first estimate for the calibration vector of each fiber.

\subsubsection{Corrections for fiber aperture losses}
\label{sec:fiber-aperture-loss-correction}

As described in \S\ref{sec:expo-data}, a table of fiber positioning offsets is part of the data product of each exposure.
We use this information to improve the calibration while also considering the mean seeing (sometimes called image quality) of the exposure, the sources surface density profiles and their measured fiber fluxes from the imaging catalogs.
We first estimate a ``flat to PSF'' correction coefficient valid for point sources only and apply it to the reported spectro-photometrically calibrated fluxes. We also determine another ``PSF to fiber'' coefficient to obtain the fiber fluxes that can be used for extended sources. The latter is not applied to the spectral fluxes but saved in the \texttt{FIBERMAP} table in the column \texttt{PSF\_TO\_FIBER\_SPECFLUX} (see \S\ref{sec:data-products}).

The focal plane PSF has been precisely characterized using the GFA images. It follows to a good approximation a Moffat profile with a parameter $\beta=3.5$~\citep{Moffat1969,Meisner2020}. We use this property, the effective fiber angular radius $r_F$ (which depends on the fiber location in the focal plane as the plate scale varies), the seeing (hereafter $s$, measured as the FWHM), and the fiber positioning offset $o$  to compute the fiber acceptance fraction for a point source that we call $F_{PSF}(s,r_F)$ in the following. It is the computed ratio of the flux entering a fiber to the total flux from a source on the focal surface. The fiber positioning offsets are used in the computation of the fiber acceptance.

The ``flat to PSF'' correction is
\begin{equation}
  c^{\ PSF}_{flat}(s,r_F) \equiv \frac{\left< F_{PSF}(s) \right>}{F_{PSF}(s,r_F)} \frac{r_F^2}{\left< r_F^2 \right>} \label{eq:c_flat_psf}
\end{equation}

where $\left< X \right>$ represents the average of $X$ over the focal plane. The second factor corrects for the variation of the fiber flat field caused by the change of solid angles of the fibers (i.e. the Jacobian of the transformation from angles to focal plane coordinates).

The ``PSF to fiber'' correction depends on the fiber acceptance fraction for an extended source, whose surface density is modeled with an exponential profile of half light radius $r_{1/2}$. We have not considered using more precise galaxy profiles for this correction because the resulting calibration was found satistfactory. We label it $F_{exp}(s,r_F,r_{1/2})$. Keeping in mind the definition of fiber flux given at the beginning of this section, this correction is

\begin{equation}
  c^{\ fiber}_{PSF}(s,r_F,r_{1/2}) \equiv \frac{F_{exp}(1\arcsec,1.5\arcsec/2,r_{1/2}) F_{PSF}(s,r_F)}{F_{exp}(s,r_F,r_{1/2})} \label{eq:c_psf_fiber}
\end{equation}

\begin{figure}
  \centering
  \includegraphics[width=0.99\columnwidth,angle=0]{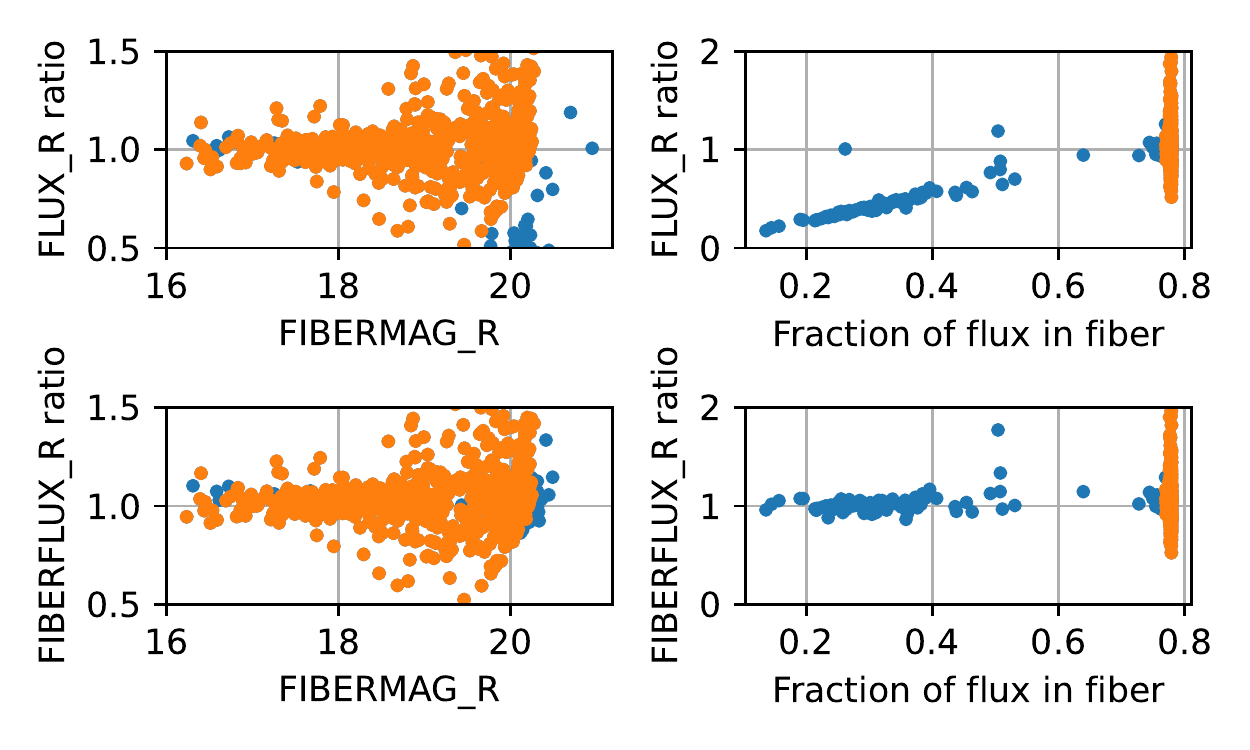}
  \caption{Ratio of spectroscopic to imaging r-band flux for the targets from several main survey dark-time tiles. The top panels compare the total flux which is only valid for point sources for DESI. The lower panels compare the fiber fluxes which are consistent between spectroscopy and imaging for both point sources (in orange) and extended sources (in blue). The x axis is the imaging fiber magnitude on the left panels and the fraction of light in the fiber aperture on the right panels.
}
  \label{fig:imaging-spectro-fiberflux-r-ratio}
\end{figure}

Figure~\ref{fig:imaging-spectro-fiberflux-r-ratio} shows the ratio between the spectroscopic flux in the DECam r-band and the same quantity from the imaging catalog. Both the total flux (\texttt{FLUX\_R}) and the fiber flux (\texttt{FIBERFLUX\_R}) are considered. For the spectroscopy, this consists in multiplying the integrated flux by the coefficient \texttt{PSF\_TO\_FIBER\_SPECFLUX}.
The figure demonstrates that the  fiber fluxes are consistent between spectroscopy and imaging for extended sources. On the bright end, the scatter in the ratio is of 6\%.

\begin{figure}
  \centering
  \includegraphics[height=0.3\columnwidth,angle=0]{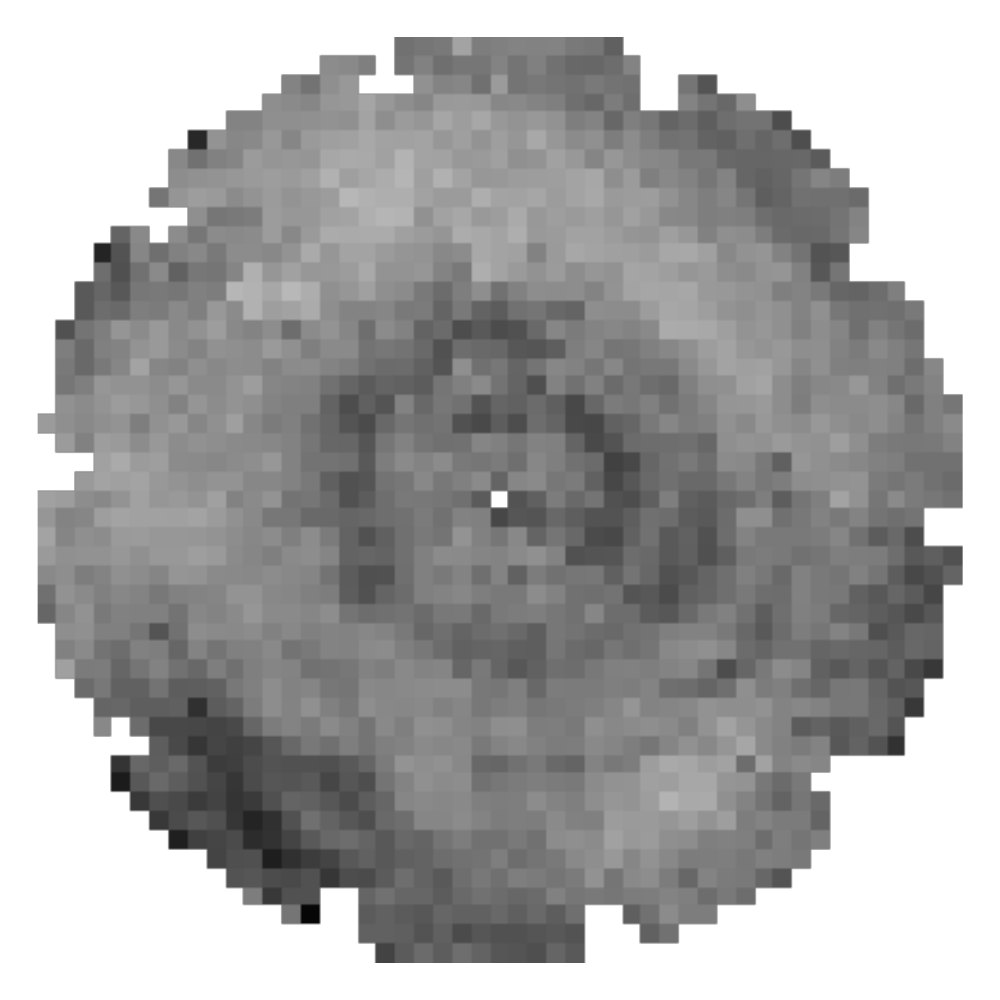}
  \includegraphics[height=0.3\columnwidth,angle=0]{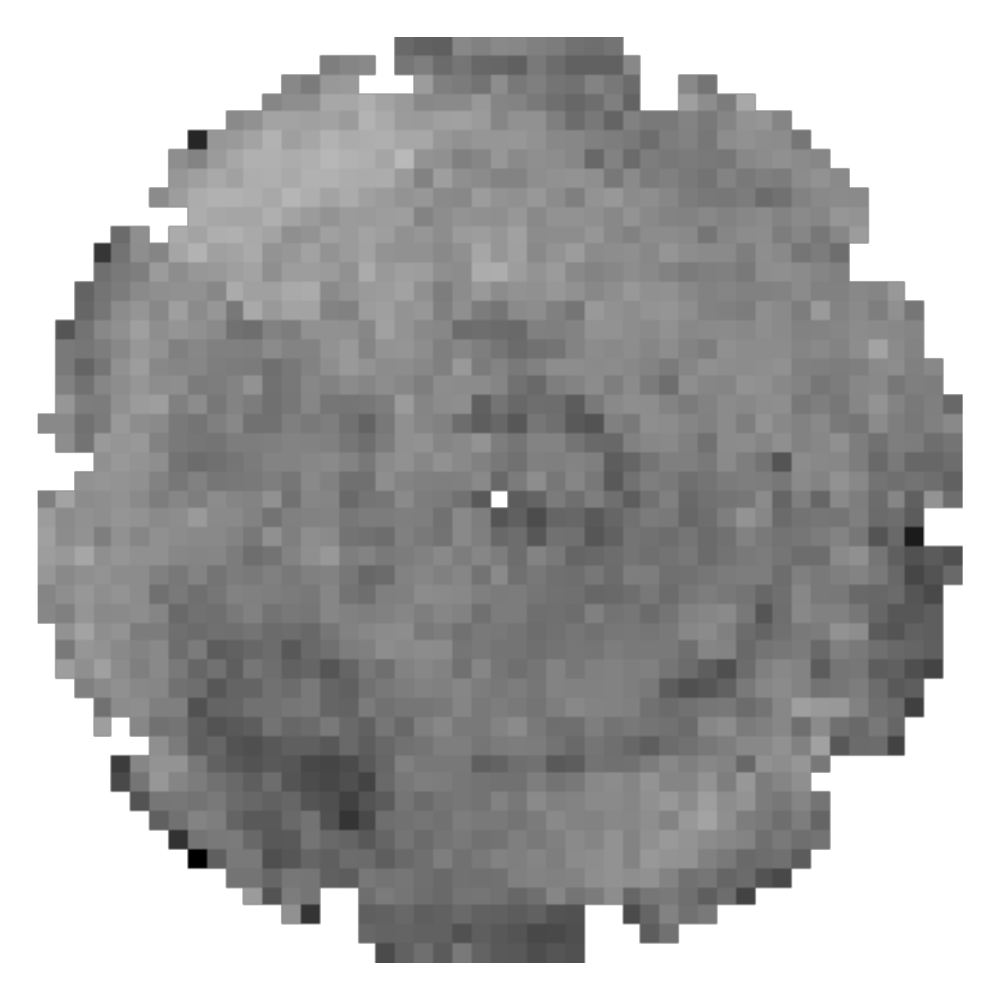}
  \includegraphics[height=0.305\columnwidth,angle=0]{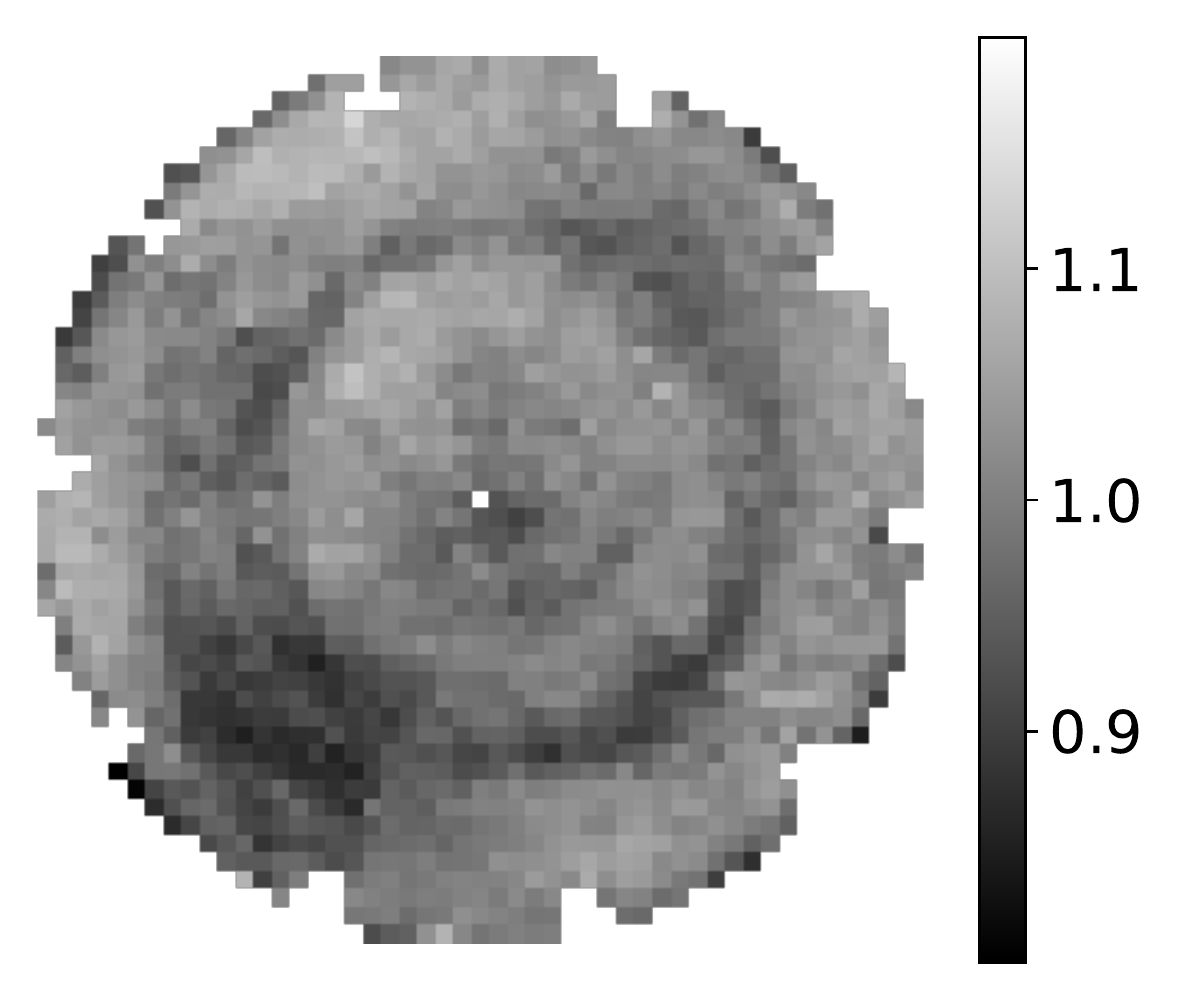}

  \caption{For left to right, average ratio of the spectroscopic to imaging flux of standard stars in the blue (4500-5500\,\AA), red (6000-7300\,\AA), and NIR (8500-9800\,\AA) as a function of focal plane coordinates.
}
  \label{fig:dflux-vs-xy}
\end{figure}

Figure~\ref{fig:dflux-vs-xy} shows color maps of the average ratio of spectroscopic to imaging fluxes for stars, in blue, red and NIR bands as a function of focal plane coordinates. While the distribution is mostly flat for the red some structures are found in the other bands. Those are due to chromatic distortions in the corrector. A variation of 10\% is typically caused by a offset of about 0.3$\arcsec$ or 20$\mu$m. This is irreducible in the sense that one has to choose which wavelength to optimize for the fiber positioning. We have chosen for DESI to optimize the throughput in the red. Note that this systematic calibration error is not corrected in the Early Data Release (EDR).

\subsubsection{Flux calibration testing with white dwarfs}

\begin{figure}
\centering
\includegraphics[width=\linewidth]{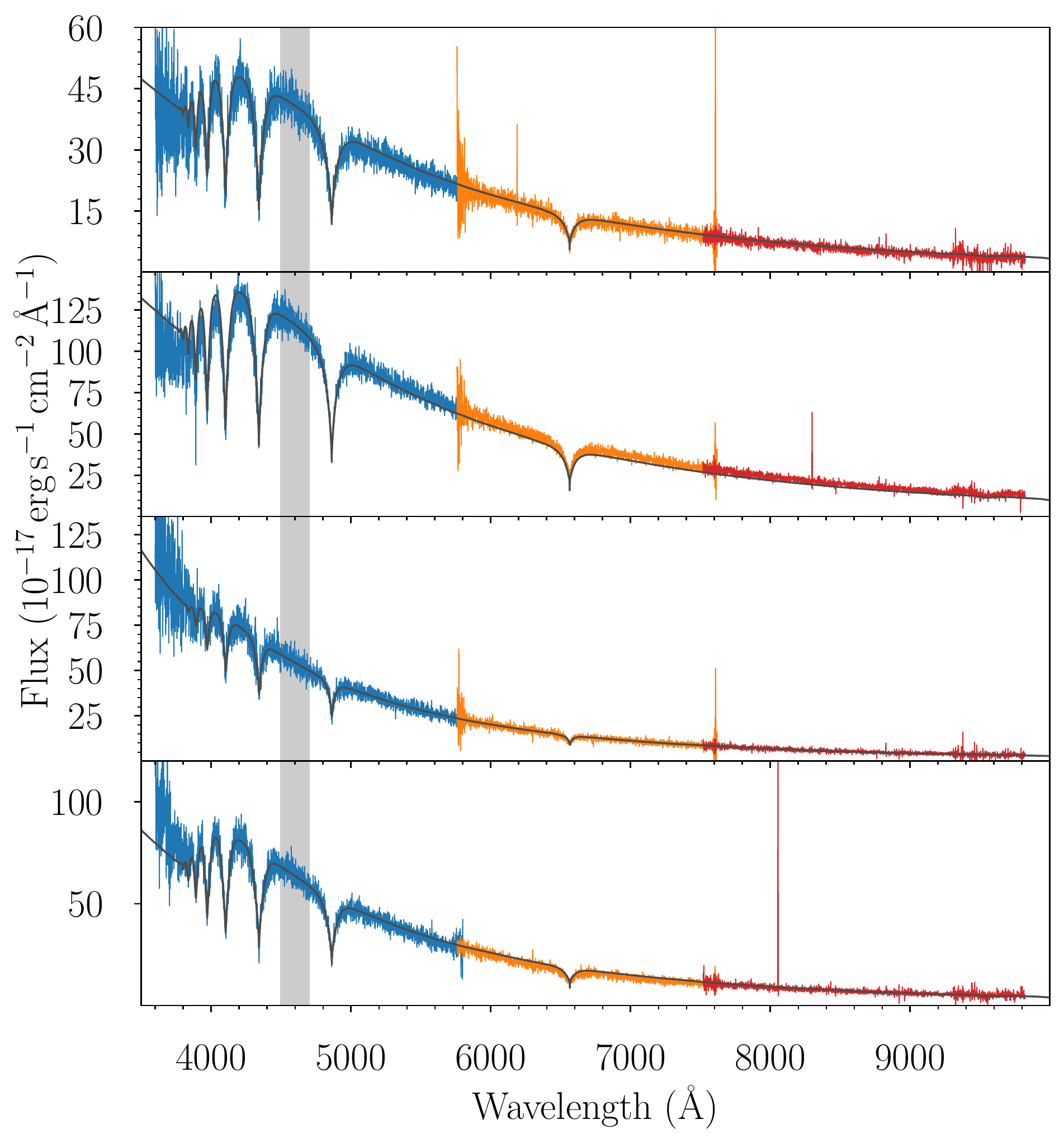}
\caption{White dwarf spectra observed by DESI showing the individual $b$- (blue), $r$- (orange), and $z$- (red) arm spectra and best fitted models from the WD pipeline~\citep{cooper22a} shown in gray, which have been scaled by a constant factor by normalizing to the region shaded in gray.}
\label{fig:wd_fits}
\end{figure}

Given the relatively simple spectra of hydrogen-dominated atmosphere white dwarfs and their largely blue-dominated spectra both in terms of spectral features and flux (Fig.~\ref{fig:wd_fits}), they serve as good tests for flux calibration and as such were observed by DESI. White dwarfs were selected in DESI based on criteria described in~\cite{cooper22a}, which was itself adapted from the selection made in \cite{gentilefusilloetal19-1}.

To test the flux calibration in DESI, we first cross-matched the observed DESI targets with the $Gaia$ DR2 catalog collated by \cite{gentilefusilloetal19-1}, which provides additional parameters beyond those in the $Gaia$ DR2 photometric catalog, including the probability, $P_{\textrm{WD}}$, that a white dwarf candidate is indeed a white dwarf. A white dwarf candidate is considered ``high-confidence’’ if $P_{\textrm{WD}} > 0.75$. This initial cross-match resulted in 4064 unique white dwarf candidates that were observed as part of the DESI EDR, with a total of 31734 exposures showing the many repeat spectra obtained.

\begin{figure}
\centering
\includegraphics[width=\linewidth]{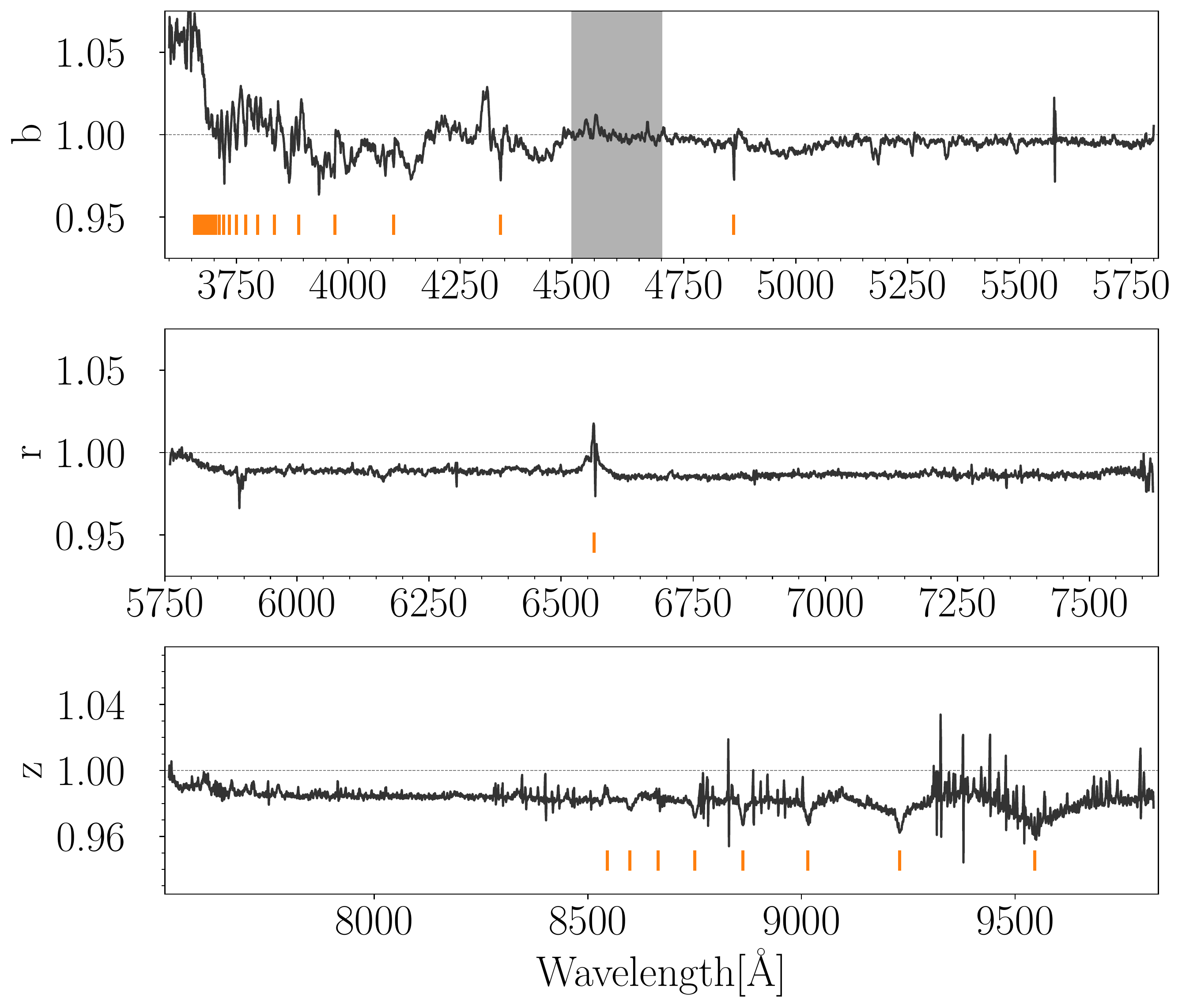}
\caption{The average residuals (black) to white dwarf spectra observed by the three spectroscopic arms of DESI and fitted by the WD pipeline~\citep{cooper22a}. The gray shaded region indicates the wavelength range over which each model was scaled to the white dwarf $b$-arm spectrum. Orange tabs denote the wavelengths of the Balmer and Paschen lines, in particular highlighting the Balmer jump around where the largest discrepancy in the residuals are observed.}
\label{fig:wd_calib_compare}
\end{figure}

To use these white dwarf candidate spectra for comparing the DESI flux calibration, we limited our sample to the highest confidence white dwarf candidates selecting $P_{\textrm{WD}} > 0.95$, spectra with a signal-to-noise ratio in the $b$-arm $> 5$, and white dwarf candidates identified as hydrogen-dominated (i.e. DA white dwarfs) based on a random forest classifier~\citep{breiman2001}. This resulted in 9854 spectra which were all fitted using the \textsc{WD} pipeline described in~\cite{cooper22a}, and examples of these fits to individual spectra are shown in Fig.\,\ref{fig:wd_fits}. These fits are applied to and scaled by a constant value to the $b$-arm DESI spectra, and do not use any information in the $r$-, or $z$-arm spectra. From these fits, we calculated a single weighted-average of the residuals from these 9854 fits which is shown for the DESI spectral range in Fig.\,\ref{fig:wd_calib_compare}.

The \textsc{WD} fits in the $b$-arm show some structure but are largely constrained within $\pm 2$\% and to wavelengths below $\simeq$\,4500\,\AA. The largest discrepancy is seen below $\simeq$\,3700\,\AA\ where a rapid change in the residual results in a $\simeq$\,6\% increase that then remains constant to the blue-edge of the spectrum.
The $r$- and $z$-arm comparisons show a slight offset of roughly $-1$\%, which is not unexpected when the spectra are fitted to the $b$-arm and normalised to a small range within that arm. There is evidence of small-scale features in the residuals that deviate from this offset, which appear to be largely constrained to H$\alpha$ in the Balmer series, and the Paschen series. Further work is needed to identify the origin of these features, as it is likely that some of these are generated from the \textsc{WD} fitting pipeline rather than the flux calibration itself. Additional improvements can also be made to the comparison by including E(B-V) values into these fits, which are currently not included.

\subsection{Cross-talk correction}
\label{sec:fibercrosstalk}

The spectrograph PSF tails extend well beyond the separation between adjacent fibers and cause a measurable contamination among neighboring spectra. As explained in \S\ref{sec:psf}, these tails were not included in the PSF model used for the extraction. The result of this omission is the presence some residual fiber cross-talk among the extracted spectra.

The PSF tails can asymptotically be described as a power law (at large radius $r$). We model them as a convolution with a kernel of the form $K(r)\propto\ r^2 \, (1+r^2)^{-(1+p/2)}$, where $r$ is the distance from the PSF center in units of pixels and $p \simeq 2.5$ the power law index (for $r \gg 1$).

\begin{figure}
  \centering
  \includegraphics[width=0.9\columnwidth,angle=0]{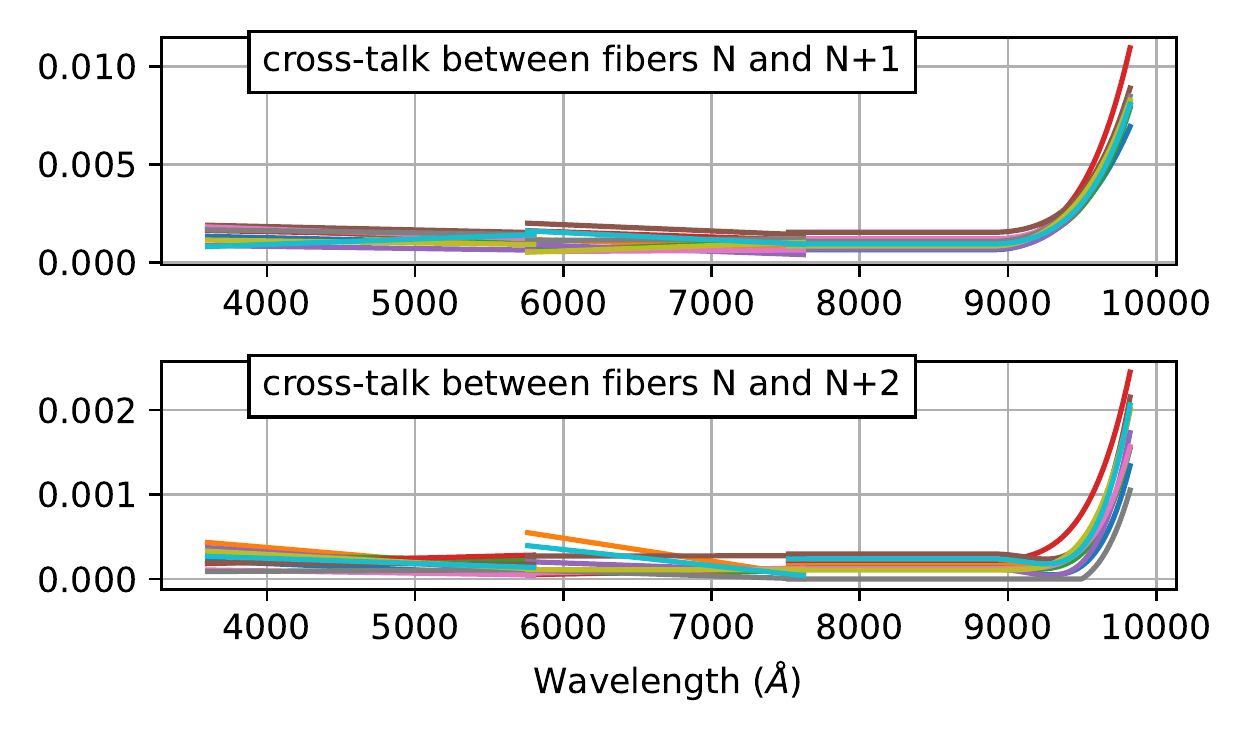}
  \caption{Measured cross-talk between fibers $N$ and $N+1$ (top) or $N+2$ (bottom) for all 30 CCD cameras, as a function of wavelength, for a central fiber ($N \sim 250$).}
  \label{fig:fibercrosstalk}
\end{figure}

We estimate the contamination of neighboring spectra using the kernel $K(r)$ and the pixel coordinates of the spectral traces in the CCD. Corrections for nearest and second nearest fibers are computed. The normalization of the kernel has been fit as a function of  the wavelength, fiber number, and fiber separation for each CCD camera independently. It has been obtained by comparing the residual signal in sky fibers after sky subtraction with a model derived from the neighboring fibers, while taking into account the variations of fiber transmission. Several hundred exposures were used to reduce the statistical uncertainties.
For the blue and red camera, the cross-talk is in the range 0.1-0.2\% for adjacent fibers. It varies slowly with wavelength and we have resorted to use the same value for each fiber of a given camera. In the NIR camera however, the cross-talk is rapidly increasing with wavelength for wavelength above 8900\,\AA. It reaches values as large as 1.5\% at the maximum wavelength of 9800\,\AA. It is also a function of the fiber location in the CCD. The cross-talk values for all cameras are shown in Figure~\ref{fig:fibercrosstalk}. The resulting contamination among fibers from the same camera is estimated from the extracted spectra and subtracted (we do not iterate the procedure because the correction is small).

\subsection{Co-addition}
\label{sec:coaddition}

For some of the delivered data products, the spectra from the same target are co-added, i.e. averaged, per camera. Co-added spectra from all of the past observations of a tile are computed as part of the daily processing and for data releases. We also provide co-added spectra of targets observed on multiple tiles (for instance Lyman-$\alpha$ QSOs), and group them per {\it Healpix} pixel on the sky (with $N_{side}=64$, corresponding to pixels of about $0.8\,\deg^2$, see~\citealt{Healpix}); see also~\S\ref{sec:data-products} for more details about the data products.

The fact that the wavelength grid is the same for all fibers in all exposures (and across cameras for their overlapping wavelength) simplifies greatly the co-addition over exposures: it is a simple weighted average per wavelength element, optionally with outlier rejection. The weights are the inverse variance of the calibrated spectra. They are the statistically optimal weights for non-variable objects. For instance exposures with poor sky transparencies are properly de-weighted as the inverse variance of the calibrated spectra scales with the square of the sky transparency. Co-added spectra still have uncorrelated fluxes because no re-sampling is involved in the process.

The resolution matrix of the co-added spectra is the weighted mean of the resolution matrices of the input spectra, with the same weights as for the flux. This is obvious if one considers that the expected values of the extracted flux $F$ are the same for all input spectra  (contrary to the recombined fluxes $\tilde F$ which depend on the resolution matrices of the input spectra, see \S\ref{sec:extraction}).
\begin{equation}
  \langle \tilde F_{coadd} \rangle =  \sum_i w_i \langle \tilde F_i \rangle = \left( \sum_i w_i R_i \right)  \langle F \rangle \equiv R_{coadd} \langle F \rangle \label{eq:ftilde_coadd}
  \end{equation}

For the co-addition of DESI main survey spectra, we use an outlier rejection threshold of 4 standard deviation as a last barrier to discard spurious spikes due to undetected cosmic ray hits. We do not co-add spectra across cameras before fitting redshifts to avoid averaging data of different resolution and losing information for the detection of sharp spectroscopic features.

\subsection{Classification and Redshift fitting}
\label{sec:redshift}

Spectral classifications and redshifts are obtained using
the Redrock\footnote{\url{https://github.com/desihub/redrock}} software.
Details of the algorithm are provided in \cite{bailey_redrock} and summarized as follows.

Redrock consists in comparing the measured spectra with a series of templates, performing for each of them a redshift scan followed by a refined fit for the best solutions.
It includes principal component analysis (PCA) templates for galaxies,
quasars, and stars.
The galaxy templates are generated from stellar population synthesis and emission-line modeling of galaxies at $0<z<1.5$.
These galaxy templates utilized a grid of theoretical, high-resolution simple stellar population models from C. Conroy (2014, private communication; see also \citealt{Conroy2018}) which are described in more detail in~\cite{bailey_redrock}.

The training sample consists of 20,000 realizations generated by the
desisim package\footnote{\url{https://github.com/desihub/desisim}}
covering restframe wavelengths 1602\,\AA\ to 11000\,\AA\ on a 0.1\,\AA\ grid and
following the expected redshift distribution and target selection color cuts for DESI
Emission Line Galaxies (10k realizations), Luminous Red Galaxies (5k),
and Bright Galaxy Survey (5k).  These 20k high resolution templates are
processed into 10 principal component eigenvectors. The Weighted Expectation Maximization Principal Component Analysis code\footnote{\url{https://github.com/sbailey/empca}} was used for this purpose (see \citealt{Bailey2012}).

Stellar templates are generated from theoretical spectral models of stars and white dwarfs, split by effective temperature $T_{\mathrm{eff}}$ to generate
5 PCA eigenvectors for each of stellar subtypes B,A,F,G,K,M, and White Dwarfs.
Cataclysmic Variables (CV) templates of 3 PCA eigenvectors are generated from archetype CV spectra used by SDSS {\it idlspec2d}~\citep{Bolton2012}.

QSO templates with 4 PCA eigenvectors are currently the same as those used by SDSS {\it idlspec2d}.

For each target at each candidate redshift, the templates are resampled to the wavelength grid of the observed data and multiplied by the resolution matrix described in \S\ref{sec:extraction}, thus providing a template matched to the resolution of each spectrum without resampling the data.
Galaxies scan redshifts $z$ on a logarithmic spacing from $0.005<z<1.7$;
Quasars scan a logarithmic redshift grid from $0.05<z<6.0$, and stars are
sampled on a linear redshift grid from $-0.002<z<0.002$ (i.e.~$\pm$600\,km\,s$^{-1}$).
$\chi^2(z)$ is measured on a finer redshift spacing around the
lowest three $\chi^2(z)$ minima for each template class.  Redrock fits a parabola to these minima to derive the final $\chi^2$ and $z$ of each
minimum.  The template with the lowest $\chi^2$ is selected as the best,
with warning flags indicating if this is within a $\Delta\chi^2<25$ of the
next-best solution, or other potential problems such as being at the edge of the scanned redshift range or failed parabola fits.
A penalty is added to the $\chi^2$ for unphysical models with a negative flux for the [\ion{O}{2}]~$\lambda \lambda 3726,3729$ doublet.
Although Redrock performs well on galaxies and stars, it misses $\sim$10-15\% of true quasars, in particular redder low redshift quasars that are not well represented by the training sample used to create the original SDSS QSO PCA templates (see~\citealt{alexander22a}).  To recover these, the spectroscopic pipeline also runs an improved version of QuasarNET \citep{Busca2018} and an \ion{Mg}{2}~$\lambda \lambda 2796,2803$ line fit to make the final QSO selection as described in \cite{chaussidon22a}. QuasarNET has been improved using additional data, and a new architecture to reduce overfitting \citep{Green2023}.


\subsection{Redshift performance}
\label{sec:redshift-perf}

Details of the performance of the pipeline and survey in terms of redshift success rate and purity are provided in a series of papers reporting on the Survey Validation (hereafter SV, see \citealt{sv22a} for an overview). We give here a short summary of those results, which are an important validation of the whole spectroscopic pipeline.

\cite{hahn22a} present the Bright Galaxy Survey, for which a redshift efficiency of 98.5\% has been obtained for the faint sample with an effective exposure time of 180~sec (see \S\ref{sec:efftime} for a definition of the effective time), with a purity of 99.5\%, with even better results for the bright sample. A comparison with the GAMA survey (DR4, see \citealt{driver2022}) is also provided with 99.7\% of redshifts in common with $| \Delta z | <0.001 (1+z)$.
\cite{zhou22a} present the Luminous Red Galaxy (LRG) sample. With a nominal effective exposure time of 1000~sec, and with quality selection cuts $\Delta \chi^2 > 15$ and $z<1.5$, they obtain a success rate of 98.9\% and a purity of 99.8\%. \cite{raichoor22a} study the Emission Line Galaxy (ELG) sample; they find that with a selection cut involving a combination of $\Delta \chi^2$ and [\ion{O}{2}] signal to noise ratio, they obtain a purity higher than 99.4\%. The ELG redshift success is however lower than for the other target classes because a significant fraction of the galaxy targets do not have bright enough emission lines or are not in the redshift range of interest. The QSO sample is studied in~\cite{chaussidon22a}. They find that the Redrock redshift code has to be complemented with a \ion{Mg}{2} line finder and a neural network classifier (QuasarNET, see~\citealt{Busca2018}) to increase the identification efficiency and avoid line confusion. A combination of the three methods give a redshift efficiency of $98.0 \pm 0.4$\% and a purity of $99.5 \pm 0.4$\%. Those figures have been derived from comparisons with truth tables obtained from deep observations and a large effort of visual inspection conducted by the collaboration \citep{lan22a,alexander22a}.

Despite those good results, the spectral classification and redshift estimation could be further improved. Redrock currently uses PCA templates to fit the data; although this provides speed and flexibility, there is no constraint that the ``best fit'' answer is physically meaningful other than penalizing negative [\ion{O}{2}] flux.  This makes Redrock overly sensitive to false fits, especially when there are data problems upstream, for instance errors in the background subtraction or flux calibration between the spectrograph arms leading to steps in the spectra. Future work may introduce updated templates based on deep DESI observations, archetypes~\citep{Cool2013,Hutchinson2016}, non-negative matrix factorization (NMF) templates~\citep{Blanton2007,Zhu2016}, or other priors focused on ensuring physically meaningful fits. Also, the current Redrock galaxy and stellar templates were generated for DESI, but the QSO templates are still the same as those used by BOSS, trained on just a few hundred QSO spectra.  Although there are many more QSO spectra available now that could be used to make improved QSO templates, initial work has found that improving the QSO performance comes at the cost of degrading the completeness of galaxy targets.  Future releases may have new QSO templates tuned to improve QSO performance while preserving the performance of galaxy templates.

\subsection{Effective exposure time}
\label{sec:efftime}

One key ingredient of the survey optimization is the definition of a spectroscopic average signal to noise ratio (hereafter SNR) and its associated effective exposure time (proportional to the SNR squared). Once this quantity is defined and a minimal threshold is set, the role of the survey operations is to achieve this minimal SNR for all of the tiles (or pointings) that compose the survey in a minimum amount of time (see~\citealt{schlafly22a}). An ideal survey would have all the tiles observed exactly at the threshold; this would both minimize the survey duration and ensure its homogeneity in terms of redshift efficiency.

Among an infinity of possible choices, we have resorted to consider for the SNR definition a quantity related to the inverse of the redshift measurement uncertainty for an ensemble of reference targets. We call this quantity TSNR (for Template SNR) and define it more precisely as a quadratic average for an ensemble of template spectra that follow a predefined redshift distribution. We have defined TSNR values for several target classes (LRG, ELG, QSO, BGS, Lyman-alpha forests) but we can use only one of them to decide on the exposure time. TSNR values are calculated for each fiber of each observation, and we use the average of the LRG TSNR values over all the fibers of a given exposure to define its effective exposure time.

\begin{figure}
  \centering
  \includegraphics[width=0.99\columnwidth,angle=0]{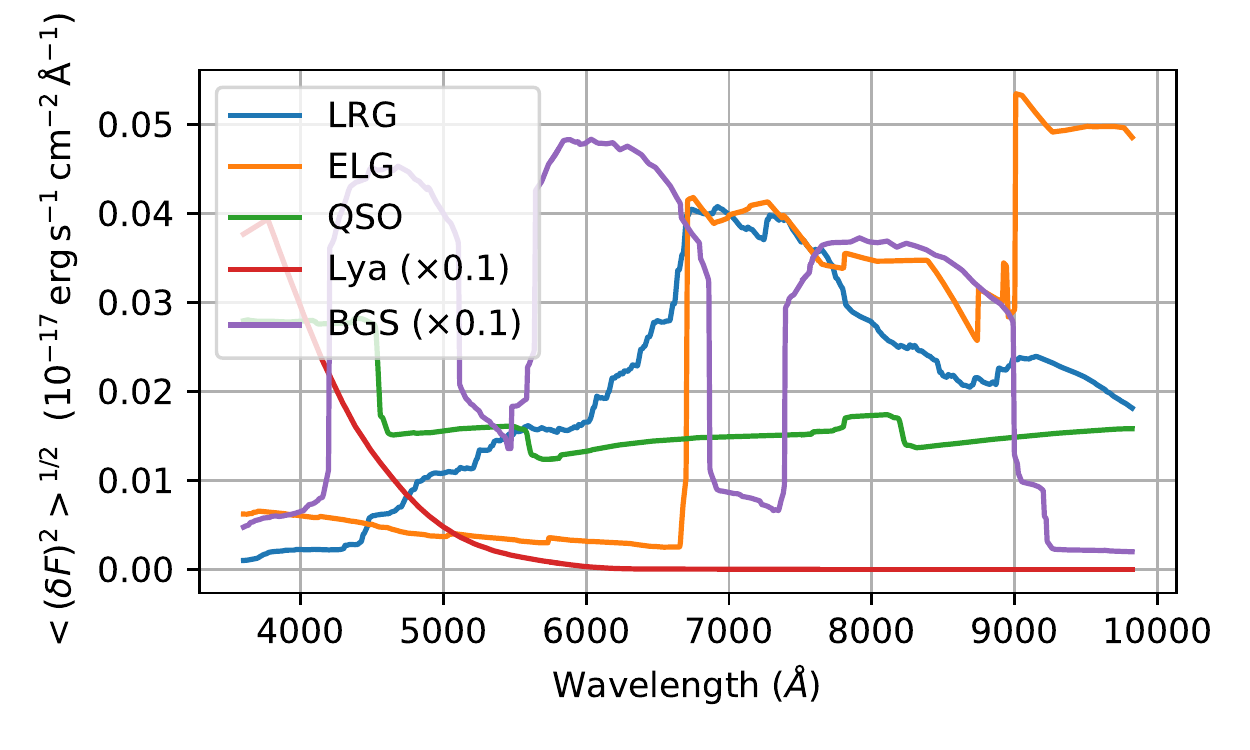}
  \caption{Templates $\sqrt{\left< (\delta F)^2 \right>}$ used to compute the TSNR values for LRG, ELG, QSO, Lyman-$\alpha$ and BGS targets (see Table~\ref{table:tsnr-zrange} for the redshift ranges).}
  \label{fig:tsnr-template-deltaf}
\end{figure}

The TSNR values are measured per camera, fiber and observation as follows:

\begin{equation}
  TSNR^2 = \sum_i  T_i^2 \left< (\delta F)^2 \right>_i / \sigma_i^2  \label{eq:tsnr}
\end{equation}

where $i$ is a spectral wavelength index, $T_i$ the calibration coefficient to convert flux to electrons in the detector (including the throughput times the exposure time), $\sigma_i$ the flux measurement uncertainty. $\left< (\delta F)^2 \right>$ is an average over an ensemble of templates that differs for each target class. $\delta F = F - med(F)$ is one template spectrum minus a median filtered version of the same spectrum (with a width of 100\,\AA).
$\delta F$ is close but different from the derivative of $F$ with respect to the redshift; it was found to be a better indicator of the redshift success rate, which is defined by a minimal $\chi^2$ difference between the best and second best solution of the redshift fit (see~\S\ref{sec:redshift}). The normalization of $\delta F$ does not play any role as it is absorbed in the conversion factor from the TSNR$^2$ values to the effective exposure time (see Eq.~\ref{eq:efftime-spec}). The magnitudes of the template spectra were arbitrarily chosen to match the peak of the magnitude distribution of the targets.
The quantity $\left< (\delta F)^2 \right>$ is shown as a function of wavelength for each target class on Figure~\ref{fig:tsnr-template-deltaf}. The redshift range considered for each target class is given in Table~\ref{table:tsnr-zrange}.

\begin{table}
  \centering
  \begin{tabular}{ll}
    Target & Redshift range \\
    \hline
    BGS & $0.13 - 0.37$ \\
    LRG & $0.68 - 0.97$ \\
    ELG & $0.80 - 1.40$ \\
    QSO & $0.56 - 1.93$ \\
    Lyman-$\alpha$ & $> 2$
    \end{tabular}
  \caption{Redshift range used to compute the term $\left< (\delta F)^2 \right>$ entering in the definition of TSNR (Template Signal to Noise Ratio) for each target class.}
\label{table:tsnr-zrange}
\end{table}

$T_i$ and $\sigma_i$ are derived from the observations. $T_i$ is the throughput term, it includes the fiber aperture losses (see \S\ref{sec:fiber-aperture-loss-correction}). For this calculation, a single reference angular size is considered for each target class\footnote{Half-light radii of 0.45\arcsec, 1\arcsec\ and 1.5\arcsec\ for ELG, LRG and BGS galaxies respectively.} but the specific plate scale and positioning error of each individual fiber, along with the specific atmospheric seeing of the observation is used. $T_i$ also contains the Galactic dust transmission. $\sigma_i$ comprises a read noise term, different for each amplifier of each CCD, with an effective number of CCD pixels derived from the PSF shape, and a Poisson noise term derived from the sky spectrum model (see \S\ref{sec:skymodelfit}), taking into account the specific flat field correction of each fiber (see \S\ref{sec:fiberflat}).

\begin{figure}
  \centering
  \includegraphics[width=0.99\columnwidth,angle=0]{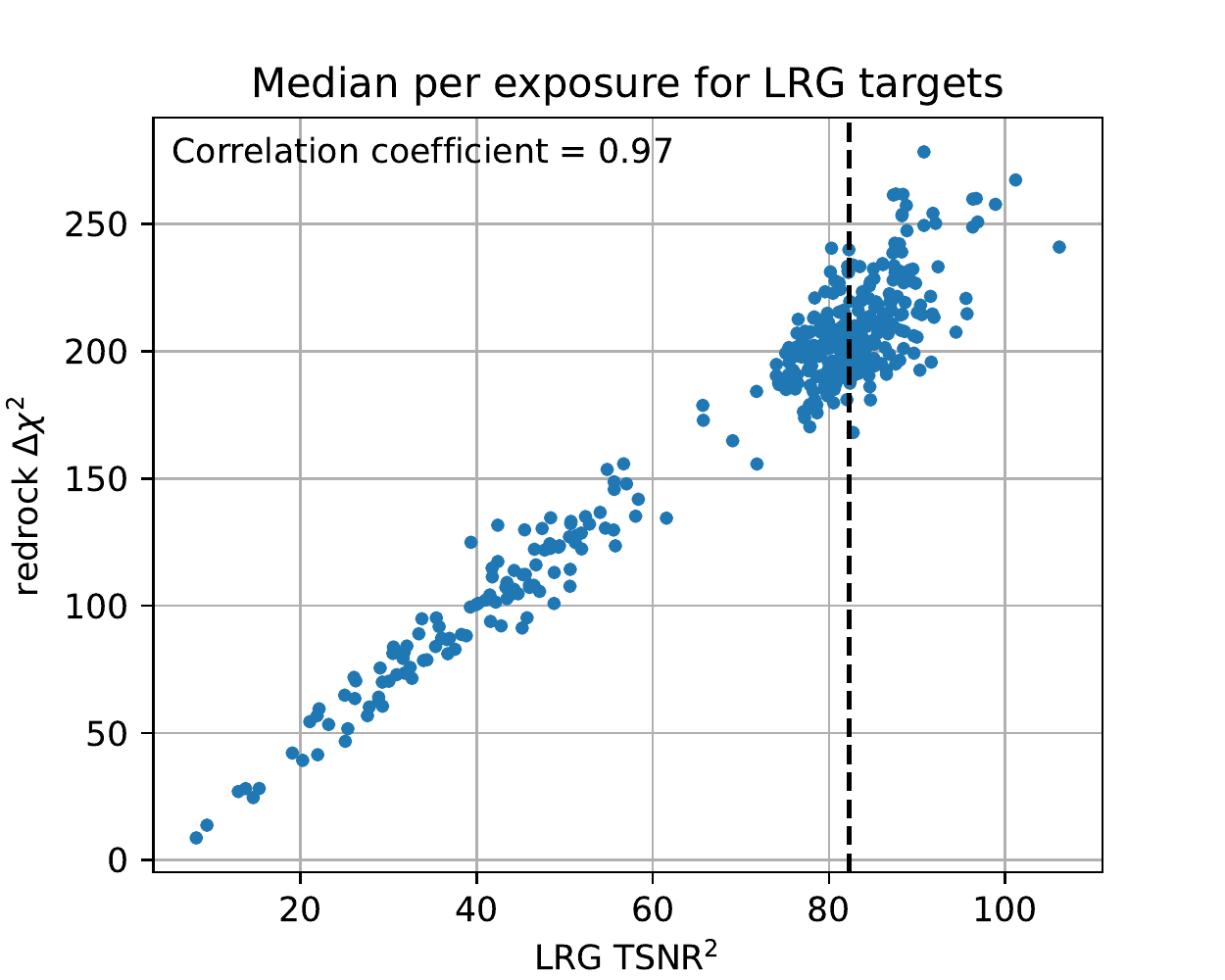}
  \caption{Comparison of the LRG TSNR$^2$ per exposure and the corresponding $\Delta \chi^2$ from Redrock. Each dot corresponds to one exposure; both the TSNR$^2$ and $\Delta \chi^2$ values are the median over the LRG target spectra observed during that exposure. The vertical dashed line indicates the TSNR$^2$ value which corresponds to an effective time of 1000 seconds.}
  \label{fig:tsnr2-vs-dchi2}
\end{figure}

The correlation between the median LRG TSNR$^2$ per exposure and the median $\Delta \chi^2$ from Redrock (see \S\ref{sec:redshift}) is shown on Figure~\ref{fig:tsnr2-vs-dchi2}. The median has been computed for the LRG targets only. A correlation coefficient of 0.97 is found, demonstrating that the TSNR values are a good indicator of the redshift success. A slightly lower correlation coefficient is obtained if the TSNR value for ELGs is used instead to the one for LRGs (the difference is due to the different weights as a function wavelength given by the values of $\left< (\delta F)^2 \right>$, the ELGs being more sensitive to the signal to noise in near infrared, see Figure~\ref{fig:tsnr-template-deltaf}).

As the signal to noise ratio increases as the square root of the exposure time if the noise is dominated by the Poisson noise from the sky background and not the read noise, it is natural to define an effective exposure time proportional to the TSNR$^2$ values.

The normalization is such that the effective exposure time corresponds to an actual exposure time when observing in nominal conditions, for a dark sky, ideal transparency, at zenith, without Galactic dust extinction, and for a median seeing of 1.1\arcsec. Based on this, we define the following {\it spectroscopic effective time}
\begin{equation}
  T_{\rm spec} = ( 12.15\,\sec )\times TSNR_{LRG}^2  \label{eq:efftime-spec}
\end{equation}

It is measured from the spectroscopic pipeline output for each observed tile, and used to verify the quality of the observations.

It is interesting to compare $T_{\rm spec}$, obtained from the spectroscopic data only, with an estimate derived from other inputs, namely the transparency and fiber acceptance derived from the GFA images in r-band, along with a model of the atmospheric transmission with airmass, and the extinction.

\begin{figure}
  \centering
  \includegraphics[width=0.99\columnwidth,angle=0]{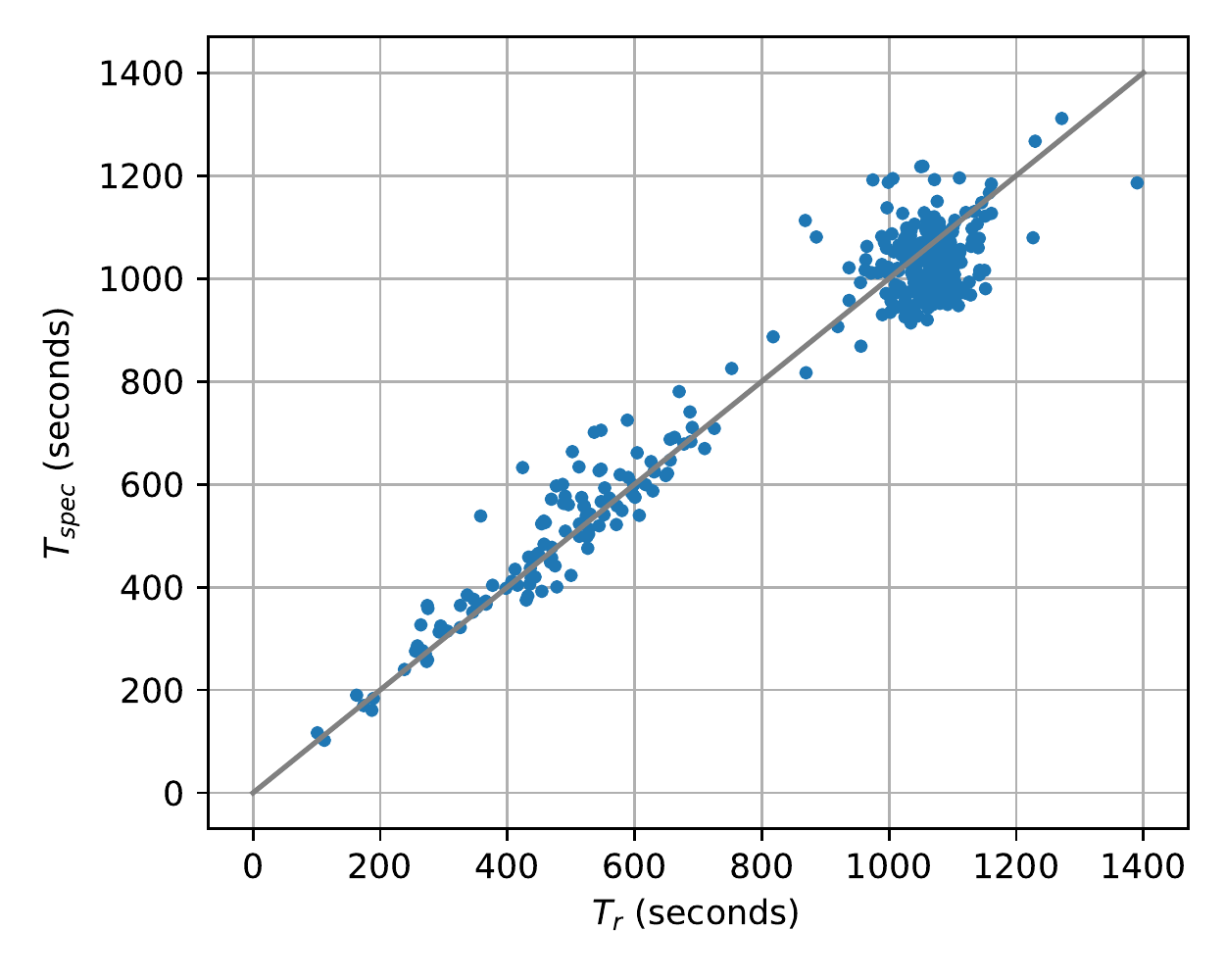}
  \caption{Effective time derived from the spectroscopic TSNR values ($T_{\rm spec}$, see Eq.~\ref{eq:efftime-spec}) as a function of the effective time $T_{\rm r}$ from Eq.~\ref{eq:efftime-r}.}
  \label{fig:teff}
\end{figure}

For this purpose we define another effective time associated with the r-band,
\begin{eqnarray}
  T_{\rm r} & \equiv & T_{exp}\left( \frac{ft}{ft_{nom}}\right)^2 \frac{s_{nom}}{s} \nonumber \\
  && \times 10^{\frac{-2}{2.5}( 0.114 (X-1) + 2.165 E(B-V))} \label{eq:efftime-r}
\end{eqnarray}
where $T_{exp}$ is the actual exposure time, $ft$ is the product of the fiber acceptance (which depends on seeing) and the sky transparency, and is obtained from the guide star images, $s$ is the sky flux measured in r-band (here we do use the spectroscopic sky measurement), $X$ the airmass, and $E(B-V)$ the reddening due to Galactic extinction. The subscript {\it nom} indicates the value in nominal conditions. $T_{\rm spec}$ and $T_{\rm r}$ values are compared in Figure~\ref{fig:teff}; they are tightly correlated.

We track both quantities $T_{\rm r}$, $T_{\rm spec}$ along with the real-time effective time derived from the exposure time calculator as part of the daily quality assessment of the observations. Another exposure is required in the rare cases where $T_{\rm spec}<850$~seconds for a tile in the main dark time program (or 153~seconds for the bright time program). Note that a tile can be composed of one or several exposures from which the spectra have been co-added (see \S\ref{sec:coaddition}).

\section{Software development methodology}
\label{sec:methodology}

The DESI data team developed the spectroscopic pipeline as open source from
the beginning, hosted on GitHub\footnote{\url{https://github.com/desihub}} with a BSD 3-clause license\footnote{\url{https://opensource.org/licenses/BSD-3-Clause}}.
This was originally motivated by the pragmatic reason that GitHub provided free hosting to open source projects but charged for closed source, and it provided a better user experience at less cost (free) than hosting and maintaining our own closed-source software repository.  Additionally, some institutions of contributing authors had burdensome procedures for publicly releasing previously proprietary packages, and it wasn't even clear that they had compatible release policies, but contributing to an already open-source package was straightforward.

The open source nature of the work also enabled us to share code with other collaborations, resulting in benefits to both. For instance eBOSS used Redrock as its redshift fitter for its final cosmology results~\citep{Ross2020,GilMarin2020,Raichoor2020}, providing testing with real-world data years before DESI observations began.  eBOSS contributed improvements back to Redrock, which accelerated DESI's readiness for initial on-sky observations.

Code contributions follow a workflow of creating a local git branch, making updates, pushing the branch to GitHub, then opening a ``pull request'' (PR) to merge the updates back into the main branch.  In most cases the PR is reviewed by someone other than the original author before merging.  We did not follow strict and extensive code review practices, but this workflow generally helped ensure that multiple people were knowledgeable about any given piece of code and helped maintain higher standards of code quality.

Packages also include unit tests that are automatically run with each PR, and were required to pass before merging.  Like the code reviews, these were also implemented at a pragmatic level, usually testing ``does the code still run without crashing?'' rather than deeply testing algorithmic correctness.  Some unit tests required large input files that were not viable to host within the testing framework at GitHub, so they are only run if the tests are run at NERSC.  A nightly cronjob at NERSC updates every package, runs all unit tests, and generates a report alerting the Data Systems manager if any tests are failing.

Unit tests were augmented with end-to-end functional tests run every night, including simulating spectra, target selection, fiber assignment, survey simulations, pixel-level data simulations, and the spectroscopic pipeline run from raw data through redshifts.  This allowed simple bugs to be caught early, freeing up developer attention for more subtle algorithmic and data quality studies.

Code tags are made after major updates and before production runs.  Since the DESI code is split across multiple packages (desispec, desitarget, specter, redrock, desiutil, ...) we require that the main branch of all packages remain compatible with each other, and the latest tags of every package remain compatible with each other.  e.g.~if a PR for package X requires a feature in a branch of package Y, that branch in Y must be reviewed and merged before the PR in X can be merged.  Quarterly software releases define a combination of code tags that are more extensively tested and confirmed to work together.

Python was selected as the primary development language to prioritize developer efficiency over raw computational efficiency.  At the same time, the core algorithms heavily leverage compiled libraries wrapped by
numpy\footnote{\url{https://numpy.org}} and
scipy\footnote{\url{https://scipy.org}} (e.g. LAPACK, BLAS), and key functions use numba\footnote{\url{https://numba.pydata.org}} just-in-time compilation.
This combination of libraries has enabled good computational performance while benefiting from the flexibility and non-expert accessibility of Python.

Most code was developed by scientist-programmers within the DESI collaboration, who had a vested interest in the quality of the outcome, rather than by professional programmers with limited domain knowledge.  Two postdocs at the
National Energy Research Scientific Computing Center (NERSC) focused on
improving the code efficiency of the algorithms (see \citealt{laurie_stephey-proc-scipy-2019})
and porting these to Graphics Processing Units (GPUs) for future machines (see \citealt{margala-gpu-2021}).

The code architecture strictly separates algorithms from the pipeline workflow wrappers that run those algorithms in parallel on multiple compute nodes for multiple input exposures.
Any individual step (e.g.~sky subtraction, flux calibration) can be run completely independently of a production workflow by reading input files, running an algorithm, and writing output files.
The parallel pipeline workflow then calls the identical functions, also reading
and writing files as the method of passing data from one pipeline step to another.
This structure enabled focused algorithm development on laptops with a deployment to High Performance Computing supercomputers using tens of thousands of cores in parallel to process hundreds of terabytes of data.
Although less efficient from an I/O perspective,
this design has been key to developer efficiency by enabling focused work and iterative debugging on individual algorithms with minimal conceptual overhead on how to run those steps.
The file-centric method of data passing also provides a natural ``checkpoint restart'' design, such that any later step of the pipeline can be run using files produced by earlier steps, without having to re-run those earlier steps.

\section{Data processing}
\label{sec:data-processing}

We address in this section the operational aspects of the data processing. We first describe the simplified processing ran during the night at Kitt Peak to provide real time information to the observers, then the offline processing ran on the NERSC computers every day to provide feedback for the survey operations, and the large reprocessing performed for data releases. We also give some details about the daily offline data quality assurance tools we have developed.
\subsection{Real time processing}


DESI exposures are automatically processed and displayed using a quality assurance tool called Nightwatch\footnote{\url{https://github.com/desihub/nightwatch}}. Its purpose is to provide
observing scientists with exposure metadata and instrument diagnostics in real time. Nightwatch handles data in three steps: (1) exposure
processing; (2) calculation of quality assurance metrics; and (3) creation of web-based plots and tables to allow visual inspection of
exposures. These steps are described in detail below.

{\bf Exposure processing}: real-time processing of exposures begins with a preprocessing step that extracts the raw CCD images from the
$b$, $r$, and $z$ cameras and masks out saturated pixels, pixels affected by cosmic rays, and in rare cases entire CCD amplifiers affected
by high read noise. For exposures without spectroscopic data, such as bias and dark frames, processing ends at this point, while
calibration and science exposures continue to spectral extraction. A quick processing routine
applies a simple boxcar extraction \citep{deBoer:1981} to estimate the spectra from fiber columns in the CCDs (Fig.\,\ref{fig:ccd-image}).
While boxcar extraction provides sub-optimal S/N compared to spectroperfectionism (Section\,\ref{sec:extraction}), its speed and
robustness make it well-suited for real-time analysis. After spectral extraction, a simplified sky subtraction is applied to science
exposures to produce sky-subtracted, flux-calibrated spectra for all fibers.

{\bf Data quality assurance (QA)}: following preprocessing and spectral extraction, several data quality metrics are computed and stored:
\begin{itemize}
  \setlength{\itemsep}{-0.5em}
  \item CCD metrics: read noise, overscan bias level, and detected cosmic rays per minute are estimated independently for the four
  amplifiers in every CCD.
  \item Camera metrics: boxcar extraction spectral trace shifts in $x$ (fiber number) and $y$ (wavelength) are estimated independently for
  the three cameras in each spectrograph.
  \item Fiber/spectra metrics: integrated and median flux per fiber (raw and calibrated), as well as SNR (raw and calibrated).
\end{itemize}

\begin{figure}
  \centering
  \includegraphics[width=\columnwidth]{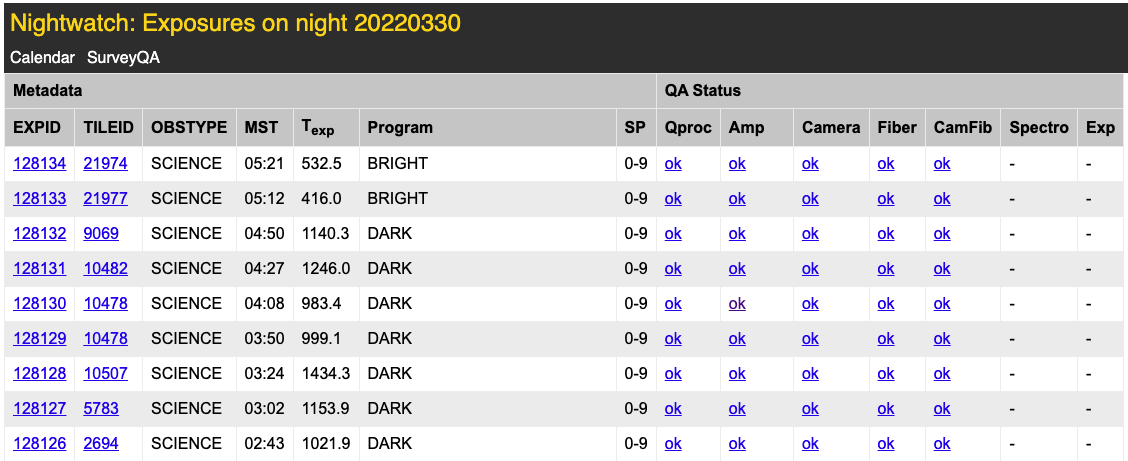}
  \caption{The random-access exposure table provided by the Nightwatch QA tool for one night of observing.}
  \label{fig:nwexp}
\end{figure}

\begin{figure}
  \centering
  \includegraphics[width=\columnwidth]{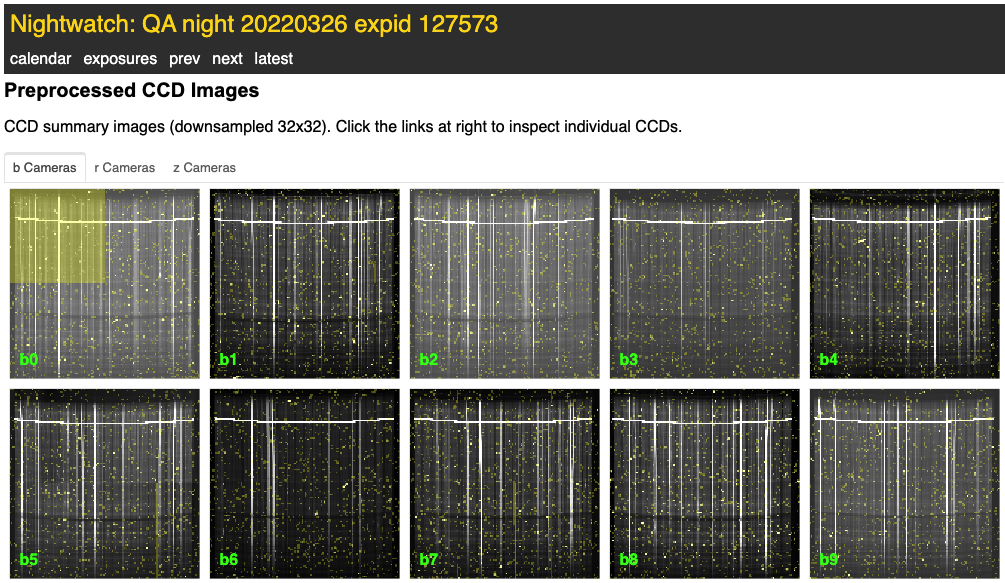}
  \caption{CCD images from the ten $b$ cameras during a dark-time exposure. An amplifier in the $b0$ camera has been masked out due to high
  overscan bias (yellow square).}
  \label{fig:nwccd}
\end{figure}

\begin{figure}
  \centering
  \includegraphics[width=\columnwidth]{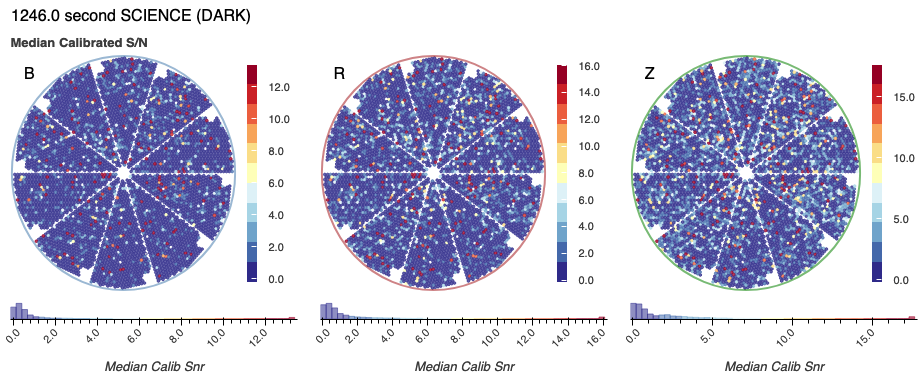}
  \caption{Focal plane SNR plots for fibers in the $b$, $r$, and $z$ cameras during a 1246~s dark-time exposure.}
  \label{fig:nwfp}
\end{figure}

{\bf QA Web interface}: after calculation of QA metrics, plots and tables are automatically generated and posted on websites hosted at
NERSC and Kitt Peak. The Nightwatch main webpage provides a calendar that allows random access to every night in DESI commissioning,
survey validation, and the main survey. For each night, a table of exposures lists exposure metadata: exposure ID, type, date, number of
spectrographs in operation, and QA status (see Fig.\,\ref{fig:nwexp}). For each exposure, users can access preprocessed CCD images with pixel
masks highlighted (Fig.\,\ref{fig:nwccd}), plots of the QA metrics, and raw and calibrated spectra. Additional focal plane metrics showing
fibers on target, fiber SNR (Fig.\,\ref{fig:nwfp}), and the accuracy of fiber positioner moves are computed on the fly and plotted. The
interface also allows interactive plotting of sky-subtracted, flux-calibrated spectra recorded in individual fibers.

The web interface provides the first look at reduced data from DESI, and is the main tool used for human validation of exposures as they
are recorded. During DESI operations, observing scientists inspect the QA plots, diagnose problems, and take appropriate action. For
real-time operations, the QA is particularly useful for recognizing problems with the CCDs, identifying broken calibration lamps or stray
light in the dome, and noticing systematic issues with positioners.

\subsection{Offline Data Pipeline Overview}

The offline pipeline is responsible for managing the processing of DESI data using the algorithms described in \S\ref{sec:algorithms-and-performance}. It is designed to operate at NERSC using the Slurm\footnote{\url{https://slurm.schedmd.com}} job scheduler system on both Cori and Perlmutter machines. While currently specialized for Slurm jobs, the code was designed to be as machine agnostic as possible, with the scheduler and machine specifications localized to several calling functions that could be upgraded in the future to work on other systems. The pipeline is tasked with two objectives: nightly processing and full dataset reprocessing. The nightly processing is the near-realtime analysis of data taken throughout a night with the goal of having processed spectra and redshifts by the next morning to inform observations the following night. The full dataset reprocessing is intended to give a self-consistent set of data processed with a tagged version of the offline pipeline, which could be used for publishable scientific analyses. These full reprocessing runs are what become internal data releases and public data releases.

Both types of data processing perform the same series of steps. Calibration data is processed first. Master bias frames are computed for a night based on available bias exposures (with null exposure time). These bias levels are then used to pre-process a dark exposure used to identify bad columns on each of the 30 cameras. Both the bias and badcolumn files are then accounted for when processing the arc lamp exposures for PSF fitting. The PSFs for each camera are then used in determining the flat fielding vectors of each fiber using the flat field exposures. Finally, the science exposures are processed using all of the calibration information. The pipeline assures that these are run in the proper order by leveraging Slurm job dependencies. For example, the individual arc lamp exposures can be processed in parallel, but they all depend on the bad columns being successfully completed prior to their execution.

Individual science exposures of the same tile are treated as independent for the extracting of spectra from the images and for the sky subtraction, but are processed together to jointly fit the standard star models that are later used for flux calibration. The flux calibration is again done individually and in parallel. Finally, redshifts are fit using all exposures of a target coadded together.


A full description of an older version of the workflow can be found in \citet{Kremin2020}, which still provides a relatively accurate description of the general pipeline even though the names and timing information have changed.

\subsubsection{Nightly Data Pipeline}

\begin{figure}
\centering
  \includegraphics[width=0.7\columnwidth,angle=0]{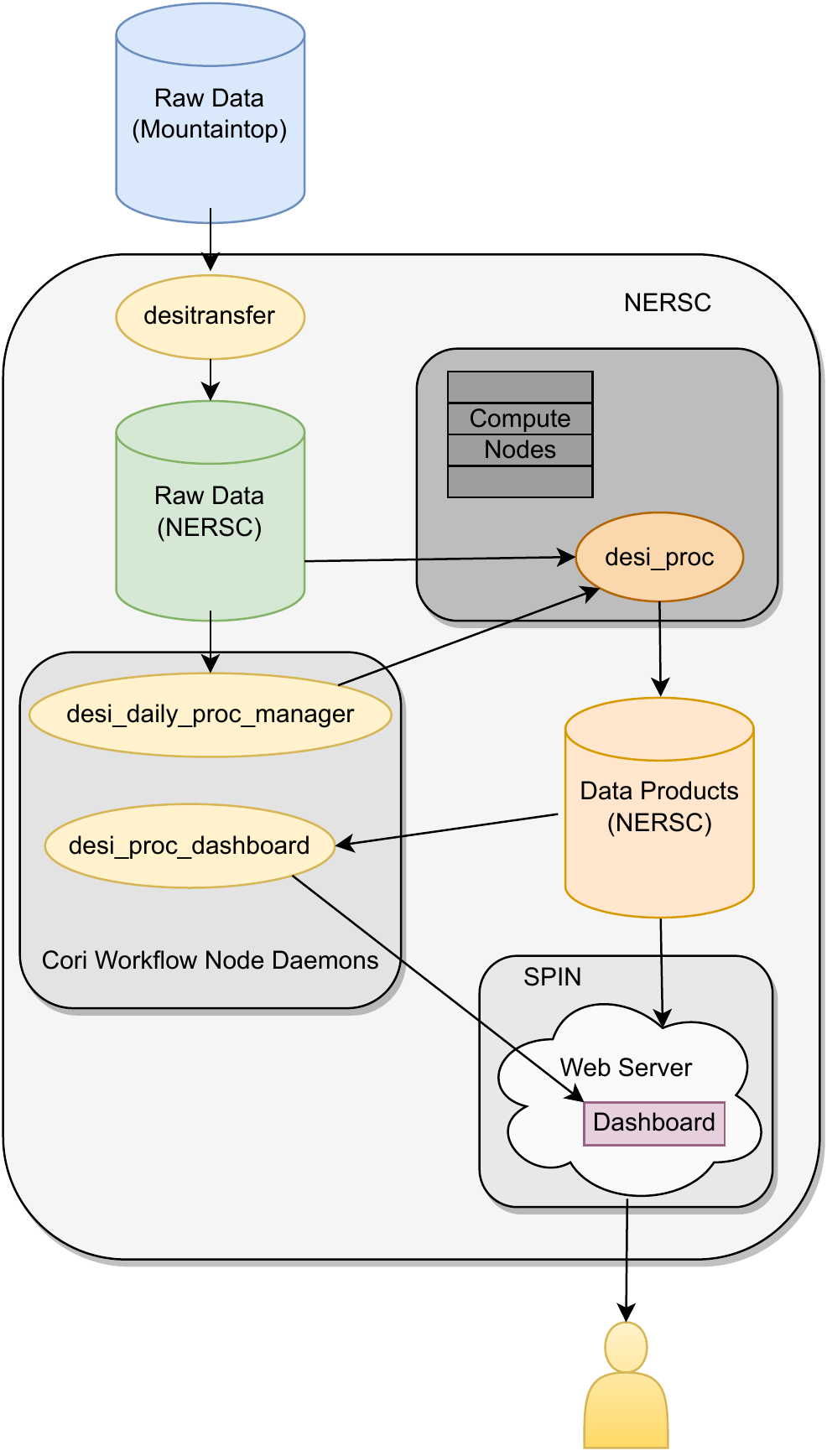}
  \caption{Flowchart depicting the path data takes from local storage on the mountaintop to processing and derived data products, and useful web services for monitoring each. Each exposure is transferred to NERSC, where a pipeline manager identifies it and submits several jobs to process the data to flux calibrated spectra and redshifts. A dashboard checks the output data and reports progress on an interactive html page for use in monitoring progress.}
  \label{fig:pipeline-flowchart}
\end{figure}

Figure~\ref{fig:pipeline-flowchart} shows the interplay between the various tools and data throughout a night of processing.

The nightly pipeline is managed by a script (\emph{desi\_daily\_proc\_manager} in Fig.~\ref{fig:pipeline-flowchart}) which identifies when new data has arrived at NERSC and submits jobs to the Slurm queue to process this data.
The jobs submitted to the queue call \emph{desi\_proc} and \emph{desi\_proc\_joint\_fit} to process single exposure and multi-exposure operations, respectively. Since the data appears chronologically, the pipeline has to be robust to hardware and software failures as well as differences in data acquisition strategy and data quality. Survey validation led to dramatic improvements in uniformity, and nightly processing is now reliably supplied by the following morning with many nights requiring no human intervention.

\begin{figure}
\centering
  \includegraphics[width=1.0\columnwidth,angle=0]{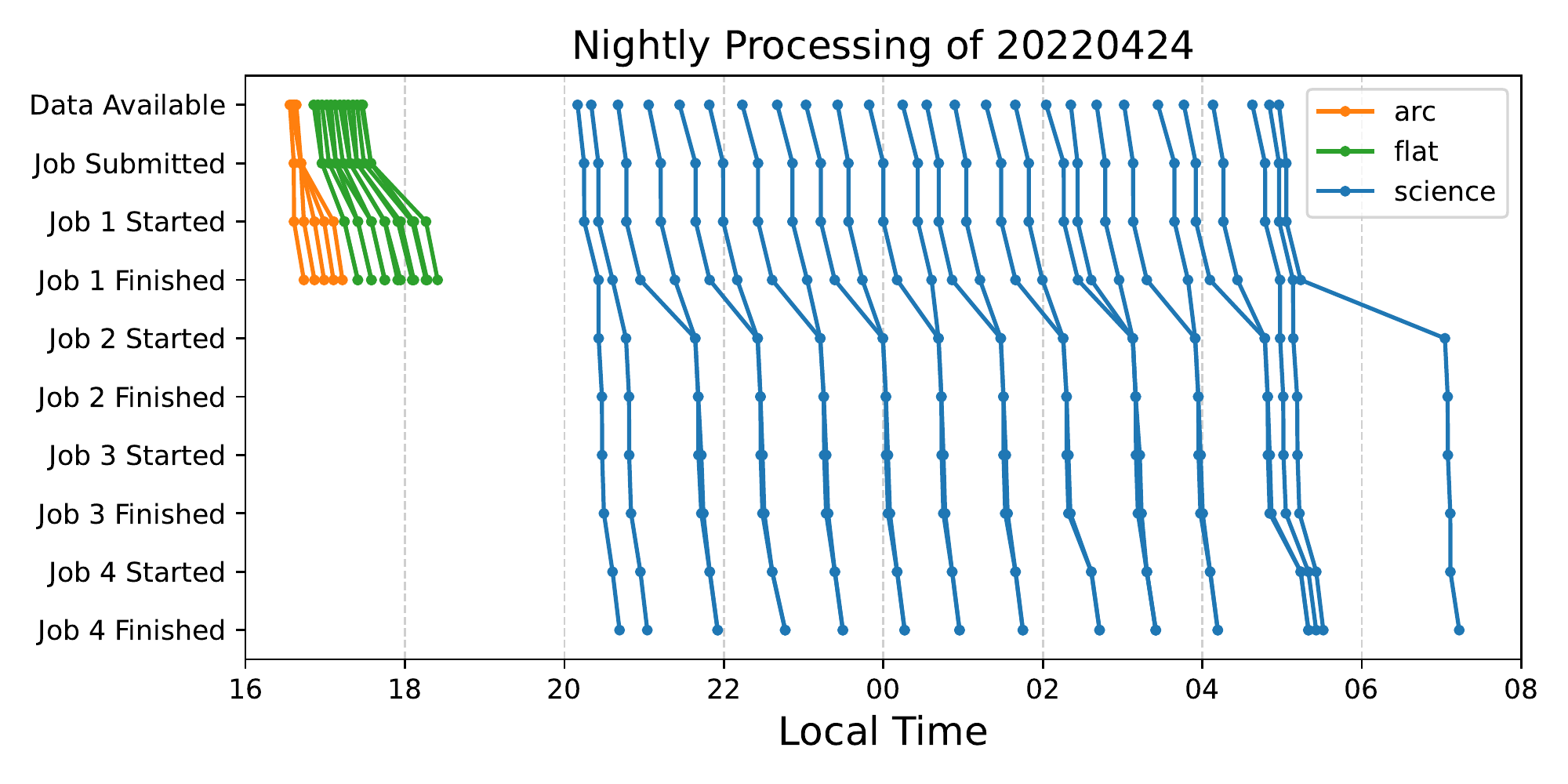}
  \caption{The speed at which data is processed throughout an example night, April 24th 2022. Orange lines are arc lamp exposures, green lines are flat exposures, and blue lines are science exposures. Science exposures are processed in four stages, with the second and fourth stages being jointly performed with all exposures for a given tile of targets. The various merging events in Job 2 are due to this joint fitting forcing earlier exposures to wait on later exposures. Per-tile jobs are submitted once data for a different tile appears or the end of the night is reached. The last tile is delayed until an end of night signal is triggered, which is the cause of the delay seen on the right-most exposure. The plot shows that on this night we had fully calibrated spectra and redshifts of targets within 20-40 minutes of the data becoming available at NERSC.}
  \label{fig:nightly-pipeline-timing}
\end{figure}

Figure~\ref{fig:nightly-pipeline-timing} shows an example night, April 24th 2022, where we successfully observed 15 tiles in 26 science exposures. The orange lines indicated arc lamp jobs (for PSF fitting), the green lines fiber flat field jobs, and the blue lines science jobs. For a given tile the joint fitting of standard stars across all exposures necessitates the waiting of some jobs for later exposures of a tile to complete. The merging of jobs seen in the middle of the night are depicting this scenario. The large delay in the rightmost exposure is a consequence of the reactive nature of the pipeline. It will not submit the joint standard star fitting until it receives new data from a different tile or the end of the night is reached. The delay in the rightmost job at the end of the night is the pipeline waiting for the end of the night when it knows no new data will arrive before submitting the final exposure for standard star fitting, flux calibration, and redshift fitting.

A table of exposure properties for a single night, called an \emph{exposure\_table}, is automatically generated as the exposures are identified by the pipeline. These tables provide per-exposure information about the type of observation, cameras available, and columns designating whether special considerations need to be made due to noisy amplifiers, failed positioning on a petal, or whether that exposure failed to meet our quality criteria and should be ignored. During the night the information is auto-populated, but can be updated as new details about specific exposures are uncovered. Version control is used as a safeguard. In addition to the repository, copies of tables used to process a specific release are kept with the data for future reference.

\subsubsection{Reprocessing Data Pipeline}

The nightly processing is run using the latest version of the pipeline on that given day. That is beneficial for providing the best outputs possible given the current understanding of the instrument and data. The implication of this is that earlier daily data are processed with less optimized versions of the pipeline. The objective of the data reprocessing is to process all data with a tagged version of the code, and utilizing knowledge of all data derived that exists for a night via the exposure tables that is not known for the realtime processing.

\begin{figure}
\centering
  \includegraphics[width=1.0\columnwidth,angle=0]{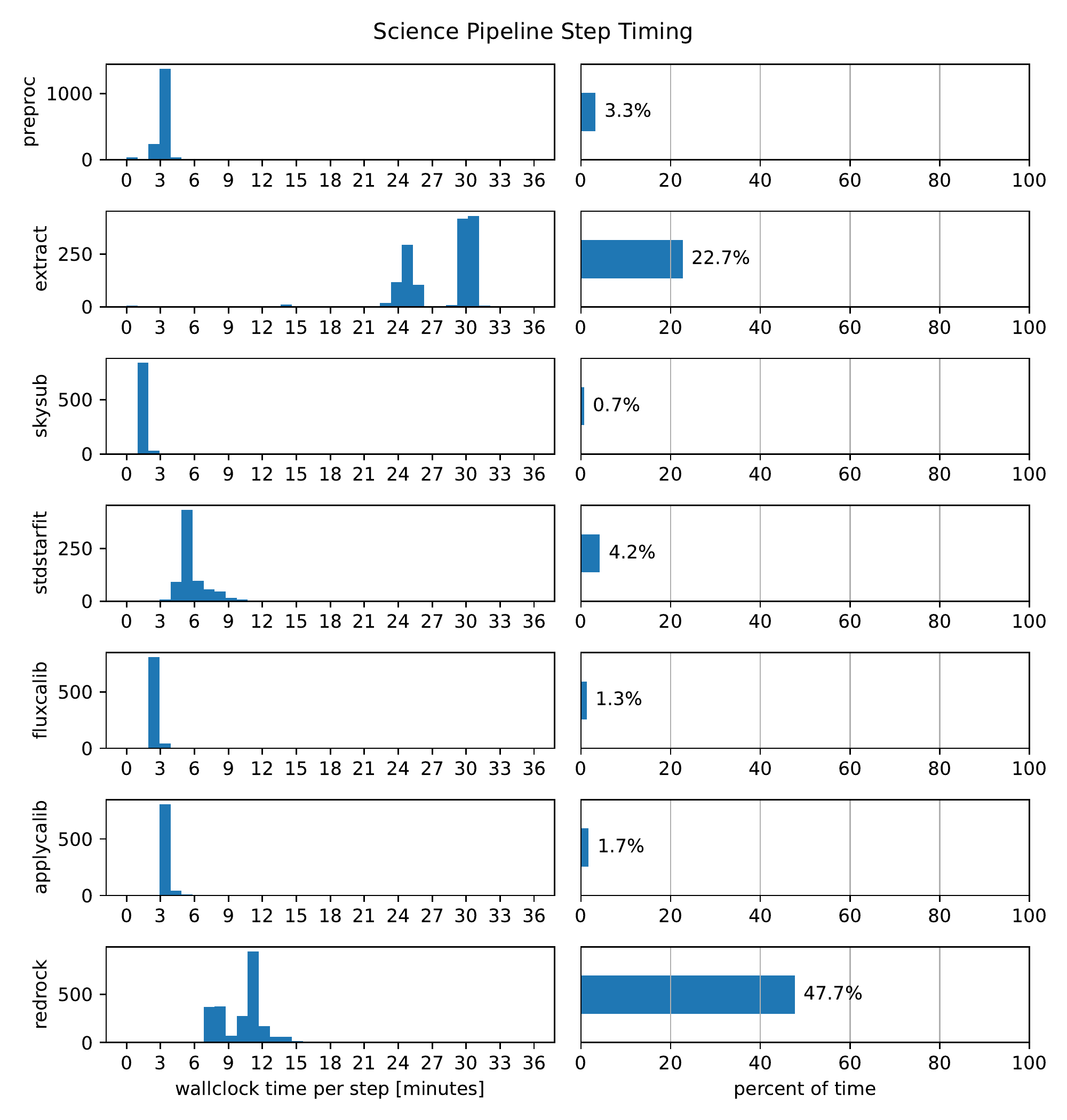}
  \caption{Distribution of wall clock times for the key science exposure processing steps for a recent internal data release. On the left is the histogram of job times in minutes. On the right is the percentage of the total computational time devoted to that step, in core-hours used, for the per-tile based science processing.}
 \label{fig:reprocessing-timing}
\end{figure}

Figure \ref{fig:reprocessing-timing} shows the performance of science exposure processing for per-tile spectra and redshifts. The left column shows the distribution of times for individual jobs for that step. The number of resources used for each step are not the same, so the right column shows the percentage of the total per-tile science processing spent in that step, computed as a function of core-hours for more accurate resource comparisons. The majority of the computational time is spent extracting the individual fibers from the images and in doing the redshift fitting of the flux calibrated spectra. For the release in question three types of redshifts were generated, which nearly triples the number of redshift function calls and time spent computing redshifts. Note also that the percentages are of the total compute time and there are steps not shown in a row.

\subsection{Offline Quality Assurance}
\label{sec:qa}

As described at the end of \S\ref{sec:observations}, DESI observations are done through tiles: we validate during daytime the bright and dark program tiles observed during the night, using a wide range of Quality Assurance (QA) checks.
Once a tile is QA-validated, we archive it (i.e. freeze it), update the spectroscopic status of the observed targets, and enable the observation of overlapping tiles in subsequent nights \citep[see][]{schlafly22a}.

We present in this subsection the principle of this QA approach, then list the performed QA. An example is provide in Appendix~\ref{sec:example-qa}.

This procedure allows us to daily monitor from an end-user point-of-view the appearance of instrumental features or specific observing conditions which impact the quality of the redshift measurement, and reduces the probability to run several nights with a lingering, unidentified issue.
Most of the time, the found issue can be solved with either an intervention on the instrument or a change in the spectroscopic pipeline, and re-running the pipeline.
In few cases, the issue cannot be solved, and the affected observed fibers are flagged as bad, i.e. are discarded.

This daily QA task has proved to be very efficient to quickly uncover issues in the first months of the DESI Main Survey.
With the survey progressing, the operations becoming more stable, and the spectroscopic pipeline more robust, those have become less frequent.
Lastly, this daily QA task is done by a member of the Survey Operations team (approximately ten members), with weekly rotations, allowing at the same time to avoid possible bias from having only a few people performing the QA, and to have a core team familiar with those aspects of the data.

The QA are performed at two levels: for each tile, and for the overall night.

For each tile, we currently look at:
\begin{itemize}
\item the redshift distribution for each target class, and compare it to the expected one;
\item the redshift $z$ as a function of the fiber number: this is efficient to identify problematic regions (e.g. a petal, a block of fibers due to some CCD feature);
\item the fiber positioning accuracy as a function of the position in focal plane: this is useful to monitor for instance that the turbulence correction\footnote{The apparent relative displacements of fixed fibers and fiducials in the fiber view camera images are used to compute a distortion map. It is attributed to air turbulence between the camera and the focal surface and is used to correct the measured fiber positions.} is effective;
\item the per-fiber spectroscopic effective time as a function of the position in focal plane;
\item various per-petal diagnoses: the read noise, the number of good positioners, the number of standard stars used by the pipeline, the r.m.s. of the standard stars r-band flux ratios, the throughput ratio to the overall tile throughput, the sky fibers throughput r.m.s. and their reduced $\chi^2$;
\item the sky position of the tile in the DESI footprint: this allows one to quickly identify if the tile falls in a particular region (e.g., close to the Galactic Plane, in a Galactic dusty region, or in the Sagittarius Stream);
\item a cutout sky image of the tile location, with an overlay of the petal geometry and the Galactic dust contours: in addition to showing possible very dusty small regions affecting some fibers, it allows one to identify if a bright star falls in the tile footprint.
\end{itemize}

And for each night, we currently monitor the following:
\begin{itemize}
\item all calibration files are present;
\item 5min \textit{DARK} image display (10 petals, 3 cameras): this allows one to visualize possible appearing CCD features;
\item CCD bad columns identification (10 petals, 3 cameras);
\item charge transfer efficiency (CTE; 10 petals, 3 cameras): a subtle effect (of the order of a few electrons) is a lower CTE in one amplifier of a given camera, for some CCD columns, leading to a discontinuity in low signal-to-noise spectra. This effect is caused by a CCD defect localised the serial register. It has been found in several CCDs and evolves with time. It can be detected with the analysis of a one second flat field image nightly taken for that purpose, where we control for each fiber the median flux (over 21 rows) above and below the CCD amplifier boundary (see example in Appendix~\ref{sec:example-qa});
\item 2D-image of sky-subtracted sky-fibers only (1 image per exposure): the (sky-subtracted) sky fibers being supposed to mostly be noise, such images are efficient to visualize remaining systematics in the data;
\item the redshift $z$ as a function of the fiber number for the sky fibers only: any systematic feature in this plot highlights possible remaining issues in the data; for instance a mis-calibration between two cameras for a petal could drive the redshift pipeline to systematically identify it as a Balmer break at a given redshift, hence creating a horizontal line in that plot;
\item various per-petal diagnoses: per-tracer n(z), per-tracer redshift success rate, new Ly-$\alpha$ identification rate for dark tiles\footnote{For the dark program, any newly observed quasar target securely identified as a quasar at $z \ge 2.1$ will be assigned for re-observation, in order to increase the signal-to-noise ratio of the spectra used for the Ly-$\alpha$ forest analysis \citep[see][]{schlafly22a}};
\item new Ly-$\alpha$ identification for dark tiles: comparison of the number of new identifications to expectation, given the tile coverage from previous observations.
\end{itemize}
Though the list of diagnoses is mostly stabilized by now, we stress that, depending on the gained experience with the data, additional ones could be added in the future (e.g., a control for out-of-focus data).\\

\section{Overview of the data products}
\label{sec:data-products}

The primary user-facing data products are sky-subtracted flux-calibrated
spectra, coadditions (hereafter 'coadds', i.e. averages) of those spectra across exposures, and tables of
spectral identification (galaxy, quasar, star) and redshift measurements.

When they are publicly released, these will be available at
\url{https://data.desi.lbl.gov}.  Each data release will contain one or more
spectroscopic pipeline runs, which are the outputs of a tagged set of code
run on a specific set of input exposures.
Survey Validation observations, taken prior to the start of the DESI main survey,
will be released in the ``Early Data Release'' (EDR) as the ``Fuji'' spectroscopic
pipeline run, which will be available at
\url{https://data.desi.lbl.gov/public/edr/spectro/redux/fuji}.
Details of the file formats and directory structure are documented at
\url{https://desidatamodel.readthedocs.io} and summarized as follows.
We first describe the directory structure by which the data are grouped
under each top-level spectroscopic pipeline run, and then summarize the
contents of the files while leaving the details of the formats to the
DESI data model online documentation.

Coadds and redshifts are provided in multiple groupings for various analysis needs.
Tile-based coadds, available under the \texttt{tiles/} directory, combine data across multiple exposures of a single tile but not across tiles.
Healpix-based coadds, available under \texttt{healpix/}, combine data across multiple tiles covering a single patch of sky, using an $N_{side}=64$ nested healpix tessalation \citep{Healpix}.
Many targets are observed on only a single tile, in which
case the coadds and redshifts are the same in both \texttt{tiles/}
and \texttt{healpix/} but are still provided
in both forms for convenience.  Some targets such as Lyman-$\alpha$ QSOs are
observed multiple times on multiple tiles, in which case the healpix-based
coadds and redshifts provide a fit that is not otherwise available in the
tile-based directories.


Although healpix-based products combine information across tiles, different
sub-programs of DESI are still processed independently so that the targeting
and observation choices of one portion of DESI do not impact the analysis
systematics of a different portion of DESI.  This organization of the output
products prioritizes homogeneity of the data quality within a program rather
than maximizing the signal-to-noise of each individual target by
combining data across programs.


On disk (and URL), these healpix-based outputs are grouped under sub-directories
\texttt{healpix/SURVEY/PROGRAM/NN/HEALPIX/}
where \texttt{HEALPIX} is the healpix number; and
\texttt{NN}=\texttt{HEALPIX}/100 to avoid having thousands of healpix directories at the same level.
See~\citealt{Myers22a} for more details about the
various surveys and programs.

Tile-based coadds and redshifts under the \texttt{tiles/} directory
also come in multiple combinations for each \texttt{TILEID},
depending upon whether the tile exposures are combined only within
a single exposure (\texttt{perexp/TILEID/EXPID/}),
a single night (\texttt{pernight/TILEID/NIGHT}),
across multiple nights (\texttt{cumulative/TILEID/LASTNIGHT}),
or in a custom combination of exposures to achieve a specific effective exposure
time (\texttt{1x\_depth/TILEID/N/} or \texttt{4x\_depth/TILEID/N/},
where \texttt{N} is an arbitrary integer group number).

Although the data are organized in multiple different sub-directories
depending upon how the exposures are combined, the files in each sub-directory
follow the same formats in all cases.  Tile-based coadds and redshifts
are split into files by spectrograph (or equivalently focal plane petal) and thus have
500 targets per file, with 5000 targets per sub-directory.  Healpix-based coadds
and redshifts have a varying number of targets per file depending upon how many
observed targets were within each healpix.  Details of the file formats
are provided at
\url{https://desidatamodel.readthedocs.io/en/latest/DESI_SPECTRO_REDUX/SPECPROD/}.

The uncoadded spectra are stored in gzip-compressed multi-HDU FITS files\footnote{\url{https://fits.gsfc.nasa.gov}}
\texttt{spectra-*.fits.gz},
with separate extensions for each of the $b,r,z$ cameras for each of the
quantities wavelength, flux, inverse variance, mask, and resolution data.  A
``fibermap'' binary table tracks which targets are assigned to which fibers
along with targeting photometry, shapes, target selection masks,
and related metadata; and per-fiber per-observation data such as the focal plane
($x,y$) location of each target and its distance from the assigned position.
An additional ``scores'' binary table records TSNR-squared values
(see \S\ref{sec:efftime})
and the median and summed counts (or flux) at various steps of the processing:
raw, fiber flat-fielded, sky subtracted, and fully calibrated spectra.

The coadd files \texttt{coadd-*.fits} (not gzip-compressed) contain the same wavelength, flux, inverse variance, and resolution data per camera, coadded across multiple input exposures for the same target.
In the coadd files, the fibermap is split into a table of information that is
the same for every exposure of a target (e.g.~photometry), plus a separate
table that contains the per-exposure information that contributed to the
coadd (e.g.~exposure IDs and the per-exposure $(x,y)$ locations).

A key feature of both the spectra and the coadds is that all
targets use the same wavelength grid, and the individual wavelength bins are
uncorrelated.  See \S\ref{sec:extraction} for more details about the validity
of the error (inverse variance) model and the uncorrelated nature of the noise. See also Appendix~\ref{sec:resampleresolution} for the best use of the resolution matrix provided in the files.

The Redrock fitted redshift files \texttt{redrock-*.fits} contain a binary table with columns listed in Table~\ref{table:redshift-columns}, as well as
propagating the \texttt{FIBERMAP} and \texttt{EXP\_FIBERMAP} HDUs from the
input coadd files.
Additional details including the $\chi^2$ vs.~$z$ scans and the best
3 fits for each SPECTYPE (GALAXY, QSO, STAR) are stored in a separate
\texttt{rrdetails-*.h5}
HDF5\footnote{\url{https://www.hdfgroup.org}} format file.

\begin{table}[h]\small
\begin{tabular}{l|l}
Z & redshift\\
ZERR & estimated error\\
ZWARN & warning mask (0=good)\\
SPECTYPE & spectral type\\
SUBTYPE &  subtype for stellar templates\\
DELTACHI2 & $\chi^2$ difference between the best\\
& and the second-best model\\
NPIXELS & number of unmasked pixels\\
& contributing to the fit\\
NCOEFF & number of template coefficients\\
COEFF & PCA template coefficients\\
CHI2 & absolute $\chi^2$
\end{tabular}
\caption{List of columns in the REDSHIFTS HDU of the
\texttt{redrock-*.fits} redshift files.}
\label{table:redshift-columns}
\end{table}

In the same directories as the core spectra, coadd, and redrock
(classification and redshift) files, the spectroscopic pipeline also
includes the outputs of several ``afterburner'' programs that are provided
by the science working groups and run automatically by the spectroscopic
pipeline.  These currently include QSO fits with the QuasarNET algorithm
(\texttt{qso\_qn-*.fits}, see \citealt{Busca2018}),
QSO fits to \ion{Mg}{2} (\texttt{qso\_mgii-*.fits}),
and fits to galaxy emission lines (\texttt{emline-*.fits}).

For end-users working with redshifts at the catalog level who don't need access to the spectra, a concatenation of the
REDSHIFTS and FIBERMAP HDUs of all of the \texttt{redrock-*.fits} files
are provided in the
\texttt{zcatalog/} subdirectory of each spectroscopic pipeline run,
\textit{e.g.}, \texttt{edr/spectro/redux/fuji/zcatalog/}.  These files are
also useful for doing an initial pre-selection of targets of interest before
reading their spectra from the individual coadd files.

In addition to the user-facing data described above, the pipeline includes
outputs for per-night fiberflat and PSF fit calibrations, preprocessed CCD
images, and intermediate files such as uncalibrated extracted spectra and
fiberflatfielded sky-subtracted but not flux-calibrated spectra.  These are
useful for QA, debugging, and specialized analyses, but are not intended for
most end-users.

Full details of all of the data formats and directory structures are
documented at \url{https://desidatamodel.readthedocs.io}.

\section{Summary}
\label{sec:summary}

We have presented the DESI spectroscopic pipeline, with all the algorithms developed to convert raw CCD images into calibrated spectra with their spectroscopic identification and redshifts. The pre-processing, spectroscopic extraction, PSF and wavelength calibration, flat-fielding, sky subtraction, and flux calibration have been described in detail. Some notable performance results are the stability of the spectrographs' PSF and relative throughput from fiber to fiber which is better than 1\% for most of the wavelength range during a night. The wavelength calibration has been tested against other surveys and the systematic error expressed in terms of radial velocity was found to be at maximum of 3\,km\,s$^{-1}$ (smaller than 0.1\,\AA), with an excess scatter of 0.8\,km\,s$^{-1}$ with respect to the expected statistical uncertainty. We are able to predict the $g-r$ colors of standard stars with a precision of 0.02 (or 2\% in flux ratio). The accuracy of the flux calibration was tested with white dwarf spectra, and the flux systematic error was found to be about 2\% in most of the wavelength range, possibly increasing to 6\% for wavelength shorter than 3700\,\AA. The scatter between the spectroscopic flux and the broadband photometric flux from the Legacy Survey was found to be of 6\% for bright objects. We also obtain consistent fiber fluxes for extended objects.

We have paid great attention to the error propagation, starting with a model of the pixel variance in the CCD images, then using the spectroperfectionism technique to provide uncorrelated spectral fluxes and their variance. The error propagation was validated with an inspection of the residuals in the sky fibers. The residual scatter in the spectral continuum was found consistent with the expected uncertainties, whereas an excess scatter of a few percent of the total flux was found on the bright sky lines. This extra scatter is included in the reported flux uncertainties.

While it was not the main purpose of this paper, we have given a overview of the redshift fitting technique used for DESI (see~\citealt{bailey_redrock} for more details). The resolution matrix retrieved from the extraction provides a well defined transform to apply to high resolution models before comparing with the data, while the uncorrelated noise in the spectra makes it easy to perform a maximum likelihood fit. The performance of the spectroscopic pipeline and survey are studied in a series of Survey Validation papers \citep{sv22a,hahn22a,zhou22a,raichoor22a,chaussidon22a,cooper22a,lan22a,alexander22a}. They demonstrate that the requirements in terms of redshift efficiency and purity are met or exceeded for all target classes.

Some new developments have been started or are envisioned to further improve the data processing.
We intend to improve the cosmic ray detection which still misses some pixels, the sky subtraction to accommodate for sky gradients in the focal plane caused by the presence of the moon or during twilight, the flux calibration as we have identified repeatable percent variations of throughput with positioner moves, the wavelength calibration with a better cross-correlation with sky lines, and finally we will improve the flux calibration guided by the systematic trends presented in the paper.

The spectral resolution matrix is a new concept in spectroscopic surveys, and some of its properties are not trivial (it is not a flux conserving convolution).
As the DESI collaboration gains experience using these spectra for science analyses, we may learn better ways to model and use the spectral resolution, potentially leading to algorithmic and format updates. We also expect to improve the Redrock spectral classification and redshift estimation with better templates and algorithms.
In addition to the changes that will improve the data quality, we have also embarked on an effort to speed-up the pipeline, porting some of our software to GPUs in order to take advantage of the new computer system at NERSC. Some of those improvements will be used for the processing of the first year data set of the DESI survey.

The algorithms presented in this paper were used to produce the {\it Fuji} data set that will be made public as part of the upcoming Early Data Release (EDR) covering DESI Survey Validation data.

\acknowledgments
The DESI collaboration is indebted to the late John Donaldson, Robin Lafever, Tammie Lavoie, and Glenn Roberts
for their many contributions to the success of this project and mourns their passing.

This research is supported by the Director, Office of Science, Office of High Energy Physics of the U.S. Department of Energy under Contract No. DE–AC02–05CH11231, and by the National Energy Research Scientific Computing Center, a DOE Office of Science User Facility under the same contract; additional support for DESI is provided by the U.S. National Science Foundation, Division of Astronomical Sciences under Contract No. AST-0950945 to the NSF's National Optical-Infrared Astronomy Research Laboratory; the Science and Technologies Facilities Council of the United Kingdom; the Gordon and Betty Moore Foundation; the Heising-Simons Foundation; the French Alternative Energies and Atomic Energy Commission (CEA); the National Council of Science and Technology of Mexico (CONACYT); the Ministry of Science and Innovation of Spain (MICINN), and by the DESI Member Institutions: \url{https://www.desi.lbl.gov/collaborating-institutions}.

The authors are honored to be permitted to conduct scientific research on Iolkam Du’ag (Kitt Peak), a mountain with particular significance to the Tohono O’odham Nation.

For more information, visit \url{https://desi.lbl.gov}.

This work has made use of data from the European Space Agency (ESA) mission Gaia (\url{https://www.cosmos.esa.int/gaia}), processed by the Gaia Data Processing and Analysis Consortium (DPAC, \url{https://www.cosmos.esa.int/web/gaia/dpac/consortium}). Funding for the DPAC has been provided by national institutions, in particular the institutions participating in the Gaia Multilateral Agreement. Source data files for the plots will be archived at \url{https://zenodo.org} (work in progress).

\vspace{5mm}
\facility{Mayall (DESI)}

\bibliographystyle{aasjournal}
\bibliography{biblio}

\begin{appendix}

\section{Emission line flux bias}
\label{sec:emission-line-bias}
The minimal variance estimator $\tilde{F}$ of the flux from an emission line extracted from a CCD image is obtained by minimizing
\begin{equation}
\chi^2 = \sum_i \sigma_i^{-2} \left( D_i - F P_i \right)^2
\end{equation}
where $i$ is a pixel index, $D_i$ the measured value in the pixel, $\sigma_i$ its uncertainty, $F$ the emission line flux and $P_i$ a model of the PSF integrated in the pixel, with the normalization $\sum_i P_i =1$.
The minimum is obtained for
\begin{equation}
  \tilde{F} = \frac{\sum_i \sigma_i^{-2} D_i P_i}{\sum_i \sigma_i^{-2} P_i^2}
\end{equation}

Replacing $D_i$ by its expectation value $F^{truth} P_i^{truth}$ (true flux and PSF profile), and assuming a constant uncertainty $\sigma_i$ (which is only the case in the limit where the flux is negligible compared to the read noise), the relative emission line flux bias due to an imperfect PSF reads
\begin{equation}
 \frac{\delta F}{F} \equiv \left<\frac{\tilde{F}}{F^{truth}}  - 1 \right> =  \frac{\sum_i P_i P_i^{truth}}{\sum_i P_i^2} -1
\end{equation}

\section{List of calibration lines}
\label{sec:arc-lamp-lines-list}

Table~\ref{table:arc-lamp-lines-list} provides a list of the arc lamp emission lines used for the wavelength calibration and PSF fit.
The wavelength are in vacuum and were obtained from the NIST Atomic Spectra Database\footnote{\url{https://physics.nist.gov/PhysRefData/ASD/lines_form.html}}.

\begin{table}[h]
  \centering
  \begin{tabular}{lllllll}
3261.995(Cd) & 3342.445(Hg) & 3404.629(Cd) & 3467.192(Cd) & 3468.648(Cd) & 3611.537(Cd) & 3613.903(Cd) \\
3651.198(Hg) & 3655.883(Hg) & 3664.327(Hg) & 3902.973(Hg) & 3907.478(Hg) & 4047.708(Hg) & 4078.988(Hg) \\
4159.762(Ar) & 4191.894(Ar) & 4197.597(Ne) & 4307.883(Cd) & 4340.443(Hg) & 4348.717(Hg) & 4359.560(Hg) \\
4414.229(Cd) & 4663.657(Cd) & 4679.459(Cd) & 4801.254(Cd) & 4917.440(Hg) & 5039.156(Ne) & 5087.239(Cd) \\
5117.929(Ne) & 5155.863(Ne) & 5332.260(Ne) & 5342.579(Ne) & 5402.063(Ne) & 5462.268(Hg) & 5563.770(Kr) \\
5571.836(Kr) & 5658.229(Ne) & 5677.380(Hg) & 5691.395(Ne) & 5721.117(Ne) & 5749.893(Ne) & 5766.017(Ne) \\
5771.210(Hg) & 5792.276(Hg) & 5821.769(Ne) & 5854.110(Ne) & 5883.525(Ne) & 5904.098(Ne) & 5908.066(Ne) \\
5915.272(Ne) & 5920.547(Ne) & 5946.481(Ne) & 5967.123(Ne) & 5976.282(Ne) & 5977.190(Ne) & 6031.667(Ne) \\
6076.019(Ne) & 6097.851(Ne) & 6130.146(Ne) & 6144.763(Ne) & 6165.299(Ne) & 6183.857(Ne) & 6219.001(Ne) \\
6268.229(Ne) & 6306.533(Ne) & 6336.179(Ne) & 6384.756(Ne) & 6404.018(Ne) & 6440.249(Cd) & 6508.325(Ne) \\
6534.687(Ne) & 6600.775(Ne) & 6653.929(Ne) & 6680.120(Ne) & 6718.897(Ne) & 6754.698(Ar) & 6873.185(Ar) \\
6909.370(Hg) & 6931.379(Ne) & 6967.352(Ar) & 7025.987(Ne) & 7034.352(Ne) & 7053.236(Ne) & 7061.054(Ne) \\
7069.167(Ar) & 7083.854(Hg) & 7093.815(Hg) & 7149.012(Ar) & 7175.915(Ne) & 7247.163(Ne) & 7274.940(Ar) \\
7355.319(Ar) & 7374.149(Ar) & 7386.014(Ar) & 7440.947(Ne) & 7474.496(Ne) & 7490.934(Ne) & 7505.935(Ar) \\
7516.721(Ar) & 7537.849(Ne) & 7546.121(Ne) & 7589.502(Kr) & 7603.638(Kr) & 7637.208(Ar) & 7687.361(Kr) \\
7696.658(Kr) & 7856.984(Kr) & 7945.365(Ne) & 7950.362(Ar) & 8008.359(Ar) & 8016.990(Ar) & 8061.721(Kr) \\
8192.308(Kr) & 8233.896(Xe) & 8349.740(Xe) & 8368.046(Ne) & 8378.661(Ne) & 8379.909(Ne) & 8419.473(Ne) \\
8420.740(Ne) & 8426.963(Ar) & 8465.682(Ne) & 8497.693(Ne) & 8511.211(Kr) & 8523.783(Ar) & 8593.618(Ne) \\
8637.019(Ne) & 8656.760(Ne) & 8670.325(Ar) & 8681.878(Ne) & 8684.306(Ne) & 8774.066(Ne) & 8776.460(Kr) \\
8779.161(Kr) & 8783.034(Ne) & 8786.166(Ne) & 8821.832(Xe) & 8856.298(Ne) & 8864.750(Xe) & 8868.191(Ne) \\
8921.950(Ne) & 8931.145(Kr) & 8933.282(Xe) & 8954.709(Xe) & 9047.930(Xe) & 9125.471(Ar) & 9165.167(Xe) \\
9170.040(Xe) & 9222.590(Ne) & 9224.111(Ne) & 9227.030(Ar) & 9354.800(Kr) & 9356.787(Ar) & 9375.880(Ne) \\
9377.330(Xe) & 9515.987(Xe) & 9660.435(Ar) & 9687.980(Xe) & 9720.830(Xe) & 9754.435(Kr) & 9787.186(Ar) \\
9802.384(Xe) \\

\end{tabular}
\caption{List of the arc lamp emission lines used for the wavelength calibration and PSF fit. Wavelength are in \AA\ in vacuum. Ambiguous lines have been discarded, but some blended lines have been kept as they are correctly handled during the PSF fit. Most faint lines have been ignored except in the wavelength regions where brighter lines are missing or when they are blended with other lines and could bias the fit.}
\label{table:arc-lamp-lines-list}
\end{table}

\section{Spectro-perfectionism and convolution}
\label{sec:spectroperfconvolution}

We discuss in the section how the transformation $\tilde F = R F$ introduced in \S\ref{sec:extraction}
can be interpreted as a convolution {\it if} the pixel noise is constant (which is not the case with real data,
because of the contribution from the Poisson noise of the signal) and {\it if} the PSF is also constant.

\subsection{Interpretation of the covariance matrix of $F$}

With the first assumption of a constant noise, Eq.~\ref{eq:extract-f} simplifies to $ C^{-1} F = A^T p$ , with $C^{-1} = A^T A$.

Each row or column of $C^{-1}$ is the auto-correlation function of the $PSF$, centered on the diagonal, and sampled on a regular grid given by the position of the flux points. We call in the following $\Delta$ the period of this grid, in units of CCD pixels. We have
\begin{equation}
C^{-1}_{j,k} \equiv c_{|j-k|} =  \left[ PSF \star PSF \right](\Delta \times |j-k|) \label{eq:2.1}
\end{equation}

The matrix $C^{-1}$ is a finite order (and symmetric) Toeplitz matrix (same coefficients for each descending diagonal of the matrix). Finite order means that $c_j=0$ for $j>k$ (with $k$ finite; we assume that the PSF has a finite extent).
When the dimension of the matrix becomes large, which corresponds to the extraction of a very large set of flux points, this matrix can be approximated as a circulant matrix, with differences only on the top right and bottom left corners (corresponding to the edges of the spectrum we want to extract)\footnote{This is presented rigorously in {\it Toeplitz and Circulant Matrices: A review} by R. M. Gray, \url{http://www-isl.stanford.edu/~gray/toeplitz.pdf}}.

The eigenvectors and eigenvalues of a circulant matrix are related to the discrete Fourier transform.
$$
C^{-1} = \mathcal{F}_n^{-1} diag(\mathcal{F}_n c) \mathcal{F}_n \ \ \mathrm{with} \  \  \mathcal{F}_{n,(j,k)} \equiv e^{-2 \pi i (jk)/n}
$$

In other words, the eigenvalues of $C^{-1}$ is the Fourier transform of one of the rows of $C^{-1}$ that we labeled $c$ in  Eq.~\ref{eq:2.1}, which is the auto-correlation of the PSF sampled on a grid of period $\Delta$. We label this quantity $\mathcal{F}\left( \left[ PSF \star PSF \right]_{|\Delta}\right)$. It is related to the power spectrum of the original PSF, but with aliasing. 

The matrix operation $C^{-1} F$ is asymptotically a ``circular convolution'' process with a kernel $c$.
Far from the edges of the spectrum, it is to a good approximation a simple convolution, $ c * F$.

\subsection{Interpretation of $R$ as a convolution kernel}

Since $C^{-1}$ is asymptotically a circulant matrix, the same is true for its square root matrix $Q$ and the resolution matrix\footnote{The resolution matrix $R$ is equal to $Q$ normalized row by row as explained in \S\ref{sec:extraction-2d} Eq.~\ref{eq:resolution-matrix}, which in our present case has no effect, as all rows have the same sum.} $R$.
The eigenvalues of $R$ are the square root of those of  $C^{-1}$.
In addition, the operation $\widetilde{F} = R F$ is a convolution,
with a kernel that we will call $r$, which is one row of $R$, and the inverse Fourier transform of the eigenvalues of $R$. In consequence we have
\begin{equation}
\mathcal{F}(r) = \sqrt{ \mathcal{F}\left( \left[ PSF \star PSF \right]_{|\Delta}\right) } \label{eq:Fr}
\end{equation}

This identification of the $R$ matrix of spectro-perfectionism as a convolution is quite accurate in the center of the extracted spectra and only becomes invalid on the edges.

\subsection{Interpretation of the extraction}

The vector $A^Tp$ is the cross-correlation of the PSF with the pixel values sampled at $\Delta$.
We can write it $(PSF \star p)_{|\Delta}$. Similarly, we have the coefficients of one row of $C^{-1}$
 as $c=(PSF \star PSF)_{|\Delta}$.

 The full spectro-perfectionism algorithm, in Fourier space, {\it omitting the $\mathcal{F}$ labels in the following}, is the convolution - deconvolution process
\begin{eqnarray}
\tilde F &=& r  \frac{( PSF^*  p )_{|\Delta}}{c} = r  \frac{( PSF^*  p )_{|\Delta}}{( PSF^*  PSF )_{|\Delta}} =  \frac{( PSF^*  p )_{|\Delta}}{\sqrt{(PSF^*  PSF )_{|\Delta}}} \label{eq:ftilde2}
\end{eqnarray}

We invite the interested reader to also see~\cite{Bolton2012b} for more discussions on this topic.

\section{Resampling the Resolution Matrix to alternate model wavelengths}
\label{sec:resampleresolution}

The spectral extraction examples in Section~\ref{sec:extraction} used an
input model wavelength grid that was exactly aligned with the output
extraction grid with a 0.8\,\AA\ wavelength step, thus simplifying the
process of convolving the input model with the square resolution matrix
$R$, which is tied to the extraction wavelength grid.

In practice, input models may be on a higher resolution or differently aligned
wavelenth grid than the output grid (default 0.8\,\AA\ for DESI).
Resampling the model to the extraction grid before applying the resolution is incorrect;
applying the resolution should happen before any model wavelength rebinning.
It is worth noting that this issue is not unique to the spectro-perfectionism technique; it applies to any astronomical spectroscopy with unresolved components.

\begin{figure}  \includegraphics[width=0.55\columnwidth,angle=0]{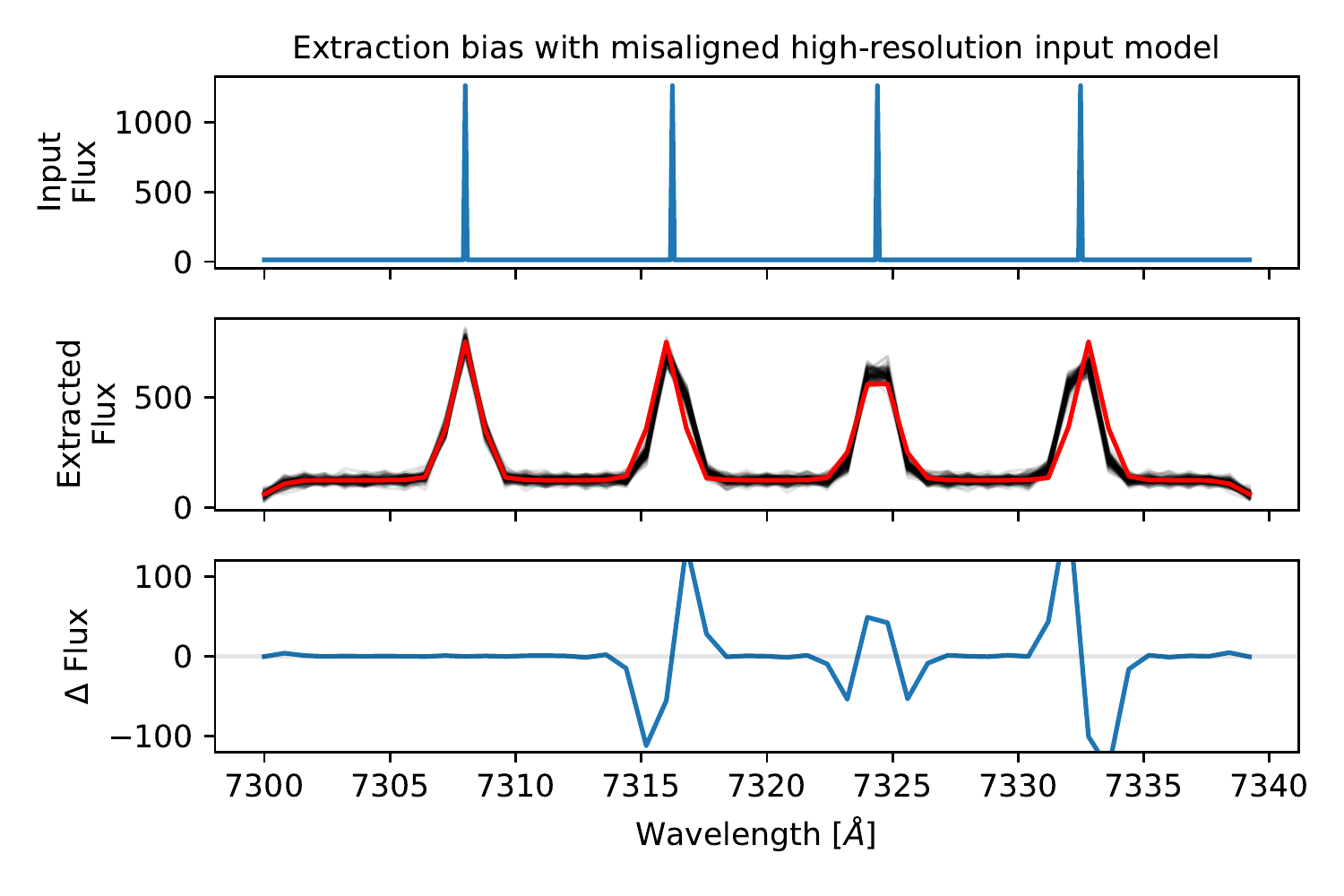}
  \centering
  \caption{Extraction model biases from an unresolved input model that is
  unaligned with the output spectral extraction wavelength grid.
  Top panel: input model, with the left-most input emission line aligned
  with the output wavelength grid, and other emission lines with offsets.
  Middle panel: input model resampled to output wavelength grid before
  multiplying by the resolution matrix $R$ (red) and 100 extractions with
  different noise realizations (black).
  Bottom panel: mean difference between extractions and resampled
  resolution convolved model.
  }
  \label{fig:exbias-offset-input}
\end{figure}

If the input model has features that are unresolved by the extraction wavelength grid,
rebinning before resolution can lead to large data vs.~model biases as shown in
Figure~\ref{fig:exbias-offset-input}.  The leftmost input emission line
is exactly aligned with the output wavelength grid, leading to no bias in the extractions.
The other 3 lines have offsets of 33\%, 50\%, and 67\% of the 0.8\,\AA\
extraction wavelength step size.  The extracted spectra have significant differences when compared to
the (incorrectly) rebinned then resolution convolved model.

\begin{figure}  \includegraphics[width=0.55\columnwidth,angle=0]{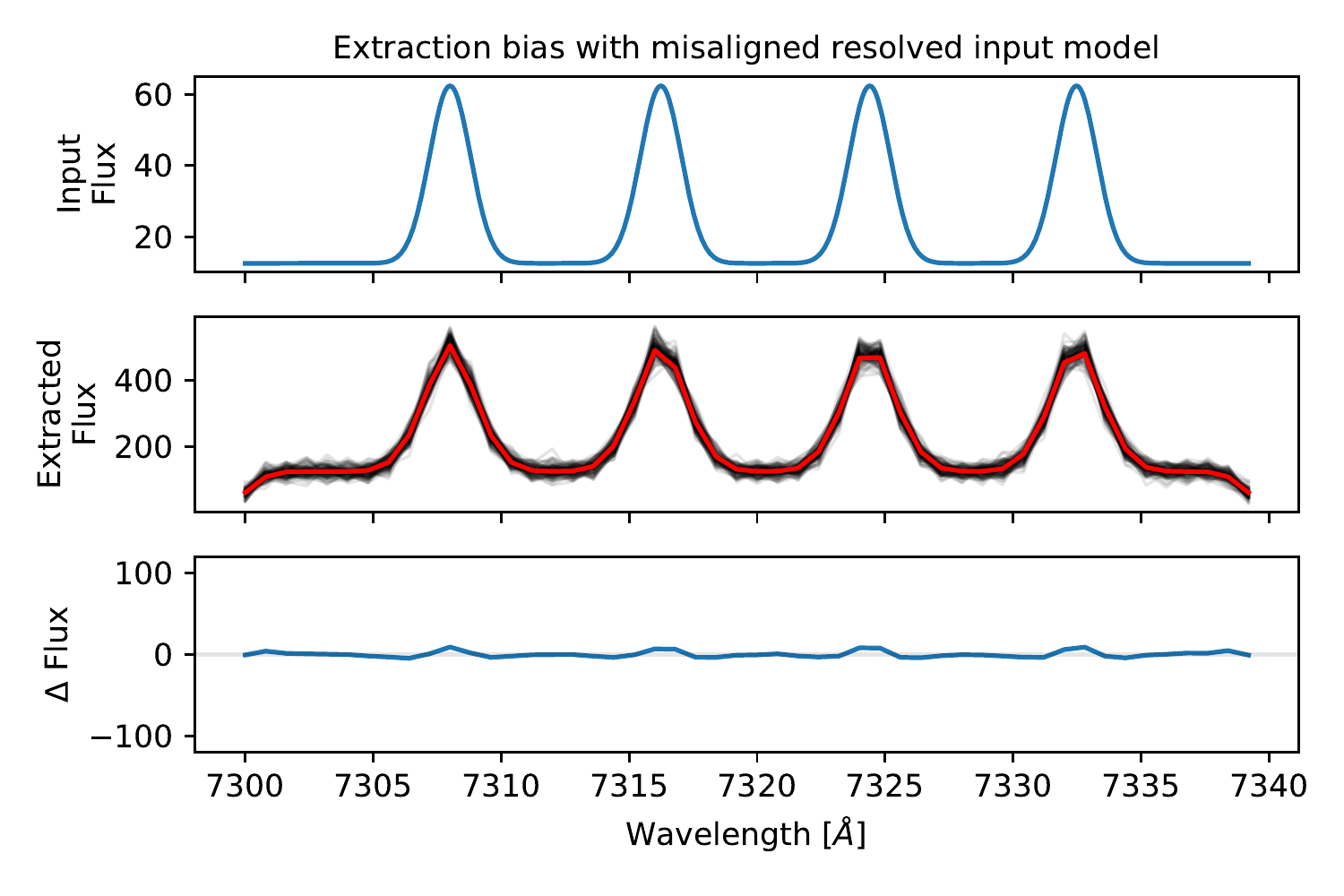}
  \centering
  \caption{Like Figure~\ref{fig:exbias-offset-input}, except that the input
  model emission lines have a 0.8\,\AA\ Gaussian $\sigma$ dispersion and thus
  are nominally resolved by the extraction wavelength grid.  Despite the
  offset centroids, the bias is much smaller than the unresolved case.}
  \label{fig:exbias-resolved-input}
\end{figure}

On the other hand, if the input model is resolved by the extraction wavelength
grid even before considering instrumental resolution, these discrepancies
are much smaller, as shown in Figure~\ref{fig:exbias-resolved-input}.  In
this case, the input model has a Gaussian $\sigma$ of 0.8\,\AA\ (one extraction
pixel), equivalent to a velocity dispersion of 16\,km\,s$^{-1}$ for a $z=1$ object.
Even though the centroids of the emission lines are still offset relative to
the extraction wavelength grid, the data vs.~model differences are much smaller.

For simplicity and algorithmic efficiency, the Redrock algorithm for
classification and redshift estimation (Section~\ref{sec:redshift}) takes
advantage of this and rebins its redshifted template spectra to the output
wavelength grid prior to convolving with $R$ for comparing to data.  Since
the internal velocity dispersion of main survey quasar and galaxy targets
are resolved by the 0.8\,\AA\ wavelength grid, the resulting biases are minimal
when comparing the model to the data.

\begin{figure}  \includegraphics[width=0.55\columnwidth,angle=0]{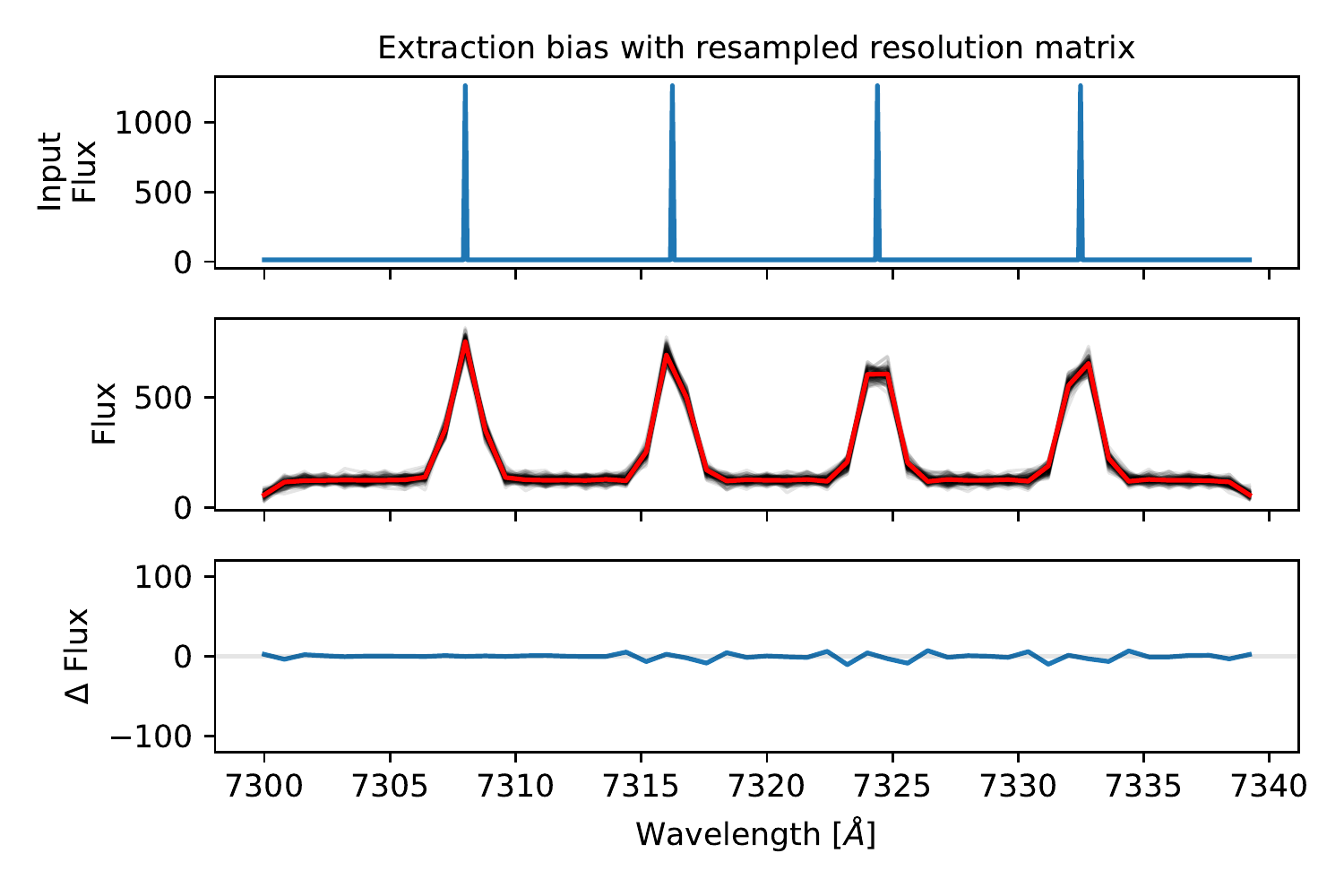}
  \centering
  \caption{Like Figure~\ref{fig:exbias-offset-input}, except that the
  resolution matrix $R$ rows are resampled to match the input wavelength
  grid, thus performing the resolution convolution and the rebinning in a
  single step, significantly reducing the extraction bias.}
  \label{fig:exbias-resample-R}
\end{figure}

However, if one needs to work with an input model that is unresolved by the
default 0.8\,\AA\ wavelength grid (e.g.~stellar models with sharp absorption
features), one needs to resample the resolution matrix to
apply the resolution prior to rebinning to the output wavelength grid.

Indeed, in the original definition of $R$ (equation~\ref{eq:resolution-matrix}),
 the square matrix applies to an array mapping the input spectrum on
same wavelength grid as the output spectra (see for instance the model-to-data comparison in equation~\ref{eq:extract-chi2-1D}).
To accomodate a higher resolution model, one can use
a cubic spline to interpolate each row of $R$ into a higher resolution
wavelength grid matching the input model, resulting in a
non-square resolution matrix $R^\prime$ that takes a high-resolution input
model $m$, applies the resolution, and rebins to the output wavelength grid.
Figure~\ref{fig:exbias-resample-R} shows the
improvement from using this resampled resolution matrix $R^\prime$.
Although there are still some artifacts, it is much improved
compared to Figure~\ref{fig:exbias-offset-input} which rebinned the model
prior to applying the resolution.

The DESI spectroscopic pipeline algorithms do not use this $R$ resampling
procedure as part of the data processing, but we document this procedure
here as an issue to consider for analyses comparing models to DESI spectra.

\section{Offline Quality Assurance Example}
\label{sec:example-qa}

\begin{figure}
	\begin{center}
		\begin{tabular}{cc}
			\multicolumn{2}{c}{\includegraphics[width=0.95\columnwidth]{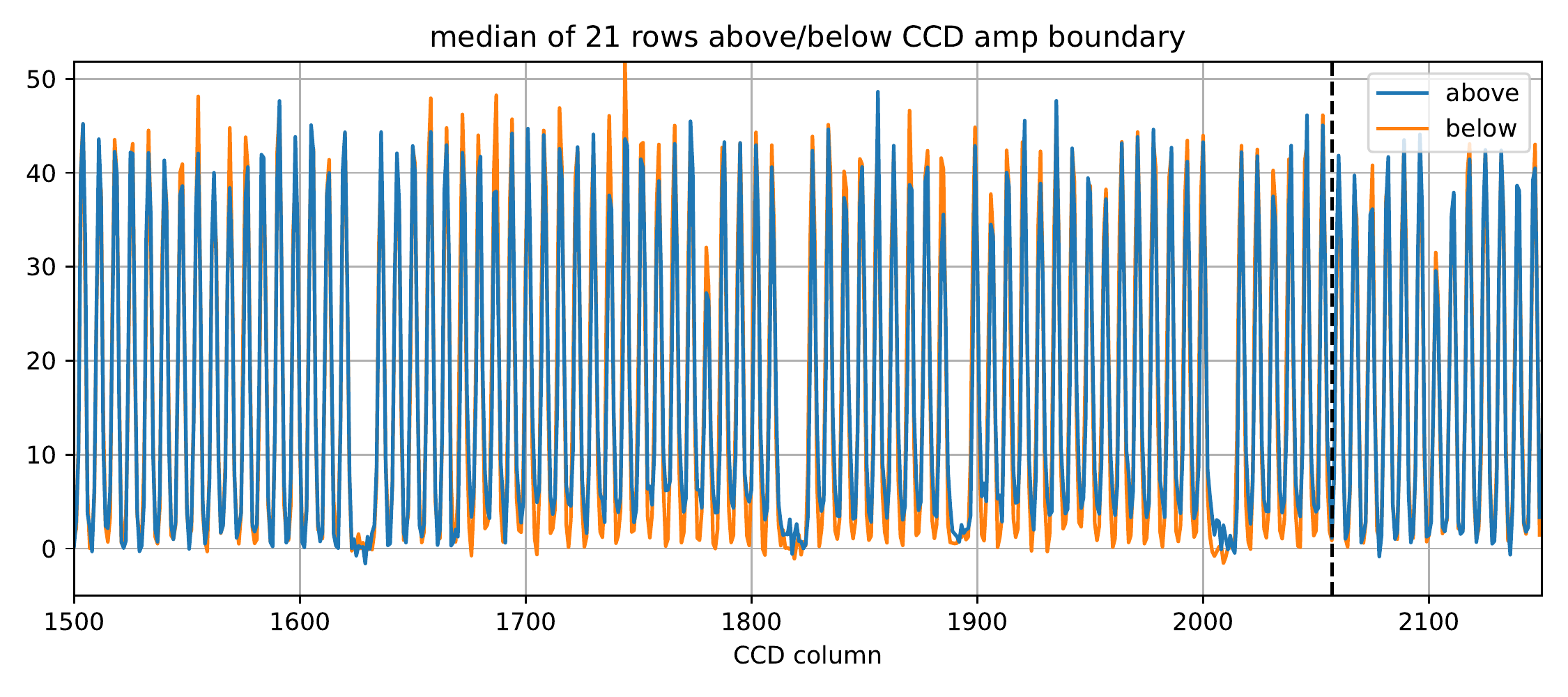}}\\
			\includegraphics[width=0.45\columnwidth]{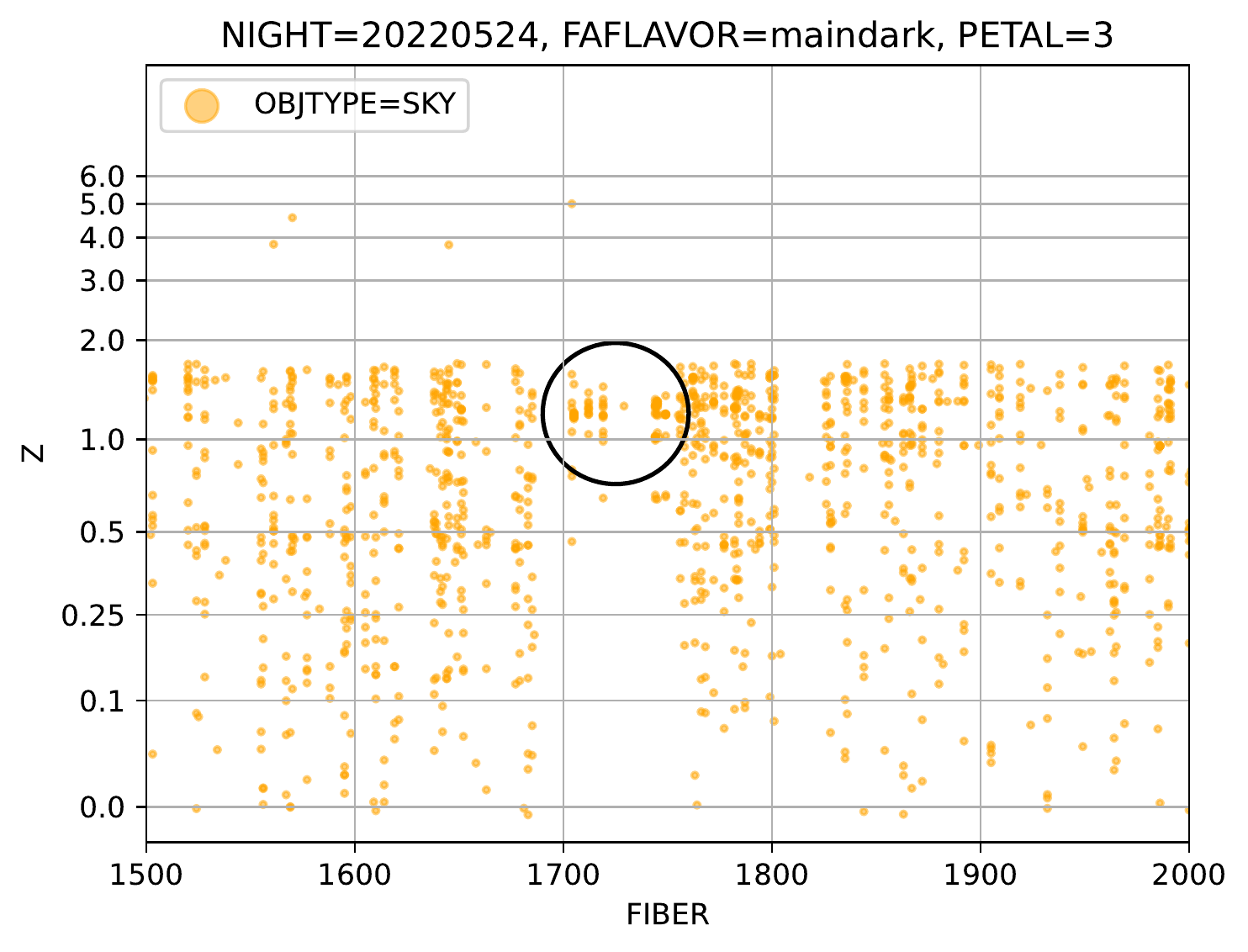} &
			\includegraphics[width=0.45\columnwidth]{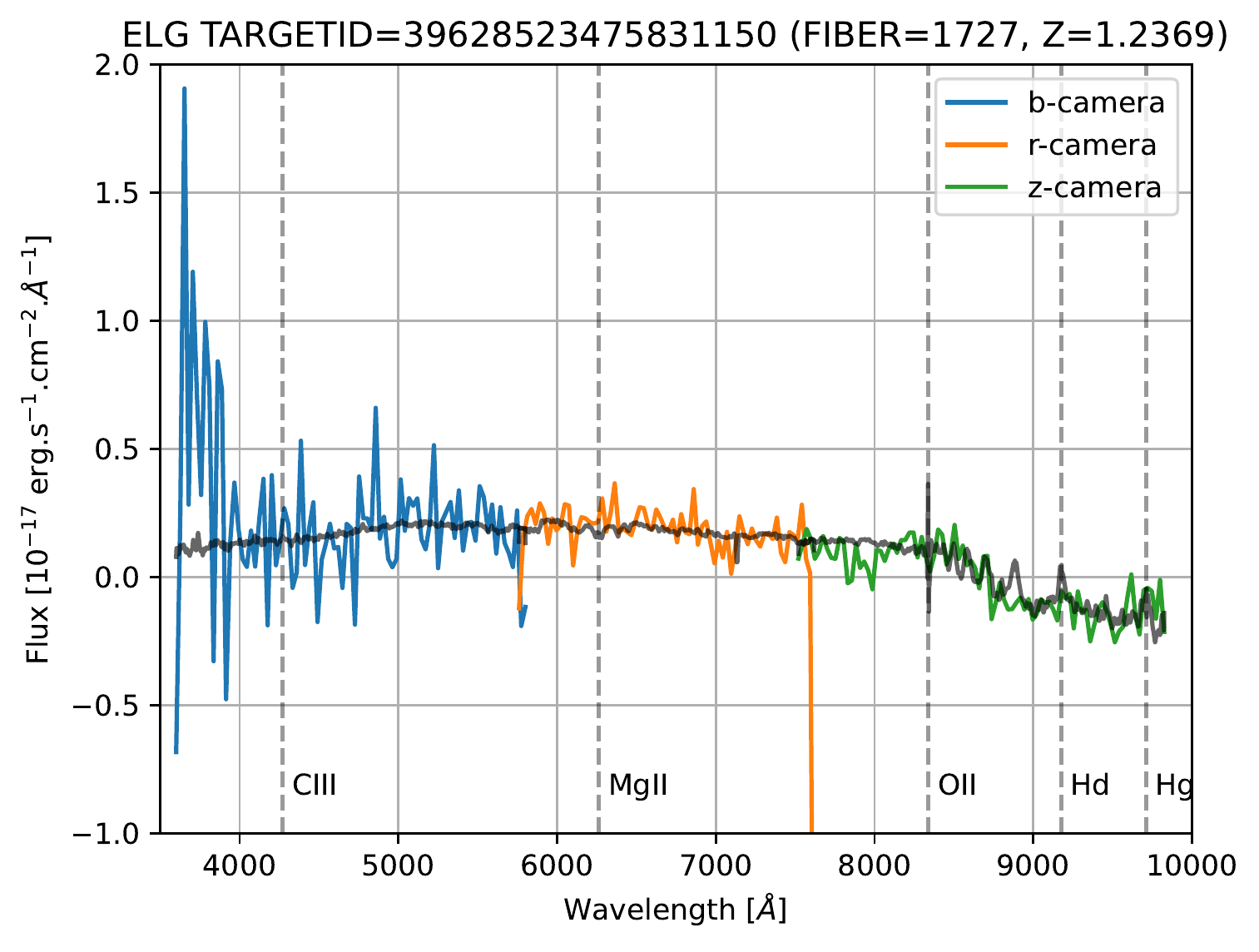}\\
		\end{tabular}
		\caption{
			Example of a CTE issue for the petal 3 for one night identified by the QA. From top left to bottom right : median counts over the 21 rows just above/below the CCD amplifier boundary for a flat field exposure, redshift vs. the fiber number for petal 3 (sky-subtracted) sky fibers for all dark tiles from that night, an ELG spectrum drawn from one of the impacted fibers. See text for more details.
		}
		\label{fig:qa-cte}
	\end{center}
\end{figure}

We illustrate in Figure~\ref{fig:qa-cte} a charge transfer efficiency (CTE) issue for the petal 3 for one night (May, 24, 2022), and its impact on redshifts (see \S\ref{sec:qa} on Quality Assurance for context).
The top panel shows a CTE issue for the C amplifier of the z3-camera: it displays the median counts values over the 21 rows just above/below the amplifier boundary along the wavelength direction (see Figure~\ref{fig:ccd-layout} for the CCD image layout with the amplifiers).
The electrons in the amplifier C are not properly transferred for the CCD columns 1670 to 2057 (last column read with this amplifier)  which approximately correspond to fibers 1705 to 1749. This results in an offset of a few electrons in the extracted spectra from those fibers.
As a consequence those $\sim$45 fibers read from the z3-camera amplifiers A and C have an artificial discontinuity at the wavelength corresponding to the A and C amplifier boundary ($\sim 8750$\,\AA), identified by the redshift pipeline as a Balmer break at redshift $\sim$1.2 for low signal-to-noise spectra.
This is visible in the bottom left plot showing for that petal the redshifts as a function of the fiber number for the (sky-subtracted) sky fibers only, with all those fibers having a best-fit redshift of $\sim$1.2.
Lastly, the bottom right plot shows an ELG spectrum drawn from one of those fibers, where one can see the artificial flux decrease redwards of $\sim 8750$\,\AA, interpreted as an (inverted) Balmer break by the redshift pipeline.

For CTE issue, the obtained data usually cannot be fixed, hence the corresponding fibers are declared as "bad" if the effect is strong; however, the issue can be resolved for later observations, with adjusting the CCD serial clock voltages.

\end{appendix}


\end{document}